\documentclass[longauth]{aa}
\usepackage{graphicx}
\usepackage[colorlinks=true, linkcolor=blue, citecolor=blue, urlcolor=blue]{hyperref}
\usepackage{txfonts}
\usepackage{lscape}
\usepackage{epstopdf}
\usepackage{CJK}
\usepackage{float}
\usepackage[absolute]{textpos}
\usepackage{amsmath}
\usepackage{mathrsfs}
\usepackage{adjustbox}
\usepackage[T1]{fontenc}
\usepackage{orcidlink}
\usepackage[switch]{lineno}

\begin{document} 
% \linenumbers
\title{The ALMA-ATOMS survey: A sample of weak hot core candidates identified through line stacking}

\author{Zi-Yang Li\inst{1,2}\orcidlink{0009-0005-7028-0735}\and
Xunchuan Liu\inst{2}\orcidlink{0000-0001-8315-4248}\and
Tie Liu\inst{2}\orcidlink{0000-0002-5286-2564}\and
Sheng-Li Qin\inst{1}\orcidlink{0000-0003-2302-0613}\and
Paul F. Goldsmith\inst{3}\orcidlink{0000-0002-6622-8396}\and
Pablo Garc{\'i}a\inst{4,5}\and
Yaping Peng\inst{6}\orcidlink{0000-0001-5703-1420}\and
Li Chen\inst{1}\orcidlink{0009-0009-8154-4205}\and
Yunfan Jiao\inst{2,8}\orcidlink{0009-0005-9867-6723}\and
Zhiping Kou\inst{7,8,1}\and
Chuanshou Li\inst{1}\orcidlink{0000-0001-5710-6509}\and
JiahangZou\inst{1,2}\orcidlink{0009-0000-9090-9960}\and
Mengyao Tang\inst{9}\orcidlink{0000-0001-9160-2944}\and
Shanghuo Li\inst{23,24}\orcidlink{0000-0003-1275-5251}\and
Meizhu Liu\inst{13}\orcidlink{0000-0002-5789-7504}\and
Guido Garay\inst{4,14}\and
Fengwei Xu\inst{10,11}\orcidlink{0000-0001-5950-1932}\and
Wenyu Jiao\inst{2}\orcidlink{0000-0001-9822-7817}\and
Qiu-Yi Luo\inst{2}\orcidlink{0000-0003-4506-3171}\and
Suinan Zhang\inst{2}\orcidlink{0000-0002-8389-6695}\and
Qi-Lao Gu\inst{2}\orcidlink{0000-0002-2826-1902}\and
Xiaofeng Mai\inst{2}\orcidlink{0000-0002-0786-7307}\and
Yan-Kun Zhang\inst{2}\orcidlink{0000-0001-7817-1975}\and
Jixiang Weng\inst{2,8}\orcidlink{0009-0008-8439-8488}\and
Chang Won Lee\inst{18,19}\and
Patricio Sanhueza\inst{25,16,17}\orcidlink{0000-0002-7125-7685}\and
Sami Dib\inst{20}\orcidlink{0000-0002-8697-9808}\and
Swagat R. Das\inst{14}\orcidlink{0000-0001-7151-0882}\and
Xindi Tang\inst{7,15}\orcidlink{0000-0002-4154-4309}\and
Leonardo Bronfman\inst{14}\and
Prasanta Gorai\inst{21,22}\and
Ken’ichi Tatematsu\inst{17}\and
Hong-Li Liu\inst{1}\orcidlink{0000-0003-3343-9645}\and
Dongting Yang\inst{1}\and
Zhenying Zhang\inst{1,2}\orcidlink{0009-0005-4295-5010}\and
Xianjin Shen\inst{1}\orcidlink{0009-0004-3244-3508}
}

\institute{School of Physics and Astronomy, Yunnan University, Kunming 650091, People's Republic of China\\
\email{qin@ynu.edu.cn }
\and 
Shanghai Astronomical Observatory, Chinese Academy of Sciences, 80 Nandan Road, Shanghai 200030, People's Republic of China
\email{liutie@shao.ac.cn, liuxunchuan001@gmail.com}
\and
Jet Propulsion Laboratory, California Institute of Technology, 4800 Oak Grove Drive, Pasadena CA 91109, USA
\and
Chinese Academy of Sciences South America Center for Astronomy, National Astronomical Observatories, CAS, Beijing 100101, People's Republic of China
\and     
Instituto de Astronom{\'i}a, Universidad Cat{\'o}lica del Norte, Av. Angamos 0610, Antofagasta, Chile.
\and
Department of Physics, Faculty of Science, Kunming University of Science and Technology, Kunming 650500, People's Republic of China
\and
Xinjiang Astronomical Observatory, Chinese Academy of Sciences, 830011 Urumqi, People's Republic of China
\and
University of Chinese Academy of Sciences, Beijing 100049, People's Republic of China
\and
Institute of Astrophysics, School of Physics and Electronical Science, Chuxiong Normal University, Chuxiong 675000, People's Republic of China
\and
I. Physikalisches Institut, Universit{\"a}t zu K{\"o}ln, Z{\"u}lpicher Stra{\ss}e 77, 50937 K{\"o}ln, Germany
\and
Kavli Institute for Astronomy and Astrophysics, Peking University, 5 Yiheyuan Road, Haidian District, Beijing 100871, People's Republic of China
\and
Max Planck Institute for Astronomy, K{\"o}nigstuhl 17, D-69117 Heidelberg, Germany
\and
Center for Astrophysics, Guangzhou University, Guangzhou 510006, People's Republic of China
\and
Departamento de Astronom\'{i}a, Universidad de Chile, Las Condes, 7591245 Santiago, Chile
\and
Key Laboratory of Radio Astronomy, Chinese Academy of Sciences, 830011 Urumqi, People's Republic of China
\and
Department of Earth and Planetary Sciences, Institute of Science Tokyo, Meguro, Tokyo, 152-8551, Japan
\and
National Astronomical Observatory of Japan, National Institutes of Natural Sciences, 2-21-1 Osawa, Mitaka, Tokyo 181-8588, Japan
\and
Korea Astronomy and Space Science Institute, 776 Daedeokdaero, Yuseong-gu, Daejeon 34055, Republic of Korea
\and
University of Science and Technology, Korea (UST), 217 Gajeong-ro, Yuseong-gu, Daejeon 34113, Republic of Korea
\and
Max Planck Institute for Astronomy, K\"{o}nigstuhl 17, 69117, Heidelberg, Germany
\and
Rosseland Centre for Solar Physics, University of Oslo, PO Box 1029 Blindern, 0315 Oslo, Norway
\and
Institute of Theoretical Astrophysics, University of Oslo, PO Box 1029 Blindern, 0315 Oslo, Norway
\and
School of Astronomy and Space Science, Nanjing University, 163 Xianlin Avenue, Nanjing 210023, People's Republic of China
\and
Key Laboratory of Modern Astronomy and Astrophysics (Nanjing University), Ministry of Education, Nanjing 210023, People's Republic of China
\and
Department of Astronomy, School of Science, The University of Tokyo, 7-3-1 Hongo, Bunkyo, Tokyo 113-0033, Japan
}

\abstract
{Hot cores represent critical astrophysical environments for high-mass star formation, distinguished by their rich spectra of organic molecular emission lines. Nevertheless, comprehensive statistical analyses of extensive hot core samples remain relatively scarce in current astronomical research.}
{We aim to utilize high-angular resolution molecular line data from the Atacama Large Millimeter and Submillimeter Array (ALMA) to identify hot cores, with a particular focus on weak-emission candidates, and to provide one of the largest samples of hot core candidates to date.}
{We propose to use spectral stacking and imaging techniques of complex organic molecules (COMs) in the ALMA-ATOMS survey, including line identification $\&$ weights, segmentation of line datacubes, resampling, stacking and normalization, moment 0 maps, and data analysis, to search for hot core candidates. The molecules involved include CH$_3$OH, CH$_3$OCHO, C$_2$H$_5$CN, C$_2$H$_5$OH, CH$_3$OCH$_3$, CH$_3$COCH$_3$, and CH$_3$CHO. We classify cores with dense emission of CH$_3$OH and at least one molecule from the other six molecules as hot core candidates. }
{In addition to the existing sample of 60 strong hot cores from the ALMA-ATOMS survey, we have detected 40 new weak candidates through stacking. All hot core candidates display compact emission from at least one of the other six COM species. For the strong sample, the stacking method provides molecular column density estimates that are consistent with previous fitting results. For the newly identified weak candidates, all species except CH$_3$CHO show compact emission in the stacked image, which cannot be fully resolved spatially. These weak candidates exhibit column densities of COMs that are approximately one order of magnitude lower than those of the strong sample. The entire hot core sample, including the weak candidates, reveals tight correlations between the compact emission of CH$_3$OH and other COM species, suggesting they may share a similar chemical environment for COMs, with CH$_3$OH potentially acting as a precursor for other COMs. Among the 100 hot cores in total, 43 exhibit extended CH$_3$CHO emission spatially correlated with SiO and H$^{13}$CO$^{\text{+}}$, suggesting that CH$_3$CHO may form in widely distributed shock regions.
}
{The molecular line stacking technique is used to identify hot core candidates in this work, leading to the identification of 40 new hot core candidates. Compared to spectral line fitting methods, it is faster and more convenient, and allows for greater sensitivity to detect weaker hot cores.  }%


\maketitle

\section{Introduction}
\label{Intro}
   The formation of high-mass stars is critical for the structure and evolution of galaxies \citep[][]{ZinneckerYorke2007}. However, our understanding of this process remains incomplete, with many unresolved questions. Among the evolutionary stages of high-mass star formation, the hot core phase is particularly pivotal. Complex organic molecules (COMs) refer to molecules containing carbon and consisting of six or more atoms \citep[][]{2009ARA&A..47..427H}. Hot cores are characterized by rich emissions of COMs, high gas temperatures (>100 K), high gas densities ($n_{\rm H_2} = 10^5 - 10^8$ cm$^{-3}$) and compact source sizes \citep[$<0.1$ pc,][]{2000prpl.conf..299K,2005IAUS..227...59C}. 
   Most COMs in space were firstly detected in hot cores \citep[][]{2022ApJS..259...30M}. Emission lines of different COMs serve as probes of various physical and chemical components within hot cores  \citep[][]{1998ARA&A..36..317V,2020ARA&A..58..727J,2021A&A...655A..65T}. As a result, the hot-core phase offers a wealth of COMs emission lines that trace the birth environment of massive stars, providing key insights into the mechanisms of their formation.
   
   Establishing a systematic and comprehensive large sample of hot core candidates is of great importance, yet studies based on such large-scale surveys are still rare. In the single-dish era, observations of hot cores were primarily limited to case studies because of sensitivities constraints \citep[][]{1997ApJS..108..301S,2000ApJ...545..309G,2001ApJS..132..281S,2006A&A...454L..41S,2007A&A...470..639F}. Massive stars, and thus also hot cores, typically form in clusters within massive clumps \citep[][]{Dib+2010a,2023ASPC..534..275O,Zhou+2024a}. The relatively large beams of single-dish telescopes could not resolve individual hot cores, introducing biases in statistical studies. Now, millimeter/submillimeter interferometric arrays (such as SMA, NOEMA, and ALMA) which have been employed with unprecedented broad bandwidths, high sensitivities, and improved resolutions \citep[][]{2014ApJ...786...38H,2016A&A...595A.117J,2018A&A...617A.100B,2023ApJ...950...57T}, enabled more efficient line surveys towards hot cores \citep[][]{2021MNRAS.505.2801L,2022MNRAS.511.3463Q}. Dozens of new hot core candidates have been identified in recent large-scale star formation surveys, such as CoCCoa \citep[e.g.][]{2023A&A...678A.137C}, DIHCA \citep[e.g.][]{2023ApJ...950...57T} and ALMA-IMF \citep[e.g.][]{2024A&A...687A.163B}. The ATOMS project has observed 146 massive clumps in the 3 mm band with ALMA \citep{2020MNRAS.496.2790L}, providing an excellent dataset for identifying hot cores. Based on the ATOMS data, \cite{2022MNRAS.511.3463Q} compiled a catalog of 60 hot cores exhibiting emission lines from three typical COMs: C$_2$H$_5$CN, CH$_3$OCHO, and CH$_3$OH. To enable the determination of the excitation temperature, they required that at least one of the three species have multiple detected transitions. As a result, their samples focus primarily on bright hot cores due to sensitivity limitations, with less attention given to those candidates with weaker emissions. This may lead to biases in our understanding of hot-core evolution. The ATOMS sample is also covered by the ongoing follow-up ALMA-QUARKS survey \citep{2024RAA....24b5009L}, which offers higher resolution in ALMA Band 6. A complete sample of hot core candidates in Band 3 would further aid in studying the inner details of hot cores at different stages. This work focuses on providing a complete sample of hot core candidates from the ATOMS survey through spectral stacking techniques.
   
   The molecular spectral stacking technique involves aligning different spectral lines with low signal-to-noise ratios (S/Rs) and applying proper weights to create a single stacked line with an improved S/R. This technique has been employed at sub-millimeter and radio wavelengths for over 20 years \citep{2005ApJ...632L...9K,2011ApJ...730...61K,2011AJ....142...37S,2013MNRAS.433.1398D,2013AJ....146..150C,2016ApJ...822L..26B,2016MNRAS.462.1192L,2023A&A...675A.104N,2023ApJ...942...66G}. Recent line-survey studies of Orion KL have demonstrated the stacking technique's effectiveness in enhancing the S/Rs of the emission of COMs \citep[e.g., CH$_3$COCH$_3$,][]{2022ApJS..263...13L} and in detecting radio recombination lines (RRLs) from carbon and oxygen ions \citep{2023A&A...671L...1L,2024ApJS..271....3L,2024A&A...688A...7P}. This method effectively recovers weaker and previously undetected line emissions. So far, the spectral stacking technique has not yet been systematically applied to search for COMs emission from hot cores. This work aims to use this technique to identify hot cores and study the spatial distribution of COMs in 146 high-mass star-forming clumps using ATOMS Band 3 data. Additionally, we compared the characteristics of faint hot core candidates with those of the brighter hot cores, providing new insights into the differences between these two populations.

   In this work, we analyze ATOMS Band 3 data for the molecules CH$_3$OH, CH$_3$OCHO, C$_2$H$_5$CN, C$_2$H$_5$OH, CH$_3$OCH$_3$, CH$_3$COCH$_3$, and CH$_3$CHO using molecular line stacking techniques. These species were selected for spectral stacking because they all have multiple transitions covered by the ATOMS survey, ensuring a robust dataset for analysis. They have been detected in the most complex organic molecule (COM)-rich hot cores \citep{2022MNRAS.512.4419P, 2022MNRAS.511.3463Q}, making them ideal tracers for hot core chemistry. CH$_3$OH, CH$_3$OCHO, C$_2$H$_5$CN, and CH$_3$OCH$_3$ are included primarily due to their relatively strong emission, which has made them common tracers of hot core environments \citep{Bisschop2006,2023A&A...678A.137C}. Their presence in hot cores highlights the complex chemical processes occurring in these regions, with many of them acting as precursors to larger, more complex molecules. C$_2$H$_5$OH (ethanol) is included because it plays an important role as an alcohol in hot cores, providing complementary insights into the organic chemistry alongside methanol \citep{2023A&A...673A..34A}. CH$_3$COCH$_3$ (acetone), a ketone, is included for its ability to trace carbonyl chemistry, offering a distinct chemical pathway in star-forming regions \citep{McGuire2016}. CH$_3$CHO (acetaldehyde) is also considered an important tracer for shock regions, as its extended emission has been associated with shock-driven processes in star-forming regions \citep{Chengalur2003}. By selecting these species, which represent a variety of functional groups such as alcohols, aldehydes, ethers, nitriles, and ketones, we ensure a diverse chemical snapshot of hot core chemistry. This diversity is crucial for understanding the complex and varied molecular processes occurring in these environments, making these species ideal candidates for the spectral stacking technique applied in this study.
   
   The structure of the paper is as follows: Section 2 describes the ALMA data. Section 3 details the identification of hot cores and the application of the molecular line stacking technique. Section 4 explores the spatial distributions of COMs. Finally, Section 5 summarizes the key findings of this work.

\section{Data}\label{data}

A sample of 146 massive clumps was observed in the ALMA band 3 survey, the ALMA-ATOMS project (ALMA ID: 2019.1.00685.S; PI: Tie Liu). Details of ALMA observations and data reduction can be found in \citet[][]{2020MNRAS.496.2821L,2020MNRAS.496.2790L}. Observations were conducted using both the Atacama Compact 7-meter Array (ACA) and the 12-meter array (C43-2 or C43-3 configurations) from September to mid-November 2019. This work utilizes the 3 mm continuum data and the two wide-band line data cubes (SPWs 7 and 8) obtained from the 12-meter array, which have angular resolutions ranging from approximately 1.2\arcsec~to 1.9\arcsec, with the maximum recoverable angular scale ranging from 14.5\arcsec~to 20.3\arcsec~across the 146 clumps.SPWs 7 and 8 have a large bandwidth of 1\,875.00 MHz and a spectral resolution of $\sim$1.6 km s$^{-1}$. The frequency ranges of SPW 7 and SPW 8 are 97\,536 to 99\,442 MHz and 99\,470 to 101\,390 MHz, respectively, covering a wide range of COMs lines. The mean 1 $\sigma$ noise level is below 10 mJy beam$^{-1}$ per channel for line data and 0.4 mJy beam$^{-1}$ for the continuum. In this work, we select COMs lines of 7 species, including CH$_3$OH, CH$_3$OCHO, C$_2$H$_5$CN, C$_2$H$_5$OH, CH$_3$OCH$_3$, CH$_3$COCH$_3$, and CH$_3$CHO (Sec. \ref{sec_thetransitions}), to directly identify hot core candidates from line emission after stacking.

\section{Analysis and results}\label{res}
\subsection{Spectral line stacking }

\begin{figure}[!thb]
\centering
\includegraphics[width=0.99\linewidth]{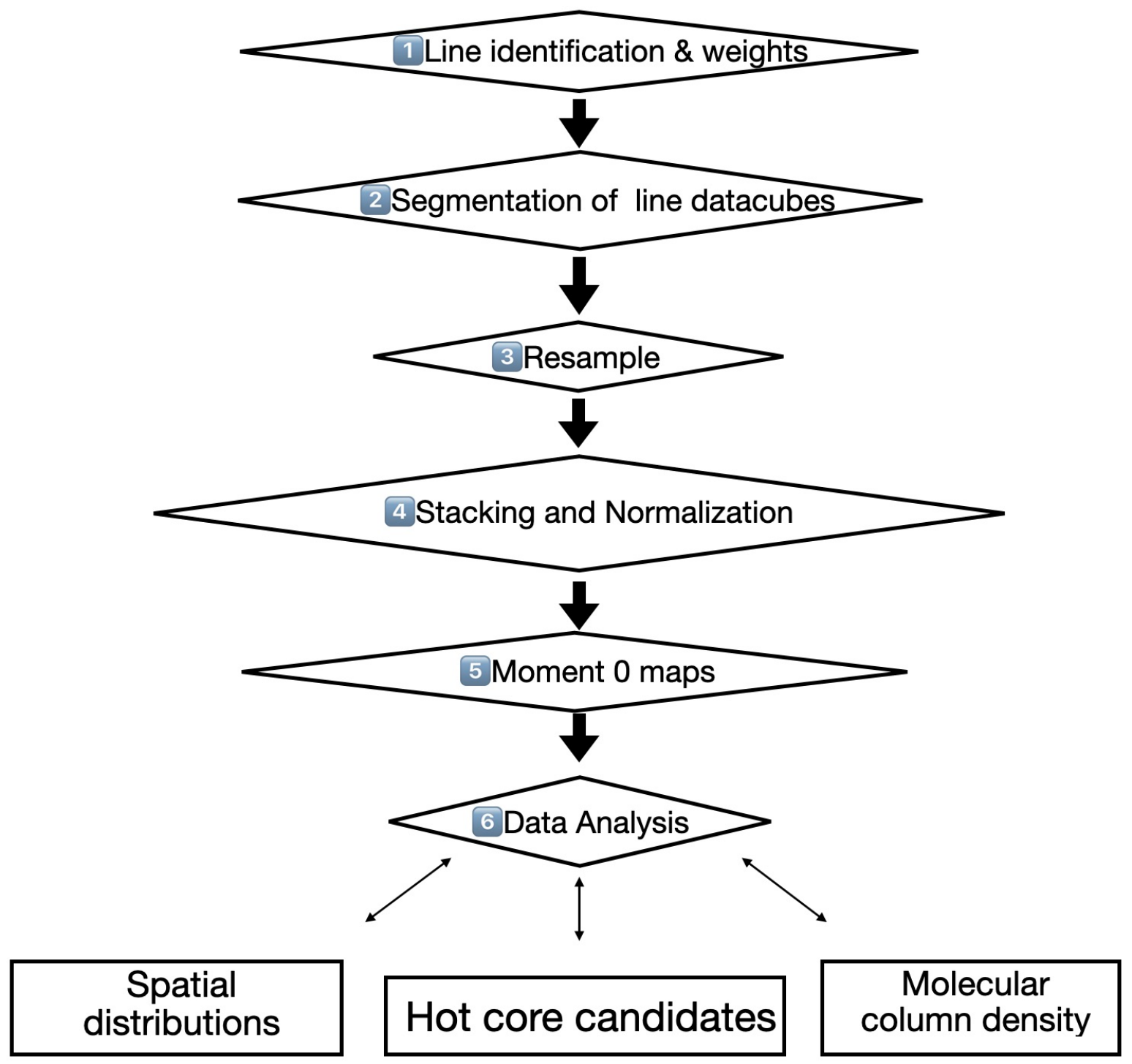}
\caption{The flowchart of the molecular spectral line stacking and hot core candidate identification.}
\label{figure_stackflow} % 设置标签，以便引用
\end{figure}

Spectral line stacking effectively enhances the S/Rs of emission from COMs. It involves combining the transitions of the same molecular specie at different rest frequencies. This technique is implemented by correcting for the rest frequency, reconstructing the coordinate axis, and applying reweighting and normalization before stacking \citep{2022ApJS..263...13L,2024ApJS..271....3L}. We applied the spectral line stacking technique to identify hot core candidates, following the procedure displayed in Figure \ref{figure_stackflow}. Below, we introduce the details of the line stacking technique.

\subsubsection{Standard spectrum of G9.62+0.19}
\label{sec_G9.62standard}
We first construct a template for hot core spectral lines using the two broad spectral windows, designated as SPW 7 and SPW 8 in the ATOMS survey, from the well-known source I18032-2032 (G9.62+0.19). Since G9.62+0.19 contains four confirmed hot cores and displays a line forest of complex organic molecules (COMs) at each core, it serves as a good template for hot cores \citep[][]{1999AAS...195.7307W,2000A&A...359L...5T,2000AAS...19713201L,2002osp..conf..225S,2004ApJ...616..301G,2005A&A...429..903L,2020MNRAS.496.2790L,2022MNRAS.512.4419P}. Figure \ref{fig:G9.62full} presents the full-band spectral lines of three hot cores (C1, C2, and C3) in G9.62+0.19. They exhibit spatially identical peaks of continuum and COM emission. Their peak spectra are extracted using CASA \citep[][]{2007ASPC..376..127M,2022PASP..134k4501C} and fitted using XCLASS\footnote{\url{https://xclass.astro.uni-koeln.de/}} \citep[][]{2017A&A...598A...7M} following \citet{2022MNRAS.512.4419P}, assuming local thermodynamic equilibrium (LTE). The critical density of the relevant transitions is typically 10$^6$ cm$^{-3}$ \citep{2001ApJ...557..736G}. The size of a hot core is very compact, with number densities generally above this threshold; therefore, the LTE assumption is valid \citep{2021MNRAS.505.2801L}. 
For CH$_{3}$OH,  CH$_{3}$OCHO, C$_{2}$H$_{5}$OH, CH$_{3}$OCH$_{3}$ and CH$_{3}$CHO, the fitted parameters from C1 are adopted as the standard parameters. CH$_{3}$COCH$_{3}$ and C$_{2}$H$_{5}$OH molecules were not detected in C1 \citep{2022MNRAS.512.4419P}. We adopt the parameters of CH$_{3}$COCH$_{3}$ and C$_{2}$H$_{5}$OH from the fits of C2 and C3, respectively.
For each species, the XCLASS fitting parameters, including source size, rotational temperature, column density, line width, as well as the core from which the spectrum is extracted, are listed in Table \ref{table1}.

A noise-free standard spectrum then modeled using XCLASS, adopting the fitted parameters (Table \ref{table1}), with the velocity offsets set to zero. The standard spectrum, after stacking (Sect. \ref{sec_stacknorm}), is adopted to calibrate the column density of the stacked spectra for the entire sample (Sect. \ref{sec_columnden}).
All the transitions adopted for stacking (Sect. \ref{sec_thetransitions}) are optically thin in the standard spectrum, and it is therefore natural to assume the same for other hot cores, especially those with weak COM emission, which are the focus of our study. Note that, in the optically thin limit, both the column density and spectral intensity will be modified by the same beam dilution factor, and their ratio will remain unchanged.

\subsubsection{Transitions and weights for stacking}
\label{sec_thetransitions}
\begin{figure*}
\centering
\includegraphics[width=0.99\linewidth]{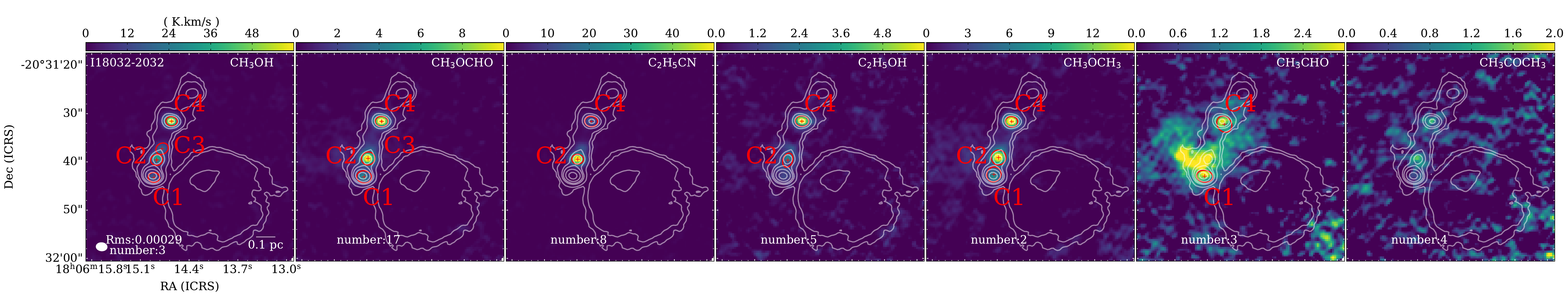}
\includegraphics[width=0.99\linewidth]{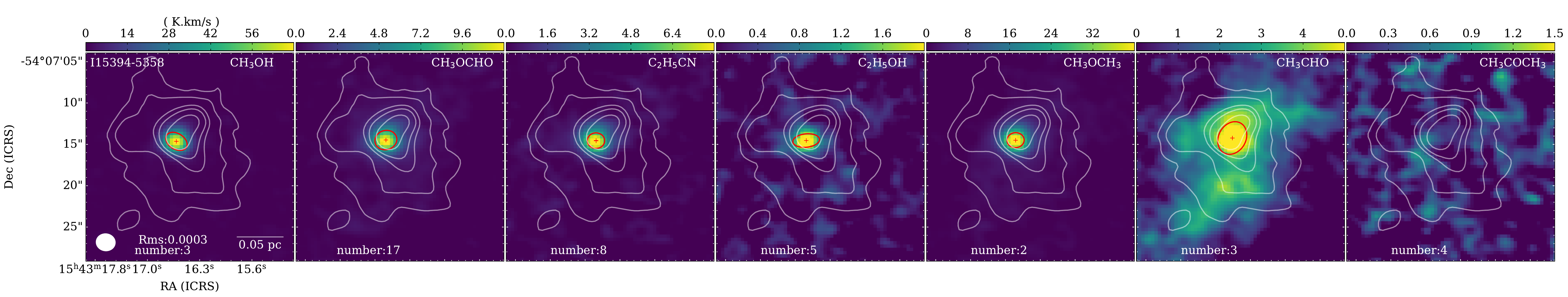} 
\includegraphics[width=0.99\linewidth]{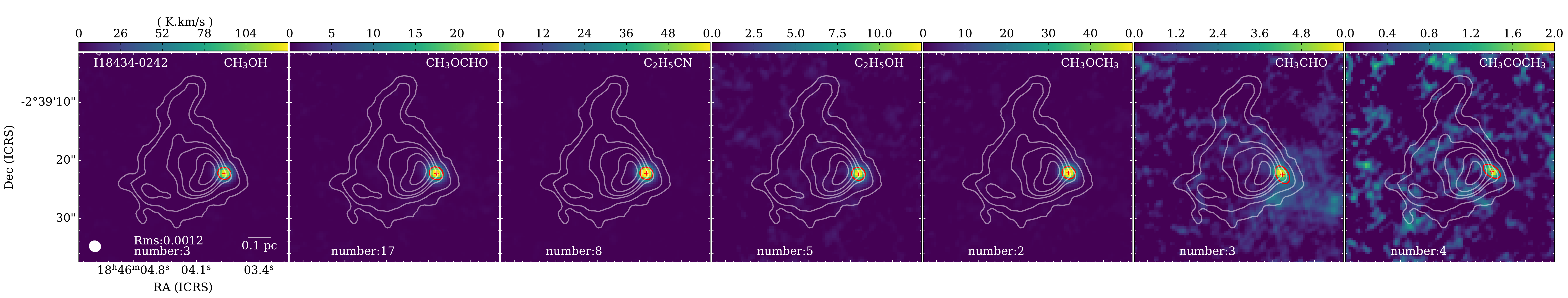}
\caption{Moment 0 maps of the stacked cubes (Sect. \ref{sec_stackmaps}) from three example sources—I18032-2032 (G9.62+0.19), I15394-5358, and I18434-0242—are shown. The contours represent the continuum emission, with levels of [5, 10, 30, 50, 100, 200] multiplied by the rms. The rms value is shown in the lower-right corner of the figure (in units of Jy beam$^{-1}$). The white filled ellipses in the lower-left corners of the left panels represent the beam of continuum emission. The red ellipses indicate the deconvolved FWHM sizes from the two-dimensional Gaussian fits to the compact cores. The images for the remaining 83 sources, which contain 94 hot cores and candidates, are presented in Fig.~\ref{fig_allthecores}.
}
\label{figure_mom0map_example} % 设置标签，以便引用
\end{figure*}

For each of the seven COM species (Sect. \ref{data}), through carefully checking the 
observed spectra and the standard spectrum, we identify the transitions that are: (1) 
detected in the hot cores of G9.62+0.19 (see upper panels in Figure \ref{figure_mom0map_example}), and (2) 
unblended with transitions of any other molecular lines. 
Note that the CH$3$OH 2(1)-2(1) transition at a rest frequency of $f{\rm rest} = 97582.898$ MHz (Table \ref{tab:all_lines}) is not included in the stacking. It has a low $E_{\rm u}$ of 20 K, which makes its emission typically optically thick and extensively distributed in high-mass star-forming regions.
All necessary molecular line information from the Jet Propulsion Laboratory (JPL\footnote{\url{http://spec.jpl.nasa.gov}}) molecular databases \citep[][]{1998JQSRT..60..883P} are accessed through XCLASS.
The rest frequencies of these unblended transitions are listed in Table \ref{table1}. 

In XCLASS, the conversion between the brightness temperature (in K) and the flux intensity of a line is conducted using the following equation:
\begin{equation}
T = 1.222 \times 10^3 \frac{I} {\nu^2 \theta_{\text{maj}} \theta_{\text{min}}},
\label{eq:1}
\end{equation}
where $I$ [mJy/beam] denotes the peak flux density, $\nu$ [GHz] is the rest frequency of the line (obtained from JPL), and $\theta_{\text{maj}}$ and $\theta_{\text{min}}$ [arcsec] represent the major and minor axis lengths of the ALMA synthesized beam at the observed frequency, respectively.
The peak brightness temperatures of these transitions
on the XCLASS-fitted noise-free spectra of  G9.62+0.19
are adopted as the weights for line stacking in following analysis. The weights are also listed in Table \ref{table1}.

\begin{figure}[!t]
\centering
\includegraphics[width=0.99\linewidth]{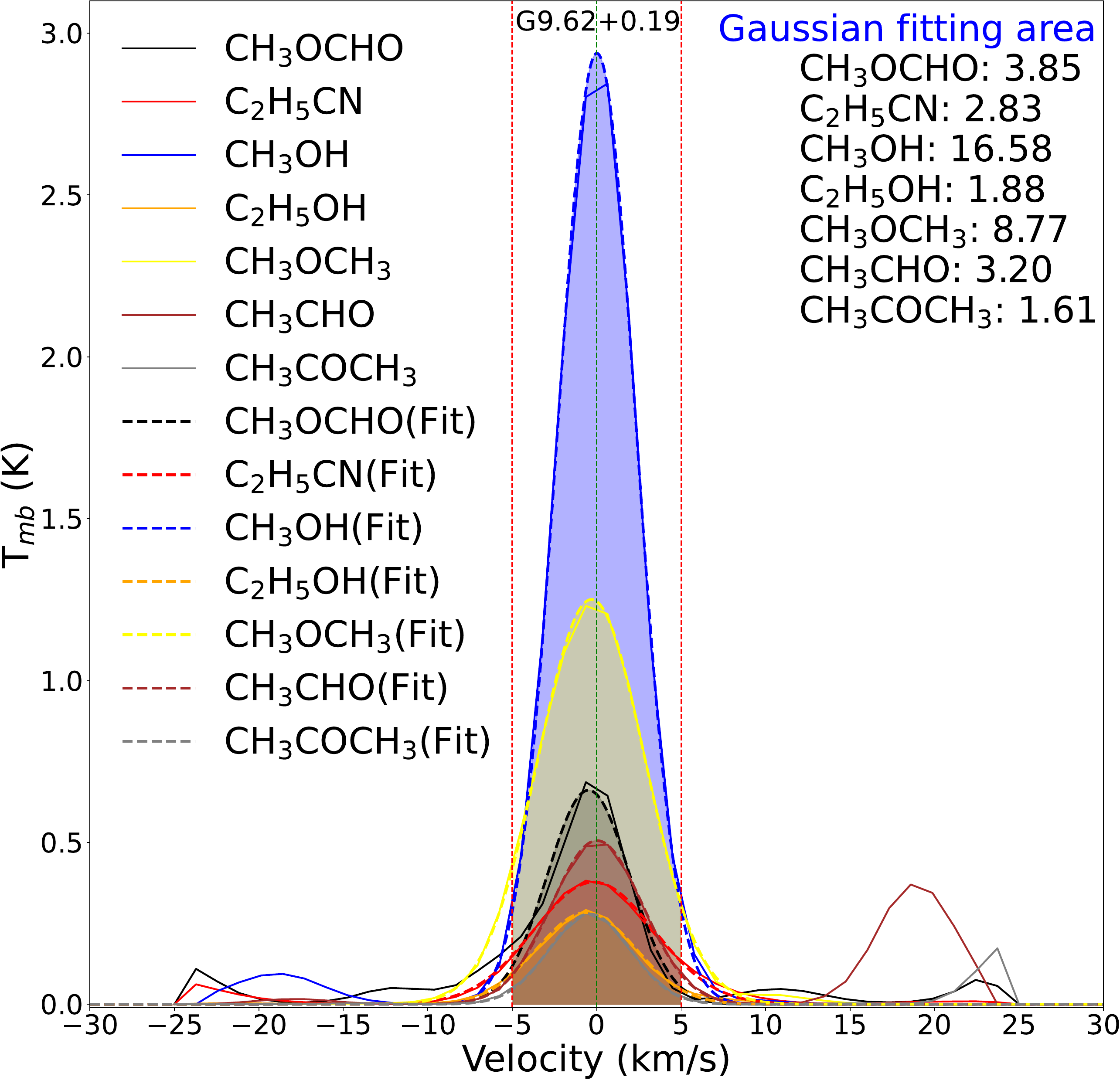}
\caption{The solid lines represent the averaged spectra of the template hot cores in G9.62+0.19 after spectral stacking. The dashed lines show Gaussian fits to the spectra. The integrated areas of the Gaussian fits (in units of K~km/s) are labeled in the upper-right corner. The vertical dotted lines indicate the velocity range of $\pm$5 km/s.}
\label{figure_stackexample} % 设置标签，以便引用
\end{figure}

\begin{figure}
\centering
\includegraphics[width=0.99\linewidth]{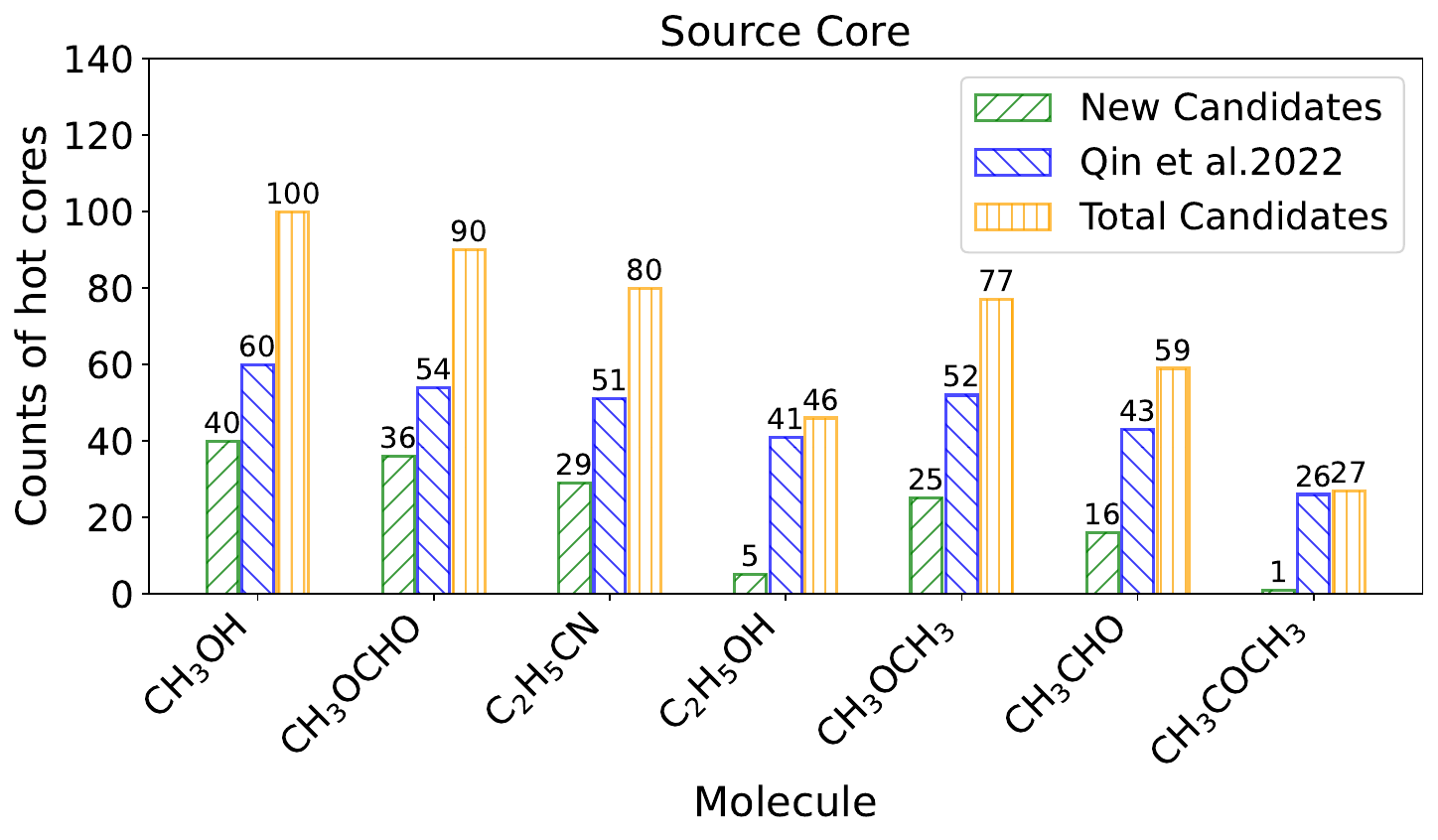}
\caption{The number of hot cores detected with different COMs. The numbers from \citet[][]{2022MNRAS.511.3463Q}, the newly detected numbers from this work, and the total are presented.
}
\label{fig_mol_counts} % 设置标签，以便引用
\end{figure}

\begin{figure*}
\centering
\begin{tabular}{@{}c@{\hspace{0.005\linewidth}}c@{\hspace{0.01\linewidth}}c@{}}
\includegraphics[width=0.32\linewidth]{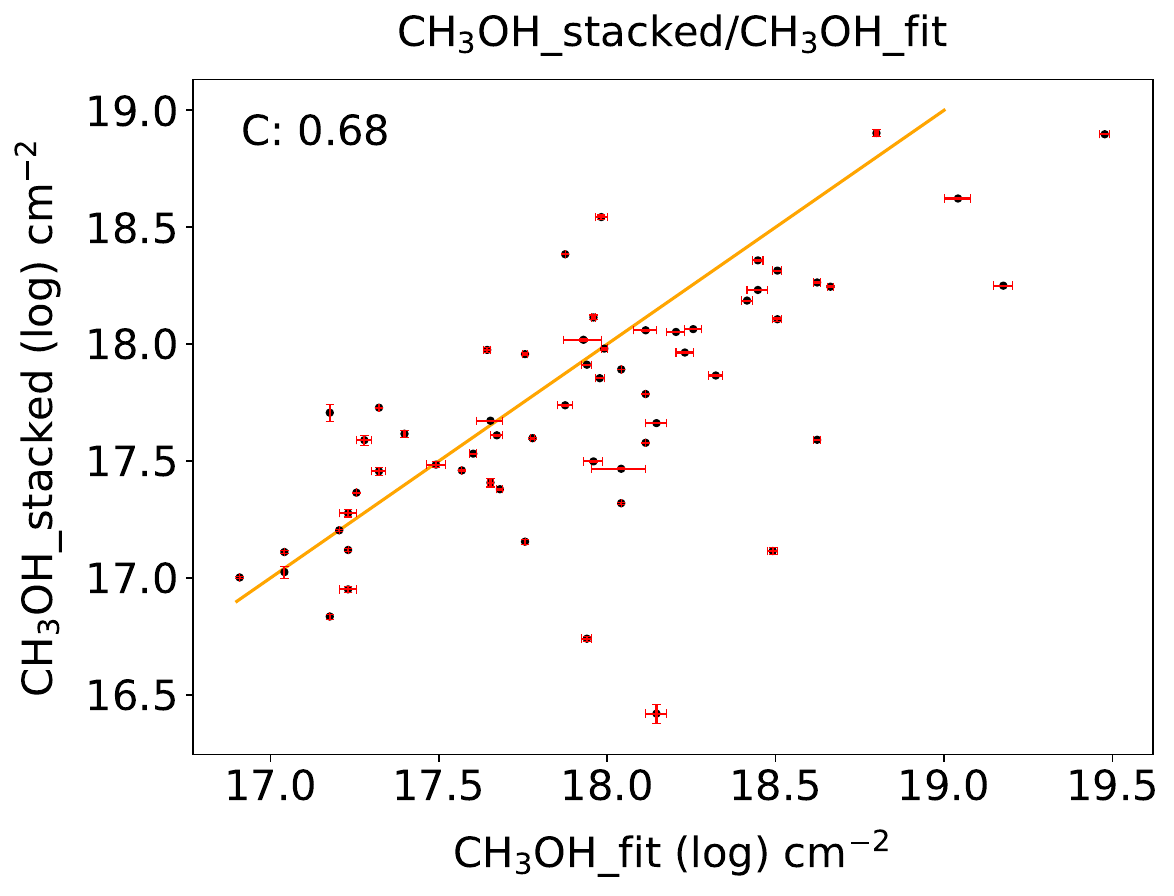} &
\includegraphics[width=0.32\linewidth]{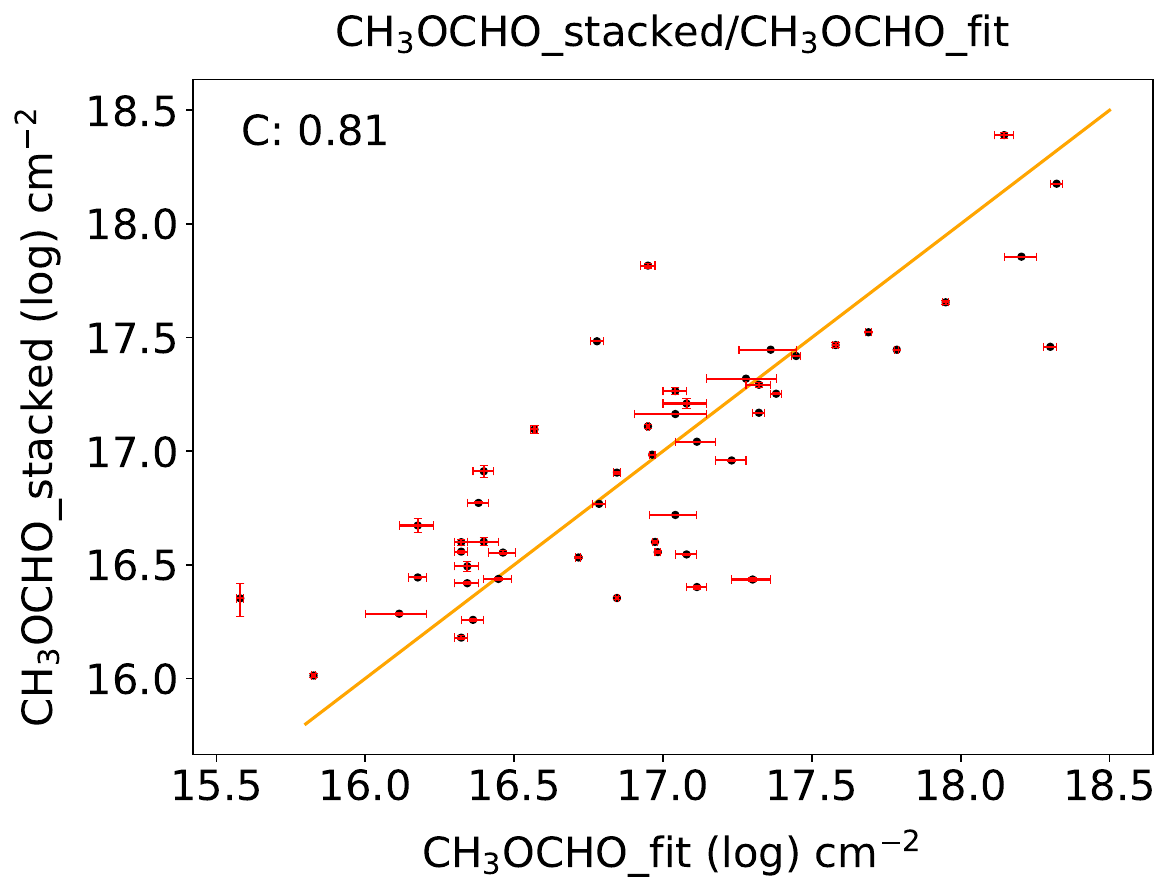} &
\includegraphics[width=0.32\linewidth]{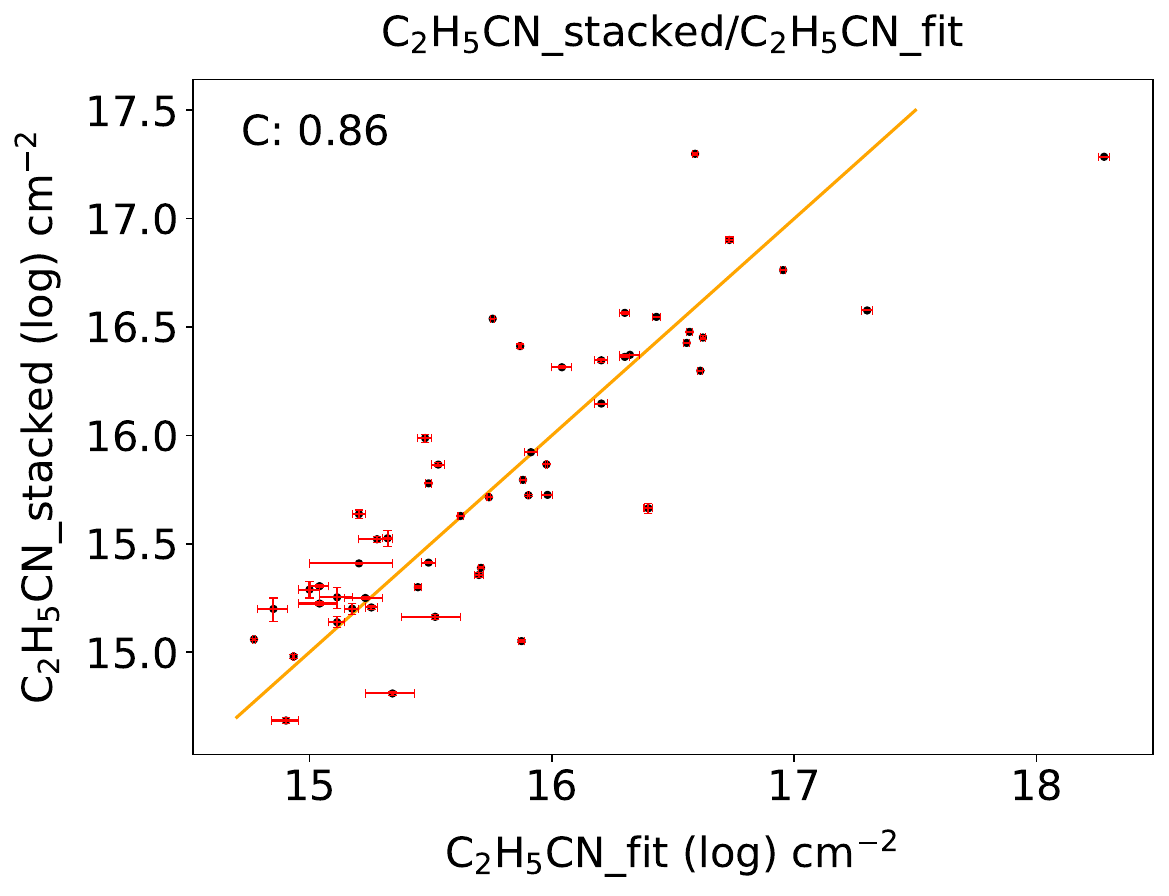} \\
\end{tabular}
\caption{Comparison of column densities derived through stacking (y-axis; Sect. \ref{sec_columnden}) and those fitted by \citet{2022MNRAS.511.3463Q} (x-axis) for the 60 brightest hot cores. The yellow line represents y = x. C denotes the correlation coefficient.}
\label{figure_compareto_qin} % 设置标签，以便引用
\end{figure*}

\begin{figure*}
\centering
\includegraphics[width=0.99\linewidth]{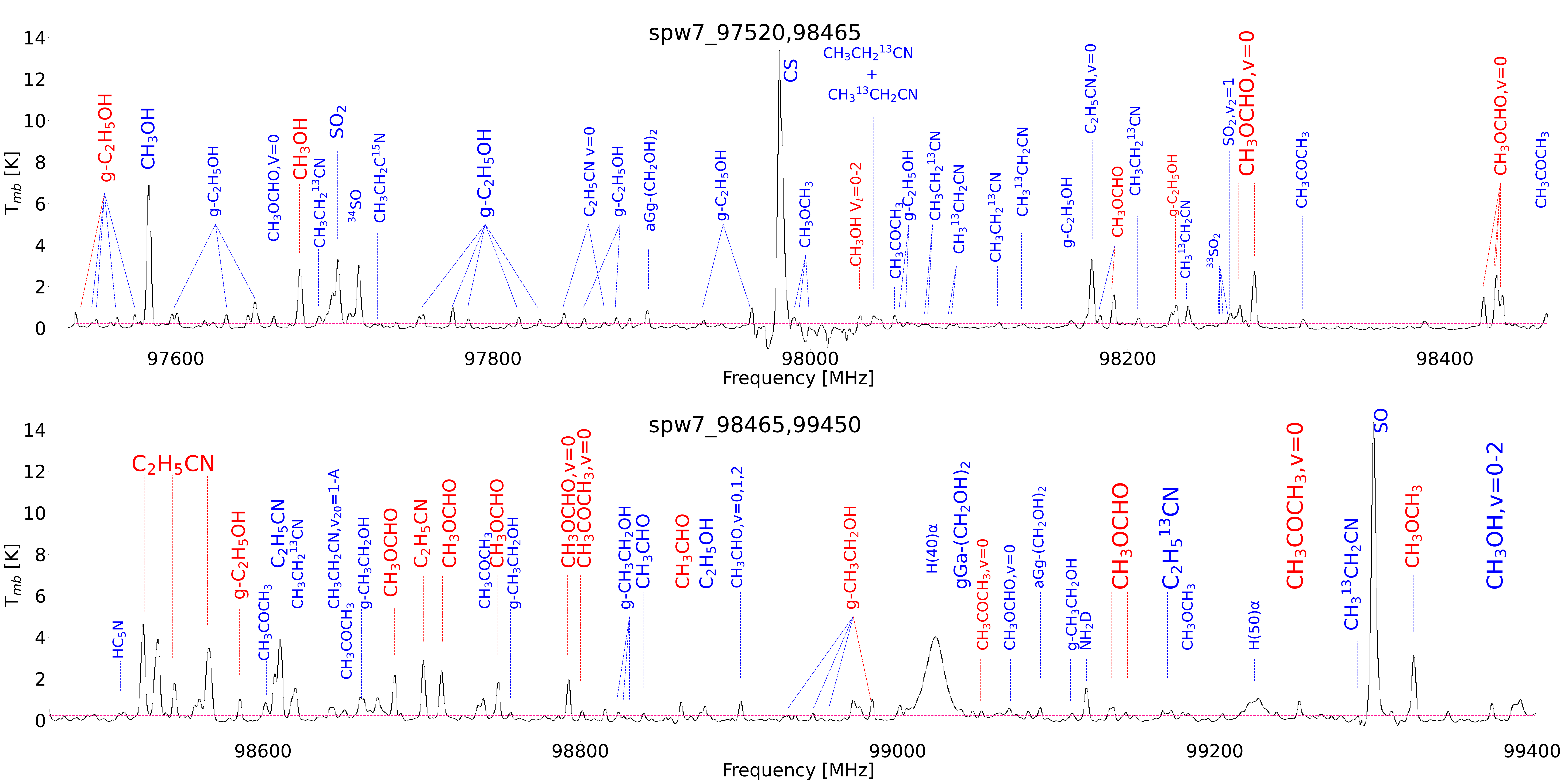}
\includegraphics[width=0.99\linewidth]{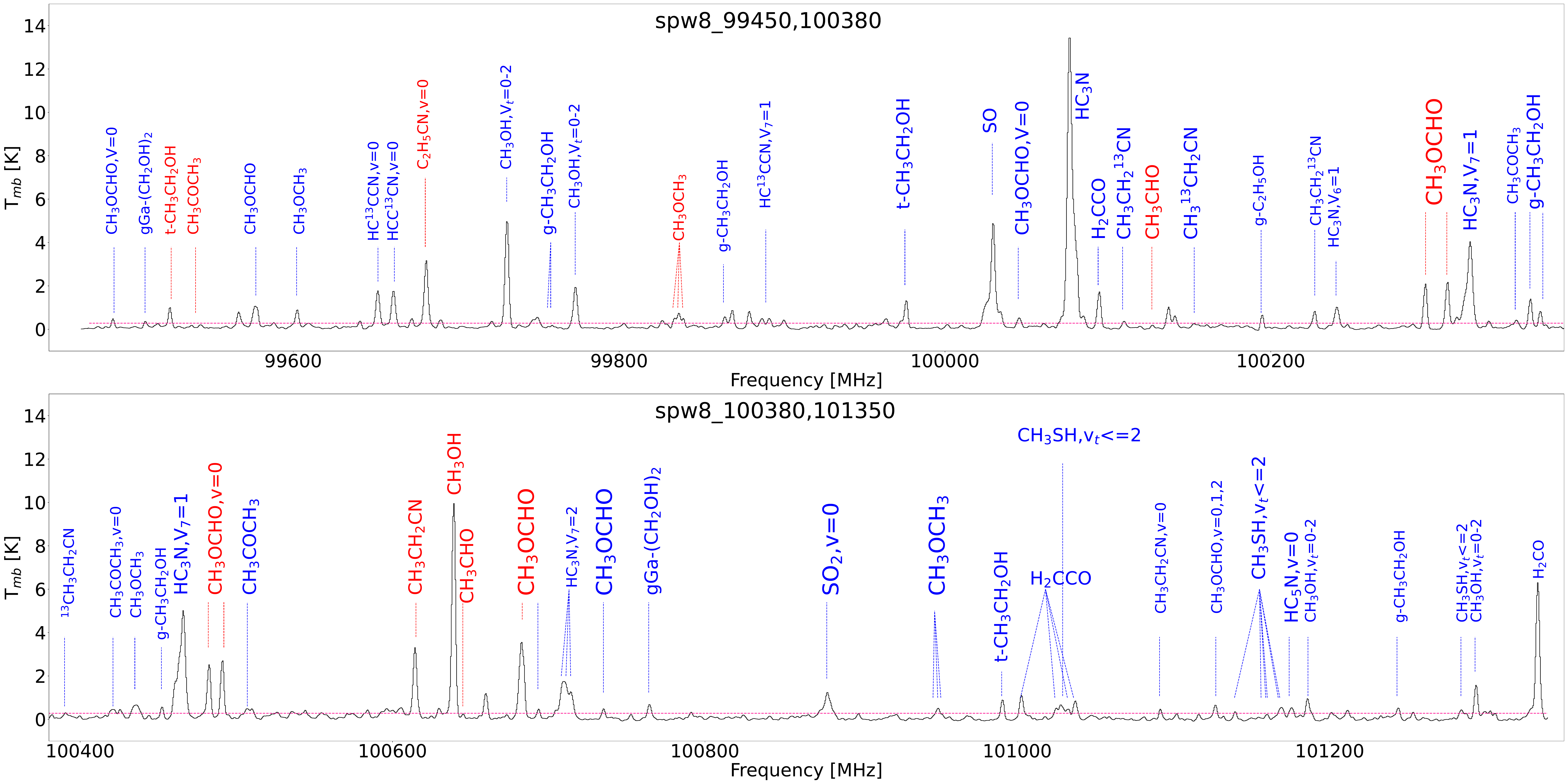}
\caption{The spectrum averaged over all 100 hot core candidates
(Sect. \ref{sec_sourcestack}). 
The transitions identified in the averaged spectrum (see Table \ref{tab:all_lines}) are labeled with the corresponding species names. The transitions selected for spectral stacking (Sect. \ref{sec_thetransitions}) in this work are marked with red labels of the species names. The 3$\sigma$ (0.2 K) noise level is indicated by horizontal pink lines.}
\label{figure_allstackspectrum} % 设置标签，以便引用
\end{figure*}

\begin{figure*}
\centering
\includegraphics[width=0.45\linewidth]{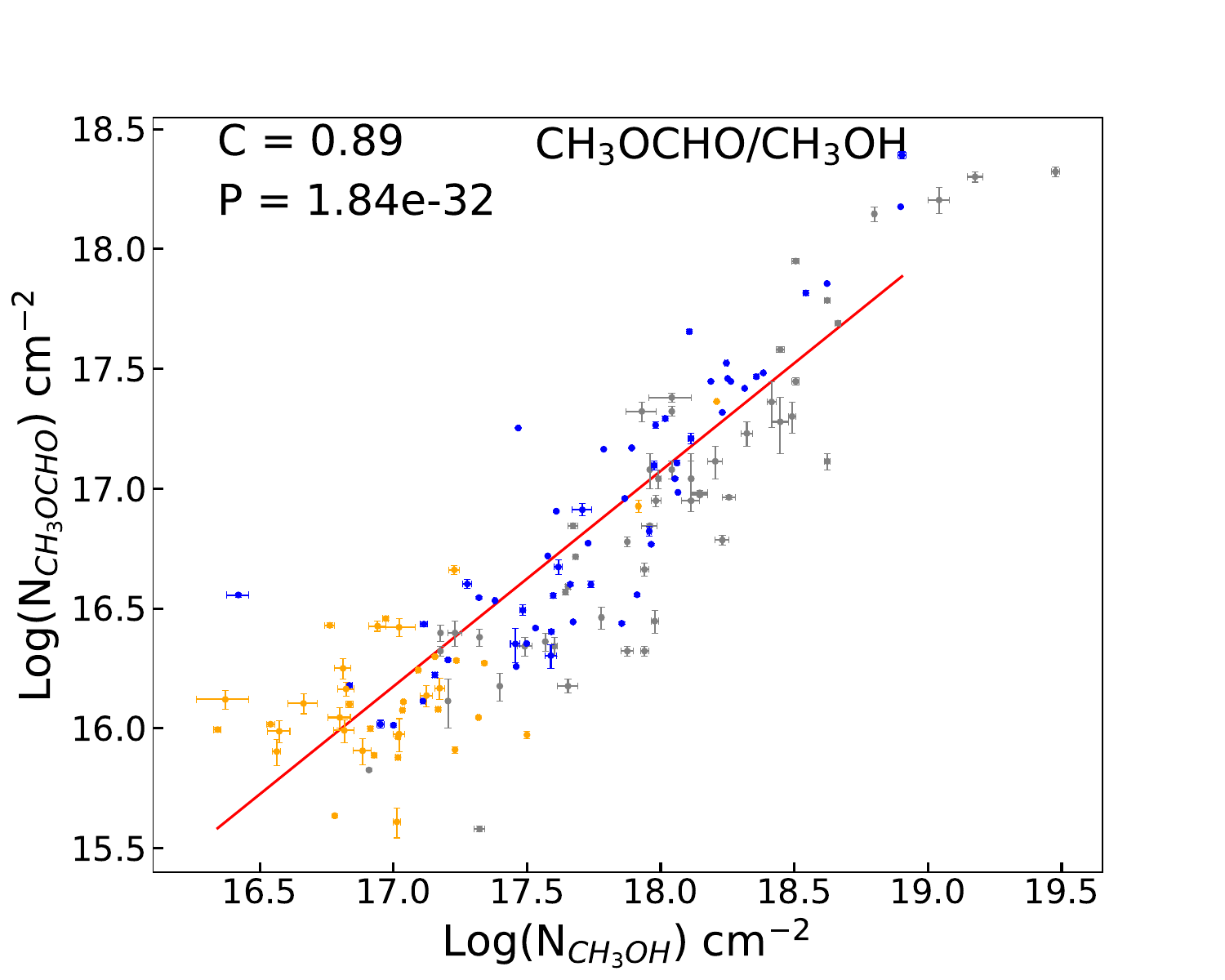}
\includegraphics[width=0.45\linewidth]{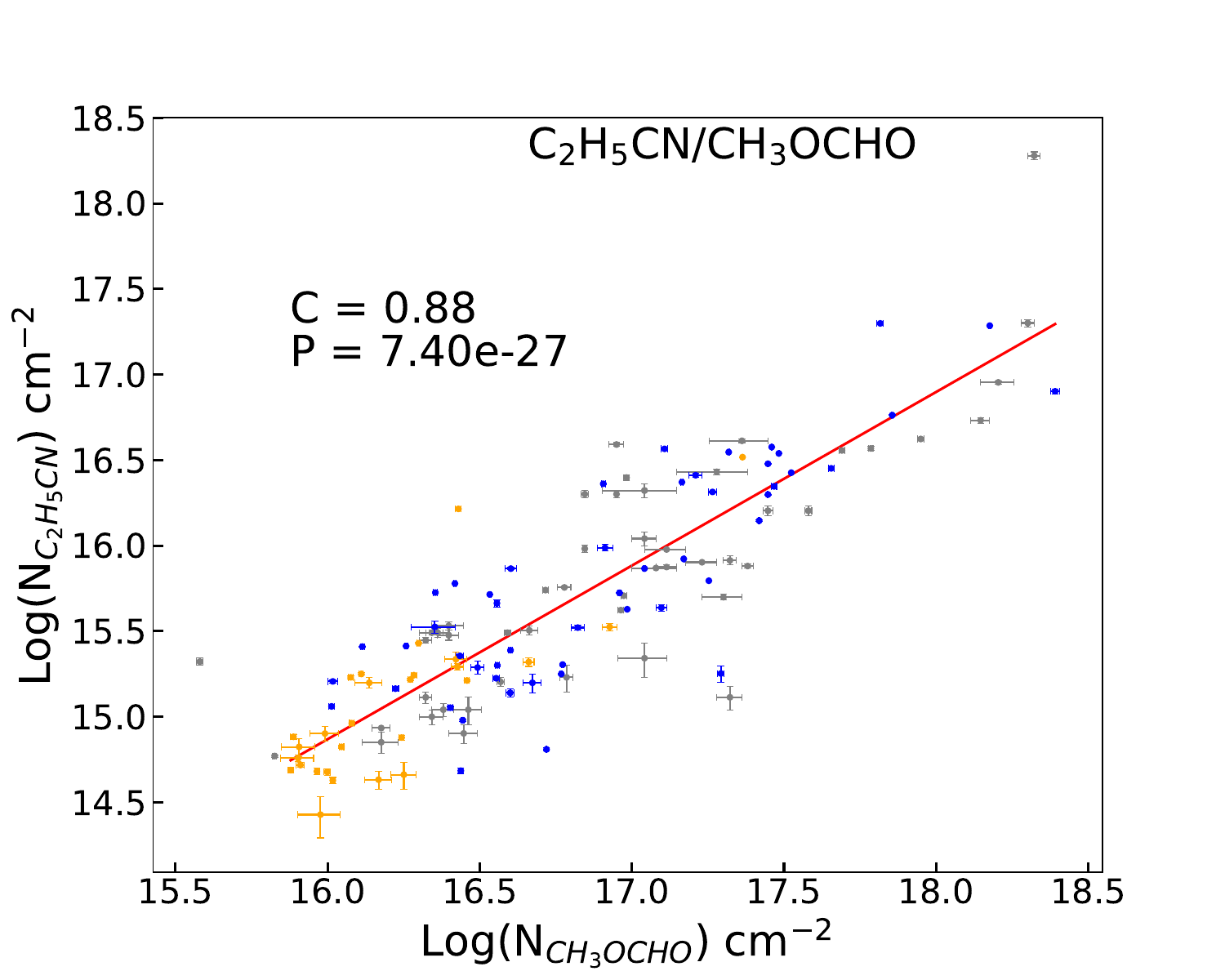}
\includegraphics[width=0.45\linewidth]{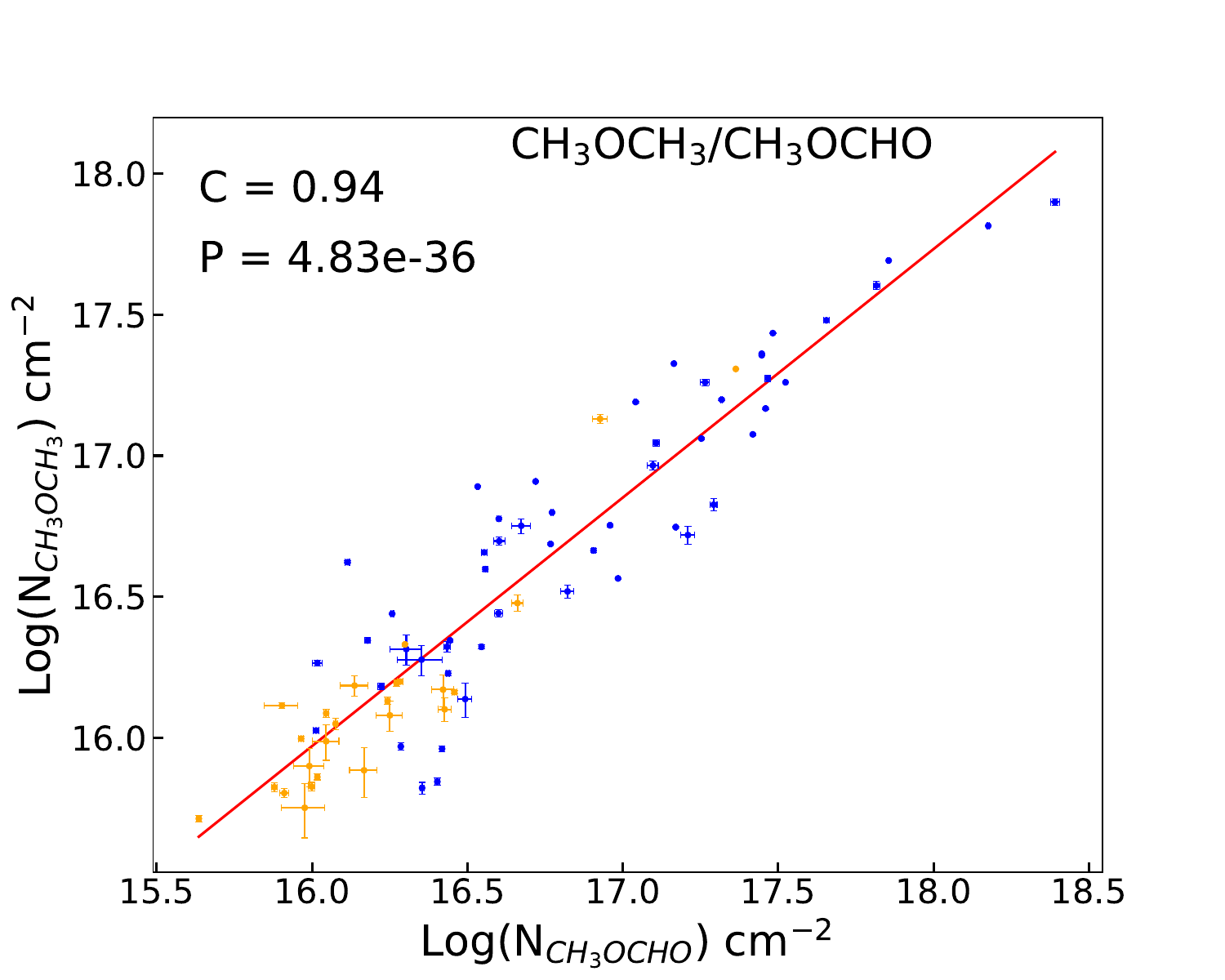}
\includegraphics[width=0.45\linewidth]{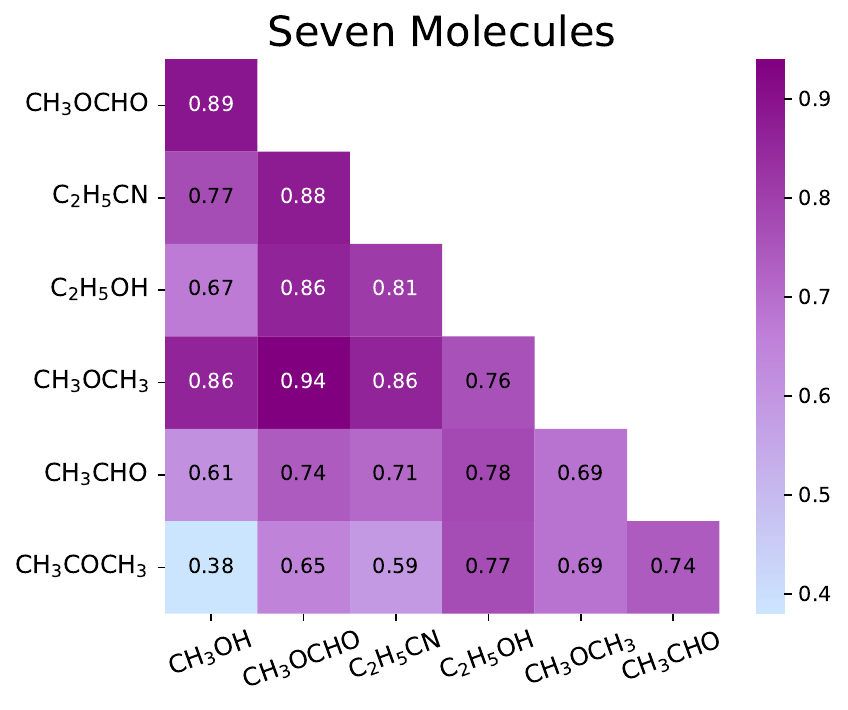}
\caption{The first three panels show the correlations of molecular column densities, with the red lines representing the linear fit in log scale. The correlation coefficients (labeled as 'C' in the upper-left corner) are displayed in each panel. Blue and yellow points represent the column densities obtained through stacking of the 60 strong hot cores and 40 weak candidates, respectively. Gray points indicate the column densities of strong hot cores fitted by \citet{2022MNRAS.511.3463Q}. The bottom-right panel displays the correlation coefficients between all species. Each cell is colored, with a deeper color representing a higher correlation coefficient. The p-value, which is also displayed in each panel, quantifies the statistical significance of the observed correlation. A smaller p-value (typically < 0.05) indicates a stronger, more statistically significant correlation, while larger p-values suggest weaker or less significant relationships between the variables.
}
\label{figure_correlations} % 设置标签，以便引用
\end{figure*}

\begin{figure}
\centering
\includegraphics[width=0.99\linewidth]{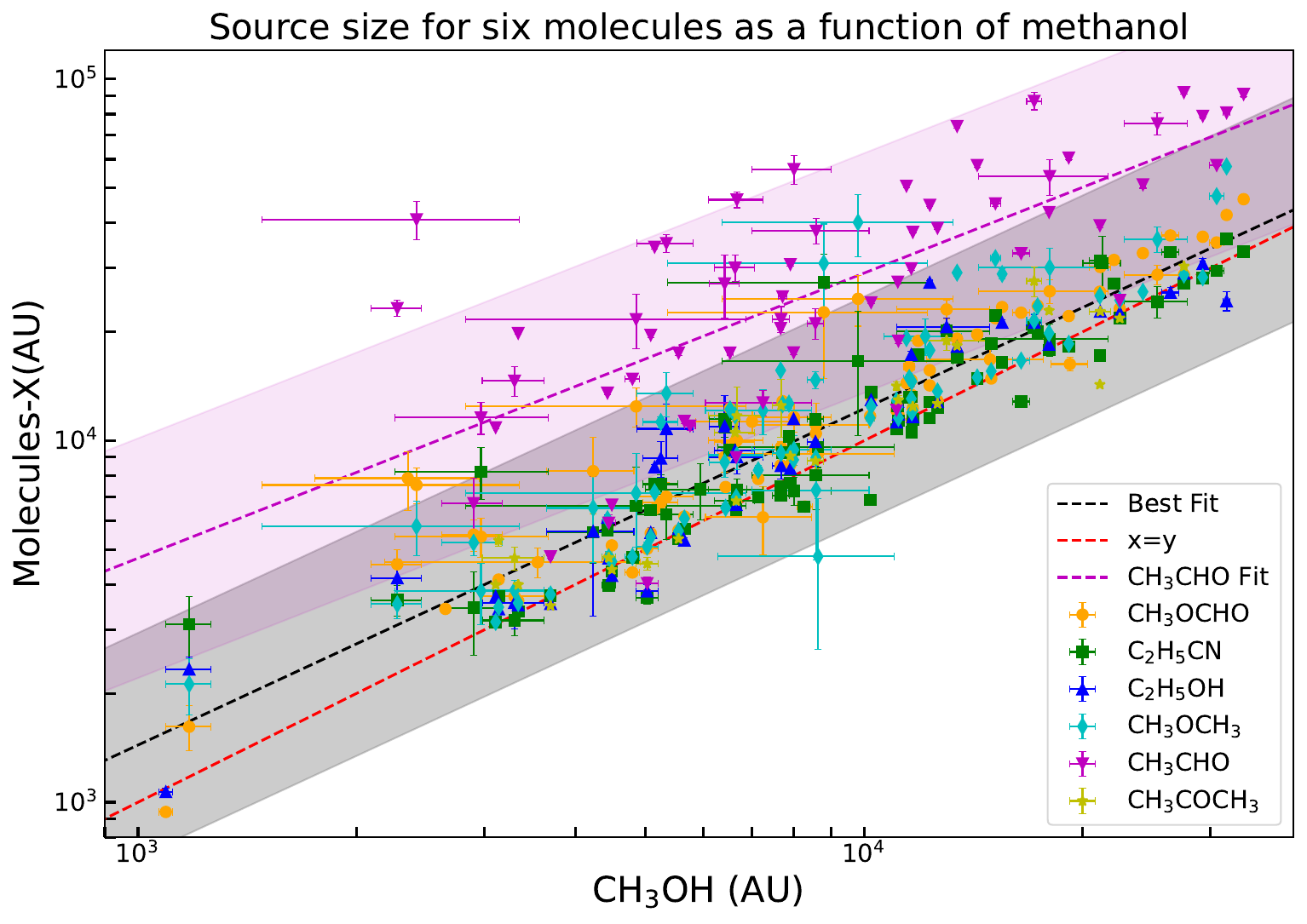}
\caption{
Relation between the beam-deconvolved sizes (in log scale) of hot core candidates for different species, with the $x$- and $y$-axis values representing the core sizes of CH$_3$OH and the other six species, respectively. The red dashed line represents $y = x$. The black dashed line shows the linear fit of all data points, except those of CH$_3$CHO. The pink dashed line shows the linear fit of the CH$_3$CHO data. The shaded regions indicate the standard deviation range of the data. }
\label{figure_coresize} % 设置标签，以便引用
\end{figure}

\begin{figure}
\centering
\includegraphics[width=0.9\linewidth]{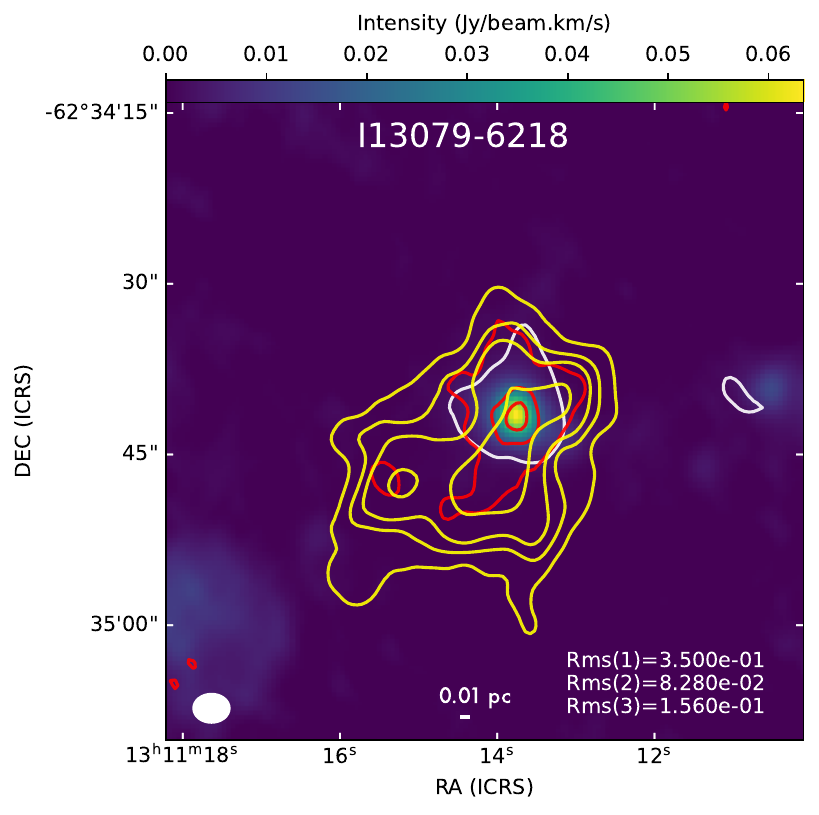}
\includegraphics[width=0.9\linewidth]{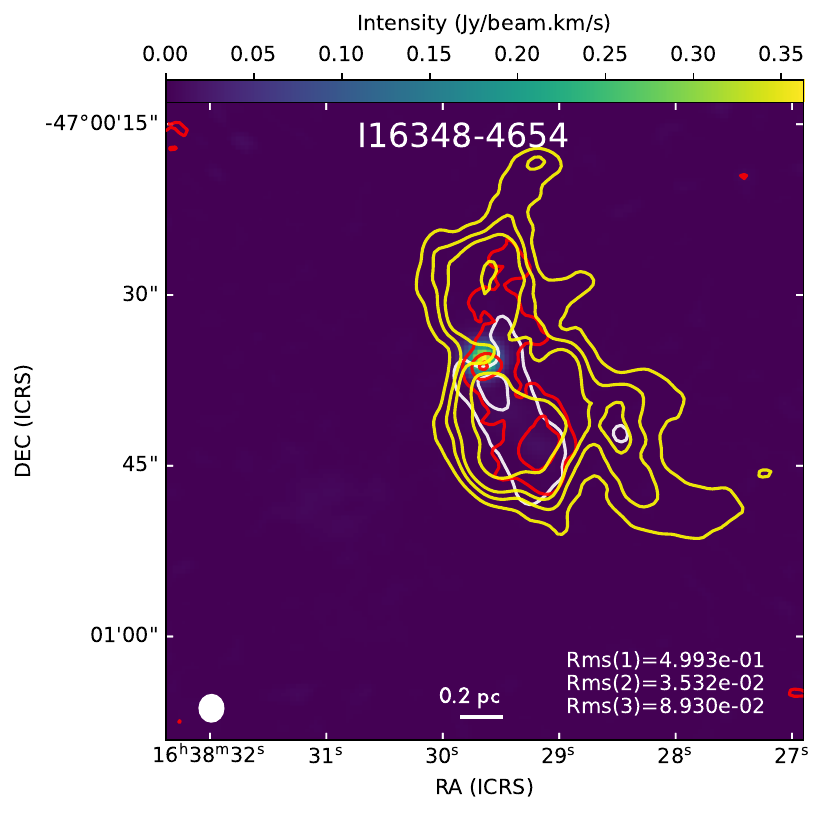}
\caption{Comparison of moment 0 maps of CH$_3$CHO (after spectral stacking), SiO and H$^{13}$CO$^{\text{+}}$. The background emission shows 3 mm continuum emission. The white, yellow and red contours are for H$^{13}$CO$^{\text{+}}$ emission, SiO, and CH$_3$CHO, respectively. Their contour levels are [5, 10, 15, 30, 50, 100, 200]$\times$Rms(1, 2, 3). Rms(1), Rms(2) and Rms(3) are shown at the lower-right corners, representing the noise values for CH$_3$CHO, SiO, and H$^{13}$CO$^{\text{+}}$, respectively, with units of K~Km~s$^{-1}$. The beam of continuum emission is placed in the lower left corner of the image.}
\label{figure_CH3CHOexample} 
\end{figure}

\begin{figure}
\centering
\includegraphics[width=0.99\linewidth]{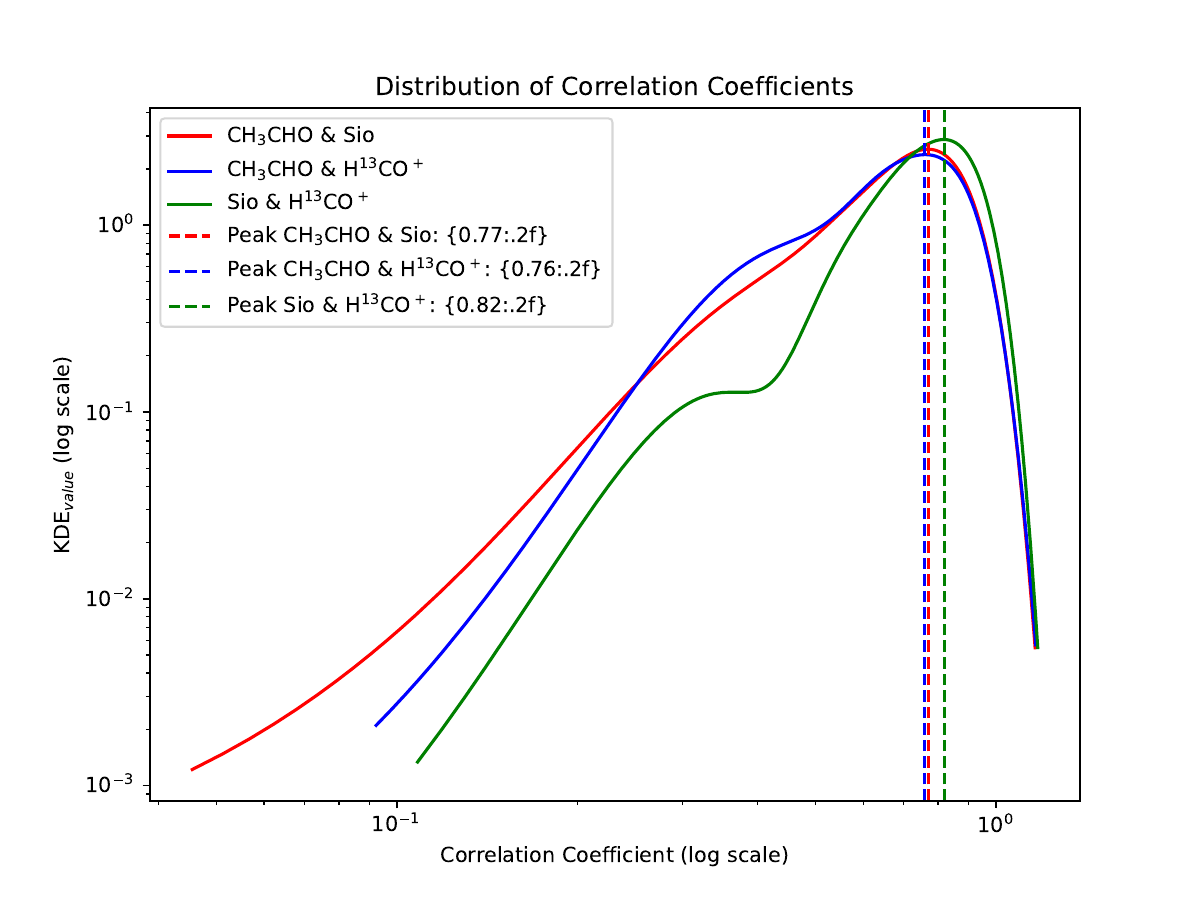}
\caption{
The kernel density estimate (KDE) smoothed distribution of correlation coefficients between the Moment 0 maps of CH$_3$CHO (after stacking), SiO (2-1), and H$^{13}$CO$^{\text{+}}$ (1-0). The KDE curves were generated using the {\it gaussian\_kde} tool from the {\it scipy} package in Python. The vertical lines indicate the peaks of the distributions.
}
\label{figure_k_kde} 
\end{figure}

\subsubsection{Resampling}
The second and third steps are to cut out and resample the data cube.
For each ATOMS source, we cut out a spectral cube with a bandwidth of \mbox{$\sim60$ km s$^{-1}$} for each transition selected above. The central frequency of the spectral cube is:
\begin{equation}
    \nu_c = \nu_0\left(1 - \frac{V_{\rm LSR}}{c}\right),
\end{equation}
where $V_{\rm LSR}$ is the systematic velocity of the source, $\nu_0$ is the rest frequency of the transition, and 
$c$ represents the speed of light. 
Next, we convert the frequency axis ($\nu$) to the velocity axis ($V$) using the relation
\begin{equation}
    V = \frac{\nu_c - \nu}{\nu_c} c. \label{eq2}
\end{equation}
Note that, for a given species, the velocity resolutions of its different transitions may vary. 
To align the velocity channels, we further resample the cut cube of each transition to a velocity resolution of 1.25 km s$^{-1}$.

\subsubsection{Spectral stacking and normalization}\label{sec_stacknorm}
In the fourth step outlined in Figure \ref{figure_stackflow}, the stacked cube ($D_{j}^{i}$) for each species ($j$) of each source ($i$) is obtained by averaging the resampled cubes ($D_{j;k}^{i}$) of different transitions (denoted as $k$; see Sect. \ref{sec_thetransitions}), weighted according to the values (denoted as $W_k$) in Table \ref{table1}, as shown below
\begin{equation}
  D_{j}^{i}  = \frac{\sum W_k D_{j;k}^{i}}{\sum W_k}.
\end{equation}
Note that if the intensities of the different transitions of the same species are proportional to those of the template spectra (Sect. \ref{sec_thetransitions}) and exhibit the same noise level, this set of weights could achieve the highest S/N for the stacked cube \citep{2022ApJS..263...13L}.
The stacked spectra of the seven COM species, obtained by applying the above procedure to the standard spectrum of G9.62+0.19 (Sect. \ref{sec_G9.62standard}), are presented in Fig. \ref{figure_stackexample}.

\subsubsection{Integrated intensity maps of stacked cubes}
\label{sec_stackmaps}
After stacking, we integrated the emission for the stacked lines within a velocity range of $\pm 5$ km s$^{-1}$. We chose this value because the typical COM line width, measured as the full width at half maximum (FWHM), for the strong sample of hot cores is approximately 5 km s$^{-1}$ \citep{2020MNRAS.496.2790L}. As shown in Figure \ref{figure_stackexample}, the adopted velocity range for integration covers the majority of the emission in the stacked spectra of G9.62+0.19 while avoiding blending.

Figure \ref{figure_mom0map_example} shows the integrated intensity maps (Moment 0 maps) of the stacked cubes of the seven species for I18032-2032, I15394-5358, and I18434-0242. I18032-2032 (G9.62+0.19) contains four hot cores, labeled as C1, C2, C3, and C4, which were previously identified by \citet{2022MNRAS.512.4419P} and \citet{2022MNRAS.511.3463Q}. This source has the largest number of hot cores in the sample. Its gas kinetic temperature and column density serve as a reference for typical hot cores. I15394-5358 and I18434-0242 are examples of sources with newly identified hot core candidates in this work. The Moment 0 maps of the seven species for the other sources that exhibit at least one compact core of CH$_3$OH are displayed in Figure \ref{fig_allthecores}.

\subsection{Hot core candidates}
\subsubsection{Identification}
We identify hot molecular core (HMC) candidates from the Moment 0 maps of the stacked cubes. An HMC candidate is visually identified by the presence of strong ($\gtrsim 5\sigma$), compact CH$_3$OH emission. Here, a compact core is defined as one with a round morphology, where its central brightest part is not resolved or is only partly resolved under the current spatial resolution (Sect. \ref{data}).
In total, we identified 100 HMC candidates, 60 of which are identical to the strong sample previously identified by \citet{2022MNRAS.511.3463Q}, and the remaining 40 are weak candidates newly identified in this work.

These HMC candidates are typically associated with compact emission in other COMs as well. Figure \ref{fig_mol_counts} shows the statistics of line detection for all 100 HMC candidates. All 40 newly identified HMC candidates exhibit compact emission of CH$_3$OCHO or C$_2$H$_5$CN. Specifically, 36 show CH$_3$OCHO emission, 29 display C$_2$H$_5$CN emission, 5 reveal C$_2$H$_5$OH emission, 25 demonstrate CH$_3$OCH$_3$ emission, 16 present CH$_3$CHO emission, and only one core shows CH$_3$COCH$_3$ emission. All 60 hot cores previously identified by \citet{2022MNRAS.511.3463Q} exhibited stronger CH$_3$OH emission and more individual COM line transitions than the newly identified weak hot core candidates. In total, 100 hot cores show CH$_3$OH emission, 90 show CH$_3$OCHO emission, 80 show C$_2$H$_5$CN emission, 46 show C$_2$H$_5$OH emission, 77 show CH$_3$OCH$_3$ emission, 59 show CH$_3$CHO emission, and 27 show CH$_3$COCH$_3$ emission.

\subsubsection{Two-dimensiona Gaussian fitting}
Two-dimensional Gaussian fitting is applied to the 100 cores on the Moment 0 maps of CH$_3$OH using the CASA {\it imfit} procedure. 
The beam-deconvolved fitting parameters are adopted, including the source size ($\theta_{\rm source}$), the position angle ($PA$), the peak flux of the integrated intensity ($I_{\rm peak}$; in units of K km s$^{-1}$), and the total flux of the integrated intensity ($I_{\rm integrated}$; in units of K km s$^{-1}$ arcsec$^2$). Here, $\theta_{\rm source}=\sqrt{ab}$, where $a$ and $b$ represent the deconvolved major and minor FWHM axes of the cores.
The sizes of molecular CH$_3$OH emission range from 1092 to 46884 AU for sources at different distances.
The fitted peak positions, $\theta_{\rm source}$, $PA$, $I_{\rm peak}$, and $I_{\rm integrated}$ of CH$_3$OH are summarized in Table \ref{tab:CH3OH}.

The same fitting procedure was applied for the other six COM species (if detected). Note that for some hot cores, the emission of CH$_3$CHO displays extended components, making the fitting of CH$_3$CHO less reliable. The extended emission of CH$_3$CHO will be further discussed in Sect. \ref{sec_extent_CH3CHO}.
The fitted parameters of the six species are summarized in Tables \ref{tab:threespecies1} and \ref{tab:threespecies2}.

\subsubsection{Column density} \label{sec_columnden}
G9.62+0.19 is adopted as the calibration source to estimate the column densities of hot core candidates from their stacked emission.
For each species, we integrate the standard spectrum of G9.62+0.19 (after stacking for each species; see Sects. \ref{sec_G9.62standard} and \ref{sec_stacknorm}) within a velocity range of $\pm 5$ km s$^{-1}$ (Fig. \ref{figure_stackexample}) to obtain the standard integrated intensity ($I_{\rm cali}$) of the stacked spectrum for that species. This allows us to quickly determine the intensity of each species by focusing on the relevant spectral features. The column density of the G9.62+0.19 hot-core value, provided in Sect. \ref{sec_G9.62standard} and listed in Table \ref{table1}, is adopted as the standard column density (denoted as $N_{\rm cali}$) for the corresponding species.
To evaluate the column density of each species, we use the following conversion formula: 
\begin{equation} \label{eq_cali}
N = \frac{N_{\rm cali}}{I_{\rm cali}}I. 
\end{equation} Here, $I$ represents the integrated intensity of the stacked spectrum, and this formula allows us to scale the species' column density relative to the standard value.

Applying Eq. \ref{eq_cali}  to the Moment 0 map of a stacked cube results in a column density map for the corresponding species. In Sect. \ref{sec_stackmaps}, we have already obtained the beam-deconvolved peak value ($I_{\rm peak}$) of the Moment map using 2D Gaussian fitting, enabling us to directly calculate the beam-deconvolved peak column density
using Eq. \ref{eq_cali}.
The derived column densities of different species are listed in Tables \ref{tab:CH3OH},
\ref{tab:threespecies1} and \ref{tab:threespecies2}.

In Figure \ref{figure_compareto_qin}, we compare the beam-deconvolved column densities (Sect. \ref{sec_columnden}) calculated using the spectral stacking method with those derived by \citet{2022MNRAS.511.3463Q} through LTE fitting for the 60 strong hot cores. The results are generally in good agreement, supporting the reliability of column densities estimated through stacking. The details of the hot cores in G9.62+0.19 remain unclear, and we cannot confirm whether different hot core candidates exhibit similar distributions of emitting regions and excitation conditions. The discrepancy in the column densities primarily arises from possible differences in excitation temperatures and spatial patterns of COM emission in G9.62+0.19.
Although the transitions chosen for stacking are optically thin, unresolved optically thick regions, which cannot be distinguished at the current resolution, may also contribute to the discrepancy. Nonetheless, the good agreement between the XCLASS fitting and the stacking conversion suggests comparable excitation conditions for different hot core candidates.

The column densities of the seven molecules span a striking range, covering one to three orders of magnitude. CH$_{3}$OH stands out with the highest column densities, ranging from 1.4$\times$10$^{18}$ to 4.4$\times$10$^{20}$ cm$^{-2}$. In comparison, CH$_{3}$OCHO and CH$_{3}$OCH$_{3}$ show column densities approximately half an order of magnitude lower, from 5.3$\times$10$^{17}$ to 2.8$\times$10$^{20}$ cm$^{-2}$ and from 1.0$\times$10$^{18}$ to 1.4$\times$10$^{20}$ cm$^{-2}$, respectively. C$_{2}$H$_{5}$OH, CH$_{3}$CHO, and CH$_{3}$COCH$_{3}$ exhibit column densities roughly one order of magnitude lower than CH$_{3}$OH, ranging from 1.3$\times$10$^{17}$ to 5.9$\times$10$^{19}$ cm$^{-2}$. C$_{2}$H$_{5}$CN, with the lowest values, spans a range from 1.6$\times$10$^{15}$ to 7.1$\times$10$^{18}$ cm$^{-2}$---roughly two orders of magnitude lower than CH$_{3}$OH.
This wide variability in column densities highlights the diverse physical conditions or evolution stages among the 100 hot core candidates under investigation.

\section{Discussion}
\label{dis}
\subsection{Source-stacking spectrum of hot-cores }
\label{sec_sourcestack}
For each of the 100 hot core candidates, we extract the full-band spectra of SPW 7 and SPW 8 at the peak location of CH$_3$OH (after stacking).  First, these frequency axis of these spectra are corrected 
through
\begin{equation}
    v' = \left(1+\frac{V_{\rm LSR}}{c}\right)v.
\end{equation}
Here, $v'$ is the corrected frequency axis.
We then resample the spectra to have aligned 
channels with a channel width of 0.49 MHz.
A source-stacking template spectrum is obtained by averaging these spectra with equal weighting.
The stacked spectrum is shown in Figure \ref{figure_allstackspectrum}. 
Thanks to the improved signal-to-noise ratio (S/N) from source stacking, the rich emission lines of Complex Organic Molecules (COMs) can be identified.
We fit the template spectrum with the emissions of species already identified from G9.62+0.19 (\cite{2020MNRAS.496.2790L} and \cite{2022MNRAS.512.4419P}) using XCLASS. The rest frequencies, transitions, and state temperatures of these molecules are compiled in Table \ref{tab:all_lines}.
The source-stacking spectrum can serve as a template for HMC studies, providing a reference for the rapid identification of molecular species in the same band. In addition to the identified transitions, there are plenty of line features yet to be identified in the ATOMS sample. We do not attempt to identify them in this work, which focuses on spectral stacking of the ATOMS hot core candidates. Instead, the source-stacking spectrum offers a valuable template for hot-core research in future studies.
The source-stacking technique can also enhance sensitivity for molecular identification in follow-up surveys, such as the ALMA-QUARKS survey in Band 6 \citep{2024RAA....24b5009L}.

\subsection{Correlations of complex molecules}

Figures \ref{figure_correlations} and \ref{fig:Ncorr_more} show the correlations between the column densities  of molecular pairs for the seven molecules, including both the strong hot cores and the weak candidates.
One of the most significant features is that the column densities of weak hot core candidates lie at the lower end of the linear correlation trend for different COM species across the entire sample. This supports the validity of estimating column density through stacking (Sect. \ref{sec_columnden}) and may suggest that the similarity in excitation conditions observed in hot cores could also be applicable to the weak hot core candidates.

We summarized the correlation coefficients of different molecular pairs in the lower-right panel of Figures \ref{figure_correlations}.
Among them, three molecular pairs exhibited strong correlations (with correlation coefficients close $\gtrsim 0.9$): CH$_3$OH versus CH$_3$OCHO, CH$_3$OCHO versus C$_2$H$_5$CN, and CH$_3$OCHO versus CH$_3$OCH$_3$ (Figure~\ref{figure_correlations}).
Their correlation coefficients are 0.89, 0.88, and 0.94, respectively. 
CH$_3$OH, CH$_3$OCHO, and C$_2$H$_5$CN demonstrated consistently strong correlations in their column densities.

The strong correlation between CH$_3$OCHO and CH$_3$OCH$_3$ has been observed in low \citep[][]{2024MNRAS.533.1583L}, intermediate \citep[][]{2018A&A...618A.145O}, and high-mass star-forming regions \citep[e.g.,][]{2007A&A...465..913B,2020A&A...641A..54C,2024MNRAS.533.1583L}. This close relationship can be attributed to a common precursor, CH$_3$O \citep[][]{2006A&A...457..927G,2008ApJ...682..283G,2013ApJ...765...60G,2016ChemRev...Oberg}, or alternatively, CH$_3$OCH$_3$ may act as a precursor to CH$_3$OCHO \citep[][]{2015MNRAS.449L..16B}. Moreover, both species are strongly correlated with CH$_3$OH (Fig. \ref{figure_correlations}), supporting the hypothesis that their precursor, CH$_3$O, is likely produced through the photodissociation of CH$_3$OH in a hot-core environment. As an isomer of CH$_3$OCH$_3$, C$_2$H$_5$OH also exhibits a strong correlation (0.86) with CH$_3$OCHO. This suggests that the chemical environments for forming the two isomers (C$_2$H$_5$OH and CH$_3$OCH$_3$) in hot cores share similarities.

The weakest correlation occurs between CH$_3$OH and CH$_3$COCH$_3$. This can be attributed to the distinct formation and excitation mechanisms of these molecules. CH$_3$OH is abundant in hot cores and may form primarily through grain-surface reactions \citep{2009ARA&A..47..427H}. In contrast, CH$_3$COCH$_3$ (acetone) is a more complex molecule that forms through both gas-phase reactions and surface reactions on dust grains \citep{1987A&A...180L..13C,2022ApJ...941..103S}. The differences in their formation pathways could lead to varying physical conditions in hot cores, which may result in the observed weak correlation.

The strong correlation between the nitrogen-bearing molecule C$_2$H$_5$CN and the oxygen-bearing molecule CH$_3$OCHO confirms that both are excellent tracers of hot cores, with the hot core environment governing their generation and/or the excitation of their emission. C$_2$H$_5$CN also shows a strong overall correlation with CH$_3$OCH$_3$, with a correlation coefficient of 0.86. However, different spatial distributions between the nitrogen-bearing C$_2$H$_5$CN and oxygen-bearing species are observed in some sources (e.g., IRAS 17158-3901, 17160-3707, and 18032-2032; see Fig. \ref{fig_allthecores}). Therefore, whether the nitrogen and oxygen differentiation is common in our sample of high-mass star-forming regions remains uncertain due to the limited angular resolution in our observations \citep{2022MNRAS.511.3463Q}. In future studies utilizing higher-resolution and more sensitive data from the ALMA-QUARKS project \citep{2024RAA....24b5009L}, a more comprehensive analysis of the differentiation between nitrogen and oxygen species in hot cores will be achievable.

\subsection{Spatial distributions of complex molecular line emission}
\label{sec_extent_CH3CHO}
The emission lines of complex organic molecules (COMs), with the exception of CH$_3$CHO, generally exhibit compact emission, concentrated around the average position of the peak emission of the seven molecules.
In contrast, CH$_3$CHO displays more extended emission in certain sources, such as I8032-2032 and I15394-5358, as shown in Figure \ref{figure_mom0map_example}. 
The sizes of the emission regions for the seven molecules were measured in each source and are compared in Figure \ref{figure_coresize}. The deconvolved sizes of the emission regions for CH$_3$OH, CH$_3$OCHO, C$_2$H$_5$CN, C$_2$H$_5$OH, CH$_3$OCH$_3$, and CH$_3$COCH$_3$ range from approximately 940 to 57\,306 AU, while the deconvolved sizes for CH$_3$CHO range from 4\,033 to 91\,884 AU.

We compare the integrated intensity maps of the stacked cube of CH$_3$CHO with those of SiO (2-1) and H$^{13}$CO$^{\text{+}}$ (1-0), both of which typically exhibit extended emission in the ATOMS sample. In 43 sources, CH$_3$CHO shows emission that is as extended as that of SiO and H$^{13}$CO$^{\text{+}}$, as demonstrated in Figure \ref{figure_CH3CHOexample}. To quantify this, we calculate the correlation between the integrated intensity maps of CH$_3$CHO, SiO, and H$^{13}$CO$^{\text{+}}$ for these 43 sources. The distribution of the correlation coefficients is shown in Figure \ref{figure_k_kde}. We find that CH$_3$CHO exhibits a strong correlation with H$^{13}$CO$^{\text{+}}$, with correlation coefficients exceeding 0.8. This suggests that CH$_3$CHO is likely widely distributed throughout some high-mass star-forming regions, similar to H$^{13}$CO$^{\text{+}}$, a common tracer of dense gas. In particular, for several sources (I13079-6218, I13134-6242, I13140-6226, I16071-5142, I16272-4837, I16318-4724, I16348-4654, and I18507+0121), CH$_3$CHO shows a strong correlation with SiO, with correlation coefficients greater than 0.8. This finding supports the possibility that CH$_3$CHO could be formed in shock-processed regions, a hypothesis that warrants further investigation in future studies \citep{Chengalur2003}.

\section{Conclusions} \label{con}

We conducted a systematic survey of hot molecular cores in 146 high-mass star forming regions using the ATOMS Band 3 data through molecular spectral stacking technique. The complex molecules used in this study are CH$_{3}$OH, CH$_{3}$OCHO, C$_{2}$H$_{5}$CN, C$_{2}$H$_{5}$OH, CH$_{3}$OCH$_{3}$, CH$_{3}$CHO, and CH$_{3}$COCH$_{3}$. The primary findings of this work are summarized as follows:

(1) We identified 100 hot core candidates using spectral stacking techniques, which shows strong and compact COMs emissions. Among them, 60 hot cores were previously identified by \citet[][]{2022MNRAS.511.3463Q}.  The other 40 are newly identified in this work.

(2) We estimated the column densities for seven molecules at the peak positions of CH$_{3}$OH emission. Among the seven molecules, CH$_{3}$OH has the highest column density.  The column densities of CH$_{3}$OCHO and CH$_{3}$OCH$_{3}$ are approximately half an order of magnitude lower than that of CH$_{3}$OH.  C$_{2}$H$_{5}$OH, CH$_{3}$CHO, and CH$_{3}$COCH$_{3}$ are an order of magnitude lower, while C$_{2}$H$_{5}$CN is about two orders of magnitude lower than CH$_{3}$OH.

(3) A tight correlation between the column densities of CH$_{3}$OCHO and CH$_{3}$OCH$_{3}$ (correlation coefficient of 0.94) is found in our hot core sample.  Strong correlations are also witnessed between the pairs of CH$_{3}$OCHO/CH$_{3}$OH, CH$_{3}$OCH$_{3}$/CH$_{3}$OH, and C$_{2}$H$_{5}$OH/CH$_{3}$OCHO.  These chemical links suggest that CH$_{3}$OH serves as a precursor for several COMs.

(4)  CH$_3$CHO exhibits significantly
extended emission in 43 out of the 100 hot core candidates. The extended emission features of CH${3}$CHO in these 43 sources are similar to those of SiO and H$^{13}$CO$^{\text{+}}$. This suggests that CH$_{3}$CHO is widely distributed and may be formed in shock regions within some high-mass star-forming clumps.

Overall, this study significantly expands the sample of hot core candidates through the spectral stacking method, providing a reliable approach for identifying molecular species in high-mass star-forming regions. The method serves as a valuable tool for future investigations into molecular distributions and formation processes in these environments.

\section*{Acknowledgements}

\begin{acknowledgements}
This work has been supported by the National Key R\&D Program of China (No.\ 2022YFA1603100). 
X.L. acknowledges the support of the Strategic Priority Research Program of the Chinese Academy of Sciences  under Grant No. XDB0800303.
T.L.\ acknowledges support from the National Natural Science Foundation of China (NSFC), through grants No.\ 12073061 and No.\ 12122307, the Tianchi Talent Program of Xinjiang Uygur Autonomous Region.
S.-L. Qin is supported by National Natural Science Foundation of China (NSFC) through grant No.12033005.
This research was carried out in part at the Jet Propulsion Laboratory, California Institute of Technology, under a contract with the National Aeronautics and Space Administration (80NM0018D0004).
Y.P. Peng acknowledges support from NSFC through grant No. 12303028. 
L.B. and G.G. acknowledge support by the ANID BASAL project FB210003.
C.W.L. acknowledges support from the Basic Science Research Program through the NRF funded by the Ministry of Education, Science and Technology (NRF- 2019R1A2C1010851) and from the Korea Astronomy and Space Science Institute grant funded by the Korea government (MSIT; project No. 2024-1-841-00).
PS was partially supported by a Grant-in-Aid for Scientific Research (KAKENHI Number JP22H01271 and JP23H01221) of JSPS.
H.-L. Liu is supported  by Yunnan Fundamental Research Project (grant No. 202301AT070118,
202401AS070121), and by Xingdian Talent Support Plan--Youth Project. 
\end{acknowledgements}

\section*{Data Availability}
The derived data underlying this article are available in thearticle and in its online supplementary material on Zenodo.

\bibliographystyle{aa}
\bibliography{ATOMS_stack_HCS}

\onecolumn
\appendix

\begin{table*}
\centering
\caption{Parameters of molecular transitions}
\label{table1}
\begin{tabular}{ccccccccc}
\hline
\hline
molecules & $\theta_{\rm source}$ & $T_{k}$ & $N_{\rm s}$  & $ \Delta V $ &$N_{\rm trans}$  &frequency & weights&database \\
 & $\prime\prime$ & K & $\times$ 10$^{15}$ cm$^{-2}$  & \text{km/s} && GHz & &   \\
\hline
 &  &  &  &   & &97.678803 & 0.22&JPL \\
CH$_3$OH & 1.6 & 100 & 200  & 5 &3& 98.030648 & 0.10&JPL \\
(C1)&  &  &  &    &  & 100.638872 & 3.87&JPL \\
\hline
&&&&&&98.190658&0.52&JPL\\
&&&&&&98.270501&0.23&JPL\\
&&&&&&98.278921&0.95&JPL\\
&&&&&&98.424207&0.67&JPL\\
&&&&&&98.431803&0.94&JPL\\
&&&&&&98.435802&0.74&JPL\\
&&&&&&98.682615&0.81&JPL\\
&&&&&&98.712001&0.65&JPL\\
CH$_3$OCHO&1.2&123&32&5&17&98.747906&0.73&JPL\\
(C1)&&&&&&98.792289&0.73&JPL\\
&&&&&&99.133272&0.24&JPL\\
&&&&&&99.135762&0.17&JPL\\
&&&&&&100.294604&0.95&JPL\\
&&&&&&100.308179&0.79&JPL\\
&&&&&&100.482241&1.01&JPL\\
&&&&&&100.490682&1.14&JPL\\
&&&&&&100.681545&1.31&JPL\\
\hline
&&&&&&98.523872&3.17&JPL\\
&&&&&& 98.533987&0.42&JPL\\
&&&&&& 98.544164&0.69&JPL\\
C$_2$H$_5$CN&1.4&140&1.2&4.5&8&98.559927&0.12&JPL\\
(C3)&&&&&&98.566615&1.59&JPL\\
&&&&&&98.701070&1.63&JPL\\
&&&&&&99.681461&1.76&JPL\\
&&&&&&100.614281&1.86&JPL\\
\hline
&&&&&&97.535908&0.23&JPL\\
&&&&&&98.230313&0.32&JPL\\
C$_2$H$_5$OH&1.6&100&9&4&5&98.583898&0.28&JPL\\
(C1)&&&&&&98.983548&0.29&JPL\\
&&&&&&99.524091&0.29&JPL\\
\hline
CH$_3$OCH$_3$&1.6&95&30&4&2&99.324430&1.38&JPL\\
(C1)&&&&&&99.836443&0.32&JPL\\
\hline
&&&&&&98.863314&0.67&JPL\\
CH$_3$CHO&1.6&100&3.8&5&3&100.127164&0.10&JPL\\
(C1)&&&&&&100.645229&0.10&JPL\\
\hline
&&&&&&98.800398&0.29&JPL\\
CH$_3$COCH$_3$&1.4&140&20&5&4&99.052559&0.37&JPL\\
(C2)&&&&&&99.256107&0.16&JPL\\
&&&&&&99.542604&0.13&JPL\\
\hline
\end{tabular}
\flushleft{
Notes: Column 1 lists the molecule names. Columns 2 to 5 provide the best-fit molecular parameters from XCLASS, as used in \citet{2022MNRAS.512.4419P}, including beam-deconvolved source size, rotational temperature, and column density. Column 6 shows the number of transitions ($N_{\rm trans}$) listed here for stacking. Column 7 contains the rest frequencies of the transitions, and Column 8 indicates the weights. All molecules are sourced from the JPL molecular database.}
\end{table*}

\section{Additional tables} \label{app:A}
Table \ref{tab:all_lines} shows the molecular line parameters for lines marked in Figure \ref{figure_allstackspectrum}.  Table \ref{tab:CH3OH} lists line parameters of CH$_3$OH. Table \ref{tab:threespecies1}-\ref{tab:threespecies2} lists the Physical parameters of CH$_3$OCHO C$_2$H$_5$CN C$_2$H$_5$OH CH$_3$OCH$_3$ CH$_3$CHO CH$_3$COCH$_3$.

\begin{longtable}{cccccc}
\caption{Identified transitions from stacked spectra of 100 hot cores} 
\label{tab:all_lines} \\
\hline
\multicolumn{1}{c}{Species} & \multicolumn{1}{c}{Transition} & \multicolumn{1}{c}{Frequency (MHz)}& \multicolumn{1}{c}{E$_{low}$ (K)} & \multicolumn{1}{c}{E$_{up}$ (K)} & \multicolumn{1}{c}{database}\\
\hline 
\endfirsthead

\multicolumn{6}{c}%
{{\tablename\ \thetable{} -- continued from previous page}} \\
\hline
\multicolumn{1}{c}{Species} & \multicolumn{1}{c}{Transition} & \multicolumn{1}{c}{Frequency (MHz)}& \multicolumn{1}{c}{E$_{low}$ (K)} & \multicolumn{1}{c}{E$_{up}$ (K)} & \multicolumn{1}{c}{database}\\
\hline 
\endhead

\hline \multicolumn{6}{r}{{Continued on next page}} \\
\endfoot

\hline
\endlastfoot
g-CH$_3$CH$_2$OH & 23(1,23)-23(0,23),v$_t$=1-0 & 97536.849 & 277.00973 & 281.69073& JPL \\
g-CH$_3$CH$_2$OH & 29(1,28)-29(2,28),v$_t$=1-0 & 97546.875 &415.6889&420.37038& JPL \\
g-CH$_3$CH$_2$OH & 26(0,26)-26(1,26),v$_t$=1-0 & 97549.692 & 336.03248 & 340.71410& JPL   \\
g-CH$_3$CH$_2$OH & 24(1,24)-24(0,24),v$_t$=1-0& 97562.811 &295.9081  &300.59035&  JPL  \\
g-CH$_3$CH$_2$OH& 20(1,20)-20(0,20),v$_t$=1-0& 97574.005 & 224.96472 & 229.64750& JPL  \\
CH$_3$OH,v$_t$=0-2& 2(1)--1(1)-v$_t$=0& 97582.798 & 16.88107 &   21.56428& JPL  \\
g-CH$_3$CH$_2$OH& 25(1,25)-25(0,25),v$_t$=1-0& 97600.390 & 315.58168 &320.26573&  JPL \\
g-CH$_3$CH$_2$OH &27(0,27)-27(1,27),v$_t$=1-0 & 97631.545 &357.25532&361.94086 & JPL\\
g-CH$_3$CH$_2$OH& 19(1,19)-19(0,19),v$_t$=1-0& 97649.502 & 209.16632 &213.85273  & JPL\\
CH$_3$OCHO,v=0& 10(4,7)-10(3,8)E&97651.270&38.48649&43.17298 & JPL\\
CH$_3$$^{13}$CH$_2$CN& 11(2,10)-10(2,9)&97672.018&27.81476&32.50231& CDMS \\
CH$_3$OH,v$_t$=0-2& 21(6)$^{+}$-22(5)$^{+}$,v$_t$=0& 97677.684 & 724.58548 &729.27330& JPL  \\
CH$_3$OH,v$_t$=0-2& 21(6)$^{+}$-22(5)$^{+}$,v$_t$=0& 97678.803 & 724.58548 &729.27330 & JPL \\
CH$_3$CH$_2$$^{13}$CN& 11(2,10)-10(2,9)&97691.544&27.92943&32.61792 & CDMS \\
CH$_3$OCHO& 10(4,7)-10(3,8)A&97694.260&38.47152&43.16008 & JPL\\
g-CH$_3$CH$_2$OH&27(1,27)-27(0,27),v$_t$=1-0& 97698.530 &357.25374&361.9425& JPL  \\
SO$_2$,v=0&7(3,5)-8(2,6)& 97702.334 &43.14608&47.83503 & JPL \\
g-CH$_3$CH$_2$OH&28(0,28)-28(1,28),v$_t$=1-0& 97708.888 &379.25309&383.94235 & JPL  \\
$^{34}$SO& 3(2)-2(1)& 97715.390 & 4.40350 &9.09307 & JPL \\
CH$_3$CH$_2$C$^{15}$N& 11(1,10)-10(1,9)& 97724.982& 24.62166&29.31175& CDMS  \\
g-CH$_3$CH$_2$OH & 28(1,28)-28(0,28),v$_t$= 1-0& 97755.610 &379.25194&383.94344 & JPL\\
g-CH$_3$CH$_2$OH&18(1,18)-18(0,18),v$_t$=1-0& 97774.307&194.14241&198.83481&  JPL\\
g-CH$_3$CH$_2$OH& 29(0,29)-29(1,29),v$_t$=1-0& 97784.113 & 402.02563 &406.71850 & JPL \\
 g-CH$_3$CH$_2$OH &29(1,29)-29(0,29),v$_t$=1-0&97815.987&402.02491&406.71931& JPL \\
 g-CH$_3$CH$_2$OH &30(1,29)-30(2,29),v$_t$=1-0 & 97828.953 & 439.74655 & 444.44157& JPL \\
 CH$_3$CH$_2$CN,v=0 &19(3,16)-19(2,17) & 97844.699 & 87.31028 & 92.00606 & JPL \\
 g-CH$_3$CH$_2$OH &30(0,30)-30(1,30),v$_t$=1-0 & 97857.476 & 425.57281 & 430.2692&  JPL \\
CH$_3$CH$_2$CN,v=0&34(4,31)-33(5,28)& 97875.099 & 269.49378&274.19102 & JPL \\
g-CH$_3$CH$_2$OH&51(4,48)-51(3,48),v$_t$=1-0 & 97877.456 & 425.57252 & 430.26987&JPL   \\
g'Ga-(CH$_2$OH)$_2$& 11(0,11),v=0-10(0,10),v=1& 97896.734 & 26.14890 &30.84718& CDMS \\
g-CH$_3$CH$_2$OH  &31(0,31)-31(1,31),v$_t$=1-0 & 97932.445 & 449.89433 &454.59432 & JPL \\
g-CH$_3$CH$_2$OH& 17(1,17)-17(0,17),v$_t$=1-0& 97962.834 & 179.89299 &184.59443 & JPL \\
CS v=0& 2-1& 97980.953 & 2.35124 &7.05355 & CDMS\\
CH$_3$OCH$_3$ & 16(3,14)-15(4,11)AA& 97990.568 & 131.90171& 136.60449& JPL  \\
CH$_3$OCH$_3$ & 16(3,14)-15(4,11)EE& 97993.397 & 131.90156& 136.60448&  JPL \\
CH$_3$OCH$_3$ & 16(3,14)-15(4,11)EA& 97996.186 & 131.90156& 136.60461 &  JPL\\
CH$_3$OH,v$_t$=0-2& 21(6)$^{+}$-22(5)$^{+}$,v$_t$=0& 98030.648 & 884.31758 &889.02228& JPL  \\
CH$_3$OH,v$_t$=0-2& 21(6)$^{+}$-22(5)$^{+}$,v$_t$=0& 98030.686 & 884.31758 &889.02228 & JPL \\
CH$_3$CH$_2$$^{13}$CN& 11(6,6)-10(6,5)& 98032.851 & 63.53867 &68.24354& CDMS \\
CH$_3$$^{13}$CH$_2$CN& 11(7,5)-10(7,4)&98039.642&76.59192&81.29711& CDMS \\
CH$_3$$^{13}$CH$_2$CN& 11(6,6)-10(6,5)& 98040.582 & 62.52260 &67.22784 & CDMS\\
CH$_3$CH$_2$$^{13}$CN& 11(8,4)-10(8,3)&98041.506&94.62149&99.32677 & CDMS\\
CH$_3$$^{13}$CH$_2$CN& 11(8,4)-10(8,3)& 98045.570& 92.81596	 &97.52144 & CDMS\\
(CH$_3$)$_2$CO,v=0&17(6,11)-17(5,12)EE&98052.399&105.88158&110.58732&JPL \\
CH$_3$$^{13}$CH$_2$CN& 11(5,7)-10(5,6)& 98052.963&50.61247&55.3183& CDMS \\
(CH$_3$)$_2$CO,v=0&17(7,11)-17(6,12)EE&98053.535&105.88158&110.58738& JPL\\
CH$_3$CH$_2$$^{13}$CN& 11(9,3)-10(9,2)& 98053.746& 113.47391 &118.17978 & CDMS\\
g-CH$_3$CH$_2$OH&33(0,33)-33(1,33),v$_t$=1-0&98056.249&500.86098&505.56691& JPL  \\
g-CH$_3$CH$_2$OH&31(1,30)-31(2,30),v$_t$=1-0&98060.630&464.57912&469.28526&  JPL \\
(CH$_3$)$_2$CO,v=0 & 24(19,5)-24(18,6)AA& 98064.051 & 255.32964 & 260.03595	& JPL \\
CH$_3$CH$_2$$^{13}$CN& 11(4,8)-10(4,7)&98072.716&41.31648&46.02325 & CDMS \\
CH$_3$CH$_2$$^{13}$CN& 11(4,7)-10(4,6)&98074.617&41.31648	&46.02335& CDMS \\
CH$_3$$^{13}$CH$_2$CN& 11(4,8)-10(4,7)& 98087.343&40.86599&45.57347&  CDMS \\
CH$_3$$^{13}$CH$_2$CN& 11(4,7)-10(4,6)& 98089.683&40.86614&45.57373	&  CDMS \\
g-CH$_3$CH$_2$OH&33(1,33)-33(0,33),v$_t$=1-0&98091.912&500.85954&505.56718& JPL \\
CH$_3$CH$_2$$^{13}$CN& 11(3,9)-10(3,8)&98117.415&33.53883&38.24775& CDMS \\
CH$_3$$^{13}$CH$_2$CN& 11(3,9)-10(3,8)&98134.857&33.28704 & 37.99680& CDMS  \\
g-CH$_3$CH$_2$OH&34(0,34)-34(1,34),v$_t$=1-0&98163.428&527.50324&532.21431& JPL \\
$^{13}$CH$_3$CH$_2$CN&11(1,10)-10(1,9)&98165.345&24.71966&29.43083& CDMS \\
CH$_3$CH$_2$CN,v=0& 11(2,10)-10(2,9)&98177.574&28.04770&32.75945& JPL \\
CH$_3$OCHO& 8(7,1)-7(7,0)E&98182.336&49.06992&53.78191& JPL \\
CH$_3$OCHO&8(7,2)-7(7,1) A&98190.658&49.05165&53.76403& JPL \\
CH$_3$OCHO,v=0&8(7,2)-7(7,1)E&98191.460&49.05007&53.76249& JPL \\
CH$_3$CH$_2$$^{13}$CN& 11(3,8)-10(3,7)&98206.212&33.54458&38.25777& CDMS \\
g-CH$_3$CH$_2$OH&16(1,16)-16(0,16),v$_t$=1-0&98230.313&166.41734&171.13162& JPL \\
CH$_3$$^{13}$CH$_2$CN& 11(3,8)-10(3,7)& 98237.791&33.29366&38.00836& CDMS \\
$^{33}$SO$_2$& 2( 2,0)-3(1,3),F=7/2-7/2&98257.864&7.80442&12.58934& JPL \\
$^{33}$SO$_2$& 2(2,0)-3(1,3),F=5/2-7/2& 98258.125&7.80442&12.58936& JPL \\
$^{33}$SO$_2$& 2(2,0)-3(1,3),F=3/2-5/2& 98260.702&7.80428&12.58934& JPL \\
$^{33}$SO$_2$& 2(2,0)-3(1,3),F=5/2-5/2& 98260.888&7.80428&12.58934& JPL \\
$^{33}$SO$_2$&2(2,0)-3(1,3),F=7/2-9/2& 98263.784&7.80413&12.58934& JPL \\
SO$_2$ v$_2$=1&16(2,14)-15(3,13)& 98264.696&894.7382&899.54462& JPL \\
$^{33}$SO$_2$& 2(2,0)-3(1,3),F=1/2-3/2& 98266.360&7.80399&12.58932& JPL \\
CH$_3$OCHO,v=0&8(6,2)-7(6,1)E&98270.501&40.43544&45.15166 & JPL \\
g-CH$_3$CH$_2$OH&32(1,31)-32(2,31),v$_t$=1-0&98274.012&490.18517&494.90156& JPL \\
CH$_3$OCHO&8(6,3)-7(6,2)E&98278.921&40.41775&45.13436& JPL \\
CH$_3$OCHO,v=0&8(6,3)-7(6,2)A&98279.762&40.41587&45.13253 & JPL\\
(CH$_3$)$_2$CO,v=0&5(4,1)-4(3,2)EE & 98310.644 & 8.24271 &12.96085 & JPL \\
CH$_3$OCHO,v=0&8(5,3)-7(5,2)E&98424.207&33.13398&37.85757 & JPL\\
CH$_3$OCHO,v=0&8(5,4)-7(5,3)E&98431.803&33.11858&37.84254 & JPL\\
CH$_3$OCHO,v=0&8(5,4)-7(5,3)A&98432.760&33.11340&33.11340 & JPL\\
CH$_3$OCHO,v=0&8(5,3)-7(5,2)A&98435.802&33.1134&37.83755 & JPL\\
g-CH$_3$CH$_2$OH&33(1,32)-33(2,32),v$_t$=1-0&98440.501&516.56629&521.29066 & JPL\\
(CH$_3$)$_2$CO,v=0& 16(6,11)-16(5,12)EA& 98462.960 &90.89896&95.62441& JPL\\
CH$_3$CH$_2$$^{13}$CN&12(2,11)-12(0,12)&98511.757&33.00202&37.72987&  CDMS\\
HC$_5$N,v=0 & J=37-36&98512.524&85.10464& 89.83253&  CDMS \\
CH$_3$CH$_2$CN,v=0& 11(6,6)-10(6,5)&98523.872&63.66683& 68.3952& JPL  \\
CH$_3$CH$_2$CN,v=0&11(7,4)-10(7,3)&98524.672&78.10684& 82.83525 &  JPL \\
g-CH$_3$CH$_2$OH&36(15,21)-37(14,24),v$_t$=0-0 & 98526.369 & 880.52397 & 885.25247& JPL  \\
CH$_3$CH$_2$CN,v=0&11(8,3)-10(8,2)&98532.084&94.75812& 99.48689& JPL\\
CH$_3$CH$_2$CN,v=0&11(5,7)-10(5,6)&98533.987&51.44253& 56.17139& JPL\\
CH$_3$CH$_2$CN,v=0&11(9,2)-10(9,1) & 98544.164 & 113.61591 & 118.34525& JPL  \\
CH$_3$CH$_2$CN,v=0&11(10,1)-10(10,0)& 98559.927 & 134.67474 & 139.40484& JPL  \\
CH$_3$CH$_2$CN,v=0&11(4,7)-10(4,6)&98566.615&41.43885&46.16927& JPL \\
g-CH$_3$CH$_2$OH&15(1,15)-15(0,15),v$_t$=1-0 & 98585.095 &153.71488&158.44619& JPL \\
(CH$_3$)$_2$CO,v=0&16(5,11)-16(4,12)EE&98600.720&90.83522&95.56728& JPL\\
(CH$_3$)$_2$CO,v=0&16(6,11)-16(5,12)EE&98600.976&90.83522&95.56729& JPL\\
CH$_3$OCHO,v=0&8(3,6)-7(3,5)E&98606.856&22.52723&27.25959 & JPL\\
CH$_3$CH$_2$CN,v$_20$=1-A&11(8,4)-10(8,3)&98609.424&630.75816&635.49070 & CDMS\\
CH$_3$OCHO,v=0&8(3,6)-7(3,5)A&98611.163&22.51141&27.24397 & JPL\\
CH$_3$CH$_3$$^{13}$CN& 28(2,26)-28(2,27)&98617.186&176.41923&181.15214& CDMS \\
CH$_3$CH$_2$CN,v$_{20}$=1-A& 11(4,7)-10(4,6)& 98644.223 & 577.94276& 582.67697& CDMS  \\
(CH$_3$)$_2$CO,v=0&5(5,1)-4(4,1)EE&98651.514&9.29114&14.02564&  JPL \\
g-CH$_3$CH$_2$OH& 41(0,41)-41(1,41),v$_t$=1-0& 98662.313 & 735.65738 & 740.3924&  JPL \\
CH$_3$OCHO,v=0& 8(4,5)-7(4,4)A& 98682.615 & 27.14928 & 31.88527 &  JPL\\
CH$_3$CH$_2$CN,v=0& 11(3,8)-10(3,7)& 98701.070 &33.66512&38.40200&  JPL\\
CH$_3$OCHO,v=0&8(4,5)-7(4,4)E & 98712.001 & 27.15877 & 31.89617&  JPL \\
CH$_3$$^{13}$CH$_2$CN& 58(3,55)-57(5,52)& 98738.158 & 742.51819	 & 747.2569&  CDMS \\
(CH$_3$)$_2$CO,v=0&16(5,11)-16(4,12)AA&98738.572&90.77163&95.5103&   JPL \\
(CH$_3$)$_2$CO,v=0&16(6,11)-16(5,12)AA&98738.836&90.77163&95.51032&   JPL  \\
CH$_3$OCHO,v=0&8(4,4)-7(4,3)E & 98747.906 &27.17172&31.91085 &    JPL  \\
g-CH$_3$CH$_2$OH&30(2,29)-30(1,29),v$_t$=1-0& 98755.450 &439.7254&444.46489 &   JPL  \\
CH$_3$OCHO,v=0&8(4,4)-7(4,3)A&98792.289&27.15187&31.89312 &    JPL  \\
(CH$_3$)$_2$CO,v=0&5(5,0)-4(4,0)EE&98800.890&9.35286&14.09453 &   JPL  \\
g-CH$_3$CH$_2$OH& 36(1,35)-36(2,35),v$_t$=1-0& 98800.965 & 600.35198 & 605.09365&  JPL   \\
g-CH$_3$CH$_2$OH& 29(2,28)-29(1,28),v$_t$=1-0& 98823.696 & 415.6594 &420.40217&  JPL   \\
g-CH$_3$CH$_2$OH&33(2,32)-33(1,32),v$_t$=1-0&98827.119&516.55651&521.29943&	JPL  \\
g-CH$_3$CH$_2$OH&43(0,43)-43(1,43),v$_t$=1-0&98831.275&802.08216&806.82529&	 JPL \\
CH$_3$OCHO,v=0&11(4,8)-11(3,9)E&98839.522&44.96972&49.71324&  JPL  \\
CH$_3$O$^{13}$CHO&13(3,11)-13(2,12),v$_t$=1-1& 98841.974 &  243.24587& 247.98951&  JPL   \\
g-CH$_3$CH$_2$OH&13(4,10)-12(5,8),v$_t$=1-0&98844.461&152.19122&156.93498& JPL 	  \\
CH$_3$CHO,v=0,1,2& 5(1,4)-4(1,3)E,v$_t$=0& 98863.314 & 11.84423 & 16.58889 & JPL \\
g-CH$_3$CH$_2$OH&32(2,31)-32(1,31),v$_t$=1-0&98869.224&490.16877&494.91372	&  JPL  \\
CH$_3$OCHO,v=0&11(4,8)-11(3,9)A&98875.228&44.95548&49.70071&  JPL  \\
g-CH$_3$CH$_2$OH&13(1,13)-13(0,13),v$_t$=1-0 & 98878.281 & 130.62739 & 135.37277 & JPL \\
g-CH$_3$CH$_2$OH&17(2,15)-16(3,13),v$_t$=0-1&98881.085&186.50787&191.25339&	 JPL \\
CH$_3$CHO,v=0,1,2&5(1,4)-4(1,3)A,v$_t$=0 & 98900.945 & 11.76683 & 16.51330 & JPL \\
g-CH$_3$CH$_2$OH&35(2,34)-35(1,34),v$_t$=1-0&98931.034&571.64322&576.39113&	 JPL \\
g-CH$_3$CH$_2$OH&48(0,48)-48(1,48),v$_t$=1-0&98947.000&981.65344&986.40212&	 JPL \\
g-CH$_3$CH$_2$OH&45(0,45)-45(1,45),v$_t$=1-0&98957.368&871.59426&876.34343&	 JPL \\
g-CH$_3$CH$_2$OH& 14(1,14)-14(0,14),v$_t$=1-0& 98983.548 & 141.78518 & 146.53562& JPL  \\
H$\alpha$& H(40)$\alpha$& 99022.953 & 0 & 0 &  \\
g'Ga-(CH$_2$OH)$_2$&9(8,1),v=1-8(8,0),v=0 & 99040.148 & 49.14071 & 53.89386 & CDMS \\
(CH$_3$)$_2$CO,v=0&15(4,11)-15(3,12)EE&99052.510&76.65875&81.41249& JPL  \\
(CH$_3$)$_2$CO,v=0&15(5,11)-15(4,12)EE&99052.559&76.65875&81.41249&  JPL  \\
CH$_3$CH$_2$CN,v=0&32(3,29)-32(2,30)&99070.600&235.17675&239.93136 &  JPL  \\
CH$_3$OCHO,v=0& 19(13,6)-20(12,9)A& 99071.877 & 219.42373 & 224.17840 &  JPL  \\
g-CH$_3$CH$_2$OH&13(4,9)-12(5,7),v$_t$=1-0&99109.251&130.78436&135.54083 &  JPL \\
NH$_2$D&5(2,4)0a-5(1,4)0s&99118.819&256.70395&261.46088 &  CDMS \\
CH$_3$CH$_2$CN,v=0&40(2,38)-39(4,35)& 99120.712 & 356.13503 & 360.89205&  JPL   \\
g-CH$_3$CH$_2$OH&38(1,37)-38(2,37),v$_t$=1-0&99126.548&660.06807&664.82537	& JPL   \\
CH$_3$OCHO,v=0&9(0,9)-8(1,8)E&99133.272&20.15427&24.91189&  JPL  \\
CH$_3$OCHO,v=0& 9(0,9)-8(1,8)A& 99135.762 & 20.13557 & 24.89331 &  JPL   \\
g-CH$_3$CH$_2$OH& 27(2,26)-27(1,26),v$_t$=1-0& 99143.725 & 369.84484 & 374.60296	&  JPL \\
(CH$_3$)$_2$CO,v=0&20(18,2)-20(17,3)AA & 99170.837 & 186.89173 &191.65115&  JPL \\
CH$_3$CH$_2$$^{13}$CN& 11(2,9)-10(2,8)&99172.521&28.09691&32.85647&   CDMS \\
CH$_3$OCH$_3$ & 25(6,19)-24(7,18)AA& 99183.408 & 342.11746& 346.87749 &   JPL \\
H$\beta$& H(50)$\beta$& 99225.208 & 0 & 0&     \\
g-CH$_3$CH$_2$OH& 39(1,38)-39(2,38),v$_t$=1-0& 99227.093 & 691.08735 & 695.84947 &  JPL   \\
CH$_3$CH$_2$CN,v=0& 15(2,14)-15(1,15)& 99253.446 & 51.09075 &55.85414&   JPL  \\
CH$_3$$^{13}$CH$_2$CN&11(2,9)-10(2,8)& 99279.632&27.99749&32.76219&   CDMS   \\
SO& 3(2)-2(1)& 99299.870 & 4.46004 & 9.22565 &  JPL  \\
CH$_3$OCH$_3$& 4(1,4)-3(0,3)EA& 99324.430 & 5.44718 & 10.21397 & JPL \\
CH$_3$OCH$_3$& 4(1,4)-3(0,3)EE& 99325.250 & 5.44675 & 10.21358 & JPL \\
CH$_3$OCH$_3$& 4(1,4)-3(0,3)AA& 99326.000 & 5.44631 & 10.21318& JPL  \\
g-CH$_3$CH$_2$OH&40(2,39)-40(1,39),v$_t$=1-0&99359.031&722.87824&727.64669&   JPL \\
CH$_3$OH,v$_t$=0-2&15(-6)-14(-7)E2,v$_t$=1 & 99374.341 & 766.24359 & 771.01278 & JPL  \\
CH$_3$OCHO,v=0& 28(4,24)-27(6,21)E&99488.215&252.30506&257.07972 &  JPL \\
g'Ga-(CH$_2$OH)$_2$&10(2,8),v=0-9(2,7),v=1 & 99509.149 &24.98148&29.75714& CDMS	 \\
t-CH$_3$CH$_2$OH&17(3,14)-17(2,15)& 99524.091 & 136.55382 & 141.33020&  JPL   \\
t-CH$_3$CH$_2$OH&19(3,16)-19(2,18),v$_t$=0-1&99537.190&222.51075&227.28776& JPL \\
(CH$_3$)$_2$CO,v=0&8(5,3)-7(6,1)EE & 99539.469 & 24.74207 & 29.51919& JPL  \\
CH$_3$OCHO&15(2,13)-15(2,14),v$_t$=1-1 & 99576.874 & 259.15463 & 263.93326& JPL  \\
CH$_3$OCH$_3$&23(10,14)-24(9,15)AE& 99602.815 &386.88014&391.6603&  JPL \\
HC$^{13}$CCN,v=0&J=11-10,F=12-11 & 99651.856 & 23.91291 & 28.69542&  CDMS  \\
HCC$^{13}$CN,v=0& J=11-10,F=12-11& 99661.474 & 23.91521 & 28.69818&   CDMS \\
CH$_3$CH$_2$CN,v=0& 11(2,9)-10(2,8)& 99681.461 & 28.21791 & 33.00183&  JPL \\
CH$_3$OH,v$_t$=0-2&6(1)-5(0)E1v$_t$=1 & 99730.940 & 335.30648 & 340.09278& JPL   \\
g-CH$_3$CH$_2$OH&43(3,41)-43(2,41),v$_t$=1-0&99756.292&842.17762&846.96514& JPL  \\
g-CH$_3$CH$_2$OH&39(3,37)-39(2,37),v$_t$=1-0&99758.844&708.30639&713.09403&  JPL  \\
CH$_3$OH,v$_t$=0-2& 20(3)-21(4)E1,v$_t$=1& 99772.834 & 897.53554 & 902.32386&  JPL   \\
CH$_3$OCH$_3$& 14(2,13)-13(3,10)AA& 99833.611 & 95.81537 & 100.6066 & JPL \\
CH$_3$OCH$_3$& 14(2,13)-13(3,10)EE& 99836.443 & 95.81537 & 100.60674&  JPL \\
CH$_3$OCH$_3$& 14(2,13)-13(3,10)EA& 99839.269& 95.81552 & 100.60702&  JPL  \\
g-CH$_3$CH$_2$OH&25(2,24)-25(1,24),v$_t$=1-0&99864.418&327.11643&331.90914 & JPL \\
HC$^{13}$CCN,v$_7$=1& J=11-10,l=1e& 99887.929 & 339.85342 & 344.64725&  CDMS  \\
t-CH$_3$CH$_2$OH& 18(3,15)-18(2,16)& 99975.883 & 152.05022 & 156.84828&  JPL  \\
SO &4(5)-4(4) & 100029.640 & 33.77481 & 38.57544 &   JPL \\
HC$^{13}$CCN,v$_7$=1 &J=11-10,l=1f & 100032.511 & 339.88823	 & 344.68901& CDMS   \\
CH$_3$CH$_2$CN,v=0& 18(3,15)-18(2,16)& 100034.425 & 78.96421& 83.76508&  JPL  \\
CH$_3$CHO,v=0,1,2&30(4,27)-31(1,30)A,v$_t$=1 & 100044.626 & 667.08480 & 671.88616&  JPL  \\
CH$_3$OCHO,v=0&J=11-10,F=10-9&100076.382&24.01478&28.81766 &  CDMS  \\
HC$_3$N,v=0&J=11-10,F=10-10 & 100078.078 & 24.01478 & 28.81774 & CDMS  \\
CH$_3$OCHO,v=0&9(1,9)-8(1,8)A&100080.542&20.13557&24.93865 &  JPL  \\
H$_2$CCO& 5(1,5)-4(1,4)& 100094.514 & 22.65974 & 27.46349 &   CDMS  \\
CH$_3$CH$_2$$^{13}$CN&11(1,10)-10(1,9) & 100109.732 & 25.19876 & 30.00330&  CDMS   \\
CH$_3$SH,v$_t$$\leq$2&4(1,3)-3(1,2)A,v$_t$=0& 100110.219 &12.28191&17.08648 &  CDMS  \\
CH$_3$$^{13}$CH$_2$CN& 11(1,10)-10(1,9)& 100155.824 & 25.18782 & 29.99458 &  CDMS  \\
g-CH$_3$CH$_2$OH& 6(1,6)-5(1,5),v$_t$=0-0& 100194.326 & 70.14976 & 74.95830 &   JPL   \\
CH$_3$CH$_2$$^{13}$CN& 54(3,51)-55(2,54)& 100233.513 & 647.25327 & 652.06375 &   CDMS  \\
HC$_3$N,v$_6$=1& J=11-10,l=1e& 100240.584 & 741.72884 & 746.53960 &  CDMS  \\
CH$_3$OCHO,v=0& 8(3,5)-7(3,4)E& 100294.604 & 22.60061 & 27.41396&  JPL  \\
CH$_3$OCHO,v=0& 8(3,5)-7(3,4)A& 100308.179 & 22.58421 & 27.39821 &  JPL  \\
HC$_3$N,v$_7$=1& J=11-10,l=1e& 100322.411 & 344.91918 & 349.73387 &  CDMS  \\
(CH$_3$)$_2$CO,v=0&8(2,6)-7(3,5)EE&100350.304&19.74566&24.56169& JPL  \\
t-CH$_3$CH$_2$OH&16(3,13)-16(2,14) & 100358.958 & 121.88975 & 126.70619	& JPL  \\
g-CH$_3$CH$_2$OH& 6(1,6)-5(1,5),v$_t$=1-1& 100365.052 & 74.81957 & 79.63630	& JPL  \\
g-CH$_3$CH$_2$OH&23(2,22)-22(3,20),v$_t$=0-1&100372.258&282.82811&287.64519&  JPL \\
$^{13}$CH$_3$CH$_2$CN&32(9,24)-33(8,25)&100390.820&305.85345	&310.67142&  CDMS	 \\
(CH$_3$)$_2$CO,v=0&8(3,6)-7(2,5)EA& 100421.172 &19.83875&24.65818& JPL \\
CH$_3$OCH$_3$&22(5,18)-21(6,15)AA& 100434.200 &261.05091&265.87096&  JPL  \\
CH$_3$OCH$_3$&22(5,18)-21(6,15)EE& 100435.500 &261.05076&265.87088&   JPL\\
g-CH$_3$CH$_2$OH&24(2,23)-24(1,23),v$_t$=1-0&100452.072&306.90763&311.72854&  JPL  \\
CH$_3$OCH$_3$&6(2,5)-6(1,6)EA& 100460.520 &19.88565&24.70697& JPL \\
CH$_3$OCH$_3$&6(2,5)-6(1,6)EE& 100463.040 &19.88522&24.70666 & JPL\\
CH$_3$OCH$_3$&6(2,5)-6(1,6)AA& 100465.700 &19.88479&24.70636&  JPL\\
HC$_3$N,v$_7$=1&J=11-10,l=1f & 100466.175 & 344.95371 & 349.77530 & CDMS \\
CH$_3$OCHO,v=0& 8(1,7)-7(1,6)E& 100482.241 & 17.95698 & 22.77934 & JPL \\
CH$_3$OCHO,v=0& 8(1,7)-7(1,6)A  &100490.682& 17.93943 & 22.76220 & JPL \\
(CH$_3$)$_2$CO,v=0&8(3,6)-7(2,5)AA&100507.065&19.64336&24.46692& JPL\\
CH$_3$CH$_2$CN,v=0&11(1,10)-10(1,9) &100614.281  & 25.32107 & 30.14977&	JPL   \\
CH$_3$OH,v$_t$=0-2& 13(2)-12(3)E1,v$_t$=0& 100638.872 & 228.77717 & 233.60704 & JPL \\
CH$_3$OCHO,v=0& 9(0,9)-8(0,8)E& 100681.545 & 20.08003 & 24.91196 & JPL \\
CH$_3$OCHO,v=0& 9(0,9)-8(0,8)A& 100683.368 & 20.06133 & 24.89334 & JPL \\
CH$_3$OCHO,v=0& 5(3,3)-5(1,4)E& 100694.666 & 10.00750 & 14.84005 & JPL \\
HC$_3$N,v$_7$=2&J=11-10,l=0 & 100708.784 & 665.85595 & 670.68918 &  CDMS\\
HC$_3$N,v$_7$=2&J=11-10,l=2e& 100711.064 & 669.12915 & 673.96249 & CDMS \\
HC$_3$N,v$_7$=2& J=11-10,l=2f& 100714.395 & 669.12958 & 673.96308 & CDMS \\
CH$_3$OCHO,v=0& 12(1,11)-12(0,12)E& 100734.805 & 42.43432 & 47.26880  & JPL\\
g'Ga-(CH$_2$OH)$_2$&15(3,12),v=0-15(2,14),v=1& 100764.618 &59.93594&64.77185&CDMS	 \\
SO$_2$,v=0&2(2,0)-3(1,3)& 100878.105 &7.74345&12.58481&	 JPL\\
CH$_3$OCH$_3$& 19(4,16)-18(5,13)AA& 100946.880 & 190.97366 & 195.81832 &JPL  \\
CH$_3$OCH$_3$& 19(4,16)-18(5,13)EE& 100949.040 & 190.97338 & 195.81814 & JPL \\
CH$_3$OCH$_3$& 19(4,16)-18(5,13)EA& 100951.970 & 190.97323 & 195.81813 & JPL\\
t-CH$_3$CH$_2$OH&8(2,7)-8(1,8) & 100990.102 & 30.32608 & 35.17281&  JPL \\
H$_2$CCO& 5(3,3)-4(3,2)& 101002.361 & 127.08442 & 131.93174	&  CDMS \\
H$_2$CCO&5(2,4)-4(2,3) & 101024.416 & 61.89194 & 66.74032 & CDMS \\
CH$_3$SH,v$_t$$\leq$2&4(-1,4)-3(-1,3)E,v$_t$=0&101029.743& 11.84107&16.68976& CDMS \\
H$_2$CCO& 5(2,3)-4(2,2)& 101032.235 & 61.89223 & 66.74099 & CDMS \\
H$_2$CCO& 5(0,5)-4(0,4)& 101036.630 & 9.69845 & 14.54742 & CDMS \\
CH$_3$CH$_2$CN,v=0&10(1,10)-9(0,9)& 101091.676 &19.21144&24.06305& JPL  \\
CH$_3$CHO,v=0,1,2&23(3,21)-22(4,18)E,v$_t$=1& 101127.345 &476.34494&481.19826&  JPL   \\
CH$_3$SH,v$_t$ $\leq$ 2&4(0,4)-3(0,3)A,v$_t$=0&101139.150&7.28213&12.13607 &  CDMS  \\
CH$_3$SH,v$_t$ $\leq$ 2&4(0,4)-3(0,3)E,v$_t$=0&101139.655&8.70797&13.56194&   CDMS  \\
CH$_3$SH,v$_t$ $\leq$ 2&4(3,1)-3(3,0)E,v$_t$=0&101156.878&46.19799&51.05279&   CDMS  \\
CH$_3$SH,v$_t$ $\leq$ 2&4(-2,3)-3(-2,2)A,v$_t$=0&101159.328&26.40604&31.26096 &  CDMS  \\
CH$_3$SH,v$_t$ $\leq$ 2&4(-3,2)-3(-3,1)E,v$_t$=0&101159.992&47.53779&52.39273 &  CDMS \\
CH$_3$SH,v$_t$$\leq2$& 4(3,1)-3(3,0)A,v$_t$=0& 101160.658 & 47.69145 & 52.54643 &   CDMS  \\
CH$_3$SH,v$_t$$\leq2$& 4(-3,2)-3(-3,1)A,v$_t$=0& 101160.694 &47.69145&52.54643 &  CDMS  \\
CH$_3$SH,v$_t$$\leq2$&4(-2,3)-3(-2,2)E,v$_t$=0&101167.158&24.76094&29.61623 &   CDMS \\
CH$_3$SH,v$_t$$\leq2$& 4(2,2)-3(2,1)E,v$_t$=0& 101168.302 & 25.41429 & 30.26963&  CDMS \\
HC$_5$N,v=0 & J=38-37&101174.677&89.83249& 94.68814 & CDMS\\
CH$_3$CH$_2$C$^{15}$N&12(1,12)-11(1,11)&101175.642&27.87922&32.73492& CDMS \\
CH$_3$SH,v$_t$$\leq2$&4(2,2)-3(2,1)A,v$_t$=0 & 101179.816 & 26.40648 & 31.26237& CDMS \\
CH$_3$OH,v$_t$=0-2&6(-2)-6(1)E2,v$_t$=0& 101185.453 & 69.80042 &74.65653 & JPL\\
g-CH$_3$CH$_2$OH&23(2,22)-23(1,22),v$_t$=1-0&101243.633&287.46785&292.32675 & JPL \\
CH$_3$SH,v$_t$$\leq2$&4(1,3)-3(1,2)E,v$_t$=0& 101284.366 &13.47193&18.33285 & CDMS \\
CH$_3$OH,v$_t$=0-2& 7(-2)-7(1)E2,v$_t$=0& 101293.415 & 86.05131 & 90.91260 &  JPL \\
H$_2$CO&6(1,5)-6(1,6) & 101332.991 &  82.70100& 87.56419 & CDMS \\

\hline 
\end{longtable}
\footnotesize 
\textbf{Notes:} The table presents the spectral lines resulting from the stacked observations of 100 hot core canditates. It includes information on the transitions, frequency, upper and lower energy level in K. The labeling of these molecular lines is based on G9.62+0.19.

\begin{tiny}
\begin{longtable}{ccccccccccccc}
\caption{Line parameters of CH$_3$OH} \label{tab:CH3OH} \\
\hline
\hline
\multicolumn{1}{c}{ID}&\multicolumn{1}{c}{Source} & \multicolumn{1}{c}{RA} & \multicolumn{1}{c}{DEC} & \multicolumn{1}{c}{Distance}&\multicolumn{1}{c}{$\theta_{\rm source}$} & \multicolumn{1}{c}{PA} &&\multicolumn{1}{c}{CH$_3$OH} \\
\hline
&&\multicolumn{1}{c}{h~m~s} &\multicolumn{1}{c}{${\circ}$~$\prime$~$\prime\prime$}&\multicolumn{1}{c}{kpc}&\multicolumn{1}{c}{$\prime\prime$}&\multicolumn{1}{c}{$^{\circ}$}&\multicolumn{1}{c}{I$_{\rm peak}$}&\multicolumn{1}{c}{I$_{\rm integrated}$}&\multicolumn{1}{c}{N}\\
&&&&&&&\multicolumn{1}{c}{K km s$^{-1}$}&\multicolumn{1}{c}{K km s$^{-1}$ arcsec$^2$}&\multicolumn{1}{c}{$\, \text{$\times$} \, \text{10$^{16}$} \, \text{cm}^{-2}$}\\
\hline 
\endfirsthead

\multicolumn{13}{c}%
{{\tablename\ \thetable{} -- continued from previous page}} \\
\hline\hline
\multicolumn{1}{c}{ID}&\multicolumn{1}{c}{Source} & \multicolumn{1}{c}{RA} & \multicolumn{1}{c}{DEC} & \multicolumn{1}{c}{Distance}& \multicolumn{1}{c}{$\theta_{\rm source}$} & \multicolumn{1}{c}{PA} &&\multicolumn{1}{c}{CH$_3$OH} \\
\hline
&&\multicolumn{1}{c}{h~m~s} &\multicolumn{1}{c}{${\circ}$~$\prime$~$\prime\prime$}&\multicolumn{1}{c}{kpc}&\multicolumn{1}{c}{$\prime\prime$}&\multicolumn{1}{c}{$^{\circ}$}&\multicolumn{1}{c}{I$_{\rm peak}$}&\multicolumn{1}{c}{I$_{\rm integrated}$}&\multicolumn{1}{c}{N}\\
&&&&&&&\multicolumn{1}{c}{K km s$^{-1}$}&\multicolumn{1}{c}{K km s$^{-1}$ arcsec$^2$}&\multicolumn{1}{c}{$\, \text{$\times$} \, \text{10$^{16}$} \, \text{cm}^{-2}$}\\
\hline 
\endhead

\hline \multicolumn{13
}{r}{{Continued on next page}} \\
\endfoot

\hline
\endlastfoot
1&I08303-4303 & 08:32:08.68 &  -43:13:45.78 &2.3& 1.26$\pm$0.12 & 168$\pm$42 & 34.30 $\pm$ 1.20 & 134.24 $\pm$ 7.04 & 414.00$\pm$14.00 \\
2&I08470-4243 & 08:48:47.79 &  -42:54:27.90 &2.1& 0.56$\pm$0.04 & 113$\pm$21 & 86.50 $\pm$ 1.10 & 215.68 $\pm$ 4.48 & 1040.00$\pm$10.00 \\
3&I09018-4816 & 09:03:33.46 &  -48:28:01.69 &2.6& 0.93$\pm$0.36 & 10$\pm$27 & 32.20 $\pm$ 1.60 & 113.28 $\pm$ 8.32 & 388.00$\pm$19.00 \\
4&I10365-5803$^N$ & 10:38:32.16 & -58:19:08.43 &2.4& 0.98$\pm$0.25 & 54$\pm$76 & 8.58 $\pm$ 0.25 & 55.68 $\pm$ 2.56 & 103.00$\pm$3.00 \\
5&I11298-6155 & 11:32:05.59 &  -62:12:25.62 &10& 2.42$\pm$0.02 & 131$\pm$3 & 13.26 $\pm$ 0.14 & 88.32 $\pm$ 0.93 & 160.00$\pm$2.00 \\
6&I11590-6452$^N$ & 12:01:36.52 &  -65:08:49.05 &0.4& 2.73$\pm$0.06 & 122$\pm$2 & 1.82 $\pm$ 0.06 & 15.36 $\pm$ 0.51 & 22.00$\pm$0.70 \\
7&I12320-6122$^N$ & 12:34:53.29 & -61:39:40.58 &3.43& 3.39$\pm$0.02 & 177$\pm$1 & 5.00 $\pm$ 0.04 & 65.22 $\pm$ 0.53 & 60.30$\pm$0.50 \\
8&I12326-6245 & 12:35:35.09 &  -63:02:31.91 &4.61& 4.58$\pm$0.02 & 126$\pm$2 & 23.90 $\pm$ 0.13 & 568.96 $\pm$ 3.04 & 288.00$\pm$2.00 \\
9&I13079-6218c1 & 13:11:13.75 &  -62:34:41.55 &3.8& 3.07$\pm$0.01 & 163$\pm$1 & 50.71 $\pm$ 0.21 & 544.96 $\pm$ 2.24 & 612.00$\pm$3.00 \\
10&I13079-6218c2$^N$ & 13:11:10.50 & -62:34:39.07 &3.8& 3.06$\pm$0.02 & 149$\pm$4 & 14.23 $\pm$ 0.14 & 150.77 $\pm$ 1.52 & 172.00$\pm$2.00 \\
11&I13134-6242 & 13:16:43.20 &  -62:58:32.30 &3.8& 3.32$\pm$0.01 & 168$\pm$1 & 64.60 $\pm$ 0.21 & 819.68 $\pm$ 2.72 & 779.00$\pm$3.00 \\
12&I13140-6226 & 13:17:15.49 &  -62:42:24.42 &3.8& 3.24$\pm$0.02 & 160$\pm$1 & 8.33 $\pm$ 0.07 & 99.60 $\pm$ 0.78 & 100.00$\pm$1.00 \\
13&I13471-6120 & 13:50:41.81 &  -61:35:10.67 &5.46& 2.74$\pm$0.01 & 69$\pm$1 & 31.37 $\pm$ 0.15 & 267.28 $\pm$ 1.26 & 378.00$\pm$2.00 \\
14&I13484-6100 & 13:51:58.31 &  -61:15:41.50 &5.4& 2.65$\pm$0.02 & 90$\pm$2 & 10.70 $\pm$ 0.13 & 85.06 $\pm$ 1.07 & 129.00$\pm$2.00 \\
15&I14164-6028$^N$ & 14:20:08.65 & -60:42:01.03 &3.19& 2.42$\pm$0.03 & 98$\pm$4 & 8.61 $\pm$ 0.13 & 57.31 $\pm$ 0.88 & 104.00$\pm$2.00 \\
16&I14212-6131$^N$ & 14:25:01.56 & -61:44:57.70 &3.44& 2.29$\pm$0.02 & 71$\pm$2 & 14.12 $\pm$ 0.14 & 83.76 $\pm$ 0.82 & 170.00$\pm$2.00 \\
17&I14498-5856 & 14:53:42.68 & -59:08:52.89 &3.16& 2.53$\pm$0.01 & 65$\pm$3 & 39.00 $\pm$ 0.30 & 284.46 $\pm$ 1.28 & 470.00$\pm$2.00 \\
18&I15254-5621 & 15:29:19.39 &  -56:31:22.34 &4& 2.77$\pm$0.01 & 180$\pm$1 & 96.08 $\pm$ 0.27 & 836.32 $\pm$ 2.24 & 1160.00$\pm$12.00 \\
19&I15290-5546$^N$ & 15:32:52.84 & -55:56:06.85 &6.76& 2.83$\pm$0.02 & 135$\pm$6 & 11.86 $\pm$ 0.13 & 109.18 $\pm$ 1.22 & 143.00$\pm$2.00 \\
20&I15394-5358$^N$ & 15:43:16.64 & -54:07:14.64 &1.82& 1.25$\pm$0.10 & 58$\pm$63 & 68.40 $\pm$ 1.10 & 507.84 $\pm$ 12.8 & 825.00$\pm$13.00 \\
21&I15411-5352$^N$ & 15:44:59.60 & -54:02:22.50 &1.82& 1.63$\pm$0.39 & 139$\pm$80 & 5.37 $\pm$ 0.38 & 46.88 $\pm$ 4.96 & 64.80$\pm$4.60 \\
22&I15437-5343 & 15:47:32.73 &  -53:52:38.80 &4.98& 1.29$\pm$0.06 & 115$\pm$14 & 45.39 $\pm$ 0.49 & 342.56 $\pm$ 5.92 & 548.00$\pm$6.00 \\
23&I15502-5302$^N$ & 15:54:06.53 &  -53:11:40.90 &5.8& 2.57$\pm$0.47 & 168$\pm$19 & 3.82 $\pm$ 0.48 & 52.32 $\pm$ 8.96 & 46.10$\pm$5.80 \\
24&I15520-5234 & 15:55:48.47 &  -52:43:06.75 &2.65& 1.98$\pm$0.11 & 83$\pm$59 & 78.30 $\pm$ 1.80 & 779.20 $\pm$ 25.60 & 945.00$\pm$22.00 \\
25&I15557-5215$^N$ & 15:59:40.71 & -52:23:27.89 &4.03& 2.13$\pm$0.39 & 131$\pm$37 & 5.43 $\pm$ 0.47 & 59.68 $\pm$ 7.36 & 65.50$\pm$5.70 \\
26&I15596-5301c1$^N$ & 16:03:32.11 & -53:09:29.98 &10.1& 0.97$\pm$0.34 & 164$\pm$39 & 12.34 $\pm$ 0.50 & 85.28 $\pm$ 5.60 & 149.00$\pm$6.00 \\
27&I15596-5301c2$^N$ & 16:03:32.63 & -53:09:26.41 &10.1& 0.87$\pm$0.34 & 67$\pm$75 & 8.72 $\pm$ 0.44 & 55.84 $\pm$ 4.64 & 105.00$\pm$5.00 \\
28&I16037-5223$^N$ & 16:07:38.19 & -52:31:01.48 &9.84& 1.83$\pm$0.37 & 117$\pm$45 & 7.23 $\pm$ 0.55 & 67.84 $\pm$ 7.68 & 87.20$\pm$6.60 \\
29&I16060-5146c1 & 16:09:52.64 &  -51:54:54.49 &5.3& 2.45$\pm$0.36 & 43$\pm$39 & 42.20 $\pm$ 3.50 & 520.00 $\pm$ 59.20 & 509.00$\pm$42.00 \\
30&I16060-5146c2$^N$ & 16:09:52.48 &  -51:54:55.80 &5.3& 2.29$\pm$0.28 & 177$\pm$34 & 31.50 $\pm$ 2.00 & 393.60 $\pm$ 33.60 & 380.00$\pm$24.00 \\
31&I16065-5158 & 16:10:19.99 &  -52:06:07.25 &3.98& 1.93$\pm$0.05 & 26$\pm$8 & 95.00 $\pm$ 1.00 & 922.40 $\pm$ 14.56 & 1150.00$\pm$10.00 \\
32&I16071-5142 & 16:10:59.73 & -51:50:22.85 &5.3& 1.26$\pm$0.11 & 168$\pm$20 & 79.00 $\pm$ 1.00 & 585.28 $\pm$ 14.08 & 957.00$\pm$14.00 \\
33&I16076-5134 & 16:11:26.59 &  -51:41:57.84 &5.3& 1.51$\pm$0.19 & 166$\pm$25 & 15.68 $\pm$ 0.57 & 127.68 $\pm$ 7.04 & 189.00$\pm$7.00 \\
34&I16119-5048c1$^N$& 16:15:45.69 &  -50:55:54.02 &3.1& 2.34$\pm$0.39 & 67$\pm$10 & 3.10 $\pm$ 0.30 & 38.08 $\pm$ 5.12 & 37.40$\pm$3.60 \\
35&I16119-5048c2$^N$ & 16:15:45.37 &  -50:55:53.80 &3.1& 2.26$\pm$0.37 & 115$\pm$43 & 5.22 $\pm$ 0.50 & 58.88 $\pm$ 7.84 & 63.00$\pm$6.00 \\
36&I16164-5046 & 16:20:11.08 &  -50:53:14.75 &3.57& 1.86$\pm$0.12 & 166$\pm$24 & 107.80 $\pm$ 2.60 & 1008.00 $\pm$ 35.20 & 1300.00$\pm$30.00 \\
37&I16172-5028c1 & 16:21:02.97 &  -50:35:12.60 &3.57& 2.40$\pm$0.06 & 52$\pm$4 & 75.10 $\pm$ 1.30 & 902.40 $\pm$ 20.80 & 906.00$\pm$16.00 \\
38&I16172-5028c2$^N$ & 16:20:59.67 &  -50:35:05.82 &3.57& 2.42$\pm$0.66 & 53$\pm$30 & 8.70 $\pm$ 1.30 & 109.76 $\pm$ 23.04 & 105.00$\pm$16.00 \\
39&I16272-4837c1 & 16:30:58.77 &  -48:43:53.57 &2.92& 1.13$\pm$0.11 & 122$\pm$46 & 105.90 $\pm$ 2.00 & 680.00 $\pm$ 20.80 & 1280.00$\pm$20.00 \\
40&I16272-4837c2 & 16:30:58.68 &  -48:43:51.32 &2.92& 0.98$\pm$0.10 & 106$\pm$54 & 21.20 $\pm$ 0.90 & 72.67 $\pm$ 6.42 & 256.00$\pm$11.00 \\
41&I16272-4837c3 & 16:30:57.29 &  -48:43:39.87 &2.92& 1.45$\pm$0.20 & 117$\pm$32 & 23.70 $\pm$ 1.00 & 176.00 $\pm$ 11.68 & 286.00$\pm$12.00 \\
42&I16297-4757$^N$ & 16:33:29.12 &  -48:03:43.74 &5.03& 3.27$\pm$0.09 & 70$\pm$2 & 3.03 $\pm$ 0.11 & 36.66 $\pm$ 1.36 & 36.60$\pm$1.30 \\
43&I16313-4729$^N$ & 16:34:54.42 & -47:35:37.45 &4.71& 1.03$\pm$0.43 & 6$\pm$44 & 11.02 $\pm$ 0.57 & 70.24 $\pm$ 5.76 & 133.00$\pm$7.00 \\
44&I16318-4724 & 16:35:33.96 &  -47:31:11.59 &7.68& 2.34$\pm$0.01 & 84$\pm$0 & 141.31 $\pm$ 0.26 & 877.44 $\pm$ 1.60 & 1700.00$\pm$4.00 \\
45&I16344-4658 & 16:38:09.49 &  -47:04:59.73 &12.1& 2.42$\pm$0.02 & 110$\pm$8 & 19.86 $\pm$ 0.19 & 131.38 $\pm$ 1.28 & 240.00$\pm$2.00 \\
46&I16348-4654c1 & 16:38:29.65 &  -47:00:35.67 &12.1& 2.28$\pm$0.01 & 102$\pm$2 & 147.40 $\pm$ 0.45 & 868.64 $\pm$ 2.72 & 1780.00$\pm$10.00 \\
47&I16348-4654c2$^N$ & 16:38:29.13 & -47:00:43.53 &12.1& 2.61$\pm$0.03 & 70$\pm$4 & 7.76 $\pm$ 0.19 & 60.03 $\pm$ 1.49 & 93.60$\pm$2.30 \\
48&I16351-4722 & 16:38:50.50 &  -47:28:00.68 &3.02& 2.62$\pm$0.01 & 86$\pm$1 & 200.82 $\pm$ 0.47 & 1550.72 $\pm$ 3.68 & 2420.00$\pm$10.00 \\
49&I16424-4531$^N$ & 16:46:06.00 & -45:36:43.71 &2.63& 2.45$\pm$0.02 & 99$\pm$6 & 17.28 $\pm$ 0.18 & 117.06 $\pm$ 1.20 & 208.00$\pm$2.00 \\
50&I16445-4459$^N$ & 16:48:05.14 & -45:05:08.09 &7.95& 2.18$\pm$0.02 & 74$\pm$27 & 18.18 $\pm$ 0.18 & 97.42 $\pm$ 0.96 & 219.00$\pm$2.00 \\
51&I16458-4512 & 16:49:30.04 &  -45:17:44.58 &3.56& 2.41$\pm$0.03 & 98$\pm$8 & 5.67 $\pm$ 0.14 & 37.26 $\pm$ 0.91 & 68.40$\pm$1.70 \\
52&I16484-4603 & 16:52:04.66 &  -46:08:33.85 &2.1& 2.42$\pm$0.01 & 109$\pm$1 & 76.31 $\pm$ 0.18 & 508.82 $\pm$ 1.23 & 921.00$\pm$2.00 \\
53&I16547-4247 & 16:58:17.18 &  -42:52:07.57 &2.74& 2.80$\pm$0.01 & 106$\pm$1 & 28.19 $\pm$ 0.17 & 250.72 $\pm$ 1.44 & 340.00$\pm$2.00 \\
54&I16562-3959$^N$ & 16:59:41.62 &  -40:03:43.21 &2.38& 3.47$\pm$0.06 & 75$\pm$4 & 5.13 $\pm$ 0.13 & 69.92 $\pm$ 1.76 & 61.90$\pm$1.60 \\
55&I17008-4040 & 17:04:22.91 &  -40:44:22.91 &2.38& 2.80$\pm$0.01 & 91$\pm$1 & 170.97 $\pm$ 0.46 & 1518.08 $\pm$ 4.00 & 2060.00$\pm$10.00 \\
56&I17016-4124c1 & 17:05:10.97 &  -41:29:06.95 &1.37& 2.65$\pm$0.02 & 99$\pm$1 & 10.81 $\pm$ 0.37 & 86.40 $\pm$ 2.88 & 130.00$\pm$4.00 \\
57&I17016-4124c2 & 17:05:11.20 &  -41:29:07.05 &1.37& 2.70$\pm$0.06 & 111$\pm$80 & 145.90 $\pm$ 1.30 & 1210.08 $\pm$ 10.72 & 1760.00$\pm$20.00 \\
58&I17136-3617$^N$ & 17:17:02.27 &  -36:20:50.39 &1.37& 2.65$\pm$1.02 & 74$\pm$15 & 1.95 $\pm$ 0.43 & 26.72 $\pm$ 7.84 & 23.50$\pm$5.20 \\
59&I17143-3700$^N$ & 17:17:45.47 & -37:03:12.03 &12.67& 2.00$\pm$0.20 & 46$\pm$44 & 13.98 $\pm$ 0.63 & 129.28 $\pm$ 8.32 & 169.00$\pm$8.00 \\
60&I17158-3901c1 & 17:19:20.43 &  -39:03:51.58 &3.38& 1.58$\pm$0.14 & 134$\pm$42 & 25.31 $\pm$ 0.69 & 192.80 $\pm$ 7.84 & 305.00$\pm$8.00 \\
61&I17158-3901c2 & 17:19:20.47 &  -39:03:49.20 &3.38& 1.76$\pm$0.28 & 36$\pm$11 & 8.78 $\pm$ 0.54 & 77.28 $\pm$ 6.88 & 106.00$\pm$7.00 \\
62&I17160-3707$^N$ & 17:19:27.43 &  -37:11:07.69 &10.5& 2.02$\pm$0.12 & 102$\pm$13 & 6.36 $\pm$ 0.49 & 29.28 $\pm$ 4 & 76.70$\pm$5.90 \\
63&I17175-3544 & 17:20:53.42 &  -35:46:57.72 &1.34& 3.75$\pm$0.13 & 58$\pm$8 & 662.00 $\pm$ 23.00 & 13712.00 $\pm$ 560.00 & 7990.00$\pm$280.00 \\
64&I17220-3609 & 17:25:25.22 &  -36:12:45.34 &8.01& 2.14$\pm$0.05 & 33$\pm$9 & 189.10 $\pm$ 2.10 & 1857.60 $\pm$ 28.80 & 2280.00$\pm$30.00 \\
65&I17233-3606 & 17:26:42.46 &  -36:09:17.85 &1.34& 3.32$\pm$0.07 & 85$\pm$7 & 289.60 $\pm$ 6.00 & 4936.00 $\pm$ 126.40 & 3490.00$\pm$70.00 \\
66&I17439-2845$^N$ & 17:47:09.15 & -28:46:16.41 &8& 2.76$\pm$0.04 & 102$\pm$2 & 12.19 $\pm$ 0.26 & 104.80 $\pm$ 2.24 & 147.00$\pm$3.00 \\
67&I17441-2822c1$^N$ & 17:47:20.09 &  -28:22:41.38 &8.1& 2.37$\pm$0.14 & 122$\pm$8 & 2.25 $\pm$ 0.20 & 14.40 $\pm$ 1.20 & 66.50$\pm$4.60 \\
68&I17441-2822c2 & 17:47:20.17 &  -28:23:04.74 &8.1& 2.61$\pm$0.09 & 151$\pm$3 & 42.24 $\pm$ 2.08 & 10.70 $\pm$ 1.01 & 26.30$\pm$2.50 \\
69&I17441-2822c3$^N$ & 17:47:19.92 &  -28:23:39.36 &8.1& 3.26$\pm$0.09 & 102$\pm$2 & 5.35 $\pm$ 0.22 & 64.32 $\pm$ 2.56 & 57.80$\pm$2.50 \\
70&I17589-2312$^N$ & 18:01:57.74 & -23:12:34.18 &2.97& 2.20$\pm$0.02 & 75$\pm$2 & 8.59 $\pm$ 0.13 & 47.06 $\pm$ 0.74 & 104.00$\pm$2.00 \\
71&I17599-2148$^N$ & 18:03:00.73 & -21:48:10.21 &2.99& 2.39$\pm$0.03 & 74$\pm$2 & 8.98 $\pm$ 0.16 & 58.10 $\pm$ 1.04 & 108.00$\pm$2.00 \\
72&I18032-2032c1 & 18:06:14.92 &  -20:31:43.22 &5.15& 2.22$\pm$0.01 & 81$\pm$2 & 17.31 $\pm$ 0.16 & 97.09 $\pm$ 0.91 & 209.00$\pm$2.00 \\
73&I18032-2032c2 & 18:06:14.88 &  -20:31:39.59 &5.15& 2.39$\pm$0.02 & 103$\pm$4 & 33.77 $\pm$ 0.37 & 217.92 $\pm$ 2.40 & 407.00$\pm$4.00 \\
74&I18032-2032c3 & 18:06:14.80 &  -20:31:37.26 &5.15& 2.47$\pm$0.02 & 122$\pm$3 & 19.17 $\pm$ 0.11 & 88.02 $\pm$ 0.10 & 232.00$\pm$1.00 \\
75&I18032-2032c4 & 18:06:14.66 &  -20:31:31.57 &5.15& 2.16$\pm$0.01 & 86$\pm$1 & 60.85 $\pm$ 0.26 & 321.65 $\pm$ 1.39 & 734.00$\pm$3.00 \\
76&I18056-1952 & 18:08:38.23 &  -19:51:50.31 &8.55& 2.63$\pm$0.01 & 89$\pm$0 & 654.60 $\pm$ 1.10 & 5115.00 $\pm$ 8.80 & 7900.00$\pm$10.00 \\
77&I18089-1732 & 18:11:51.45 &  -17:31:28.96 &2.5& 2.22$\pm$0.01 & 106$\pm$1 & 127.27 $\pm$ 0.33 & 713.92 $\pm$ 1.92 & 1540.00$\pm$0.00 \\
78&I18110-1854$^N$ & 18:14:00.90 & -18:53:26.21 &3.37& 2.46$\pm$0.06 & 119$\pm$3 & 5.68 $\pm$ 0.18 & 38.90 $\pm$ 1.23 & 68.50$\pm$2.20 \\
79&I18117-1753 & 18:14:39.51 &  -17:52:00.08 &2.57& 2.20$\pm$0.01 & 110$\pm$0 & 93.45 $\pm$ 0.19 & 511.74 $\pm$ 1.06 & 1130.00$\pm$0.00 \\
80&I18134-1942$^N$ & 18:16:22.12 &  -19:41:27.07 &1.25& 2.12$\pm$0.01 & 101$\pm$1 & 26.16 $\pm$ 0.19 & 133.42 $\pm$ 0.98 & 316.00$\pm$2.00 \\
81&I18159-1648c1 & 18:18:54.66 &  -16:47:50.28 &1.48& 2.10$\pm$0.01 & 101$\pm$1 & 24.29 $\pm$ 0.22 & 121.26 $\pm$ 1.09 & 293.00$\pm$3.00 \\
82&I18159-1648c2 & 18:18:54.34 &  -16:47:49.97 &1.48& 2.12$\pm$0.02 & 107$\pm$2 & 32.83 $\pm$ 0.35 & 167.52 $\pm$ 1.76 & 396.00$\pm$4.00 \\
83&I18182-1433 & 18:21:09.05 &  -14:31:47.88 &4.71& 2.17$\pm$0.01 & 108$\pm$1 & 44.32 $\pm$ 0.30 & 237.12 $\pm$ 1.60 & 535.00$\pm$4.00 \\
84&I18236-1205 & 18:26:25.79 &  -12:03:53.08 &2.17& 2.21$\pm$0.05 & 53$\pm$29 & 7.41 $\pm$ 0.22 & 40.88 $\pm$ 1.20 & 89.40$\pm$2.70 \\
85&I18264-1152$^N$ & 18:29:14.37 &  -11:50:22.88 &3.33& 3.06$\pm$0.04 & 153$\pm$1 & 10.31 $\pm$ 0.17 & 109.28 $\pm$ 1.76 & 124.00$\pm$2.00 \\
86&I18290-0924 & 18:31:44.13 &  -09:22:12.25 &5.34& 2.16$\pm$0.04 & 52$\pm$5 & 11.85 $\pm$ 0.28 & 62.93 $\pm$ 1.47 & 143.00$\pm$3.00 \\
87&I18316-0602 & 18:34:20.91 &  -05:59:42.00 &2.09& 2.12$\pm$0.01 & 52$\pm$1 & 67.68 $\pm$ 0.28 & 344.14 $\pm$ 1.41 & 816.00$\pm$3.00 \\
88&I18341-0727$^N$ & 18:36:49.95 &  -07:24:42.13 &6.04& 2.51$\pm$0.04 & 82$\pm$3 & 6.80 $\pm$ 0.15 & 48.58 $\pm$ 1.09 & 82.00$\pm$1.80 \\
89&I18411-0338 & 18:43:46.23 &  -03:35:29.77 &7.41& 2.09$\pm$0.01 & 53$\pm$3 & 26.12 $\pm$ 0.16 & 128.26 $\pm$ 0.80 & 315.00$\pm$2.00 \\
90&I18434-0242$^N$ & 18:46:03.75 & -02:39:22.21 &5.16& 2.16$\pm$0.01 & 75$\pm$1 & 134.42 $\pm$ 0.36 & 713.76 $\pm$ 1.92 & 1620.00$\pm$20.00 \\
91&I18461-0113$^N$ & 18:48:41.93 &  -01:10:02.55 &5.16& 2.30$\pm$0.04 & 55$\pm$11 & 7.02 $\pm$ 0.15 & 42.18 $\pm$ 0.91 & 84.70$\pm$1.80 \\
92&I18469-0132 & 18:49:33.05 &  -01:29:03.34 &5.16& 2.25$\pm$0.01 & 58$\pm$4 & 38.08 $\pm$ 0.17 & 218.93 $\pm$ 0.96 & 459.00$\pm$2.00 \\
93&I18479-0005$^N$ & 18:50:30.73 &  -00:01:59.18 &13& 2.57$\pm$0.02 & 58$\pm$4 & 9.03 $\pm$ 0.12 & 67.73 $\pm$ 0.88 & 109.00$\pm$1.00 \\
94&I18507+0110 & 18:53:18.56 & 01:14:58.23 &1.56& 2.88$\pm$0.02 & 136$\pm$2 & 348.00 $\pm$ 2.00 & 3262.40 $\pm$ 17.60 & 4190.00$\pm$20.00 \\
95&I18507+0121 & 18:53:18.01 & 01:25:25.56 &1.56& 2.14$\pm$0.01 & 44$\pm$2 & 152.00 $\pm$ 3.20 & 791.20 $\pm$ 2.08 & 1830.00$\pm$10.00 \\
96&I18517+0437 & 18:54:14.24 & 04:41:40.65 &2.36& 2.18$\pm$0.01 & 44$\pm$3 & 59.00 $\pm$ 0.70 & 317.87 $\pm$ 1.14 & 715.00$\pm$3.00 \\
97&I19078+0901c1 & 19:10:13.16 & 09:06:12.49 &11.11& 2.54$\pm$0.03 & 51$\pm$3 & 10.92 $\pm$ 0.20 & 77.82 $\pm$ 1.41 & 132.00$\pm$2.00 \\
98&I19078+0901c2 & 19:10:14.13 & 09:06:24.67 &11.11& 4.22$\pm$0.08 & 32$\pm$4 & 4.56 $\pm$ 0.12 & 92.00 $\pm$ 2.40 & 55.00$\pm$1.40 \\
99&I19095+0930 & 19:11:54.99 & 09:35:50.27 &6.02& 2.23$\pm$0.01 & 43$\pm$1 & 32.00 $\pm$ 0.30 & 178.70 $\pm$ 1.12 & 390.00$\pm$2.00 \\
100&I19097+0847$^N$ & 19:12:09.21 & 08:52:14.59 &8.47& 3.61$\pm$0.08 & 59$\pm$2 & 2.88 $\pm$ 0.10 & 42.50 $\pm$ 1.41 & 34.70$\pm$1.20 \\
\hline 
\end{longtable}
\end{tiny}
\footnotesize 
\textbf{Notes:} Column 1 lists the ID number. Column 2 provides the names of the hot core candidates, with a superscript `$N$' indicating the hot core candidates newly identified in this work. Columns 3 to 7 contain the parameters of the hot core candidates, including peak position, distance from the Sun, deconvolved size, and position angle. The distances from the Sun are compiled in \citet{2020MNRAS.496.2790L}. The peak positions, sizes, and position angles of the hot core candidates are derived from 2D Gaussian fitting of the CH$_3$OH integrated intensity maps. Columns 8 to 10 present the peak fluxes, integrated fluxes, and column densities.

\begin{landscape}
\begin{longtable}{ccccccccccc}
\caption{Physical parameters of CH$_3$OCHO C$_2$H$_5$CN C$_2$H$_5$OH} \label{tab:threespecies1} \\
\hline
\hline
\multicolumn{1}{c}{ID}&\multicolumn{1}{c}{Source} &&\multicolumn{1}{c}{CH$_3$OCHO}&&\multicolumn{1}{c}{C$_2$H$_5$CN}&&\multicolumn{1}{c}{C$_2$H$_5$OH} \\
\hline
&&\multicolumn{1}{c}{I$_{\rm peak}$}&\multicolumn{1}{c}{I$_{\rm integrated}$}&\multicolumn{1}{c}{N}&\multicolumn{1}{c}{I$_{\rm peak}$}&\multicolumn{1}{c}{I$_{\rm integrated}$}&\multicolumn{1}{c}{N}&\multicolumn{1}{c}{I$_{\rm peak}$}&\multicolumn{1}{c}{I$_{\rm integrated}$}&\multicolumn{1}{c}{N}\\
&&\multicolumn{1}{c}{K km s$^{-1}$}&\multicolumn{1}{c}{K km s$^{-1}$ arcsec$^2$}&\multicolumn{1}{c}{$\, \text{$\times$} \, \text{10$^{15}$} \, \text{cm}^{-2}$}&\multicolumn{1}{c}{K km s$^{-1}$}&\multicolumn{1}{c}{K km s$^{-1}$ arcsec$^2$}&\multicolumn{1}{c}{$\, \text{$\times$} \, \text{10$^{15}$} \, \text{cm}^{-2}$}&\multicolumn{1}{c}{K km s$^{-1}$}&\multicolumn{1}{c}{K km s$^{-1}$ arcsec$^2$}&\multicolumn{1}{c}{$\, \text{$\times$} \, \text{10$^{15}$} \, \text{cm}^{-2}$}\\
\hline
\endfirsthead

\multicolumn{11}{c}
{\tablename\ \thetable\ -- \textit{Continued from previous page}} \\
\hline\hline
\multicolumn{1}{c}{ID}&\multicolumn{1}{c}{Source} &&\multicolumn{1}{c}{CH$_3$OCHO}&&\multicolumn{1}{c}{C$_2$H$_5$CN}&&\multicolumn{1}{c}{C$_2$H$_5$OH} \\
\hline
&&\multicolumn{1}{c}{I$_{\rm peak}$}&\multicolumn{1}{c}{I$_{\rm integrated}$}&\multicolumn{1}{c}{N}&\multicolumn{1}{c}{I$_{\rm peak}$}&\multicolumn{1}{c}{I$_{\rm integrated}$}&\multicolumn{1}{c}{N}&\multicolumn{1}{c}{I$_{\rm peak}$}&\multicolumn{1}{c}{I$_{\rm integrated}$}&\multicolumn{1}{c}{N}\\
&&\multicolumn{1}{c}{$K$}&\multicolumn{1}{c}{$K\, \text{km/s}$}&\multicolumn{1}{c}{$\, \text{$\times$} \, \text{10$^{15}$} \, \text{cm}^{-2}$}&\multicolumn{1}{c}{$K$}&\multicolumn{1}{c}{$K\, \text{km/s}$}&\multicolumn{1}{c}{$\, \text{$\times$} \, \text{10$^{15}$} \, \text{cm}^{-2}$}&\multicolumn{1}{c}{$K$}&\multicolumn{1}{c}{$K\, \text{km/s}$}&\multicolumn{1}{c}{$\, \text{$\times$} \, \text{10$^{15}$} \, \text{cm}^{-2}$}\\

\endhead
\multicolumn{11}{c}
{\tablename\ \thetable\ -- \textit{Continued from previous page}} \\
\hline\hline
\multicolumn{1}{c}{ID}&\multicolumn{1}{c}{Source} &&\multicolumn{1}{c}{CH$_3$OCHO}&&\multicolumn{1}{c}{C$_2$H$_5$CN}&&\multicolumn{1}{c}{C$_2$H$_5$OH} \\
\hline
&&\multicolumn{1}{c}{I$_{\rm peak}$}&\multicolumn{1}{c}{I$_{\rm integrated}$}&\multicolumn{1}{c}{N}&\multicolumn{1}{c}{I$_{\rm peak}$}&\multicolumn{1}{c}{I$_{\rm integrated}$}&\multicolumn{1}{c}{N}&\multicolumn{1}{c}{I$_{\rm peak}$}&\multicolumn{1}{c}{I$_{\rm integrated}$}&\multicolumn{1}{c}{N}\\
&&\multicolumn{1}{c}{K km s$^{-1}$}&\multicolumn{1}{c}{K km s$^{-1}$ arcsec$^2$}&\multicolumn{1}{c}{$\, \text{$\times$} \, \text{10$^{15}$} \, \text{cm}^{-2}$}&\multicolumn{1}{c}{K km s$^{-1}$}&\multicolumn{1}{c}{K km s$^{-1}$ arcsec$^2$}&\multicolumn{1}{c}{$\, \text{$\times$} \, \text{10$^{15}$} \, \text{cm}^{-2}$}&\multicolumn{1}{c}{K km s$^{-1}$}&\multicolumn{1}{c}{K km s$^{-1}$ arcsec$^2$}&\multicolumn{1}{c}{$\, \text{$\times$} \, \text{10$^{15}$} \, \text{cm}^{-2}$}\\
\hline
\endhead

\hline \multicolumn{11}{r}{\textit{Continued on next page}} \\
\endfoot

\hline
\endlastfoot
1&I08303-4303 & 5.67 $\pm$ 0.39 & 49.60 $\pm$ 4.16 & 47.10$\pm$3.20 & 3.73 $\pm$ 0.46 & 19.68 $\pm$ 3.36 & 1.58$\pm$0.20 & - & - & - \\
2&I08470-4243 & 23.60 $\pm$ 0.63 & 67.20 $\pm$ 2.88 & 196.00$\pm$5.00 & 4.22 $\pm$ 0.48 & 20.48 $\pm$ 3.20 & 1.79$\pm$0.20 & 2.47 $\pm$ 0.34 & 5.18$\pm$1.25 & 11.80$\pm$1.60 \\
3&I09018-4816 & 2.42 $\pm$ 0.28 & 28.80 $\pm$ 3.84 & 20.10$\pm$2.30 & - & - & - & - & - &-  \\
4&I10365-5803$^N$ & 0.49 $\pm$ 0.08 & 8.80 $\pm$ 1.76 & 4.07$\pm$0.58 & - & - & - & - & - & - \\
5&I11298-6155 & 2.32 $\pm$ 0.03 & 28.64 $\pm$ 0.40 & 19.30$\pm$0.20 & - & - & - & - & - & - \\
6&I11590-6452$^N$ & 1.19 $\pm$ 0.03 & 7.42 $\pm$ 0.19 & 9.88$\pm$0.25 &-  &  -& - & 0.95 $\pm$ 0.04 & 7.78$\pm$0.29 & 4.54$\pm$0.19 \\
7&I12320-6122$^N$ & 0.52 $\pm$ 0.01 & 9.04 $\pm$ 0.19 & 4.32$\pm$0.08 & - & - & - & - & - & - \\
8&I12326-6245 & 2.18 $\pm$ 0.03 & 77.25 $\pm$ 1.17 & 18.10$\pm$0.20 & 6.10 $\pm$ 0.04 & 96.11 $\pm$ 0.58 & 2.59$\pm$0.02 & - & - & - \\
9&I13079-6218c1 & 17.54 $\pm$ 0.09 & 207.86 $\pm$ 1.02 & 146.00$\pm$1.00 & 55.44 $\pm$ 0.19 & 578.72 $\pm$ 1.92 & 23.50$\pm$0.10 & 3.77 $\pm$ 0.03 & 40.03$\pm$0.32 & 18.00$\pm$0.10 \\
10&I13079-6218c2$^N$ & 2.31 $\pm$ 0.05 & 23.50 $\pm$ 0.50 & 19.20$\pm$0.40 & 4.12 $\pm$ 0.12 & 37.25 $\pm$ 1.06 & 1.75$\pm$0.05 & - & - & - \\
11&I13134-6242 & 17.77 $\pm$ 0.07 & 230.82 $\pm$ 0.93 & 148.00$\pm$1.00 & 19.71 $\pm$ 0.07 & 234.59 $\pm$ 0.85 & 8.37$\pm$0.03 & 5.04 $\pm$ 0.03 & 64.88$\pm$0.42 & 24.10$\pm$0.10 \\
12&I13140-6226 & 1.24 $\pm$ 0.02 & 23.89 $\pm$ 0.45 & 10.30$\pm$0.20 & 2.70 $\pm$ 0.03 & 34.78 $\pm$ 0.42 & 1.15$\pm$0.01 & - & - &-  \\
13&I13471-6120 & 6.31 $\pm$ 0.03 & 52.85 $\pm$ 0.27 & 52.40$\pm$0.20 & 1.52 $\pm$ 0.03 & 19.87 $\pm$ 0.43 & 0.65$\pm$0.01 & - & - &-\\
14&I13484-6100 & 1.57 $\pm$ 0.03 & 23.54 $\pm$ 0.46 & 13.00$\pm$0.20 & 6.05 $\pm$ 0.05 & 51.65 $\pm$ 0.42 & 2.57$\pm$0.02 & - &-  &-  \\
15&I14164-6028$^N$ & 1.11 $\pm$ 0.02 & 20.51 $\pm$ 0.37 & 9.22$\pm$0.17 & 1.13 $\pm$ 0.05 & 10.78 $\pm$ 0.46 & 0.48$\pm$0.02 & - & - & - \\
16&I14212-6131$^N$ & 0.98 $\pm$ 0.03 & 8.21 $\pm$ 0.22 & 8.14$\pm$0.25 & 1.23 $\pm$ 0.04 & 12.50 $\pm$ 0.40 & 0.52$\pm$0.02 & - & - & - \\
17&I14498-5856 & 3.35 $\pm$ 0.04 & 33.02 $\pm$ 0.38 & 27.80$\pm$0.30 & 2.25 $\pm$ 0.04 & 22.91 $\pm$ 0.38 & 0.96$\pm$0.02 & 0.71 $\pm$ 0.02 & 10.62$\pm$0.37 & 3.39$\pm$0.10 \\
18&I15254-5621 & 11.62 $\pm$ 0.05 & 120.82 $\pm$ 0.50 & 96.50$\pm$0.40 & 10.01 $\pm$ 0.06 & 83.28 $\pm$ 0.50 & 4.25$\pm$0.03 & 8.09 $\pm$ 0.05 & 72.62$\pm$0.43 & 38.60$\pm$0.20 \\
19&I15290-5546$^N$ & 2.39 $\pm$ 0.03 & 29.06 $\pm$ 0.40 & 19.90$\pm$0.20 & 6.33 $\pm$ 0.06 & 50.51 $\pm$ 0.45 & 2.69$\pm$0.03 & - & - & - \\
20&I15394-5358$^N$ & 10.17 $\pm$ 0.58 & 129.44 $\pm$ 10.24 & 84.50$\pm$4.80 & 7.88 $\pm$ 0.38 & 79.68 $\pm$ 5.60 & 3.35$\pm$0.16 & 1.94 $\pm$ 0.15 & 23.52$\pm$2.56 & (9.27$\pm$0.72)$\times$10$^{15}$ \\
21&I15411-5352$^N$ & 2.14 $\pm$ 0.20 & 34.08 $\pm$ 4.16 & 17.80$\pm$1.70 & 1.08 $\pm$ 0.19 & 31.20 $\pm$ 6.40 & 0.46$\pm$0.08 & - & - & - \\
22&I15437-5343 & 4.79 $\pm$ 0.15 & 44.96 $\pm$ 2.08 & 39.80$\pm$1.20 & 3.24 $\pm$ 0.19 & 37.92 $\pm$ 3.04 & 1.38$\pm$0.08 & 1.76 $\pm$ 0.18 & 19.52$\pm$2.88 & 8.41$\pm$0.86 \\
23&I15502-5302$^N$ & 1.53 $\pm$ 0.14 & 25.44 $\pm$ 3.04 & 12.70$\pm$1.20 & - & - & - & - & - & - \\
24&I15520-5234 & 15.01 $\pm$ 0.65 & 193.60 $\pm$ 11.36 & 125.00$\pm$5.00 & 10.22 $\pm$ 0.45 & 151.20 $\pm$ 8.80 & 4.34$\pm$0.19 & 1.92 $\pm$ 0.20 & 35.84$\pm$4.64 & 9.17$\pm$0.96 \\
25&I15557-5215$^N$ & 1.18 $\pm$ 0.13 & 16.96 $\pm$ 2.56 & 9.80$\pm$1.08 & 1.88 $\pm$ 0.19 & 18.88 $\pm$ 2.72 & 0.80$\pm$0.08 & - & - & - \\
26&I15596-5301c1$^N$ & 1.77 $\pm$ 0.18 & 22.08 $\pm$ 3.20 & 14.70$\pm$1.50 & 1.01$\pm$0.12 & 11.20 $\pm$ 1.92 & 0.43$\pm$0.05 &  -& - &-  \\
27&I15596-5301c2$^N$ & 1.14 $\pm$ 0.18 & 10.56 $\pm$ 2.88 & 9.47$\pm$1.50 & 0.63$\pm$0.17 & 5.28 $\pm$ 2.08 & 0.27$\pm$0.07 & - & - &-  \\
28&I16037-5223$^N$ & 3.22 $\pm$ 0.16 & 44.32 $\pm$ 3.04 & 26.70$\pm$1.30 & 4.62 $\pm$ 0.22 & 47.52 $\pm$ 3.36 & 1.96$\pm$0.09 & - & - & - \\
29&I16060-5146c1 & 9.82 $\pm$ 0.58 & 262.40 $\pm$ 19.20 & 81.60$\pm$4.80 & 22.90 $\pm$ 1.00 & 497.60 $\pm$ 27.20 & 9.72$\pm$0.42 & 9.45 $\pm$ 0.55 & 216.16$\pm$15.20 & 45.10$\pm$2.60 \\
30&I16060-5146c2$^N$ & - & - & - & - & - & - &-  & - & - \\
31&I16065-5158 & 15.47 $\pm$ 0.32 & 186.08 $\pm$ 5.28 & 128.00$\pm$3.00 & 86.60 $\pm$ 1.10 & 780.48 $\pm$ 14.40 & 36.80$\pm$0.50 & 9.44 $\pm$ 0.32 & 100.16$\pm$4.80 & 45.10$\pm$1.50 \\
32&I16071-5142 & 22.13 $\pm$ 0.71 & 208.32 $\pm$ 9.76 & 184.00$\pm$6.00 & 48.59 $\pm$ 0.82 & 366.40 $\pm$ 9.60 & 20.60$\pm$0.30 & 6.58 $\pm$ 0.25 & 56.96$\pm$3.20 & 31.40$\pm$1.20 \\
33&I16076-5134 & 4.81 $\pm$ 0.20 & 52.16 $\pm$ 3.04 & 40.00$\pm$1.70 & 17.28 $\pm$ 0.30 & 129.76 $\pm$ 3.52 & 7.34$\pm$0.13 & - & - & - \\
34&I16119-5048c1$^N$ & 1.34 $\pm$ 0.13 & 14.24 $\pm$ 1.92 & 11.10$\pm$1.10 &-  & - & - & - & - & - \\
35&I16119-5048c2$^N$ & 1.17 $\pm$ 0.12 & 14.88 $\pm$ 2.08 & 9.72$\pm$1.00 & - & - & - & - & - & - \\
36&I16164-5046 & 19.51 $\pm$ 0.93 & 276.80 $\pm$ 17.60 & 162.00$\pm$8.00 & 60.80 $\pm$ 1.40 & 579.20 $\pm$ 19.20 & 25.80$\pm$0.60 & 13.68 $\pm$ 0.58 & 182.56$\pm$10.40 & 65.40$\pm$2.80 \\
37&I16172-5028c1 & 7.99 $\pm$ 0.38 & 109.12 $\pm$ 7.04 & 66.40$\pm$3.20 & 7.82 $\pm$ 0.27 & 135.04 $\pm$ 5.92 & 3.32$\pm$0.11 & 1.98 $\pm$ 0.21 & 28.00$\pm$4.00 & 9.46$\pm$1.00 \\
38&I16172-5028c2$^N$ & 3.18 $\pm$ 0.26 & 41.12 $\pm$ 4.64 & 26.40$\pm$2.20 & 5.11 $\pm$ 0.50 & 69.28 $\pm$ 9.12 & 2.17$\pm$0.21 & - & - & - \\
39&I16272-4837c1 & 54.40 $\pm$ 1.20 & 368.32 $\pm$ 13.12 & 452.00$\pm$10.00 & 66.60 $\pm$ 1.20 & 421.12 $\pm$ 12.48 & 28.30$\pm$0.50 & 8.47 $\pm$ 0.26 & 56.64$\pm$2.72 & 40.50$\pm$1.20 \\
40&I16272-4837c2 & - & - & - & - & - & - & - & - & - \\
41&I16272-4837c3 & 2.71 $\pm$ 0.45 & 41.12 $\pm$ 8.80 & 22.50$\pm$3.70 & 7.90 $\pm$ 0.66 & 74.56 $\pm$ 8.96 & 3.35$\pm$0.28 & 1.63 $\pm$ 0.29 & 15.20$\pm$3.84 & 7.79$\pm$1.39 \\
42&I16297-4757$^N$ & 0.94 $\pm$ 24.00 & 21.46 $\pm$ 0.56 & 8.00$\pm$1.99 & 1.35 $\pm$ 0.06 & 9.97 $\pm$ 0.43 & 0.57$\pm$0.03 & - & - & - \\
43&I16313-4729$^N$ & 1.65 $\pm$ 0.17 & 21.12 $\pm$ 2.88 & 13.70$\pm$1.40 & 3.73 $\pm$ 0.27 & 26.72 $\pm$ 2.88 & 1.58$\pm$0.11 & - & - & - \\
44&I16318-4724 & 25.10 $\pm$ 0.11 & 182.37 $\pm$ 0.82 & 208.00$\pm$1.00 & 82.97 $\pm$ 0.17 & 505.81 $\pm$ 1.02 & 35.20$\pm$0.10 & 11.78 $\pm$ 0.06 & 77.63$\pm$0.38 & 56.30$\pm$0.30 \\
45&I16344-4658 & 4.10 $\pm$ 0.04 & 42.72 $\pm$ 0.46 & 34.10$\pm$0.30 & 12.22 $\pm$ 0.08 & 75.41 $\pm$ 0.48 & 5.19$\pm$0.03 & 1.16$\pm$0.06 & 8.66$\pm$0.42 & 5.54$\pm$0.29 \\
46&I16348-4654c1 & 34.66 $\pm$ 0.18 & 220.67 $\pm$ 1.17 & 288.00$\pm$1.00 & 88.85 $\pm$ 0.25 & 511.92 $\pm$ 1.41 & 37.70$\pm$0.10 & 9.09 $\pm$ 0.07 & 59.30$\pm$0.43 & 43.40$\pm$0.30 \\
47&I16348-4654c2$^N$ & 3.45 $\pm$ 0.06 & 47.42 $\pm$ 0.80 & 28.70$\pm$0.50 & 3.84 $\pm$ 0.08 & 38.90 $\pm$ 0.82 & 1.63$\pm$0.03 & 1.01 $\pm$ 0.08 & 4.60$\pm$0.38 & 4.82$\pm$0.38 \\
48&I16351-4722 & 36.66 $\pm$ 0.09 & 335.84 $\pm$ 0.82 & 304.00$\pm$1.00 & 81.37 $\pm$ 0.15 & 588.61 $\pm$ 1.12 & 34.60$\pm$0.10 & 4.31 $\pm$ 0.05 & 37.62$\pm$0.46 & 20.60$\pm$0.20 \\
49&I16424-4531$^N$ & 1.34 $\pm$ 0.03 & 12.18 $\pm$ 0.29 & 11.10$\pm$0.20 & 1.57 $\pm$ 0.04 & 30.38 $\pm$ 0.78 & 0.67$\pm$0.02 & - & - & - \\
50&I16445-4459$^N$ & 2.25 $\pm$ 0.04 & 22.53 $\pm$ 0.42 & 18.70$\pm$0.30 & 3.89 $\pm$ 0.07 & 27.87 $\pm$ 0.46 & 1.65$\pm$0.03 & - & - & - \\
51&I16458-4512 & 1.82 $\pm$ 0.04 & 17.82 $\pm$ 0.40 & 15.10$\pm$0.30 & - & - & - & - & - & - \\
52&I16484-4603 & 7.06 $\pm$ 0.06 & 55.92 $\pm$ 0.45 & 58.60$\pm$0.50 & 4.19 $\pm$ 0.05 & 44.42 $\pm$ 0.54 & 1.78$\pm$0.02 & 1.34 $\pm$ 0.06 & 10.64$\pm$0.45 & 6.40$\pm$0.29 \\
53&I16547-4247 & 3.16 $\pm$ 0.04 & 34.94 $\pm$ 0.42 & 26.20$\pm$0.30 & 14.16 $\pm$ 0.06 & 118.88 $\pm$ 0.54 & 6.01$\pm$0.03 & - & - & - \\
54&I16562-3959$^N$ & - & - & - & 2.68 $\pm$ 0.04 & 23.12 $\pm$ 0.38 & 1.14$\pm$0.02 & - & - & - \\
55&I17008-4040 & 31.58 $\pm$ 0.09 & 285.36 $\pm$ 0.78 & 262.00$\pm$1.00 & 33.04 $\pm$ 0.16 & 276.27 $\pm$ 1.36 & 14.00$\pm$0.10 & 8.40 $\pm$ 0.05 & 74.16$\pm$0.42 & 40.10$\pm$0.20 \\
56&I17016-4124c1 & 3.28$\pm$0.08 & 32.03 $\pm$ 0.75 & 27.20$\pm$0.70 & 5.35$\pm$0.12 & 52.94 $\pm$ 1.22 & 2.27$\pm$0.05 & 1.36$\pm$0.05 & 19.15$\pm$0.71 & 6.50$\pm$0.24 \\
57&I17016-4124c2 &- & -& -& -& -& -& -& -& -\\
58&I17136-3617$^N$ & 1.59 $\pm$ 0.15 & 29.12 $\pm$ 3.36 & 13.20$\pm$1.20 & - & - & - & - & - & - \\
59&I17143-3700$^N$ & 5.51 $\pm$ 0.23 & 58.40 $\pm$ 3.36 & 45.80$\pm$1.90 & 4.92 $\pm$ 0.28 & 43.36 $\pm$ 3.52 & 2.09$\pm$0.12 & - & - & - \\
60&I17158-3901c1 & 3.75 $\pm$ 0.19 & 35.84 $\pm$ 2.56 & 31.10$\pm$1.60 & 4.58 $\pm$ 0.39 & 46.72 $\pm$ 5.44 & 1.94$\pm$0.17 &1.16$\pm$0.17 & 19.36$\pm$3.52 & 5.54$\pm$0.81 \\
61&I17158-3901c2 &-  &-  &-  & 3.74 $\pm$ 0.21 & 32.80 $\pm$ 2.72 & 1.59$\pm$0.09 \\
62&I17160-3707$^N$ & -&- &- & 1.57 $\pm$ 0.19 & 23.68 $\pm$ 3.68 & 0.67$\pm$0.08 & - & - & - \\
63&I17175-3544 & 296.00 $\pm$ 9.10 & 6144.00 $\pm$ 224.00 & 2460.00$\pm$80.00 & 187.90 $\pm$ 5.90 & 2486.40 $\pm$ 102.40 & 79.80$\pm$2.50 & 185.70 $\pm$ 5.00 & 2584.00$\pm$89.60 & 887.00$\pm$24.00 \\
64&I17220-3609 & 35.28 $\pm$ 0.77 & 433.76 $\pm$ 12.48 & 293.00$\pm$6.00 & 52.30 $\pm$ 1.20 & 641.60 $\pm$ 19.20 & 22.20$\pm$0.50 & 14.37 $\pm$ 0.41 & 182.40$\pm$6.72 & 68.60$\pm$2.00 \\
65&I17233-3606 & 78.90 $\pm$ 1.90 & 1486.40 $\pm$ 44.80 & 655.00$\pm$16.00 & 469.10 $\pm$ 6.60 & 6844.80 $\pm$ 121.60 & 199.00$\pm$3.00 & 12.58 $\pm$ 0.38 & 237.12$\pm$8.80 & 60.10$\pm$1.80 \\
66&I17439-2845$^N$ & 1.44 $\pm$ 0.03 & 25.52 $\pm$ 0.59 & 12.00$\pm$0.20 & 2.16 $\pm$ 0.06 & 28.32 $\pm$ 0.72 & 0.92$\pm$0.03 & - & - & - \\
67&I17441-2822c1$^N$ & 1.76$\pm$0.12 & 8.03 $\pm$ 0.53 & 14.60$\pm$1.00 & - & - & - & - & - & - \\
68&I17441-2822c2 & 4.34$\pm$0.08 & 68.24 $\pm$ 1.28 & 36.00$\pm$0.70 & 10.86$\pm$0.55 & 185.12 $\pm$ 9.44 & 4.61$\pm$0.23 & 11.58$\pm$0.20 & 103.36$\pm$1.76 & 55.30$\pm$1.00 \\
69&I17441-2822c3$^N$ & 3.24$\pm$0.07 & 76.32 $\pm$ 1.76 & 26.90$\pm$0.60 & 38.54 $\pm$ 0.29 & 735.52 $\pm$ 0.96 & 16.40$\pm$0.10 & 2.21 $\pm$ 0.14 & 25.12$\pm$1.60 & 10.60$\pm$0.70 \\
70&I17589-2312$^N$ & 0.91 $\pm$ 0.02 & 17.51 $\pm$ 0.40 & 7.56$\pm$0.17 & 1.15 $\pm$ 0.04 & 13.12 $\pm$ 0.46 & 0.49$\pm$0.02 & - & - & - \\
71&I17599-2148$^N$ & 1.43 $\pm$ 0.03 & 11.14 $\pm$ 0.24 & 11.90$\pm$0.20 & 4.00 $\pm$ 0.09 & 24.70 $\pm$ 0.53 & 1.70$\pm$0.04 & - & - & - \\
72&I18032-2032c1 & 4.23 $\pm$ 0.05 & 37.63 $\pm$ 0.45 & 35.10$\pm$0.40 & - & - & - & - & - & - \\
73&I18032-2032c2 & 9.69$\pm$0.15 & 83.63 $\pm$ 1.28 & 80.50$\pm$1.20 & 54.24$\pm$0.45 & 307.36 $\pm$ 2.56 & 23.00$\pm$0.2 & 1.74 $\pm$ 0.03 & 55.65$\pm$1.07 & 8.31$\pm$0.14 \\
74&I18032-2032c3 & - & - & - & - & - & - & - & - & - \\
75&I18032-2032c4 & 10.96 $\pm$ 0.12 & 72.70 $\pm$ 0.78 & 91.00$\pm$1.00 & 12.47 $\pm$ 0.11 & 82.51 $\pm$ 0.75 & 5.30$\pm$0.05 & 6.00 $\pm$ 0.07 & 41.98$\pm$0.50 & 28.70$\pm$0.30 \\
76&I18056-1952 & 180.93 $\pm$ 0.52 & 1500.96 $\pm$ 4.32 & 1500.00$\pm$0.00 & 454.36 $\pm$ 0.64 & 3295.04 $\pm$ 4.64 & 193.00$\pm$0.00 & 66.47 $\pm$ 0.22 & 527.36$\pm$1.76 & 318.00$\pm$1.00 \\
77&I18089-1732 & 33.66 $\pm$ 0.09 & 194.85 $\pm$ 0.54 & 280.00$\pm$1.00 & 46.82 $\pm$ 0.11 & 245.60 $\pm$ 0.64 & 19.90$\pm$0.00 & 6.71 $\pm$ 0.09 & 39.41$\pm$0.53 & 32.10$\pm$0.40 \\
78&I18110-1854$^N$ & 1.52 $\pm$ 0.05 & 6.58 $\pm$ 0.21 & 12.60$\pm$0.40 & - & - & - & - & - & - \\
79&I18117-1753 & 13.25 $\pm$ 0.07 & 86.50 $\pm$ 0.45 & 110.00$\pm$1.00 & 17.32 $\pm$ 0.11 & 101.06 $\pm$ 0.62 & 7.35$\pm$0.05 & 2.59 $\pm$ 0.06 & 12.48$\pm$0.30 & 12.40$\pm$0.30 \\
80&I18134-1942$^N$ & 1.13 $\pm$ 0.04 & 9.58 $\pm$ 0.30 & 9.39$\pm$0.33 & - & - & - & - & - & - \\
81&I18159-1648c1 & 21.55 $\pm$ 0.17 & 112.93 $\pm$ 0.88 & 179.00$\pm$1.00 & 14.69 $\pm$ 0.13 & 74.59 $\pm$ 0.66 & 6.24$\pm$0.06 & 2.95 $\pm$ 0.05 & 20.85$\pm$0.37 & 14.10$\pm$0.20 \\
82&I18159-1648c2 & 4.31 $\pm$ 0.10 & 38.05 $\pm$ 0.88 & 35.80$\pm$0.80 & 3.95 $\pm$ 0.07 & 28.11 $\pm$ 0.53 & 1.68$\pm$0.03 & 0.93 $\pm$ 0.07 & 5.65$\pm$0.40 & 4.44$\pm$0.33 \\
83&I18182-1433 & 7.13 $\pm$ 0.07 & 63.54 $\pm$ 0.64 & 59.20$\pm$0.60 & 4.75 $\pm$ 0.06 & 44.98 $\pm$ 0.59 & 2.02$\pm$0.03 & 2.46 $\pm$ 0.06 & 21.10$\pm$0.54 & 11.80$\pm$0.30 \\
84&I18236-1205 & 1.25$\pm$0.05 & 5.58 $\pm$ 0.24 & 10.40$\pm$0.40 & 3.79 $\pm$ 0.08 & 21.17 $\pm$ 0.46 & 1.61$\pm$0.03 & - & - & - \\
85&I18264-1152$^N$ & 2.11 $\pm$ 0.04 & 29.49 $\pm$ 0.54 & 17.50$\pm$0.30 & 1.78 $\pm$ 0.06 & 8.70 $\pm$ 0.30 & 0.76$\pm$0.03 & - & - & - \\
86&I18290-0924 & 2.01 $\pm$ 0.05 & 20.53 $\pm$ 0.50 & 16.70$\pm$0.40 & 3.43 $\pm$ 0.09 & 23.60 $\pm$ 0.64 & 1.46$\pm$0.04 & - & - & - \\
87&I18316-0602 & 4.35 $\pm$ 0.07 & 0.76 $\pm$ 0.01 & 36.10$\pm$0.60 & 4.70 $\pm$ 0.07 & 38.43 $\pm$ 0.61 & 2.00$\pm$0.03 & 2.07 $\pm$ 0.05 & 20.05$\pm$0.45 & 9.84$\pm$0.24 \\
88&I18341-0727$^N$ & 1.20 $\pm$ 0.03 & 17.68 $\pm$ 0.43 & 9.97$\pm$0.25 & 1.12 $\pm$ 0.05 & 17.06 $\pm$ 0.70 & 0.48$\pm$0.02 & - & - & - \\
89&I18411-0338 & 2.72 $\pm$ 0.05 & 30.76 $\pm$ 0.48 & 22.60$\pm$0.30 & 12.53 $\pm$ 0.10 & 70.16 $\pm$ 0.56 & 5.32$\pm$0.04 & 1.45 $\pm$ 0.04 & 13.42$\pm$0.35 & 6.93$\pm$0.19 \\
90&I18434-0242$^N$ & 27.77 $\pm$ 0.14 & 165.28 $\pm$ 0.83 & 231.00$\pm$1.00 & 77.55 $\pm$ 0.21 & 429.06 $\pm$ 1.17 & 32.90$\pm$0.10 & 13.14 $\pm$ 0.10 & 78.30$\pm$0.58 & 62.70$\pm$0.50 \\
91&I18461-0113$^N$ & 0.93 $\pm$ 0.02 & 71.73 $\pm$ 1.33 & 7.72$\pm$0.17 & 1.80 $\pm$ 0.06 & 22.78 $\pm$ 0.70 & 0.76$\pm$0.03 & - & - & - \\
92&I18469-0132 & 4.80 $\pm$ 0.07 & 44.16 $\pm$ 0.66 & 39.90$\pm$0.60 & 5.77 $\pm$ 0.06 & 42.80 $\pm$ 0.42 & 2.45$\pm$0.03 & 1.10$\pm$0.04 & 14.02$\pm$0.48 & 5.25$\pm$0.19 \\
93&I18479-0005$^N$ & 1.55 $\pm$ 0.03 & 22.58 $\pm$ 0.43 & 12.90$\pm$0.20 & 4.20 $\pm$ 0.07 & 31.44 $\pm$ 0.53 & 1.78$\pm$0.03 & - & - & - \\
94&I18507+0110 & 86.38 $\pm$ 0.55 & 1054.08 $\pm$ 6.72 & 717.00$\pm$5.00 & 136.37 $\pm$ 0.76 & 1205.28 $\pm$ 6.72 & 57.90$\pm$0.30 & 47.00 $\pm$ 0.30 & 392.48$\pm$2.56 & 225.00$\pm$1.00 \\
95&I18507+0121 & 33.66 $\pm$ 0.15 & 201.95 $\pm$ 0.88 & 280.00$\pm$1.00 & 70.78 $\pm$ 0.22 & 379.06 $\pm$ 1.15 & 30.10$\pm$0.10 & 16.27 $\pm$ 0.10 & 93.39$\pm$0.54 & 77.70$\pm$0.50 \\
96&I18517+0437 & 3.30 $\pm$ 0.04 & 28.16 $\pm$ 0.32 & 27.40$\pm$0.30 & 1.14 $\pm$ 0.04 & 13.36 $\pm$ 0.43 & 0.48$\pm$0.02 & 1.43 $\pm$ 0.04 & 20.74$\pm$0.51 & 6.83$\pm$0.19 \\
97&I19078+0901c1 & 2.05 $\pm$ 0.04 & 62.26 $\pm$ 0.02 & 17.04 $\pm$ 0.22 & 6.33 $\pm$ 0.07 & 69.6 $\pm$ 0.82 & 2.68 $\pm$ 0.08 & - & - & - \\
98&I19078+0901c2 & 8.14 $\pm$ 0.07 & 70.24 $\pm$ 0.61 & 67.66 $\pm$ 0.51 & 11.29 $\pm$ 0.10 & 78.34 $\pm$ 0.70 & 4.79 $\pm$ 0.12 & 1.73 $\pm$ 0.04 & 21.86 $\pm$ 0.51 & 8.28 $\pm$ 0.09 \\
99&I19095+0930 & 3.05 $\pm$ 0.06 & 34.82 $\pm$ 0.62 & 25.30$\pm$0.50 & 2.65 $\pm$ 0.05 & 23.84 $\pm$ 0.46 & 1.13$\pm$0.02 & 2.22 $\pm$ 0.05 & 22.98$\pm$0.50 & 10.60$\pm$0.20 \\
100&I19097+0847$^N$ & 1.25 $\pm$ 0.02 & 24.69 $\pm$ 0.43 & 10.40$\pm$0.20 & 1.00 $\pm$ 0.04 & 13.70 $\pm$ 0.53 & 0.43$\pm$0.02 & - & - & - \\
\hline 
\end{longtable}

\begin{longtable}{ccccccccccc}
\caption{Physical parameters of CH$_3$OCH$_3$ CH$_3$CHO CH$_3$COCH$_3$} \label{tab:threespecies2} \\
\hline\hline
\multicolumn{1}{c}{ID} &\multicolumn{1}{c}{Source} &&\multicolumn{1}{c}{CH$_3$OCH$_3$}&&\multicolumn{1}{c}{CH$_3$CHO}&&\multicolumn{1}{c}{CH$_3$COCH$_3$} \\
\hline
&&\multicolumn{1}{c}{I$_{\rm peak}$}&\multicolumn{1}{c}{I$_{\rm integrated}$}&\multicolumn{1}{c}{N}&\multicolumn{1}{c}{I$_{\rm peak}$}&\multicolumn{1}{c}{I$_{\rm integrated}$}&\multicolumn{1}{c}{N}&\multicolumn{1}{c}{I$_{\rm peak}$}&\multicolumn{1}{c}{I$_{\rm integrated}$}&\multicolumn{1}{c}{N}\\
&&\multicolumn{1}{c}{K km s$^{-1}$}&\multicolumn{1}{c}{K km s$^{-1}$ arcsec$^2$}&\multicolumn{1}{c}{$\, \text{$\times$} \, \text{10$^{15}$} \, \text{cm}^{-2}$}&\multicolumn{1}{c}{K km s$^{-1}$}&\multicolumn{1}{c}{K km s$^{-1}$ arcsec$^2$}&\multicolumn{1}{c}{$\, \text{$\times$} \, \text{10$^{15}$} \, \text{cm}^{-2}$}&\multicolumn{1}{c}{K km s$^{-1}$}&\multicolumn{1}{c}{K km s$^{-1}$ arcsec$^2$}&\multicolumn{1}{c}{$\, \text{$\times$} \, \text{10$^{15}$} \, \text{cm}^{-2}$}\\
\hline
\endfirsthead

\multicolumn{11}{c}
{\tablename\ \thetable\ -- \textit{Continued from previous page}} \\
\hline\hline
\multicolumn{1}{c}{ID} &\multicolumn{1}{c}{Source} &&\multicolumn{1}{c}{CH$_3$OCH$_3$}&&\multicolumn{1}{c}{CH$_3$CHO}&&\multicolumn{1}{c}{CH$_3$COCH$_3$} \\
\hline
&&\multicolumn{1}{c}{I$_{\rm peak}$}&\multicolumn{1}{c}{I$_{\rm integrated}$}&\multicolumn{1}{c}{N}&\multicolumn{1}{c}{I$_{\rm peak}$}&\multicolumn{1}{c}{I$_{\rm integrated}$}&\multicolumn{1}{c}{N}&\multicolumn{1}{c}{I$_{\rm peak}$}&\multicolumn{1}{c}{I$_{\rm integrated}$}&\multicolumn{1}{c}{N}\\
&&\multicolumn{1}{c}{K km s$^{-1}$}&\multicolumn{1}{c}{K km s$^{-1}$ arcsec$^2$}&\multicolumn{1}{c}{$\, \text{$\times$} \, \text{10$^{15}$} \, \text{cm}^{-2}$}&\multicolumn{1}{c}{K km s$^{-1}$}&\multicolumn{1}{c}{K km s$^{-1}$ arcsec$^2$}&\multicolumn{1}{c}{$\, \text{$\times$} \, \text{10$^{15}$} \, \text{cm}^{-2}$}&\multicolumn{1}{c}{K km s$^{-1}$}&\multicolumn{1}{c}{K km s$^{-1}$ arcsec$^2$}&\multicolumn{1}{c}{$\, \text{$\times$} \, \text{10$^{15}$} \, \text{cm}^{-2}$}\\
\hline
\endhead

\hline \multicolumn{11}{r}{\textit{Continued on next page}} \\
\endfoot

\hline
\endlastfoot
1&I08303-4303 & 16.50 $\pm$ 1.00 & 134.88$\pm$10.40 & 56.40$\pm$3.40 & 5.24 $\pm$ 0.90 & 62.40$\pm$12.48 & 6.22$\pm$1.07 & - & - & - \\
2&I08470-4243 & 19.60 $\pm$ 1.00 & 65.60$\pm$5.28 & 67.00$\pm$3.40 & - & - & - & - & - & - \\
3&I09018-4816 & 6.02 $\pm$ 0.73 & 48.64$\pm$7.36 & 20.60$\pm$2.50 & 1.43 $\pm$ 0.24 & 404.80$\pm$68.80 & 1.70$\pm$0.28 & - & - & - \\
4&I10365-5803$^N$ & - & - & - & - & - & - & - & - & - \\
5&I11298-6155 & 2.72 $\pm$ 0.08 & 20.60$\pm$0.61 & 9.30$\pm$0.27 & 1.26 $\pm$ 0.05 & 37.15$\pm$1.30 & 1.50$\pm$0.05 & - & - & - \\
6&I11590-6452$^N$ & - & - & - & - & - & - & - & - & - \\
7&I12320-6122$^N$ & 1.51 $\pm$ 0.04 & 30.90$\pm$0.90 & 5.16$\pm$0.14 & - & - & - & - & - & - \\
8&I12326-6245 & 8.04 $\pm$ 0.07 & 271.52$\pm$2.40 & 27.50$\pm$0.20 & 0.92 $\pm$ 0.02 & 76.24$\pm$1.58 & 1.09$\pm$0.02 & 3.23 $\pm$ 0.04 & 35.15$\pm$0.45 & 40.20$\pm$0.50 \\
9&I13079-6218c1 & 61.96 $\pm$ 0.27 & 712.64$\pm$3.20 & 212.00$\pm$1.00 & 4.20 $\pm$ 0.04 & 469.44$\pm$4.32 & 4.99$\pm$0.05 & 2.01 $\pm$ 0.04 & 24.26$\pm$0.46 & 24.90$\pm$0.50 \\
10&I13079-6218c2$^N$ & 4.61$\pm$0.09 & 51.92$\pm$0.99 & 15.80$\pm$0.30 & - & - & - & - & - & - \\
11&I13134-6242 & 16.31 $\pm$ 0.11 & 239.74$\pm$1.55 & 55.80$\pm$0.40 & 2.75 $\pm$ 0.03 & 323.36$\pm$3.36 & 3.27$\pm$0.04 & 3.35 $\pm$ 0.04 & 42.14$\pm$0.48 & 41.50$\pm$0.50 \\
12&I13140-6226 & 3.10 $\pm$ 0.05 & 76.46$\pm$1.20 & 10.60$\pm$0.20 & 1.99 $\pm$ 0.02 & 313.28$\pm$3.36 & 2.36$\pm$0.02 & - & - & - \\
13&I13471-6120 & 23.69 $\pm$ 0.11 & 217.06$\pm$1.02 & 81.00$\pm$0.40 & - & - & - & - & - & - \\
14&I13484-6100 & 12.24 $\pm$ 0.12 & 106.99$\pm$1.01 & 41.90$\pm$0.40 & 1.48 $\pm$ 0.03 & 192.32$\pm$3.52 & 1.76$\pm$0.04 & - & - & - \\
15&I14164-6028$^N$ & 2.90 $\pm$ 0.06 & 50.94$\pm$1.04 & 9.92$\pm$0.21 & 2.17 $\pm$ 0.04 & 151.68$\pm$2.40 & 2.58$\pm$0.05 & - & - & - \\
16&I14212-6131$^N$ & 1.86 $\pm$ 0.08 & 28.51$\pm$1.15 & 6.36$\pm$0.24 & - & - & - & - & - & - \\
17&I14498-5856 & 6.47 $\pm$ 0.10 & 57.54$\pm$0.88 & 22.10$\pm$0.30 & 1.92 $\pm$ 0.04 & 67.14$\pm$1.28 & 2.28$\pm$0.05 & - & - & - \\
18&I15254-5621 & 10.73 $\pm$ 0.10 & 126.67$\pm$1.22 & 36.70$\pm$0.30 & 7.61 $\pm$ 0.10 & 79.65$\pm$1.09 & 9.04$\pm$0.12 & 1.56 $\pm$ 0.06 & 22.22$\pm$0.83 & 19.30$\pm$0.70 \\
19&I15290-5546$^N$ & 6.26 $\pm$ 0.09 & 54.29$\pm$0.75 & 21.40$\pm$0.30 & 1.91 $\pm$ 0.03 & 29.41$\pm$0.46 & 2.27$\pm$0.04 & - & - & - \\
20&I15394-5358$^N$ & 39.50 $\pm$ 1.60 & 396.80$\pm$24.00 & 135.00$\pm$5.00 & 4.15 $\pm$ 0.29 & 796.80$\pm$57.60 & 4.93$\pm$0.34 & - & - & - \\
21&I15411-5352$^N$ & 3.51 $\pm$ 0.44 & 39.36$\pm$7.04 & 12.00$\pm$1.50 & 2.41 $\pm$ 0.28 & 130.24$\pm$16.96 & 2.86$\pm$0.33 & - & - & - \\
22&I15437-5343 & 8.06 $\pm$ 0.26 & 73.44$\pm$4.16 & 27.60$\pm$0.90 & 1.27 $\pm$ 0.29 & 50.72$\pm$13.12 & 1.51$\pm$0.34 & - & - & - \\
23&I15502-5302$^N$ & - & - & - & - & - & - & - & - & - \\
24&I15520-5234 & 27.00 $\pm$ 1.00 & 715.20$\pm$32.00 & 92.30$\pm$3.40 & - & - & - & - & - & - \\
25&I15557-5215$^N$ & 2.32 $\pm$ 0.36 & 24.64$\pm$5.44 & 7.93$\pm$1.23 & 2.01 $\pm$ 0.21 & 214.40$\pm$24.00 & 2.39$\pm$0.25 & - & - & - \\
26&I15596-5301c1$^N$ & 2.24 $\pm$ 0.45 & 52.96$\pm$12.80 & 7.66$\pm$1.54 & - & - & - & - & - & - \\
27&I15596-5301c2$^N$ & 1.65 $\pm$ 0.36 & 28.00$\pm$7.84 & 5.64$\pm$1.23 & - & - & - & - & - & - \\
28&I16037-5223$^N$ & 3.69 $\pm$ 0.36 & 64.16$\pm$8.16 & 12.60$\pm$1.20 & 2.44 $\pm$ 0.32 & 97.44$\pm$14.40 & 2.90$\pm$0.38 & - & - & - \\
29&I16060-5146c1 & - & - & - & - & - & - & 7.16 $\pm$ 0.43 & 142.08$\pm$10.56 & 88.80$\pm$5.30 \\
30&I16060-5146c2$^N$ & 11.50 $\pm$ 1.20 & 240.00$\pm$32.00 & 39.30$\pm$4.10 & - & - & - & - & - & - \\
31&I16065-5158 & 32.46 $\pm$ 0.82 & 372.96$\pm$13.12 & 111.00$\pm$3.00 & 5.04 $\pm$ 0.50 & 196.80$\pm$22.40 & 5.98$\pm$0.59 & 1.80 $\pm$ 0.26 & 30.24$\pm$5.60 & 22.30$\pm$3.20 \\
32&I16071-5142 & 53.20 $\pm$ 1.40 & 464.00$\pm$17.60 & 182.00$\pm$5.00 & 6.05 $\pm$ 0.40 & 558.40$\pm$38.40 & 7.18$\pm$0.47 & 2.20 $\pm$ 0.25 & 24.80$\pm$4.00 & 27.30$\pm$3.10 \\
33&I16076-5134 & 14.55 $\pm$ 0.49 & 130.08$\pm$6.56 & 49.80$\pm$1.70 & 2.06 $\pm$ 0.25 & 275.20$\pm$33.60 & 2.45$\pm$0.30 & - & - & - \\
34&I16119-5048c1$^N$ & 2.84 $\pm$ 0.41 & 64.64$\pm$11.20 & 9.71$\pm$1.40 & 4.43 $\pm$ 0.35 & 109.44$\pm$10.40 & 5.26$\pm$0.42 & - & - & - \\
35&I16119-5048c2$^N$ & - & - & - & - & - & - & - & - & - \\
36&I16164-5046 & 15.30 $\pm$ 1.10 & 273.60$\pm$25.60 & 52.30$\pm$3.80 & 2.69 $\pm$ 0.29 & 236.80$\pm$27.20 & 3.19$\pm$0.34 & 4.57 $\pm$ 0.36 & 71.04$\pm$7.36 & 56.70$\pm$4.50 \\
37&I16172-5028c1 & 9.64 $\pm$ 0.53 & 239.04$\pm$15.68 & 33.00$\pm$1.80 & 3.17 $\pm$ 0.37 & 144.96$\pm$18.88 & 3.76$\pm$0.44 & 4.61 $\pm$ 0.29 & 56.32$\pm$4.96 & 57.10$\pm$3.60 \\
38&I16172-5028c2$^N$ & 4.34 $\pm$ 0.55 & 36.32$\pm$7.04 & 14.80$\pm$1.90 & - & - & - & - & - & - \\
39&I16272-4837c1 & 88.30 $\pm$ 1.50 & 616.00$\pm$16.00 & 302.00$\pm$5.00 & 7.79 $\pm$ 0.83 & 260.80$\pm$32.00 & 9.25$\pm$0.99 & 9.50 $\pm$ 0.35 & 73.92$\pm$4.00 & 118.00$\pm$4.00 \\
40&I16272-4837c2 & - & - & - & - & - & - & - & - & - \\
41&I16272-4837c3 & 5.52 $\pm$ 0.67 & 68.80$\pm$11.36 & 18.90$\pm$2.30 & - & - & - & - & - & - \\
42&I16297-4757$^N$ & 3.80 $\pm$ 0.10 & 47.04$\pm$1.18 & 13.00$\pm$0.30 & 2.67 $\pm$ 0.05 & 129.60$\pm$2.24 & 3.17$\pm$0.06 & - & - & - \\
43&I16313-4729$^N$ & 4.48 $\pm$ 0.38 & 34.88$\pm$4.48 & 15.30$\pm$1.30 & 1.97 $\pm$ 0.35 & 57.44$\pm$11.84 & 2.34$\pm$0.42 & - & - & - \\
44&I16318-4724 & 46.20 $\pm$ 0.21 & 347.50$\pm$1.57 & 158.00$\pm$1.00 & 7.18 $\pm$ 0.10 & 253.28$\pm$3.36 & 8.52$\pm$0.11 & 2.61 $\pm$ 0.07 & 26.26$\pm$0.74 & 32.40$\pm$0.90 \\
45&I16344-4658 & 22.71 $\pm$ 0.16 & 140.37$\pm$1.01 & 77.70$\pm$0.50 & 3.85 $\pm$ 0.05 & 186.40$\pm$2.24 & 4.57$\pm$0.06 & - & - & - \\
46&I16348-4654c1 & 42.92 $\pm$ 0.21 & 279.20$\pm$1.39 & 147.00$\pm$1.00 & 4.69 $\pm$ 0.09 & 307.04$\pm$5.92 & 5.57$\pm$0.11 & 4.93 $\pm$ 0.08 & 35.09$\pm$0.56 & 61.10$\pm$1.00 \\
47&I16348-4654c2$^N$ & 4.25 $\pm$ 0.08 & 108.00$\pm$1.92 & 14.50$\pm$0.20 & 6.36 $\pm$ 0.06 & 321.76$\pm$3.04 & 7.55$\pm$0.07 & - & - & - \\
48&I16351-4722 & 79.56 $\pm$ 0.27 & 849.44$\pm$2.88 & 272.00$\pm$1.00 & 3.65 $\pm$ 0.05 & 428.96$\pm$5.76 & 4.33$\pm$0.06 & 4.51 $\pm$ 0.06 & 46.10$\pm$0.56 & 55.90$\pm$0.70 \\
49&I16424-4531$^N$ & 3.56 $\pm$ 0.12 & 24.90$\pm$0.86 & 12.20$\pm$0.40 & - & - & - & - & - & - \\
50&I16445-4459$^N$ & 4.55 $\pm$ 0.11 & 45.06$\pm$1.06 & 15.60$\pm$0.40 & - & - & - & - & - & - \\
51&I16458-4512 & 6.47 $\pm$ 0.15 & 49.42$\pm$1.20 & 22.10$\pm$0.50 & - & - & -  & - & - & - \\
52&I16484-4603 & 14.21 $\pm$ 0.16 & 109.02$\pm$1.20 & 48.60$\pm$0.50 & 1.41 $\pm$ 0.04 & 98.72$\pm$2.72 & 1.67$\pm$0.05 & - & - & - \\
53&I16547-4247 & 2.67 $\pm$ 0.06 & 98.56$\pm$2.08 & 9.13$\pm$0.21 & 2.44$\pm$0.04 & 154.24$\pm$2.40 & 2.90$\pm$0.05 & - & - & - \\
54&I16562-3959$^N$ & - & - & - & - & - & - & - & - & - \\
55&I17008-4040 & 34.72 $\pm$ 0.17 & 337.76$\pm$1.60 & 119.00$\pm$1.00 & 6.10 $\pm$ 0.10 & 98.90$\pm$1.54 & 7.24$\pm$0.11 & 4.77 $\pm$ 0.06 & 44.32$\pm$0.58 & 59.10$\pm$0.70 \\
56&I17016-4124c1 & 6.14$\pm$0.26 & 45.76$\pm$1.92 & 21.00$\pm$0.90 & - & - & - &  -& - & - \\
57&I17016-4124c2 & -& -& -&- &- &- & -&- &- \\
58&I17136-3617$^N$ & - & - & - & - & - & - & - & - & - \\
59&I17143-3700$^N$ & 8.77 $\pm$ 0.58 & 118.88$\pm$10.24 & 30.00$\pm$2.00 & 2.77 $\pm$ 0.24 & 124.96$\pm$11.68 & 3.29$\pm$0.28 & - & - &-  \\
60&I17158-3901c1 & 4.01 $\pm$ 0.57 & 93.44$\pm$15.68 & 13.70$\pm$1.90 & 3.28 $\pm$ 0.25 & 420.80$\pm$33.60 & 3.89$\pm$0.30 & - & - & - \\
61&I17158-3901c2 & - & - & - & - & - & - &  -& - &-  \\
62&I17160-3707$^N$ &  -& - & - & - & - &  -&  -& - &-  \\
63&I17175-3544 & 231.90 $\pm$ 6.60 & 4912.00$\pm$176.00 & 793.00$\pm$23.00 & 115.10 $\pm$ 2.90 & 1708.80$\pm$54.40 & 137.00$\pm$3.00 & 43.60 $\pm$ 1.70 & 774.40$\pm$36.80 & 540.00$\pm$21.00 \\
64&I17220-3609 & 54.90 $\pm$ 1.40 & 702.40$\pm$24.00 & 188.00$\pm$5.00 & 5.69 $\pm$ 0.42 & 788.80$\pm$59.20 & 6.76$\pm$0.50 & 3.88 $\pm$ 0.32 & 70.72$\pm$7.20 & 48.10$\pm$4.00 \\
65&I17233-3606 & 117.20 $\pm$ 3.70 & 2225.60$\pm$86.40 & 401.00$\pm$13.00 & 20.50 $\pm$ 1.10 & 552.00$\pm$33.60 & 24.30$\pm$1.30 & 19.22 $\pm$ 0.63 & 360.00$\pm$14.40 & 238.00$\pm$8.00 \\
66&I17439-2845$^N$ & - & - & - & - &  -&  -& - & - & - \\
67&I17441-2822c1$^N$ & - & - &  -& - &-  &-  & - & - & - \\
68&I17441-2822c2 & - & - &  -& - & - &  -& 18.05$\pm$0.36 & 161.60$\pm$3.20 & 224.00$\pm$4.00 \\
69&I17441-2822c3$^N$ & - & - &-  &-  &  -& - &  -& - & - \\
70&I17589-2312$^N$ & 1.95 $\pm$ 0.07 & 37.54$\pm$1.39 & 6.67$\pm$0.24 & 2.11$\pm$0.04 & 82.72$\pm$1.39 & 2.51$\pm$0.05 & - & - & - \\
71&I17599-2148$^N$ & 3.27 $\pm$ 0.14 & 28.51$\pm$1.18 & 11.20$\pm$0.50 & - &-  &-  & - & - & - \\
72&I18032-2032c1 & 6.14 $\pm$ 0.13 & 96.64$\pm$2.08 & 21.00$\pm$0.40 & 2.92 $\pm$ 0.04 & 320.00$\pm$4.32 & 3.47$\pm$0.05 & - &  -&-  \\
73&I18032-2032c2 & 13.48 $\pm$ 0.23 & 183.04$\pm$3.20 & 46.10$\pm$0.80 & - & - &-  & - & - & - \\
74&I18032-2032c3 & - & - & - & - &-  & - & - & - & - \\
75&I18032-2032c4 & 16.56 $\pm$ 0.21 & 109.39$\pm$1.41 & 56.60$\pm$0.70 & 2.22$\pm$0.04 & 71.50$\pm$1.41 & 2.64$\pm$0.05 & - & - &  -\\
76&I18056-1952 & 190.97 $\pm$ 0.42 & 1765.12$\pm$3.84 & 653.00$\pm$1.00 & 57.62 $\pm$ 0.47 & 540.16$\pm$4.48 & 68.40$\pm$0.60 & 35.68 $\pm$ 0.17 & 263.73$\pm$1.28 & 442.00$\pm$2.00 \\
77&I18089-1732 & 66.32 $\pm$ 0.25 & 382.06$\pm$1.44 & 227.00$\pm$1.00 & 2.97 $\pm$ 0.06 & 165.12$\pm$3.52 & 3.53$\pm$0.07 & 4.92 $\pm$ 0.12 & 25.57$\pm$0.61 & 61.00$\pm$1.50 \\
78&I18110-1854$^N$ & - & - & - & - & - & - & - &-  & - \\
79&I18117-1753 & 45.26 $\pm$ 0.23 & 289.60$\pm$1.41 & 155.00$\pm$1.00 & 3.35 $\pm$ 0.08 & 73.92$\pm$1.76 & 3.98$\pm$0.09 & - &-  &-  \\
80&I18134-1942$^N$ &  -& - & - & - & - & - & - & - &-  \\
81&I18159-1648c1 & 33.75 $\pm$ 0.28 & 174.18$\pm$1.44 & 115.00$\pm$1.00 & 3.75$\pm$0.06 & 229.12$\pm$3.52 & 4.45$\pm$0.07 & 2.42 $\pm$ 0.08 & 19.90$\pm$0.62 & 30.00$\pm$0.90 \\
82&I18159-1648c2 & 13.28 $\pm$ 0.17 & 82.62$\pm$1.26 & 45.40$\pm$0.60 &-  & - & - & 0.87 $\pm$ 0.04 & 12.58$\pm$0.64 & 10.80$\pm$0.50 \\
83&I18182-1433 & 18.40 $\pm$ 0.19 & 144.78$\pm$1.49 & 62.90$\pm$0.60 & 4.70 $\pm$ 0.07 & 139.36$\pm$2.08 & 5.58$\pm$0.08 &  -& - &-  \\
84&I18236-1205 & 5.38 $\pm$ 0.11 & 29.57$\pm$0.59 & 18.40$\pm$0.40 & 3.14 $\pm$ 0.07 & 26.62$\pm$0.56 & 3.73$\pm$0.08 &  -& - &  -\\
85&I18264-1152$^N$ & 3.95 $\pm$ 0.12 & 53.76$\pm$1.60 & 13.50$\pm$0.40 & - &-  &-  &  -& - &-  \\
86&I18290-0924 & 4.46 $\pm$ 0.11 & 38.85$\pm$1.20 & 15.20$\pm$0.40 &  -&  -&  -&  - & - & - \\
87&I18316-0602 & 11.57 $\pm$ 0.17 & 110.24$\pm$1.60 & 39.60$\pm$0.60 & 2.77 $\pm$ 0.04 & 132.32$\pm$2.08 & 3.29$\pm$0.05 &  -& - &-  \\
88&I18341-0727$^N$ & 1.96 $\pm$ 0.07 & 62.24$\pm$2.24 & 6.70$\pm$0.24 & 1.92 $\pm$ 0.05 & 121.92$\pm$3.20 & 2.28$\pm$0.06 & - &  -& - \\
89&I18411-0338 & 1.94 $\pm$ 0.09 & 33.36$\pm$1.55 & 6.63$\pm$0.31 & - & - & - &  -& - & - \\
90&I18434-0242$^N$ & 59.47 $\pm$ 0.25 & 336.96$\pm$1.42 & 203.00$\pm$1.00 & 4.78 $\pm$ 0.12 & 72.32$\pm$1.76 & 5.68$\pm$0.14 & 1.81 $\pm$ 0.07 & 12.91$\pm$0.50 & 22.40$\pm$0.90 \\
91&I18461-0113$^N$ & - & - & - &  -& - & - &-  & - & - \\
92&I18469-0132 & 17.47 $\pm$ 0.19 & 126.70$\pm$1.44 & 59.70$\pm$0.60 & 2.91 $\pm$ 0.05 & 110.72$\pm$1.76 & 3.45$\pm$0.06 &  -& - & - \\
93&I18479-0005$^N$ & - & - & - & 1.68 $\pm$ 0.04 & 93.44$\pm$2.24 & 1.99$\pm$0.05 &-  &-  &-  \\
94&I18507+0110 & 143.92 $\pm$ 0.97 & 1435.20$\pm$9.60 & 492.00$\pm$3.00 & 8.76 $\pm$ 0.14 & 180.00$\pm$2.88 & 10.40$\pm$0.20 & 28.85 $\pm$ 0.21 & 260.64$\pm$1.92 & 358.00$\pm$3.00 \\
95&I18507+0121 & 67.40 $\pm$ 0.30 & 401.92$\pm$1.76 & 230.00$\pm$1.00 & 3.77 $\pm$ 0.05 & 689.28$\pm$8.48 & 4.48$\pm$0.06 & 4.01 $\pm$ 0.08 & 29.66$\pm$0.59 & 49.70$\pm$1.00 \\
96&I18517+0437 & 4.93 $\pm$ 0.09 & 51.90$\pm$0.94 & 16.90$\pm$0.30 & 0.88 $\pm$ 0.02 & 212.16$\pm$5.12 & 1.04$\pm$0.02 & - & - &-  \\
97&I19078+0901c1 & - & - & - & - & - & - & 2.25 $\pm$ 0.04 & 33.47 $\pm$ 0.61 & 28.00 $\pm$ 0.42 \\
98&I19078+0901c2 & 5.49 $\pm$ 0.11 & 79.82 $\pm$ 1.6 & 18.78 $\pm$ 0.12 & - &  - & -  &  - &  - & -  \\
99&I19095+0930 & 2.04 $\pm$ 0.06 & 131.20$\pm$3.68 & 6.98$\pm$0.21 & 2.11 $\pm$ 0.03 & 361.28$\pm$5.28 & 2.51$\pm$0.04 & 1.53 $\pm$ 0.06 & 16.02$\pm$0.62 & 19.00$\pm$0.70 \\
100&I19097+0847$^N$ & 2.12 $\pm$ 0.05 & 75.20$\pm$1.76 & 7.25$\pm$0.17 & 3.59 $\pm$ 0.05 & 186.40$\pm$2.40 & 4.26$\pm$0.06 &  - & -  & -  \\

\hline

\end{longtable}
\footnotesize 
\textbf{Notes:} I$_{\rm peak}$, I$_{\rm integrated}$, N represent the peak flux value, total integrated flux value, and molecular column density of CH$_3$OCHO C$_2$H$_5$CN C$_2$H$_5$OH CH$_3$OCH$_3$ CH$_3$CHO CH$_3$COCH$_3$.
\end{landscape}

\newpage

\section{Additional figures}  \label{app:B}
 
\begin{figure}[hbp!]
\centering
{\includegraphics[width=1\linewidth]{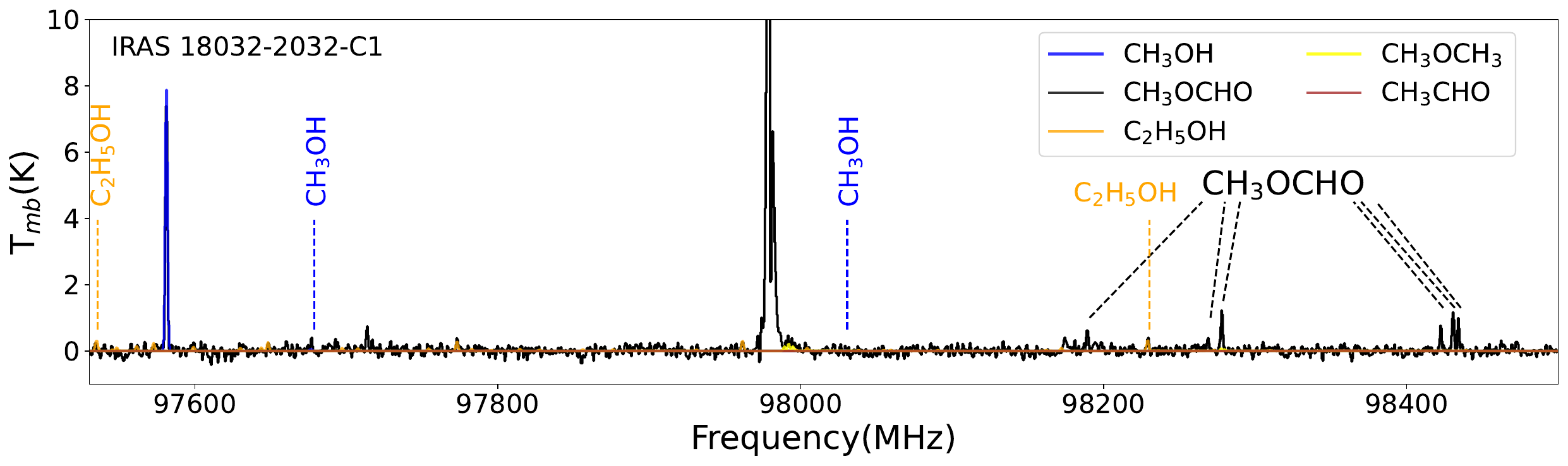}}
\quad
{\includegraphics[width=1\linewidth]{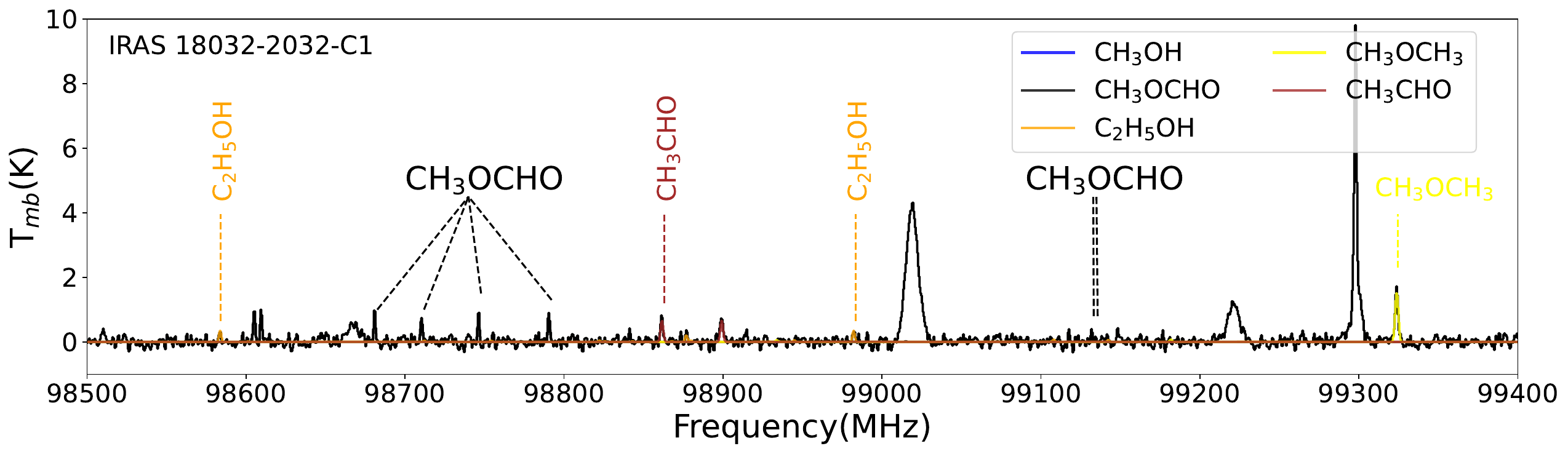}}
\quad
{\includegraphics[width=1\linewidth]{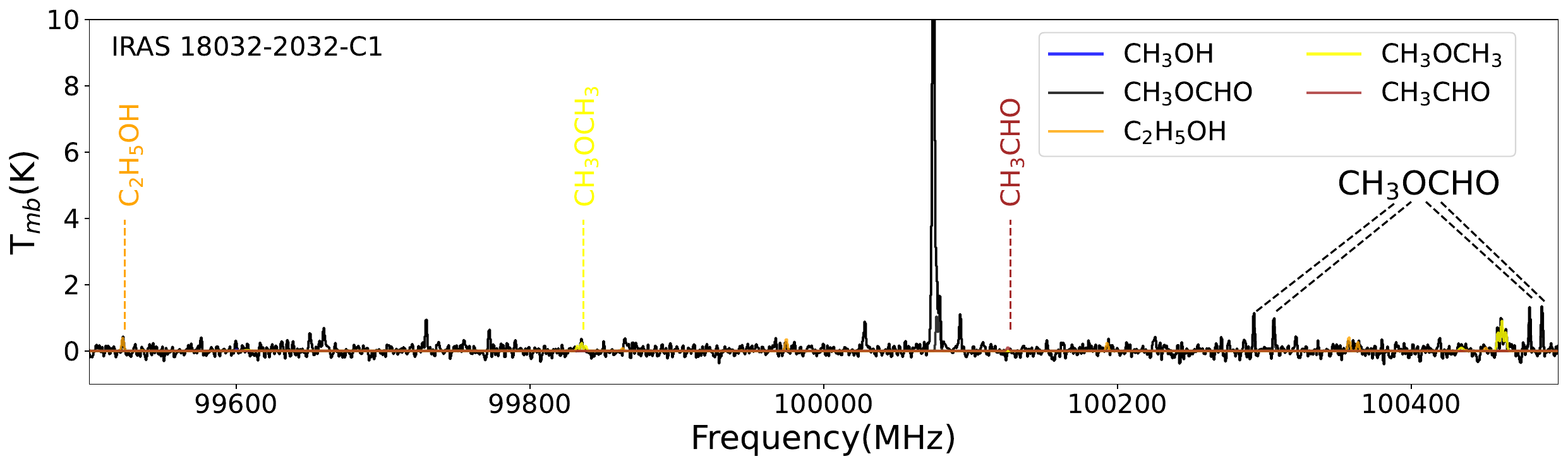}}
\quad
{\includegraphics[width=1\linewidth]{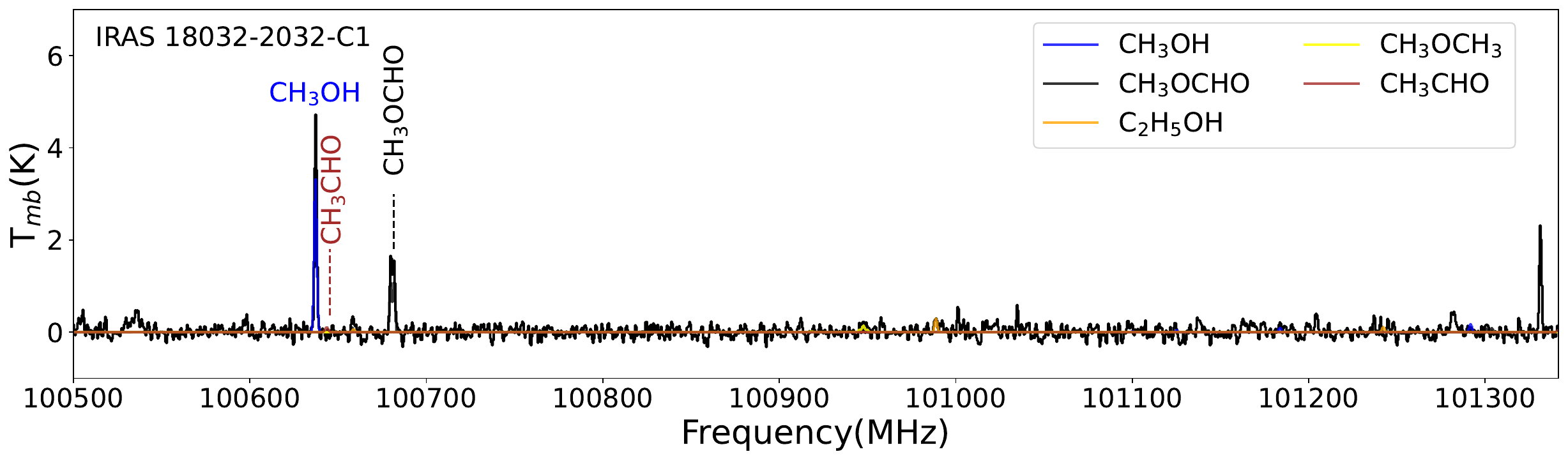}}
\quad
\caption{The figure shows the spectral lines of three cores (C1, C2, C3) in G9.62+0.19, with the molecular spectral lines in Table \ref{table1} annotated. The black line represents the observed spectrum, the red line corresponds to CH$_3$OH, the blue line represents CH$_3$OCHO, the deepskyblue line is C$_2$H$_5$OH, the green line corresponds to CH$_3$OCH$_3$, the orange line represents CH$_3$CHO, the purple line corresponds to CH$_3$COCH$_3$, and the deep pink line corresponds to C$_2$H$_5$CN.}
\label{fig:G9.62full}
\end{figure}

\clearpage
\setcounter{figure}{\value{figure}-1}
\begin{figure}
\centering
{\includegraphics[width=1\linewidth]{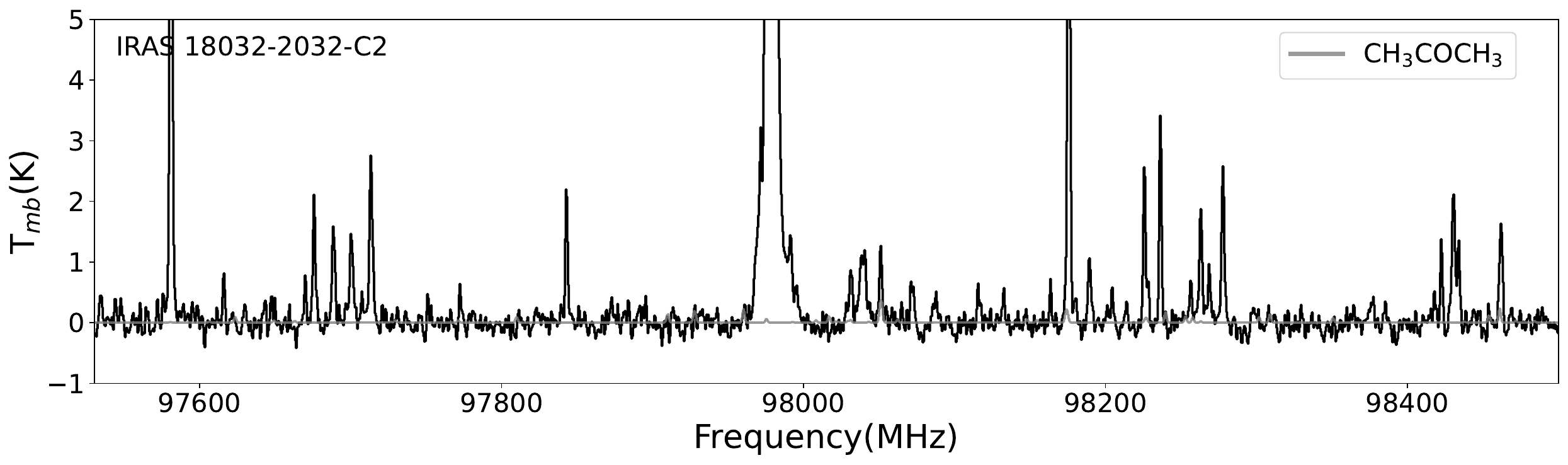}}
\quad
{\includegraphics[width=1\linewidth]{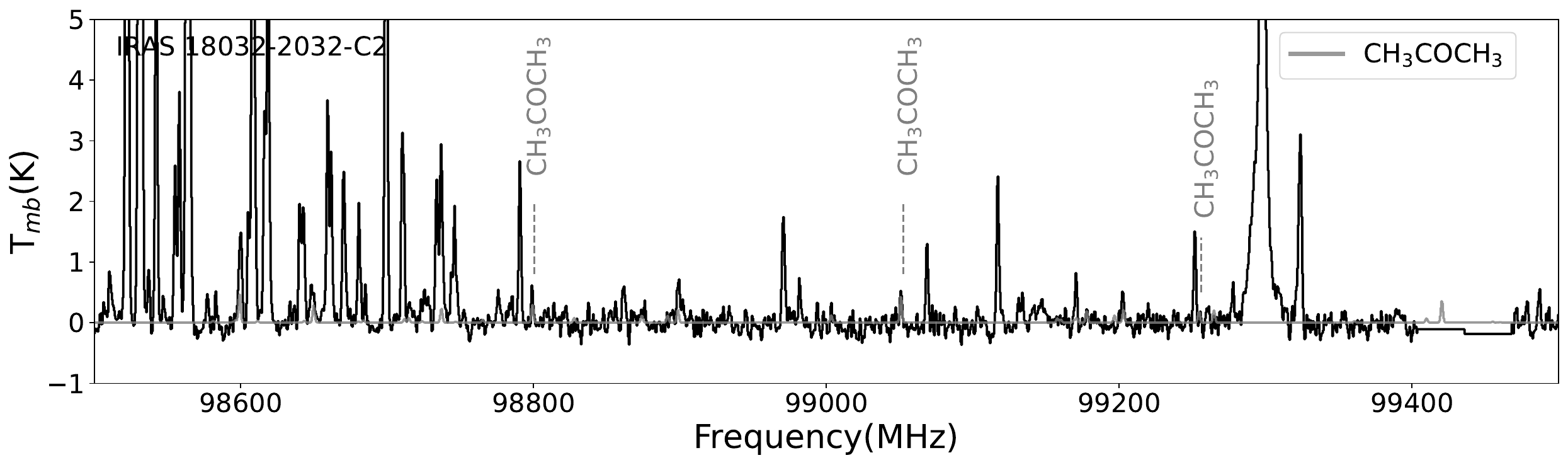}}
\quad
{\includegraphics[width=1\linewidth]{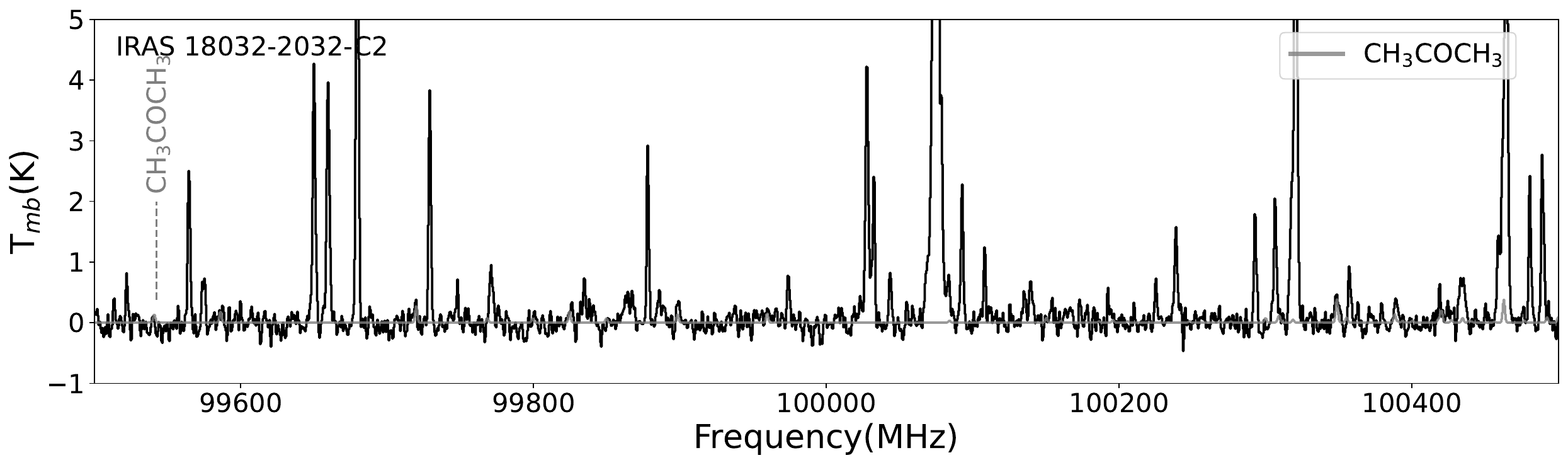}}
\quad
{\includegraphics[width=1\linewidth]{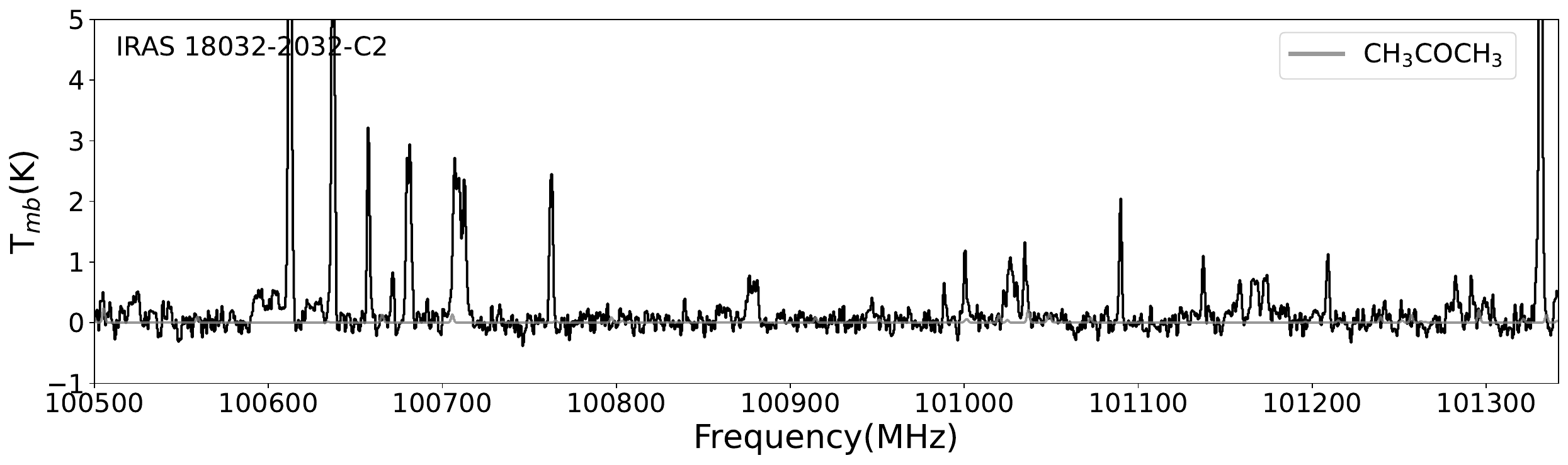}}
\quad
\caption{Continued.}
\end{figure}

\clearpage
\setcounter{figure}{\value{figure}-1}
\begin{figure}
\centering
{\includegraphics[width=1\linewidth]{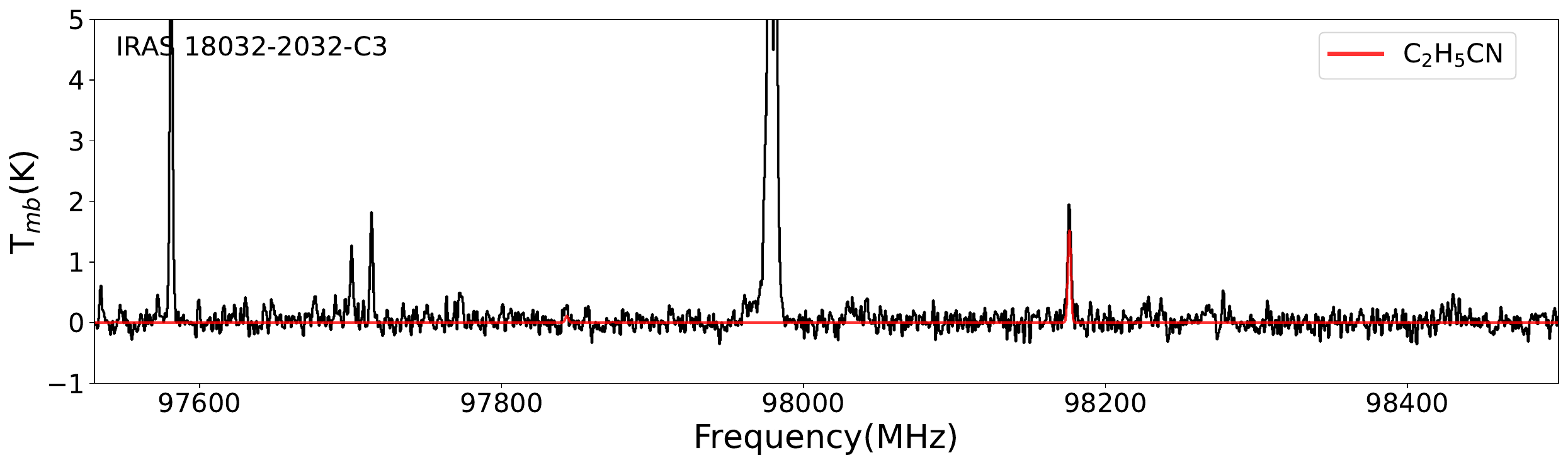}}
\quad
{\includegraphics[width=1\linewidth]{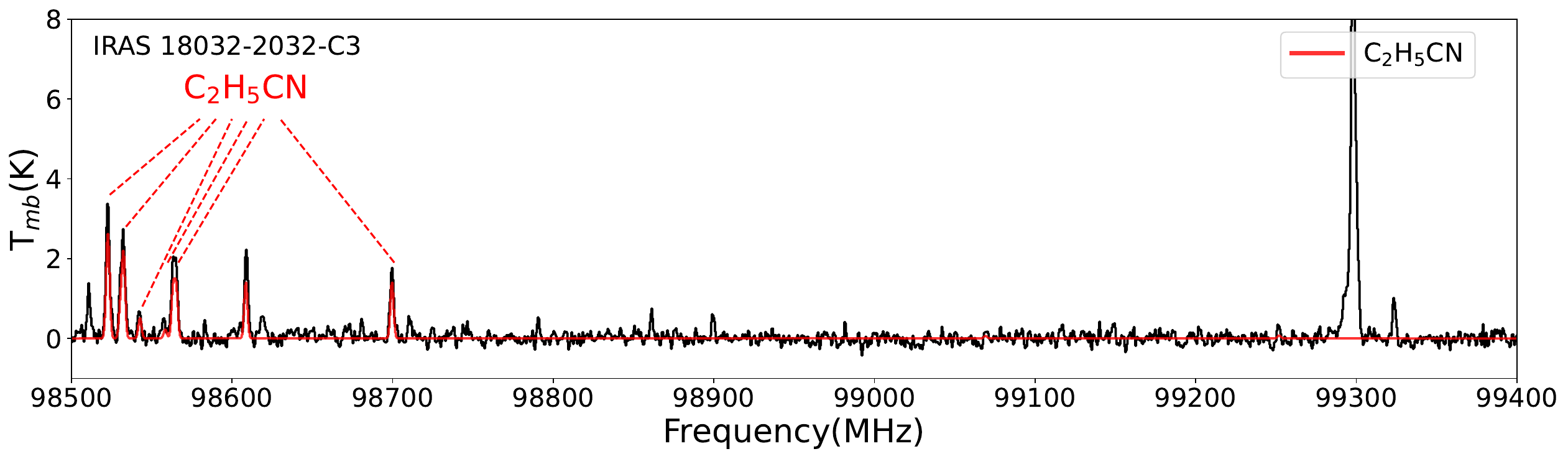}}
\quad
{\includegraphics[width=1\linewidth]{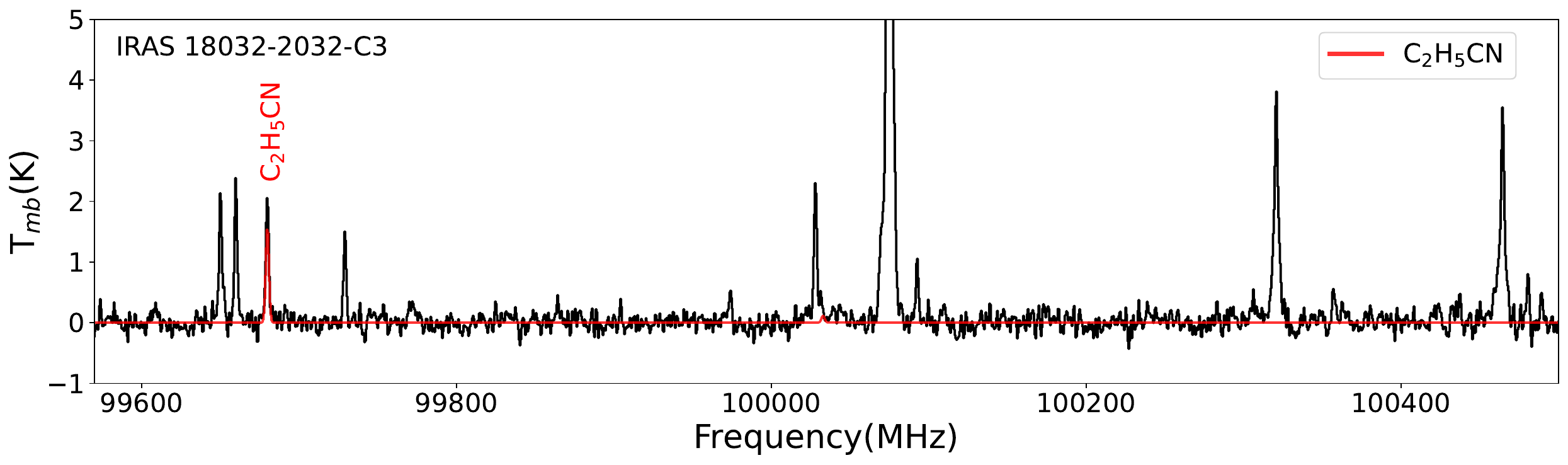}}
\quad
{\includegraphics[width=1\linewidth]{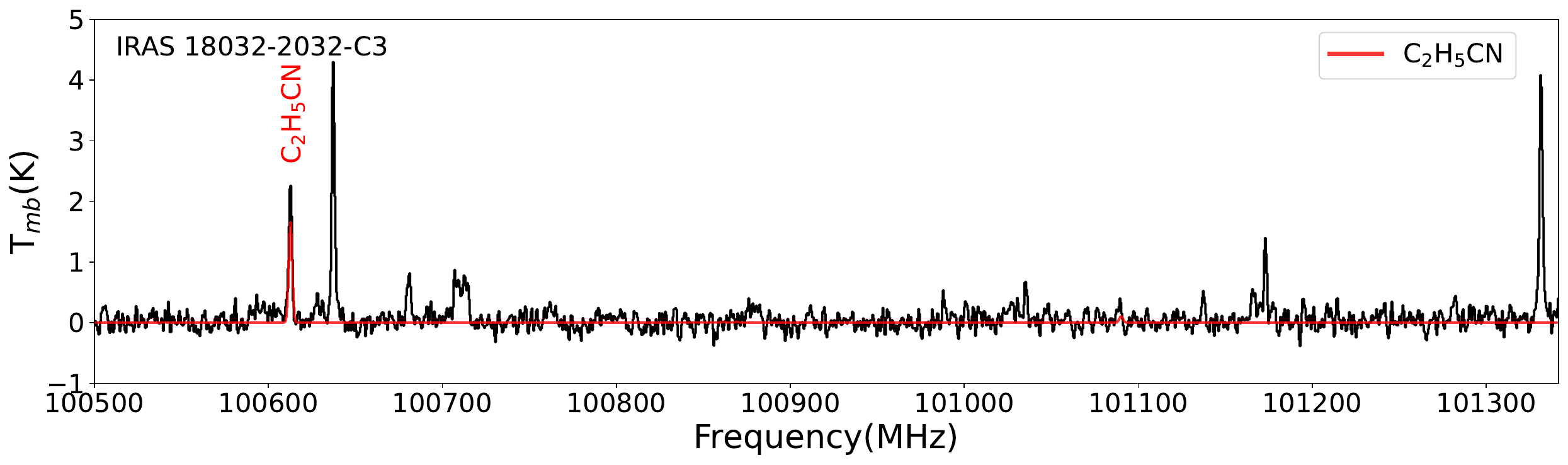}}
\quad
\caption{Continued.}
\end{figure}

\begin{figure}[hbp!]
\centering
{\includegraphics[height=5.31cm,width=6.91cm]{CH3OHandCH3OCHO.pdf}}
\quad
{\includegraphics[height=5.31cm,width=6.91cm]{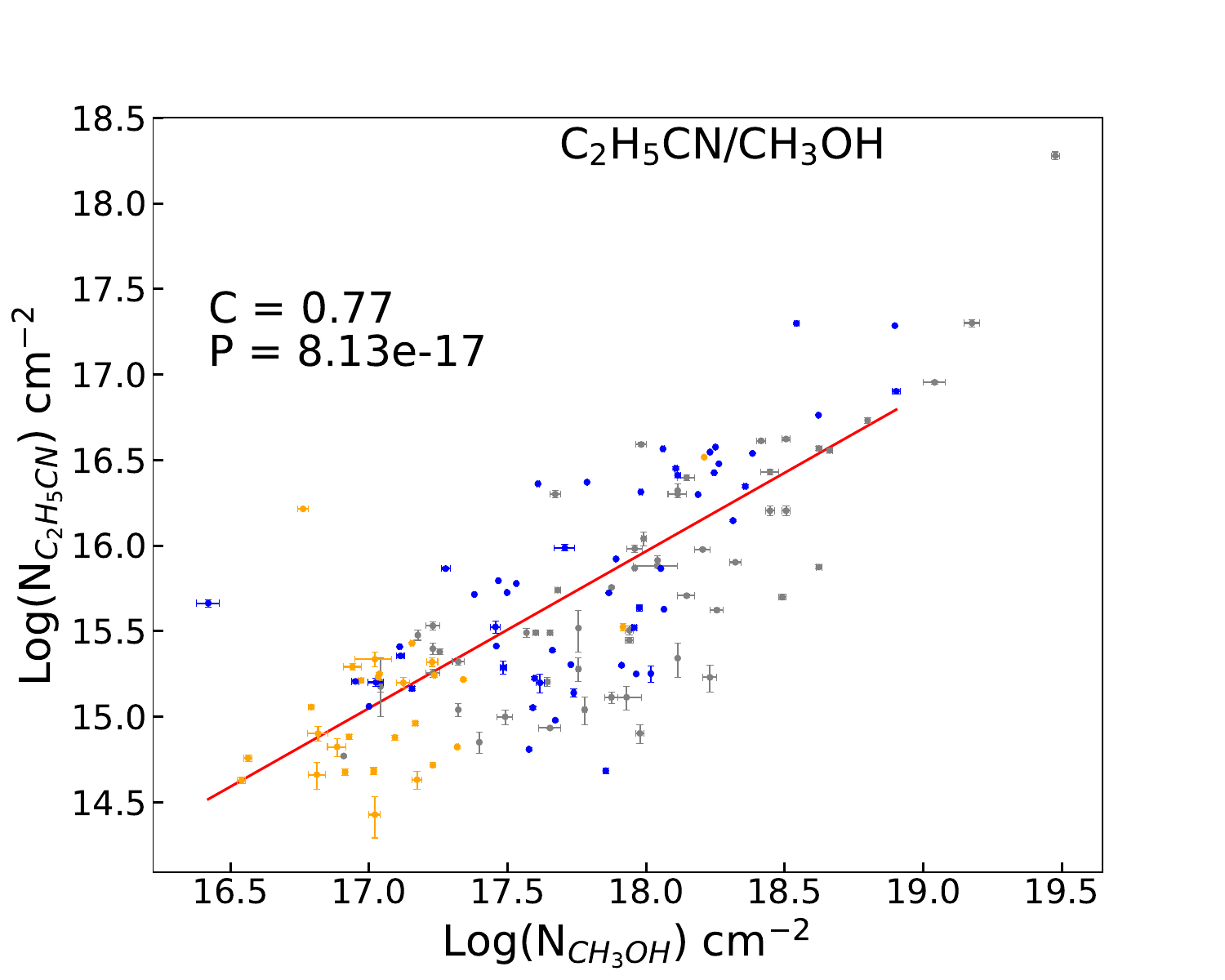}}
\quad
{\includegraphics[height=5.31cm,width=6.91cm]{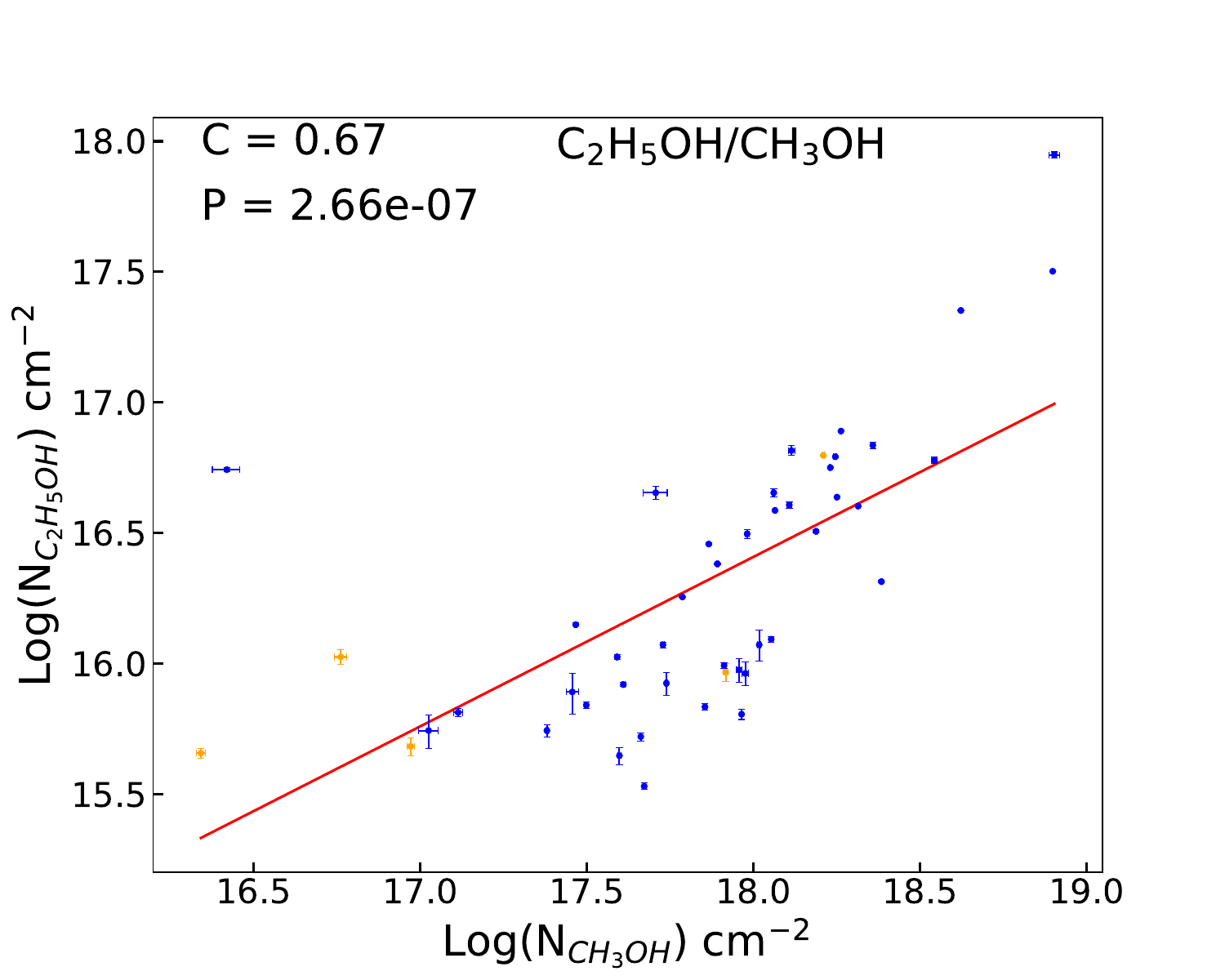}}
\quad
{\includegraphics[height=5.31cm,width=6.91cm]{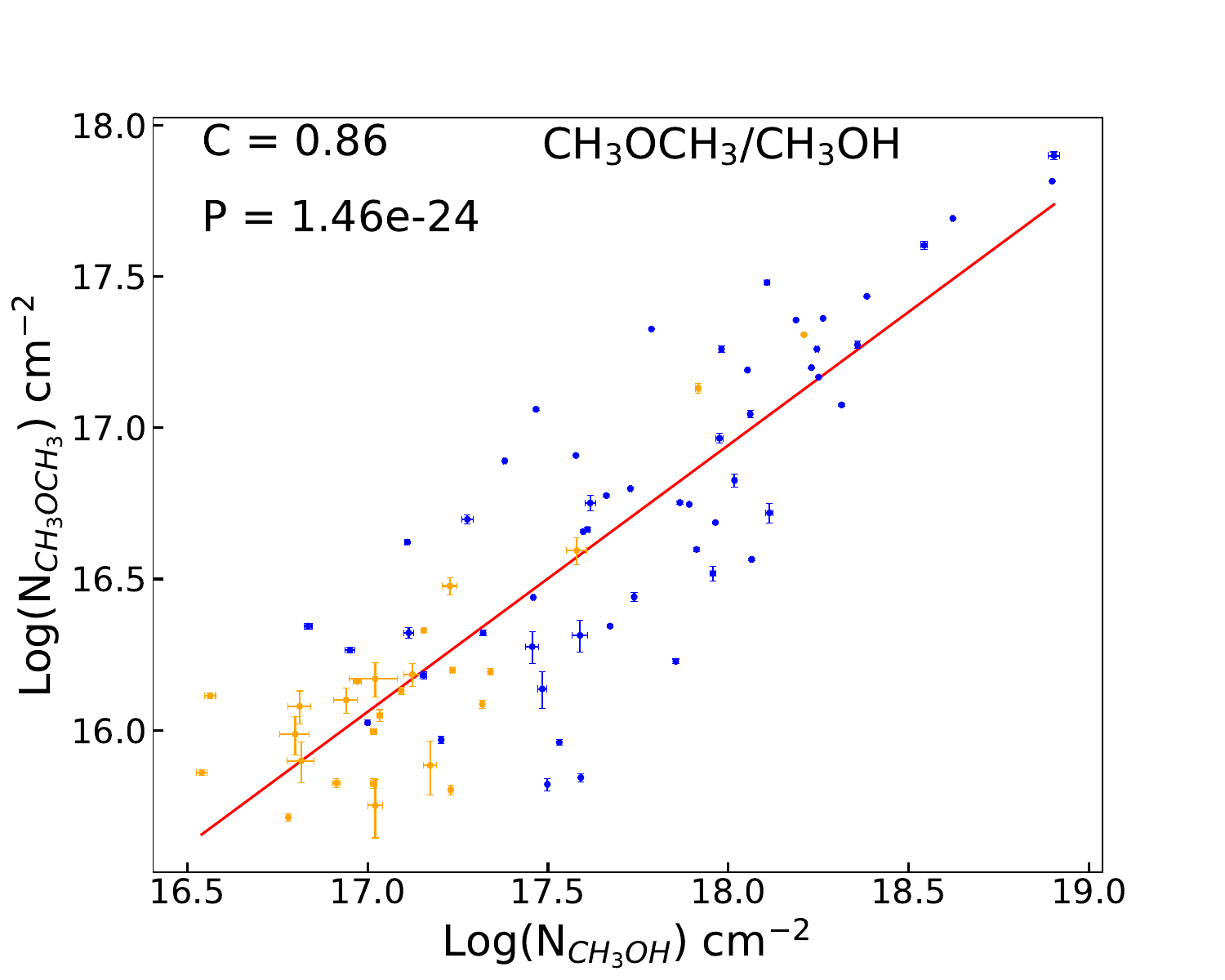}}
\quad
{\includegraphics[height=5.31cm,width=6.91cm]{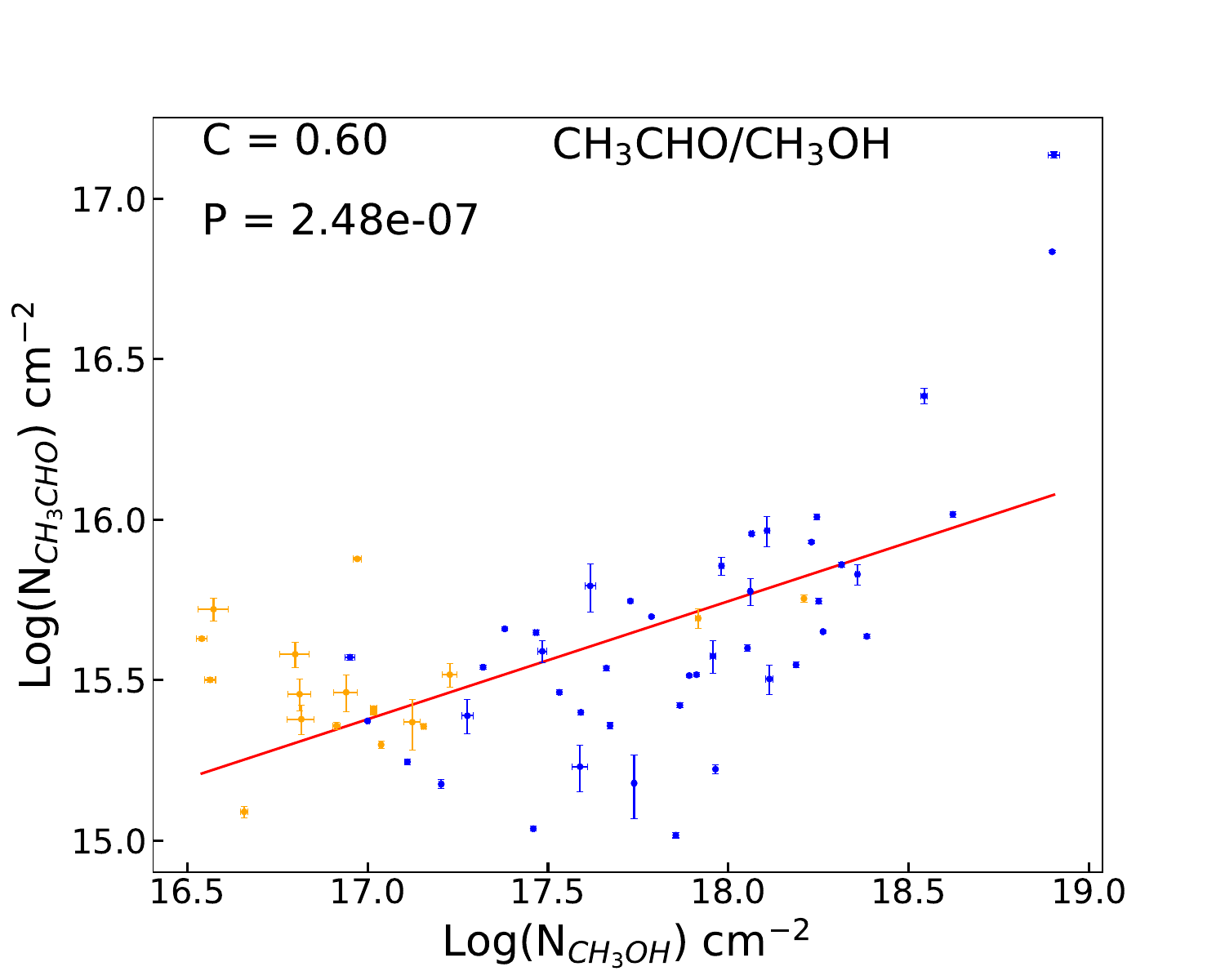}}
\quad
{\includegraphics[height=5.31cm,width=6.91cm]{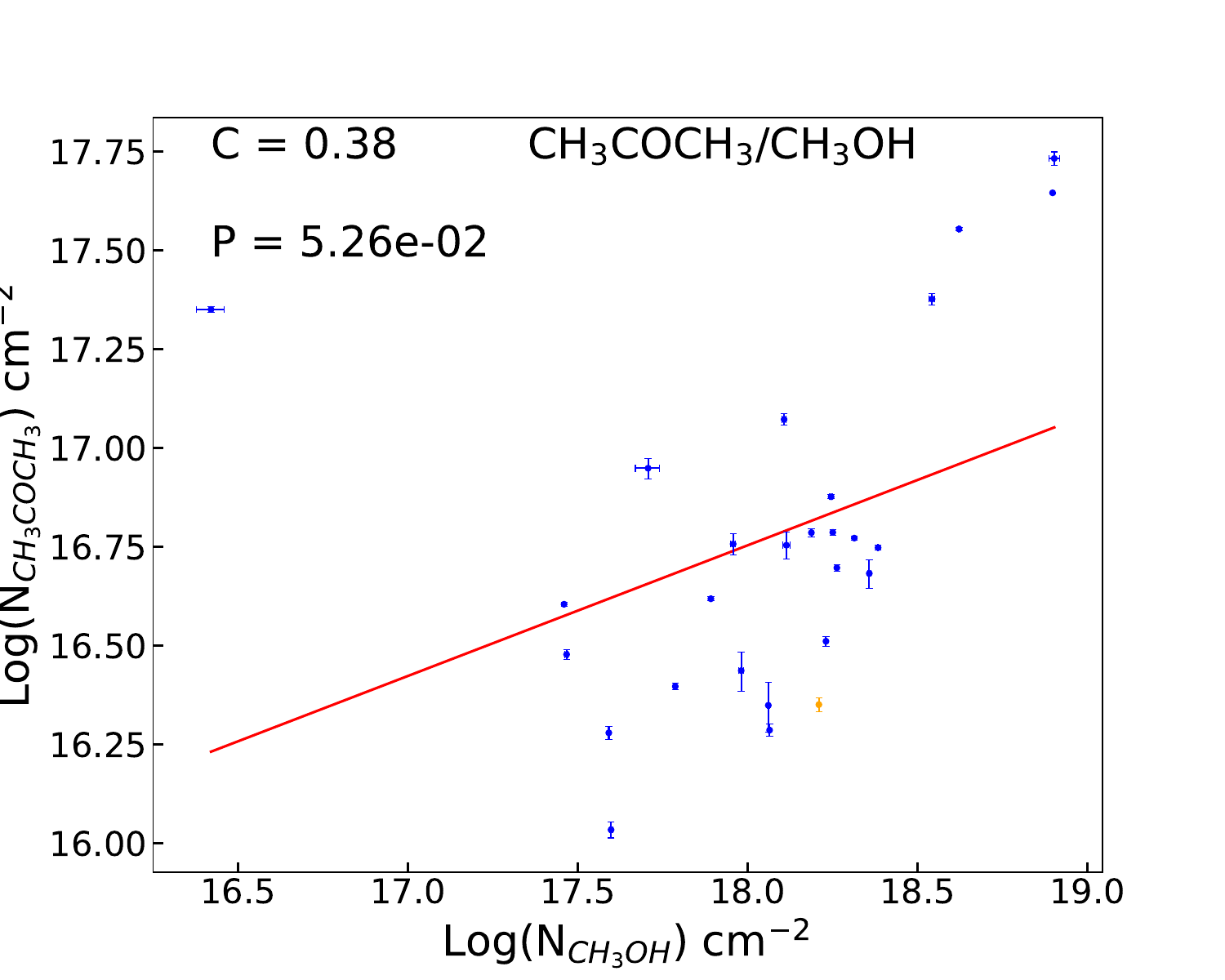}}
\quad
{\includegraphics[height=5.31cm,width=6.91cm]{CH3OCHOandC2H5CN.pdf}}
\quad
{\includegraphics[height=5.31cm,width=6.91cm]{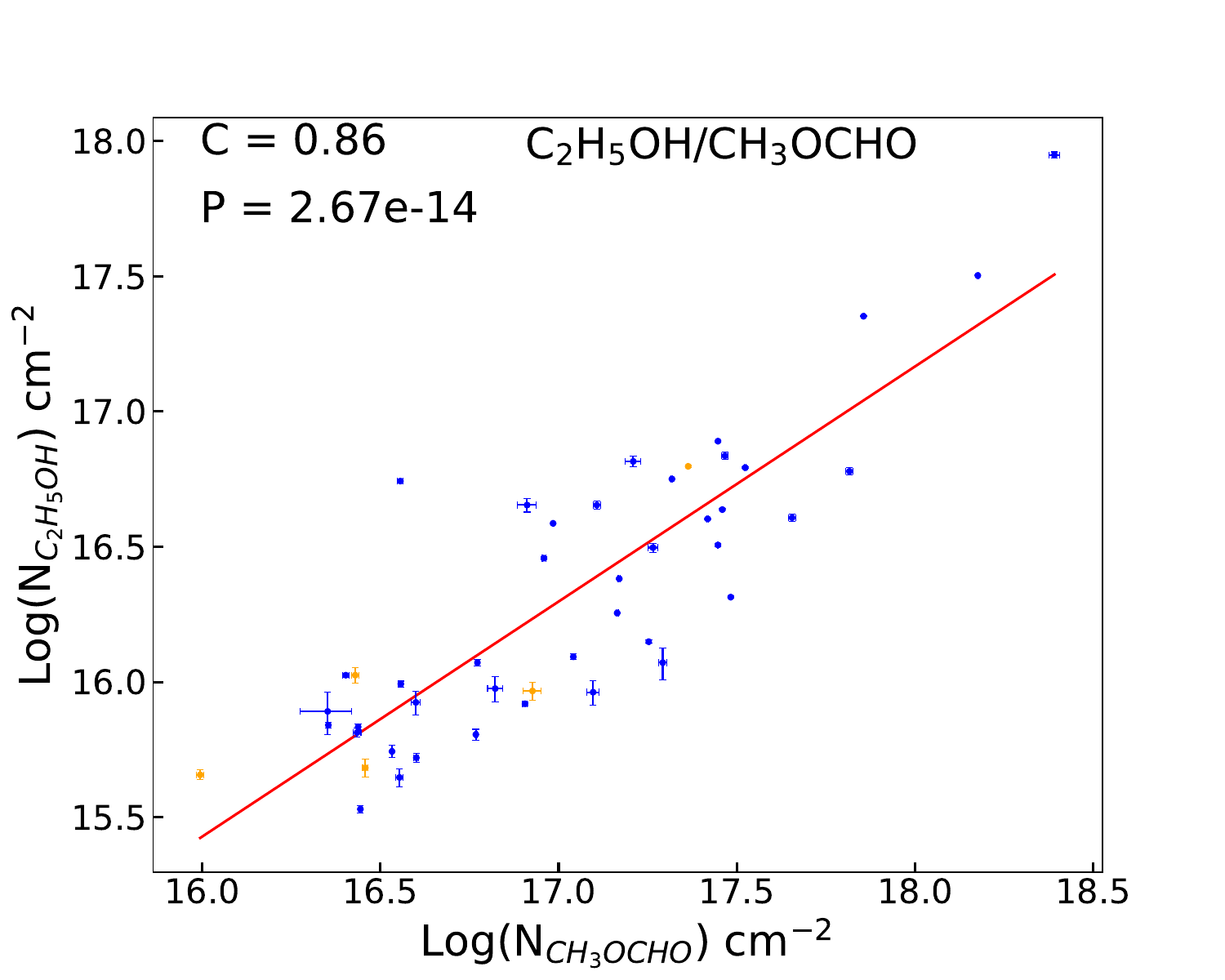}}
\quad
\caption{Correlations of molecular column densities, and 'C' denotes the correlation coefficient. The red line indicates the linear regression.}
\label{fig:Ncorr_more}
\end{figure}

\clearpage
\setcounter{figure}{\value{figure}-1}
\begin{figure}
\centering
{\includegraphics[height=5.31cm,width=6.91cm]{CH3OCHOandCH3OCH3.pdf}}
\quad
{\includegraphics[height=5.31cm,width=6.91cm]{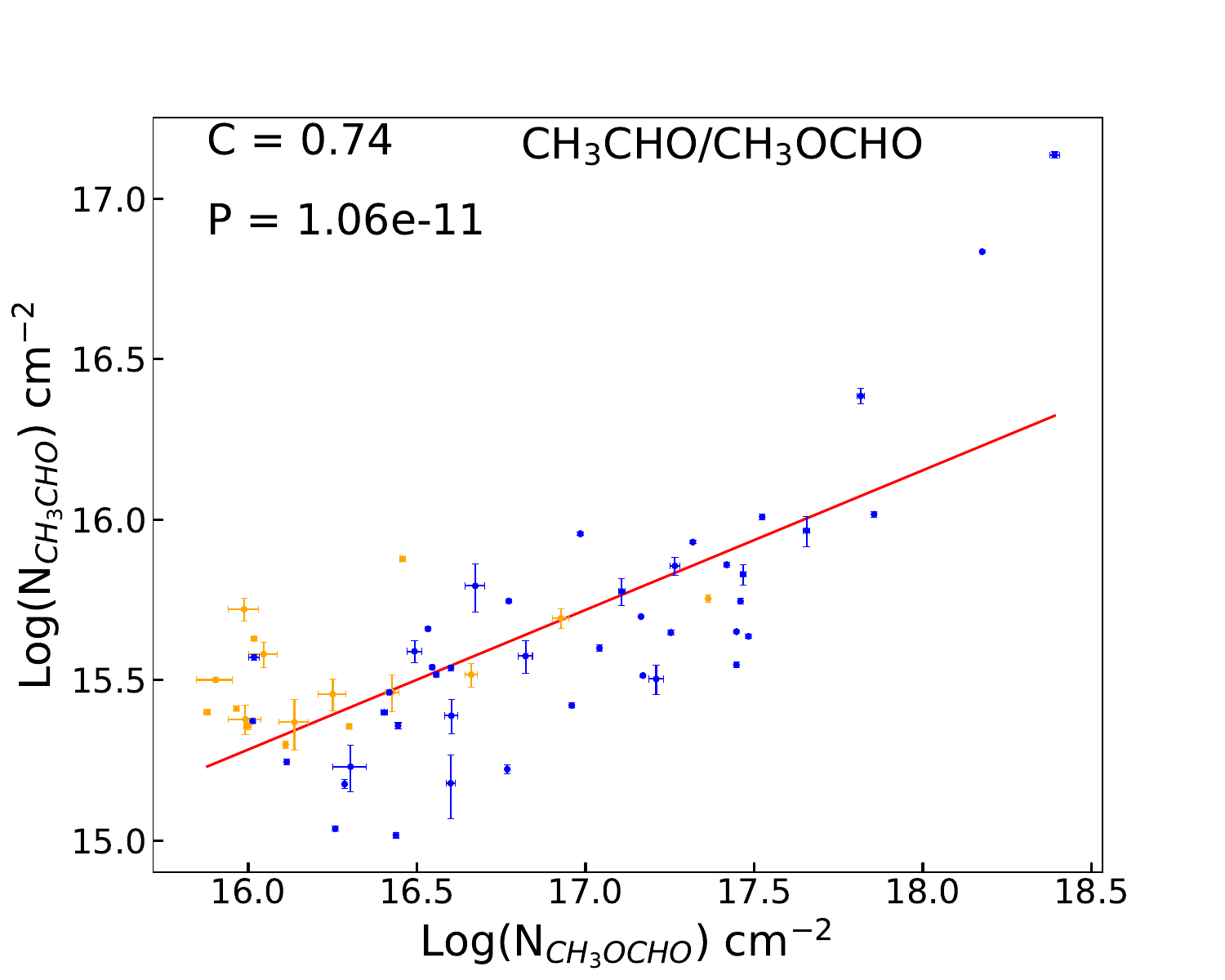}}
\quad
{\includegraphics[height=5.31cm,width=6.91cm]{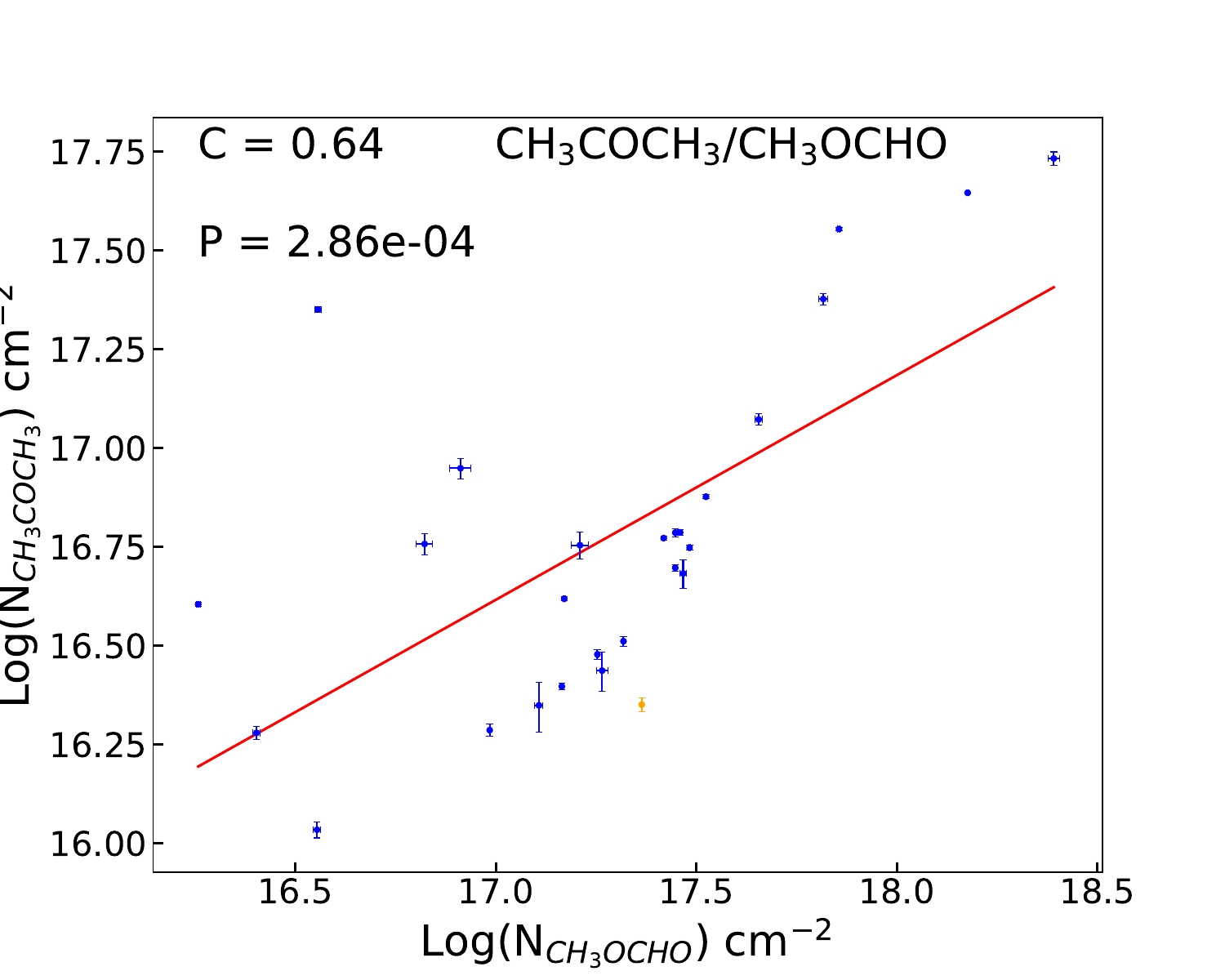}}
\quad
{\includegraphics[height=5.31cm,width=6.91cm]{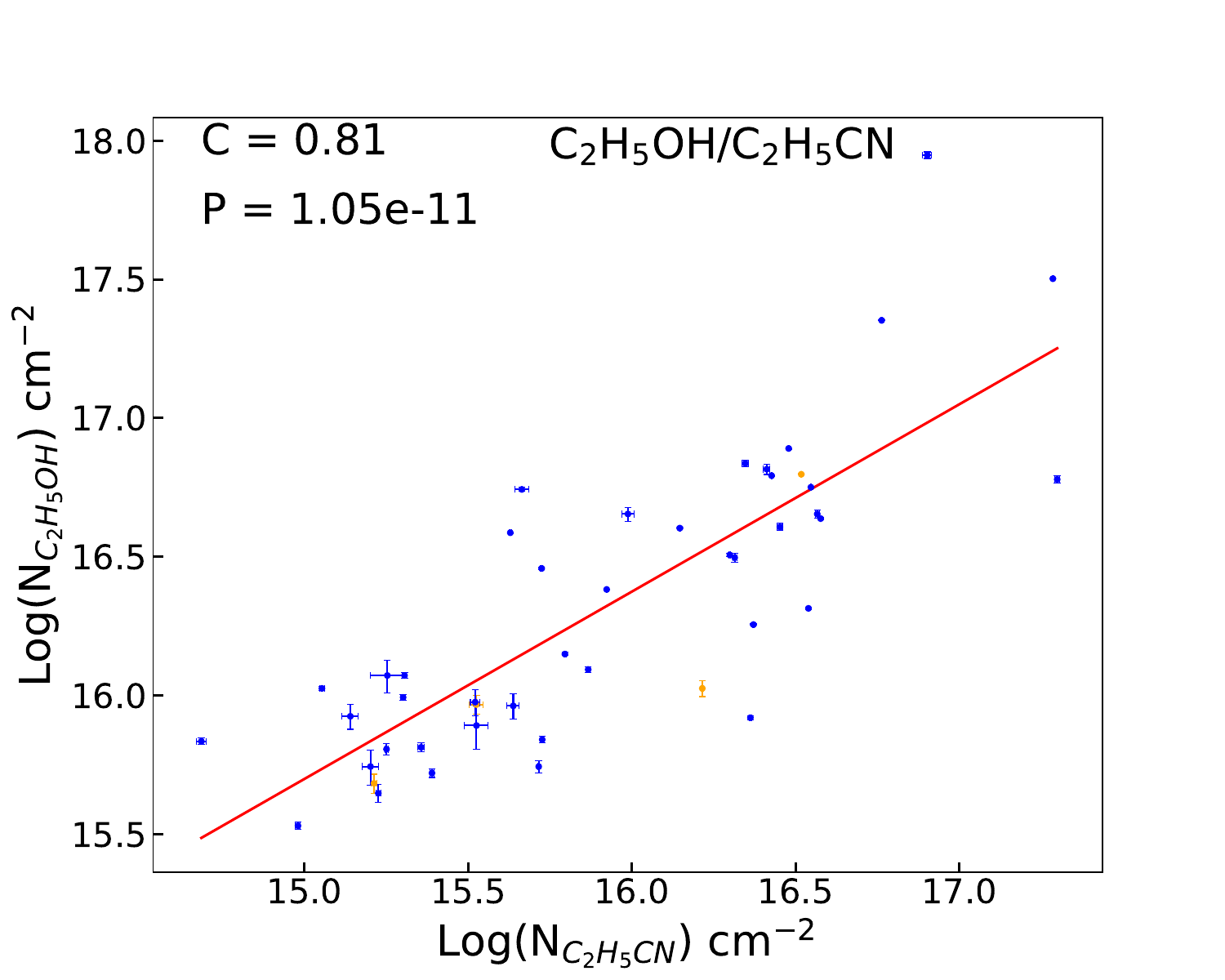}}
\quad
{\includegraphics[height=5.31cm,width=6.91cm]{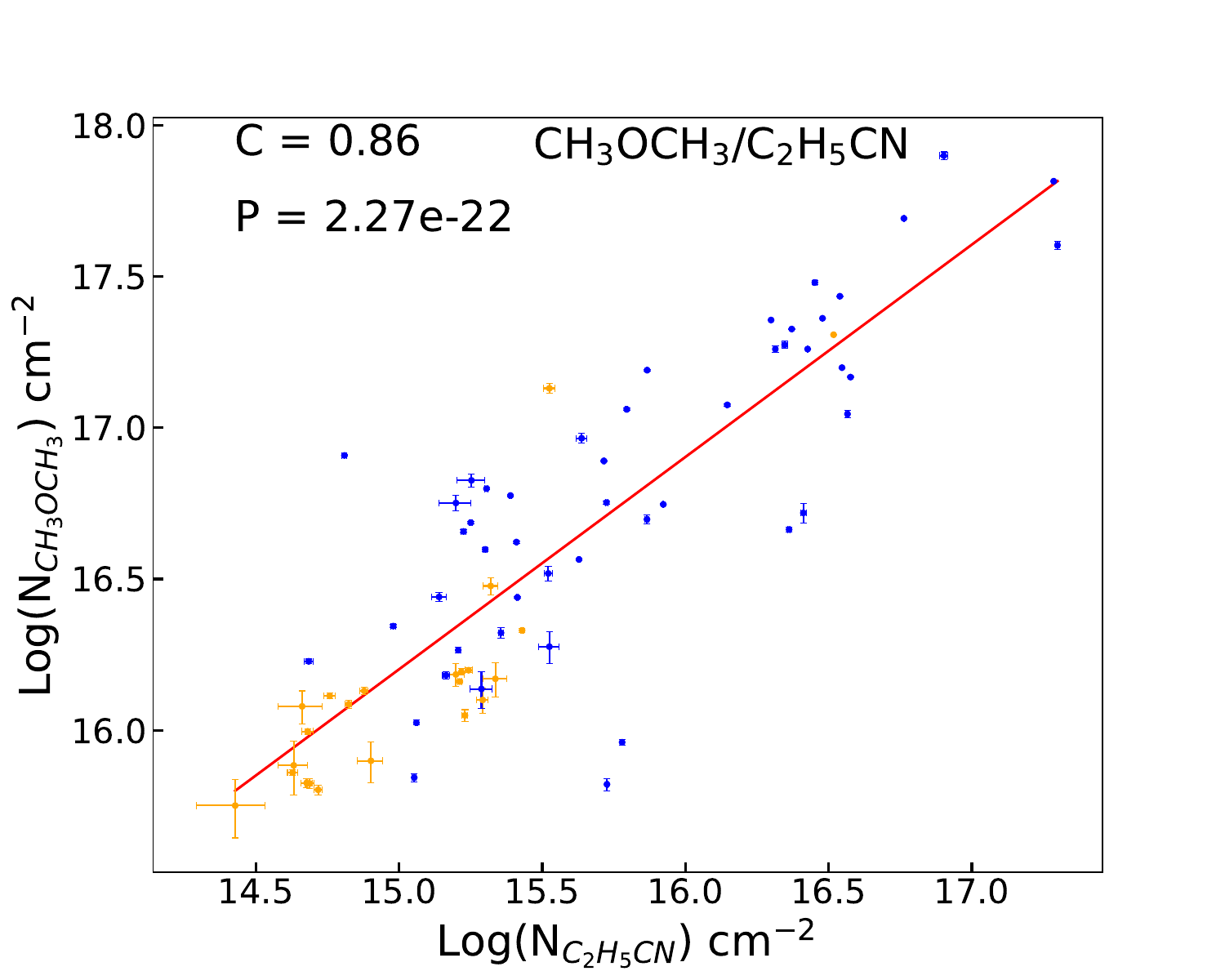}}
\quad
{\includegraphics[height=5.31cm,width=6.91cm]{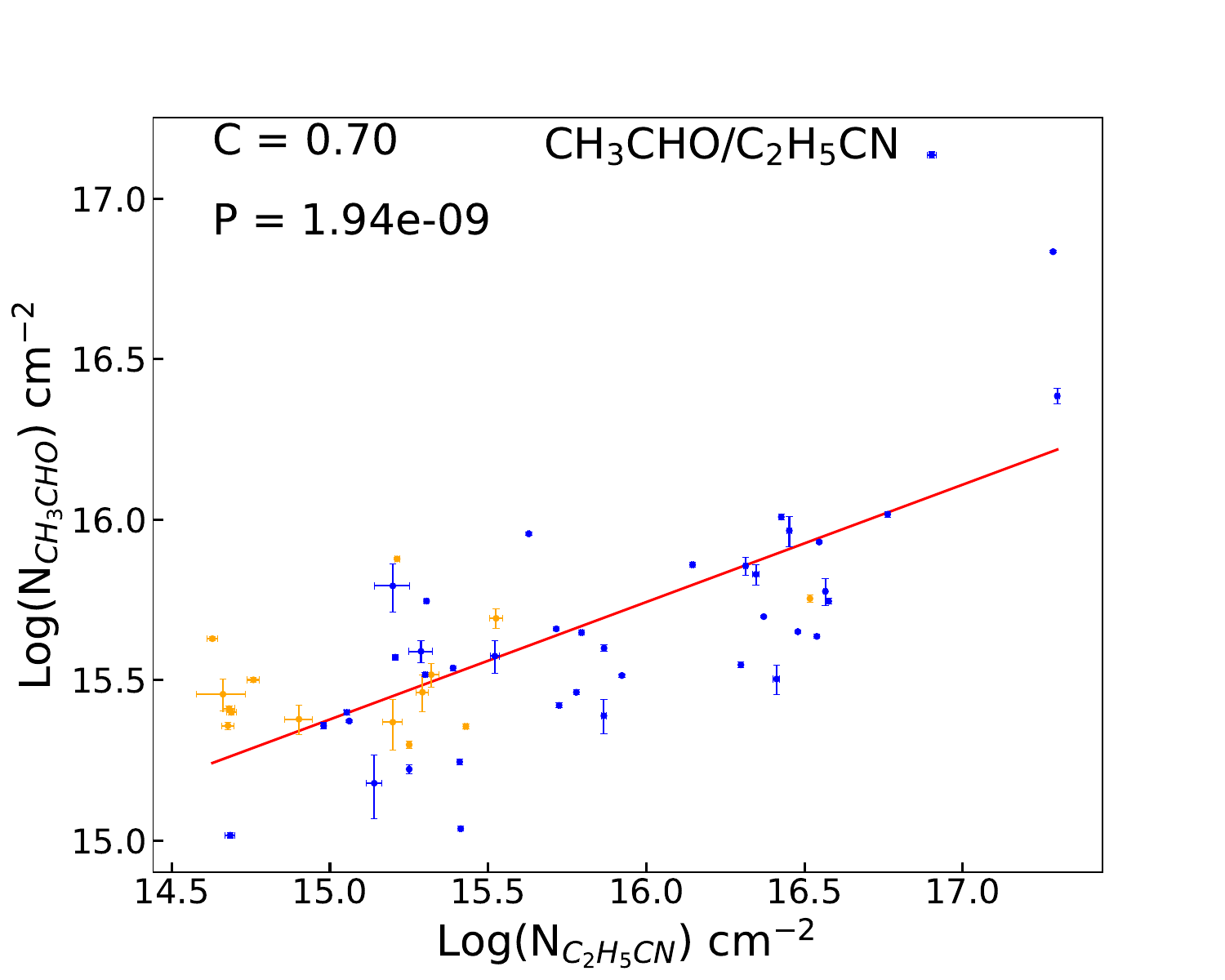}}
\quad
{\includegraphics[height=5.31cm,width=6.91cm]{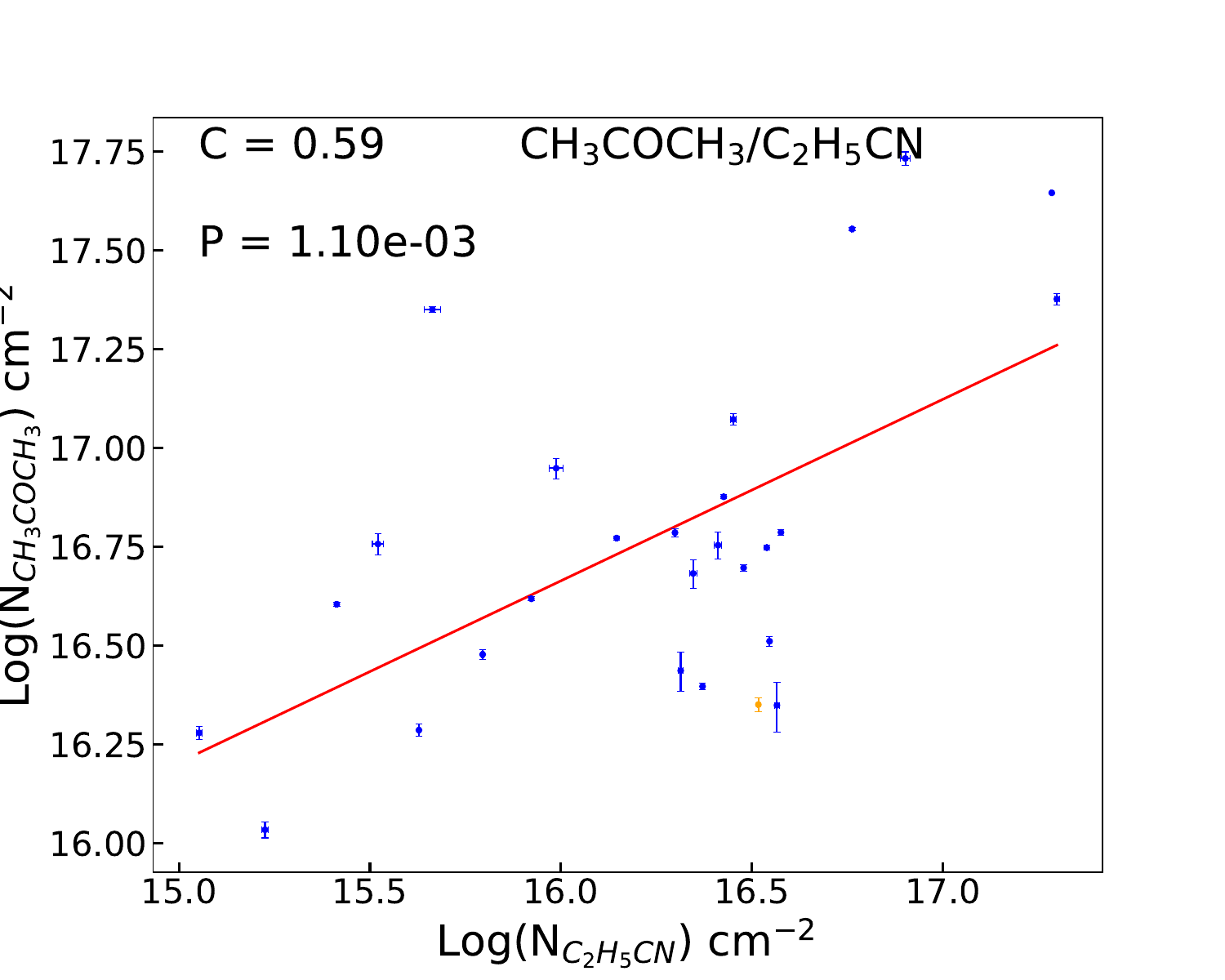}}
\quad
{\includegraphics[height=5.31cm,width=6.91cm]{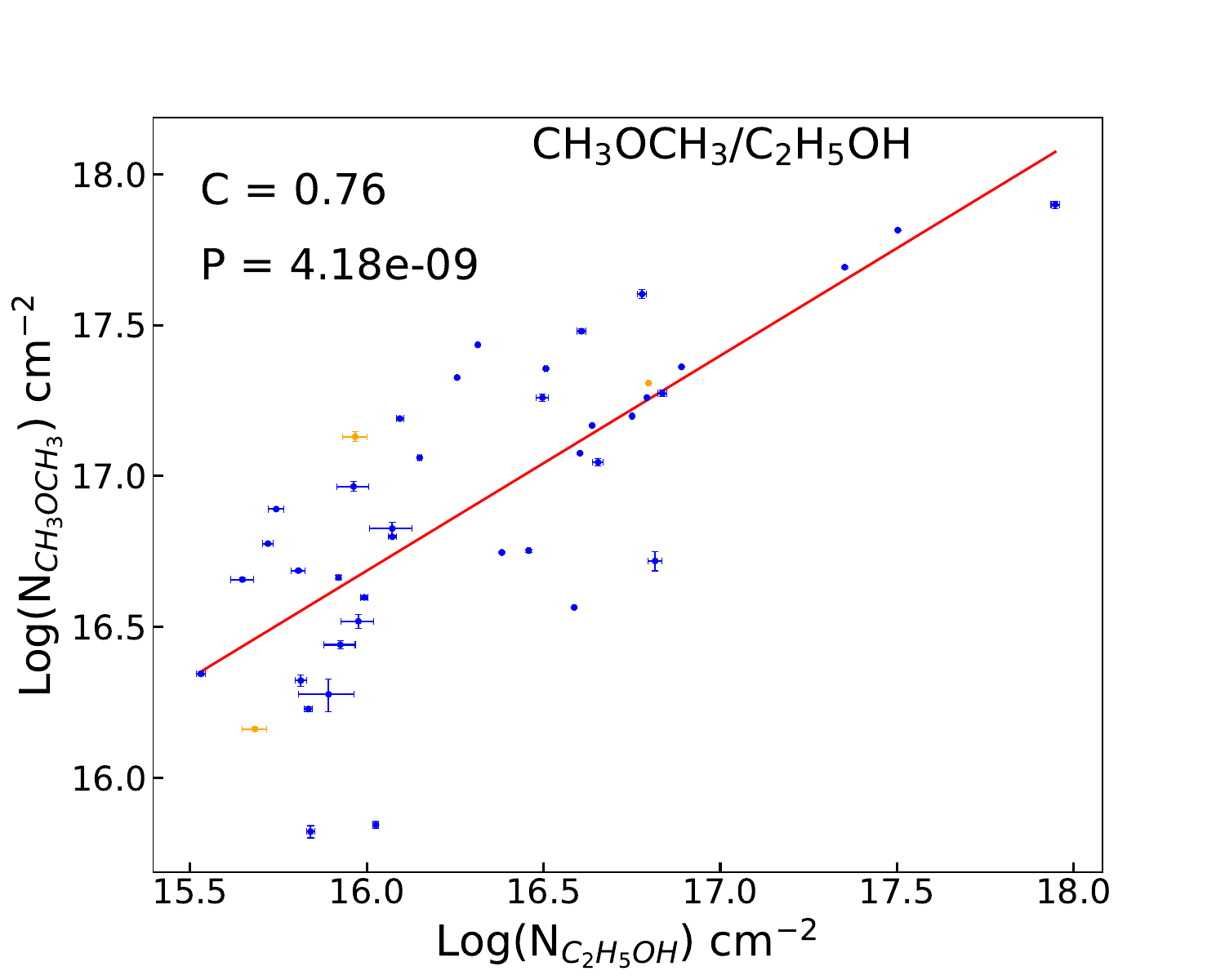}}
\quad
\caption{Continued.}
\end{figure}

\clearpage
\setcounter{figure}{\value{figure}-1}
\begin{figure}
  \centering 
{\includegraphics[height=5.31cm,width=6.91cm]{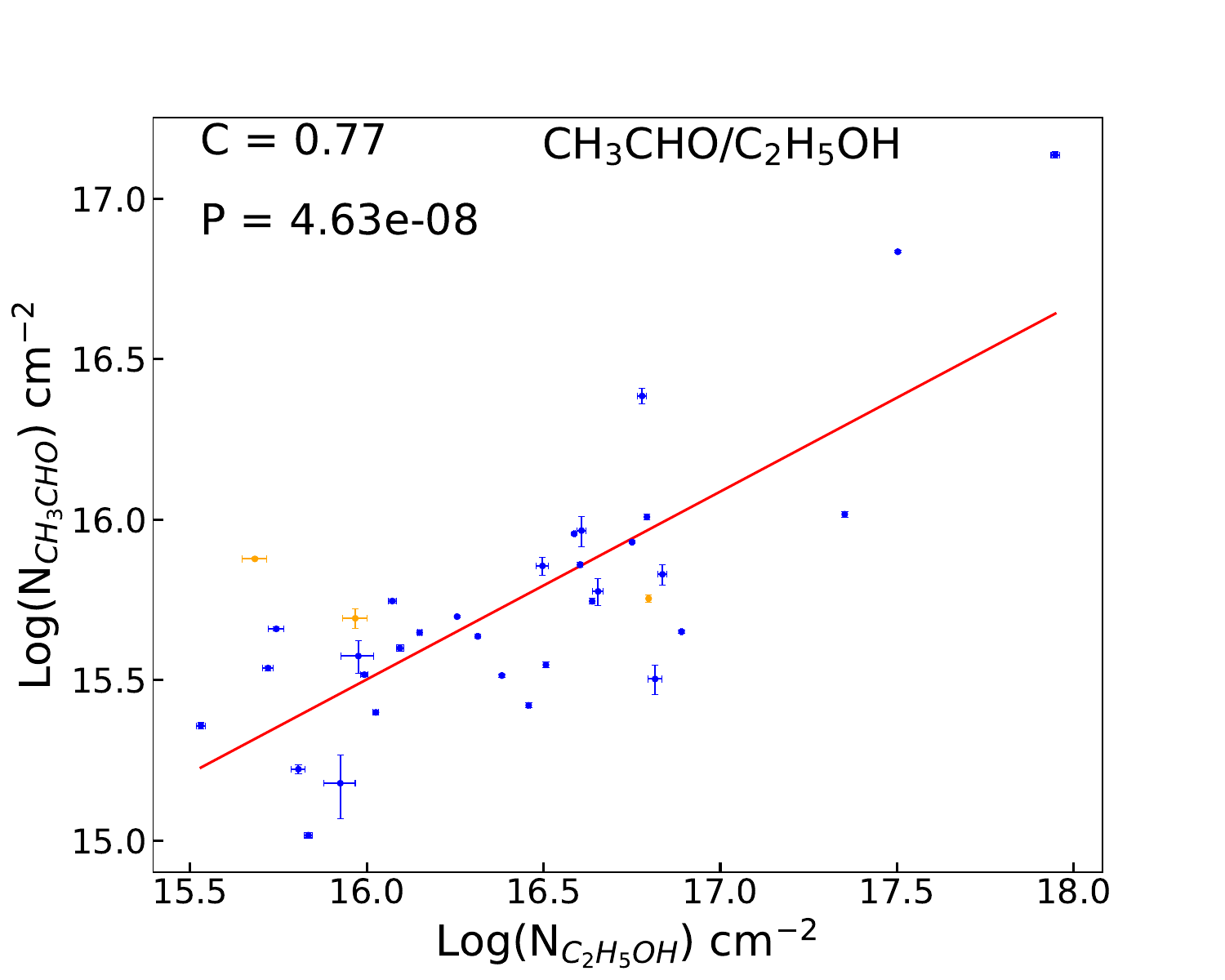}}
\quad
{\includegraphics[height=5.31cm,width=6.91cm]{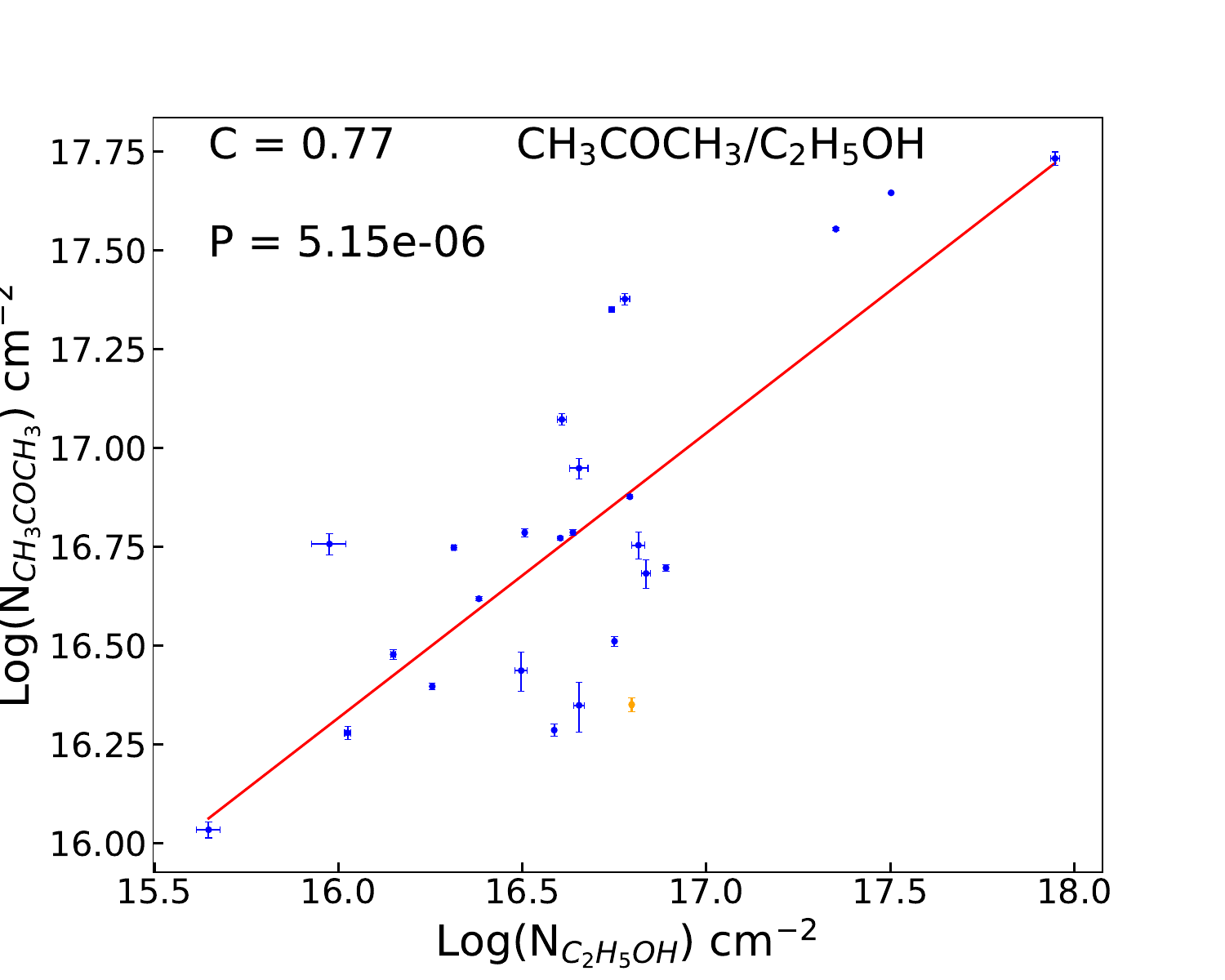}}
\quad
{\includegraphics[height=5.31cm,width=6.91cm]{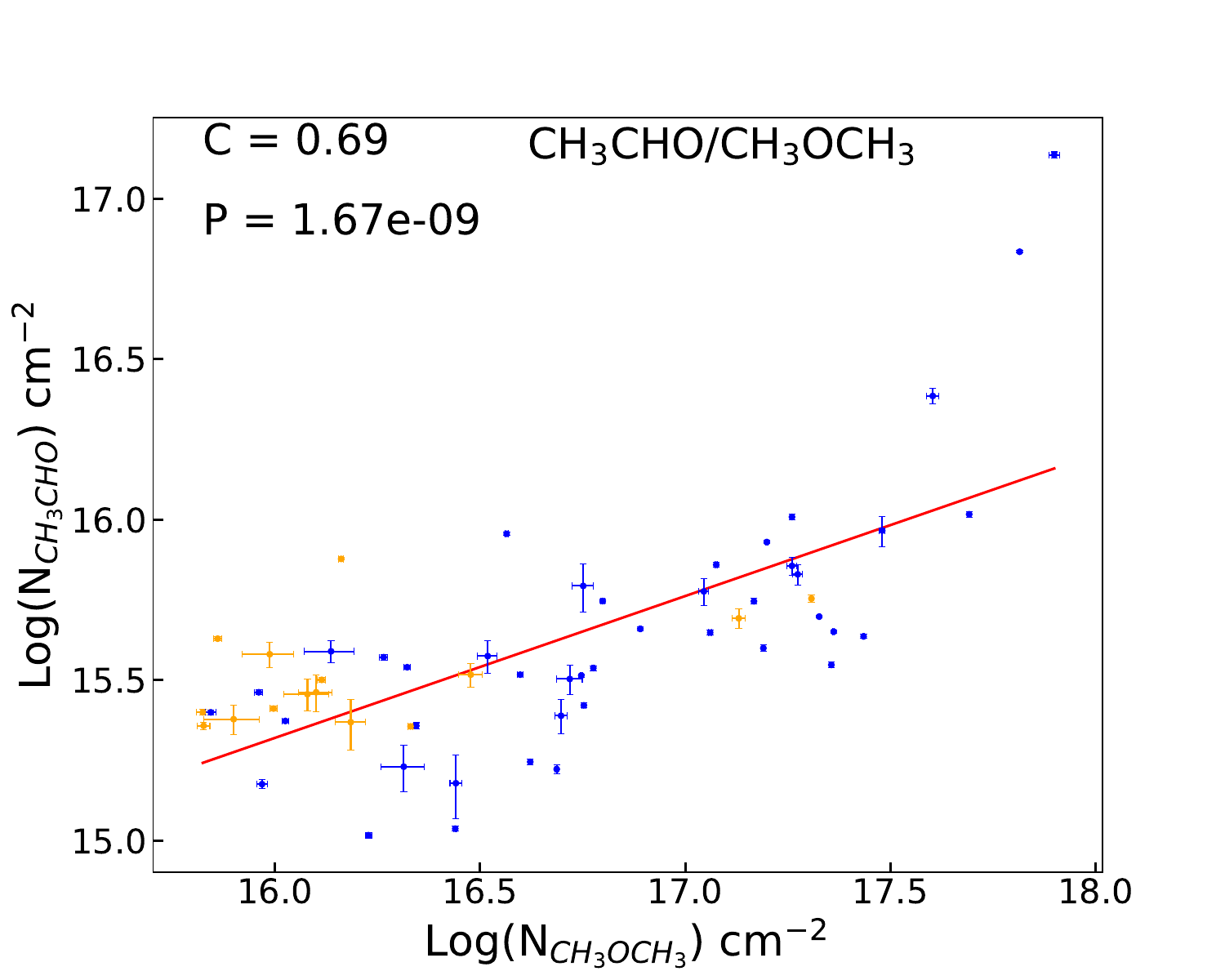}}
\quad
{\includegraphics[height=5.31cm,width=6.91cm]{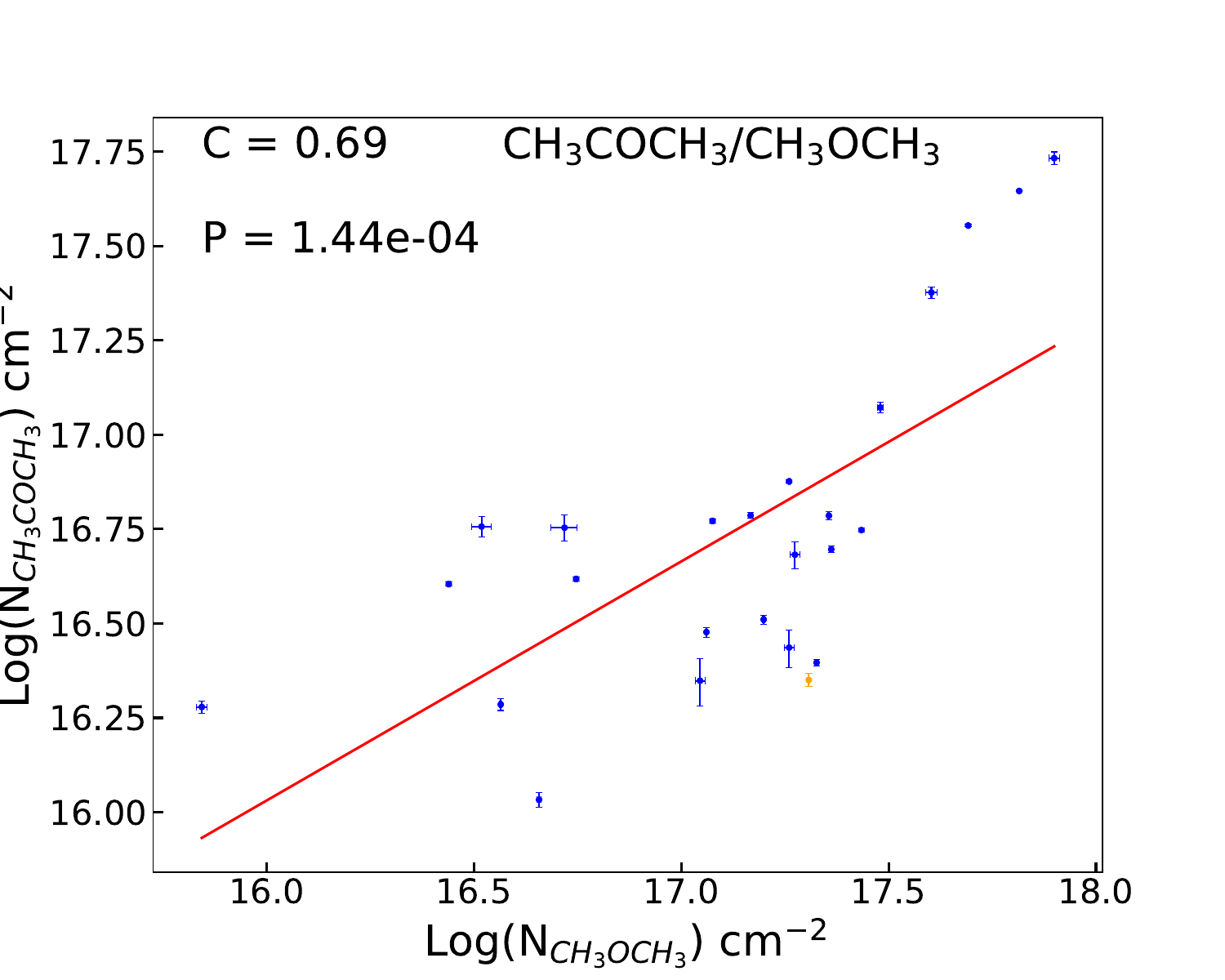}}
\quad
{\includegraphics[height=5.31cm,width=6.91cm]{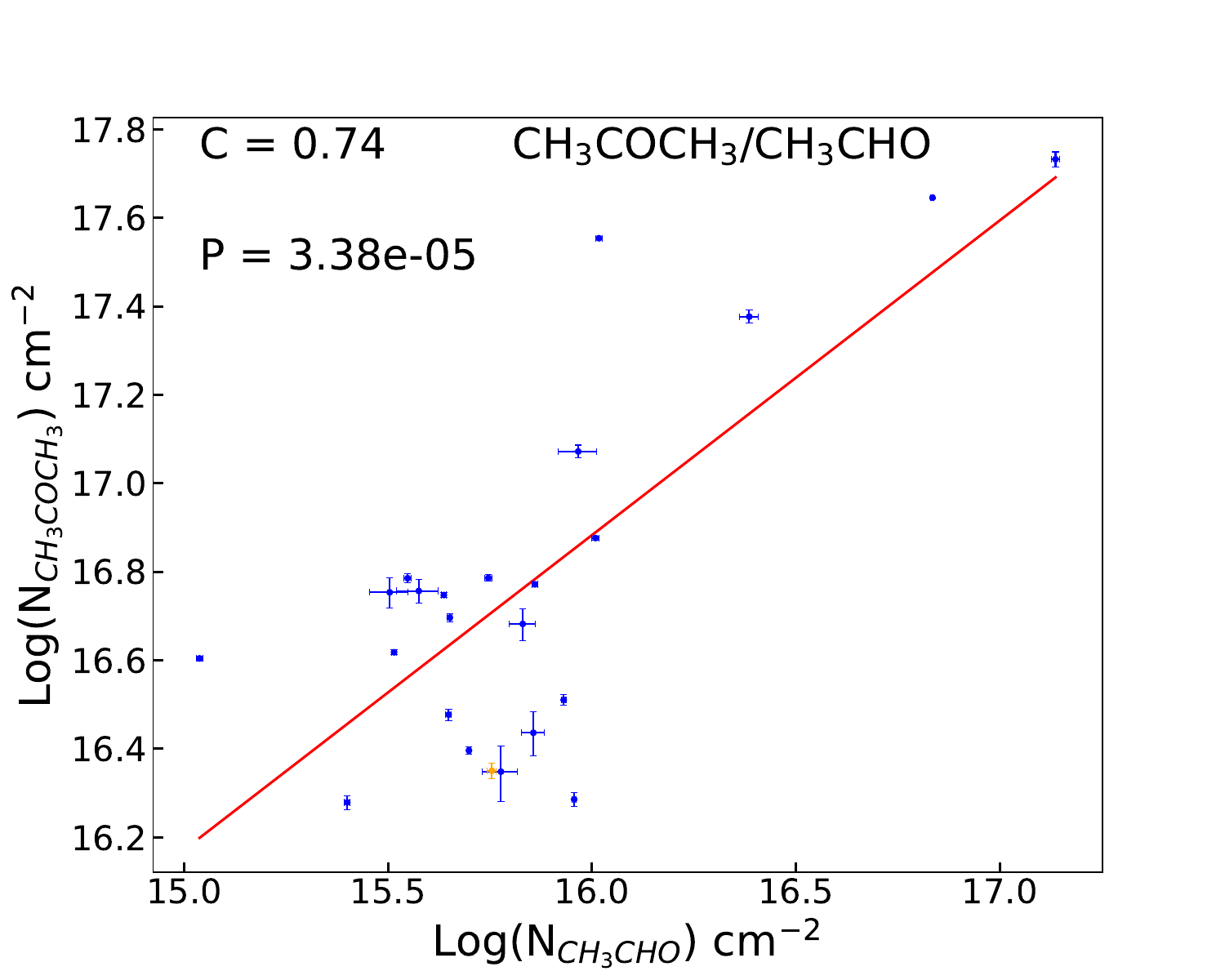}}
\quad
\caption{Continued.}
\end{figure}

\begin{figure}
\centering
{\includegraphics[width=6.5cm,height=6.5cm]{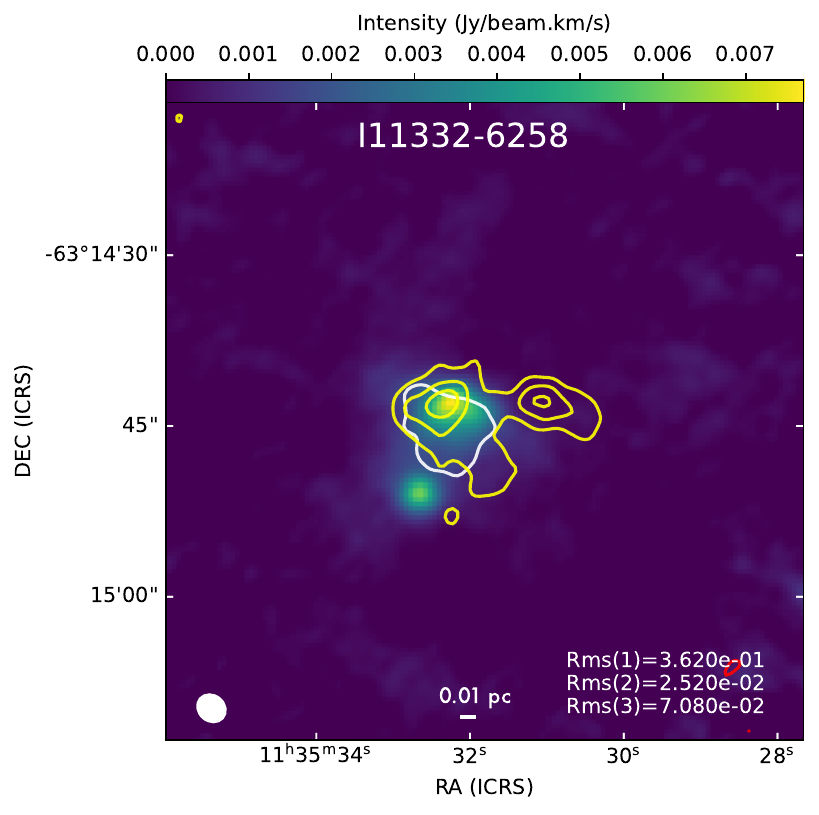}}
\quad
{\includegraphics[width=6.5cm,height=6.5cm]{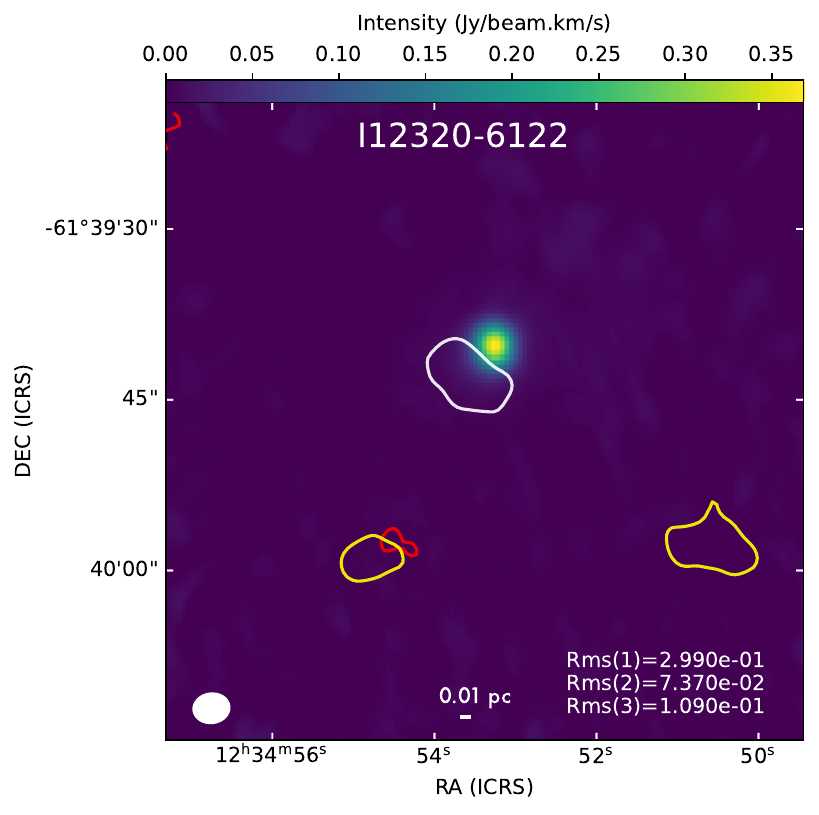}}
\quad
{\includegraphics[width=6.5cm,height=6.5cm]{H13CO_I13079-6218.pdf}}
\quad
{\includegraphics[width=6.5cm,height=6.5cm]{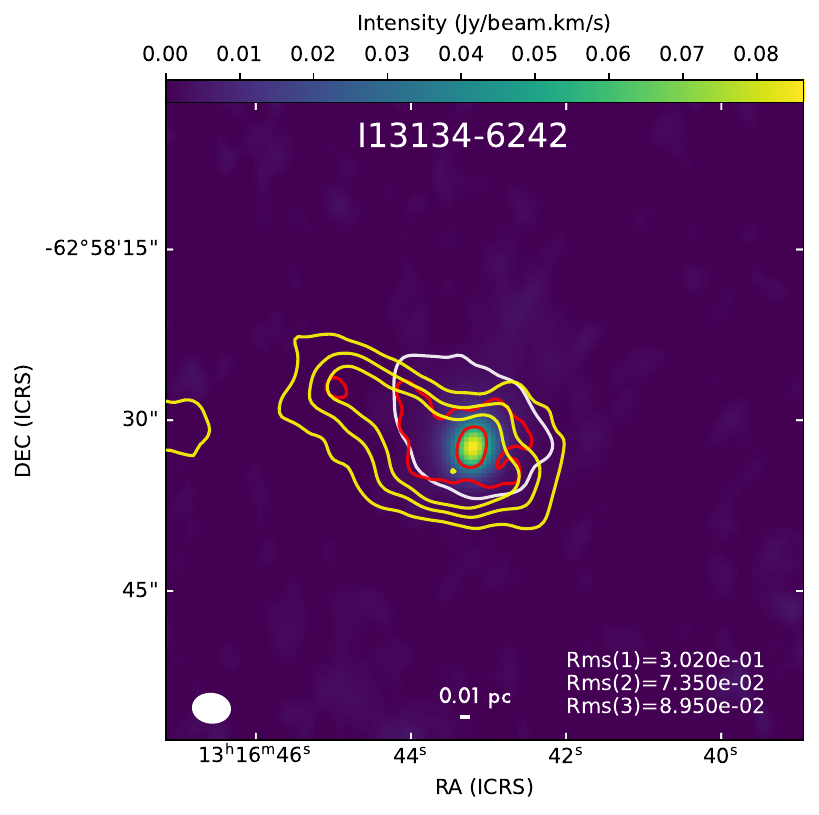}}
\quad
{\includegraphics[width=6.5cm,height=6.5cm]{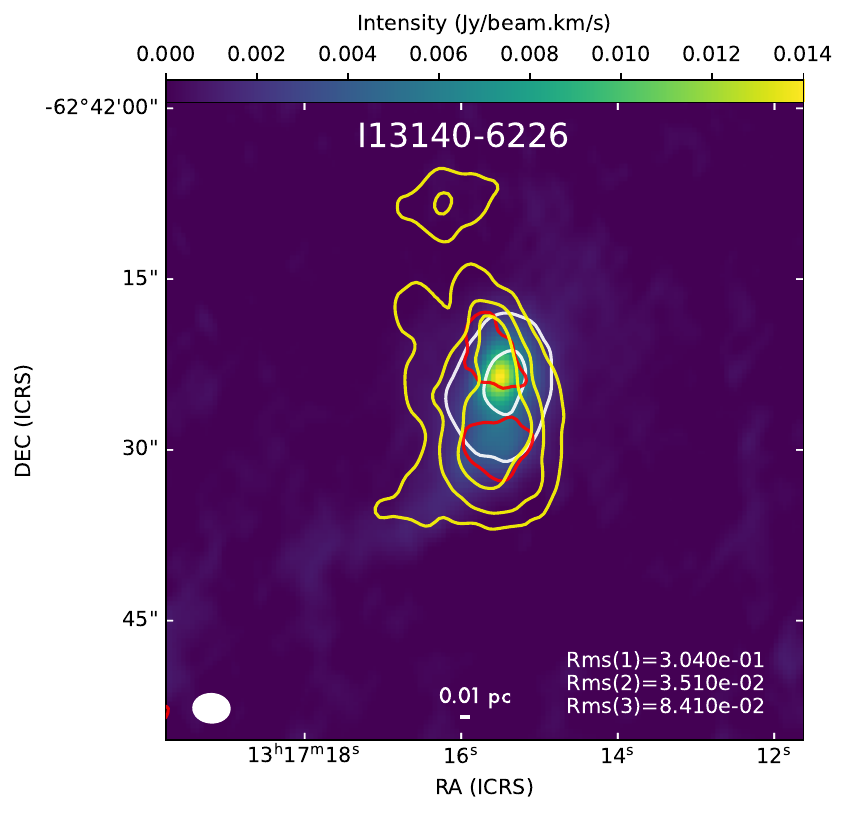}}
\quad
{\includegraphics[width=6.5cm,height=6.5cm]{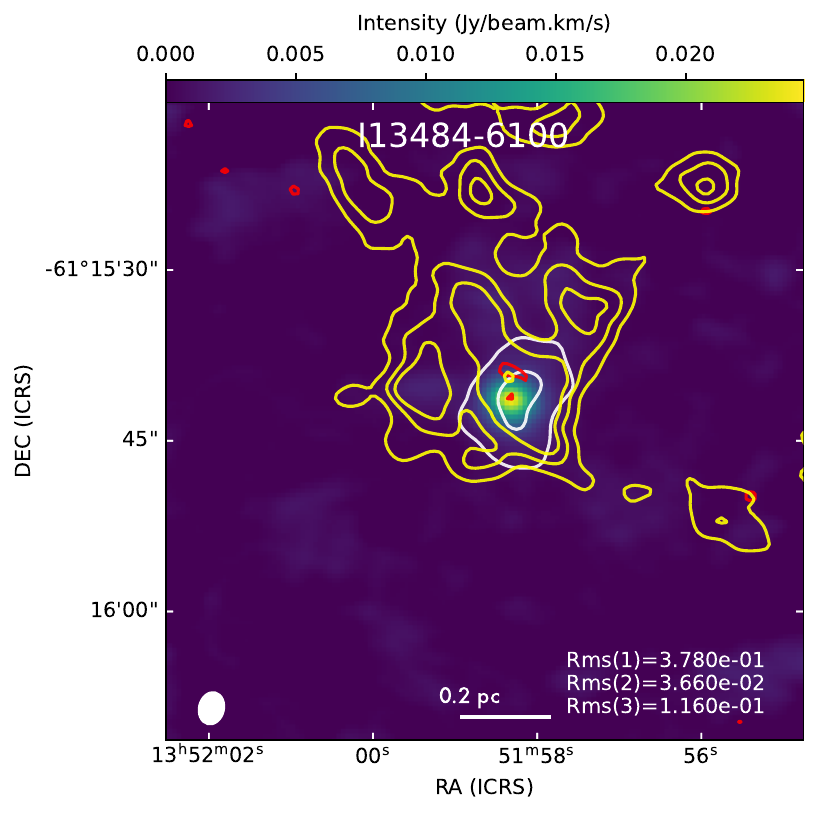}}
\caption{Comparison of moment 0 maps of CH$_3$CHO (after spectral stacking), SiO and H$^{13}$CO$^{\text{+}}$. The background emission shows 3 mm continuum emission. The white, yellow and red contours are for H$^{13}$CO$^{\text{+}}$ emission, SiO, and CH$_3$CHO, respectively. Their contour levels are [5, 10, 15, 30, 50, 100, 200]*Rms(1, 2, 3), Rms(1) represents the noise value for CH$_3$CHO, Rms(2) represents the noise value for SiO, and Rms(3) represents the noise value for H$^{13}$CO$^{\text{+}}$, with units of K Km s$^{-1}$. The beam ellipse is placed in the lower left corner of the image.}
\label{fig_allthecoresH3CHOall}
\end{figure}

\clearpage
\setcounter{figure}{\value{figure}-1}
\begin{figure}
\centering
{\includegraphics[width=6.5cm,height=6.5cm]{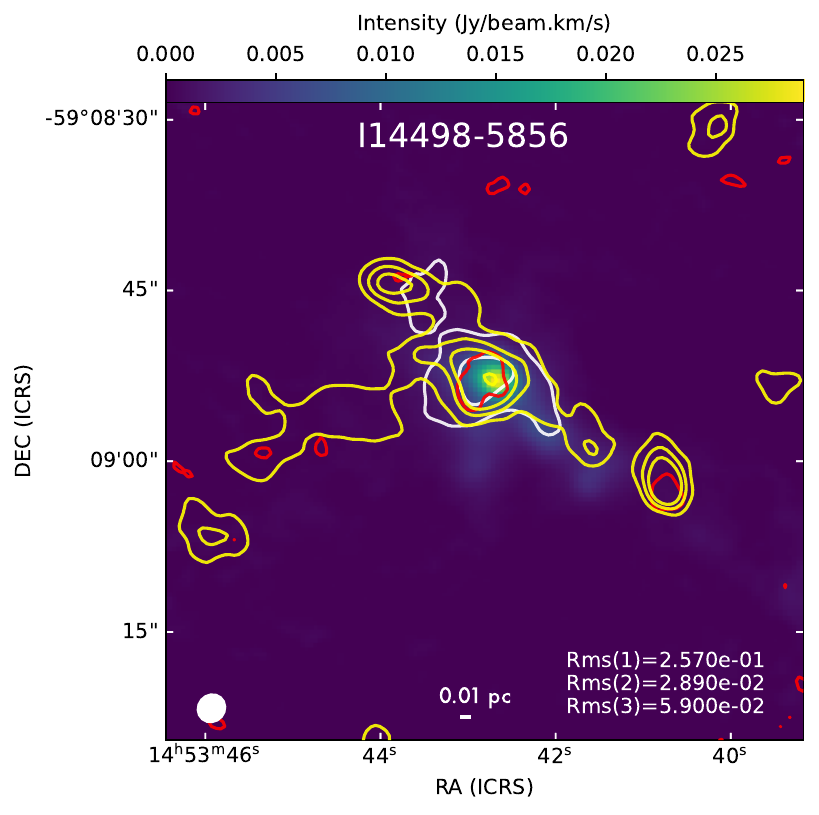}}
\quad
{\includegraphics[width=6.5cm,height=6.5cm]{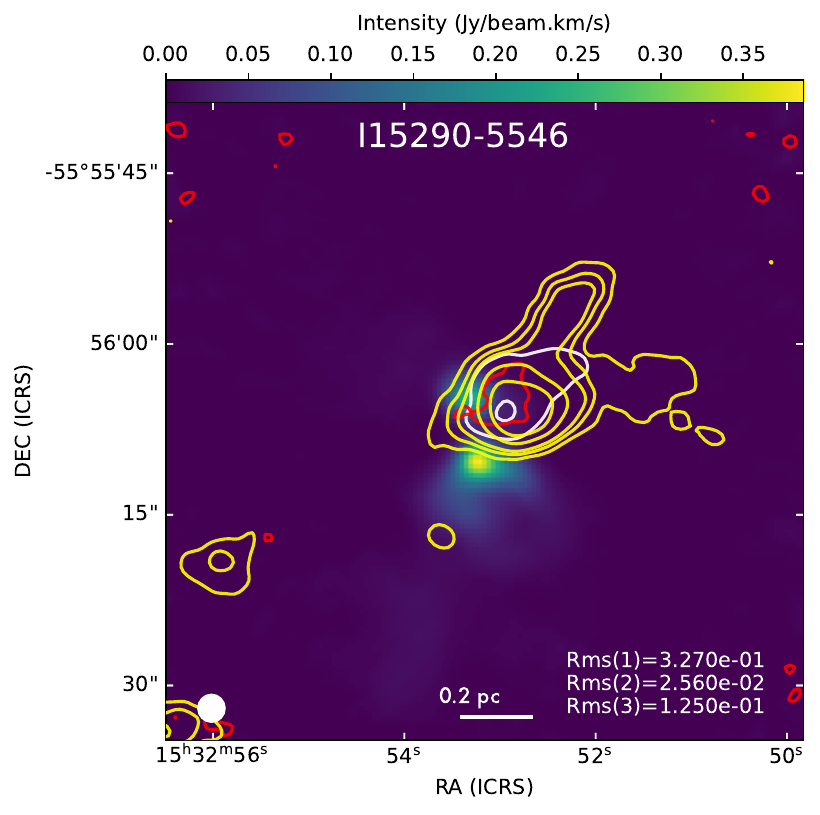}}
\quad
{\includegraphics[width=6.5cm,height=6.5cm]{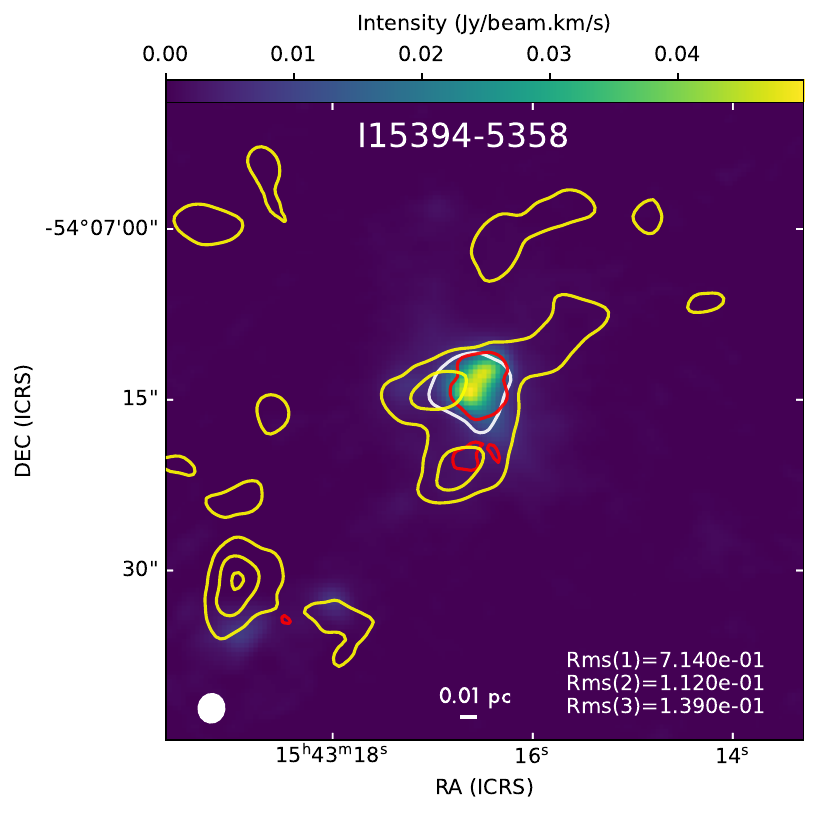}}
\quad
{\includegraphics[width=6.5cm,height=6.5cm]{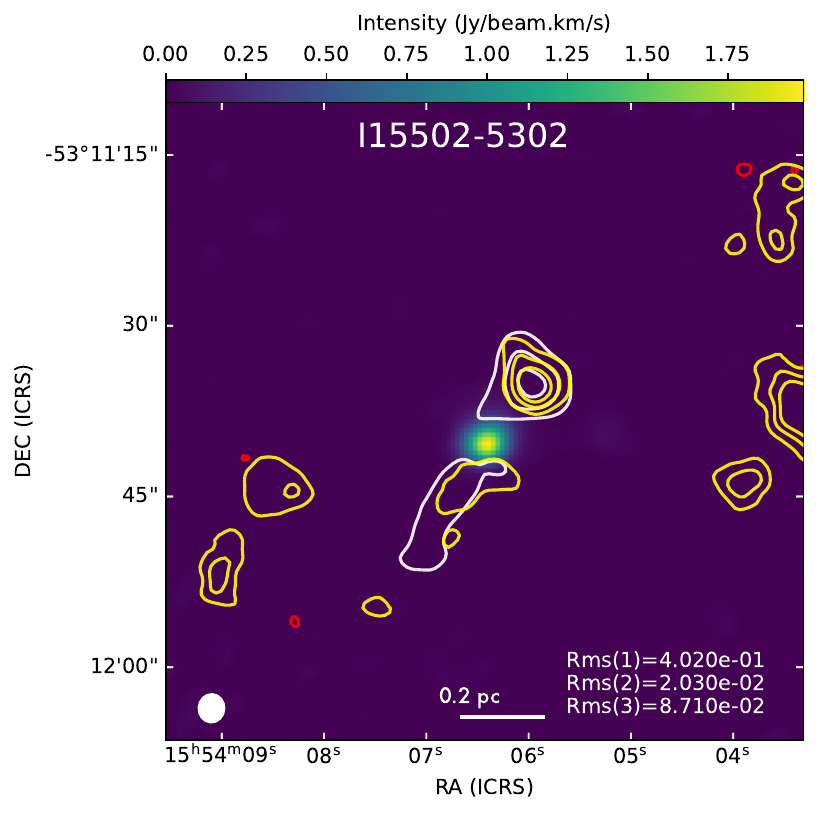}}
\quad
{\includegraphics[width=6.5cm,height=6.5cm]{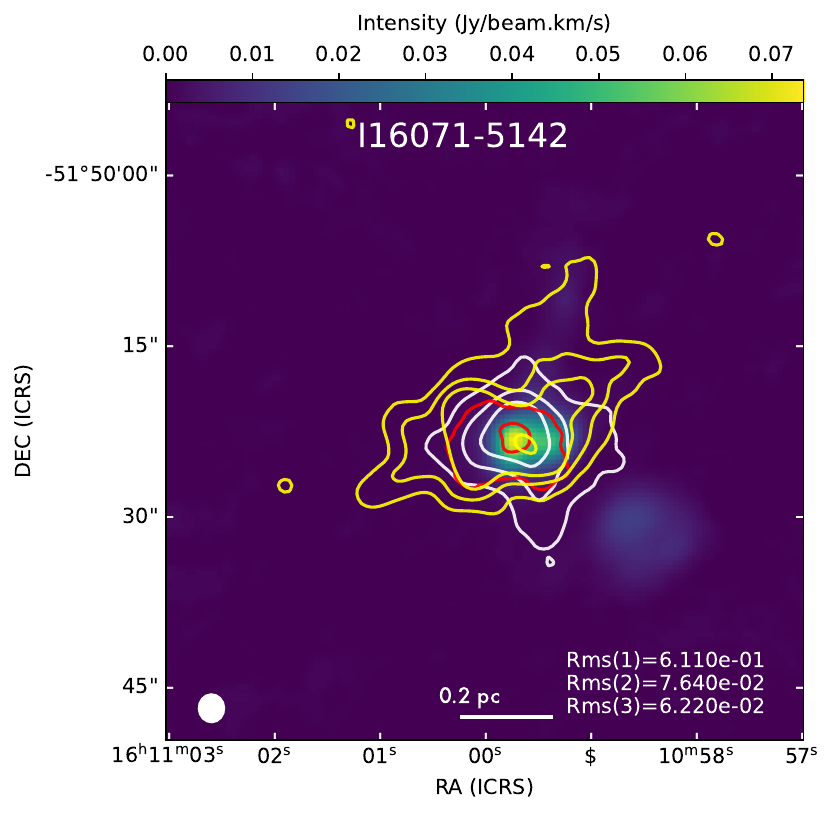}}
\quad
{\includegraphics[width=6.5cm,height=6.5cm]{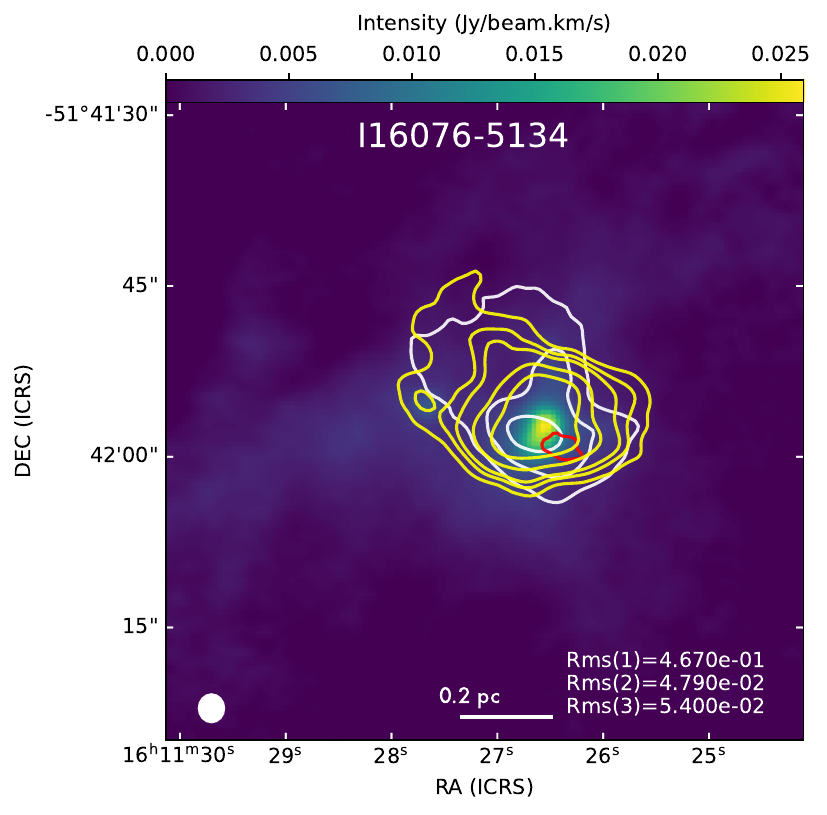}}
\caption{Continued.}
\end{figure}

\clearpage
\setcounter{figure}{\value{figure}-1}
\begin{figure}
\centering
{\includegraphics[width=6.5cm,height=6.5cm]{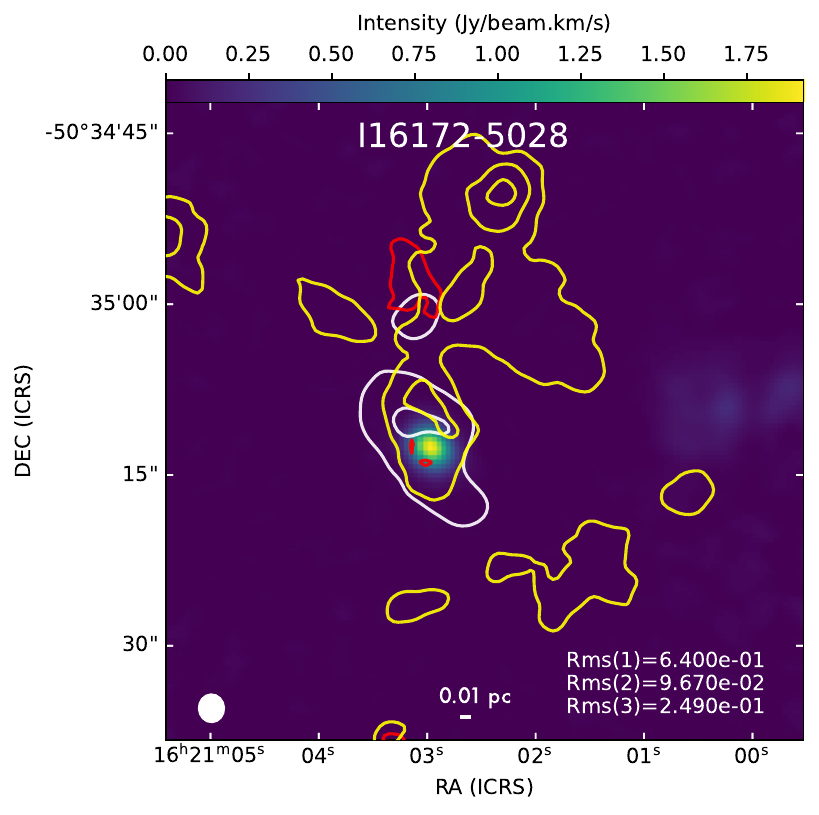}}
\quad
{\includegraphics[width=6.5cm,height=6.5cm]{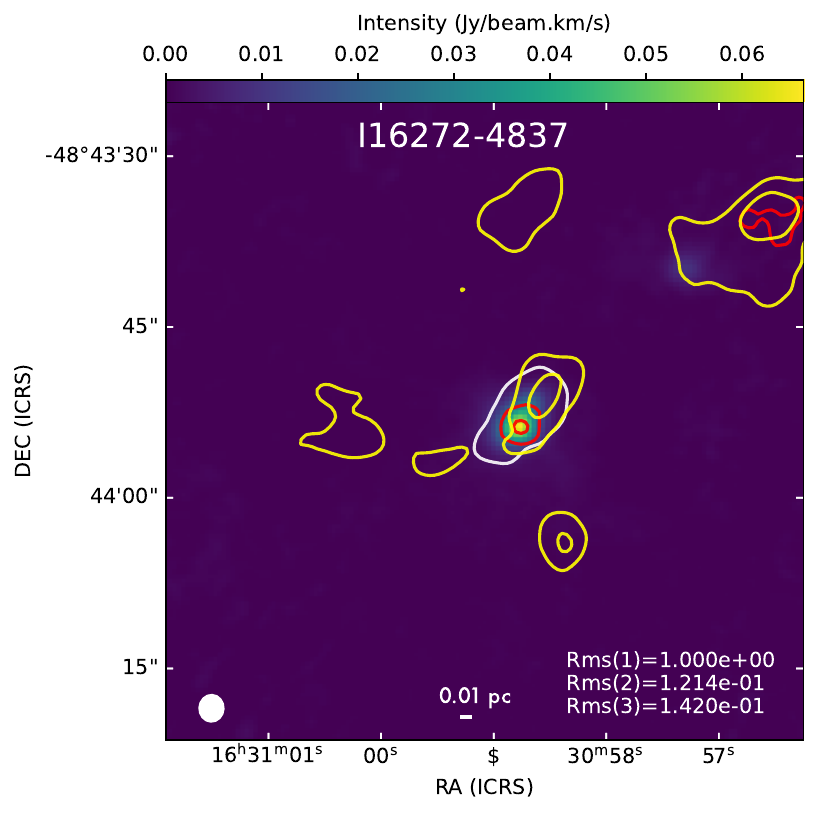}}
\quad
{\includegraphics[width=6.5cm,height=6.5cm]{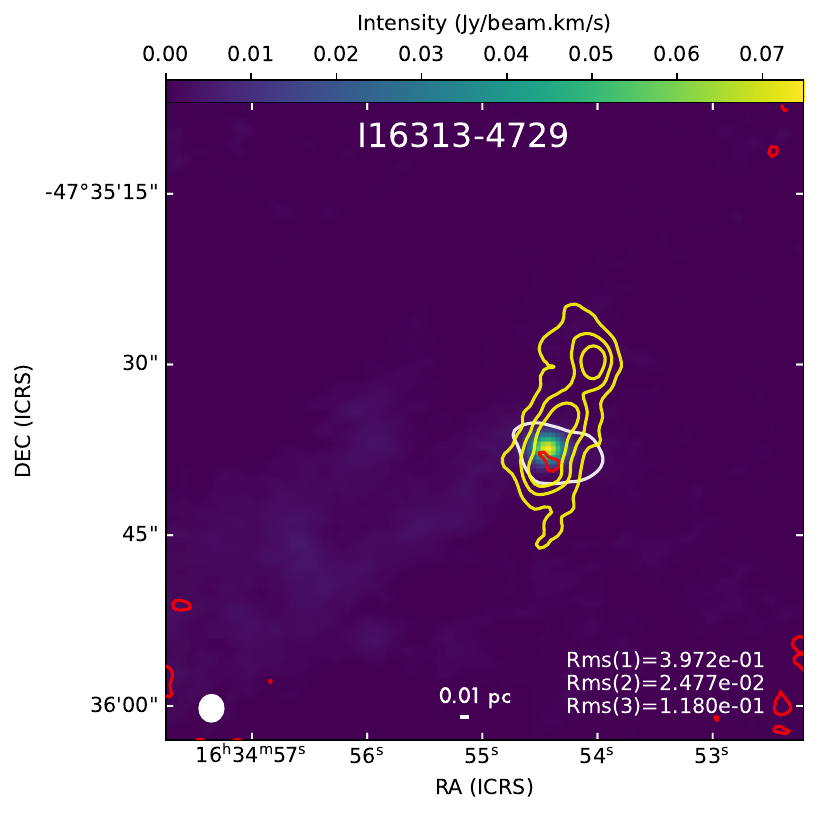}}
\quad
{\includegraphics[width=6.5cm,height=6.5cm]{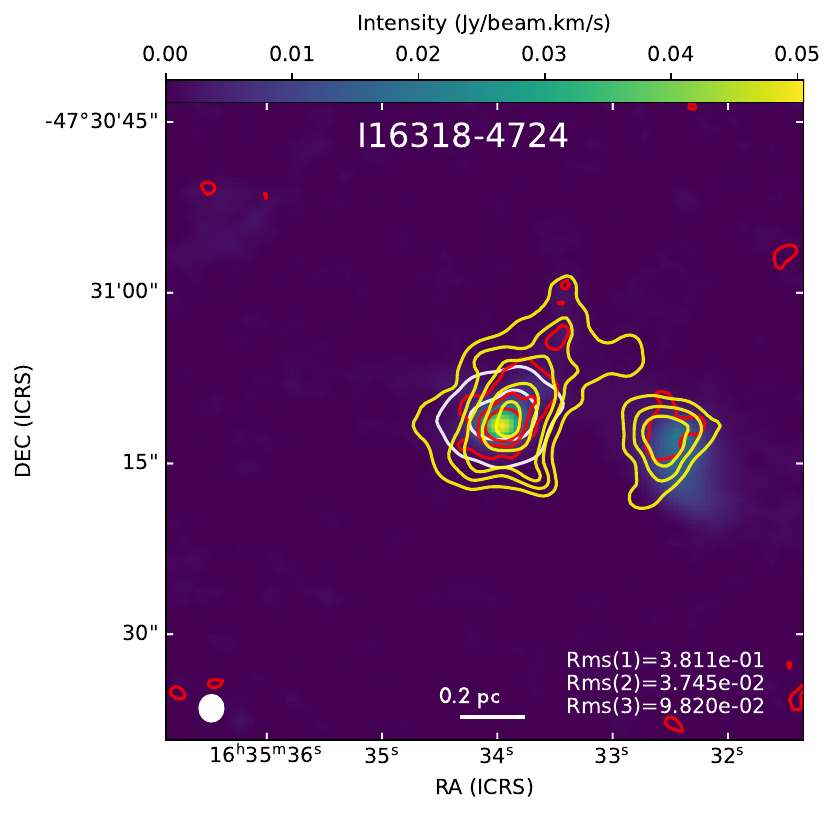}}
\quad
{\includegraphics[width=6.5cm,height=6.5cm]{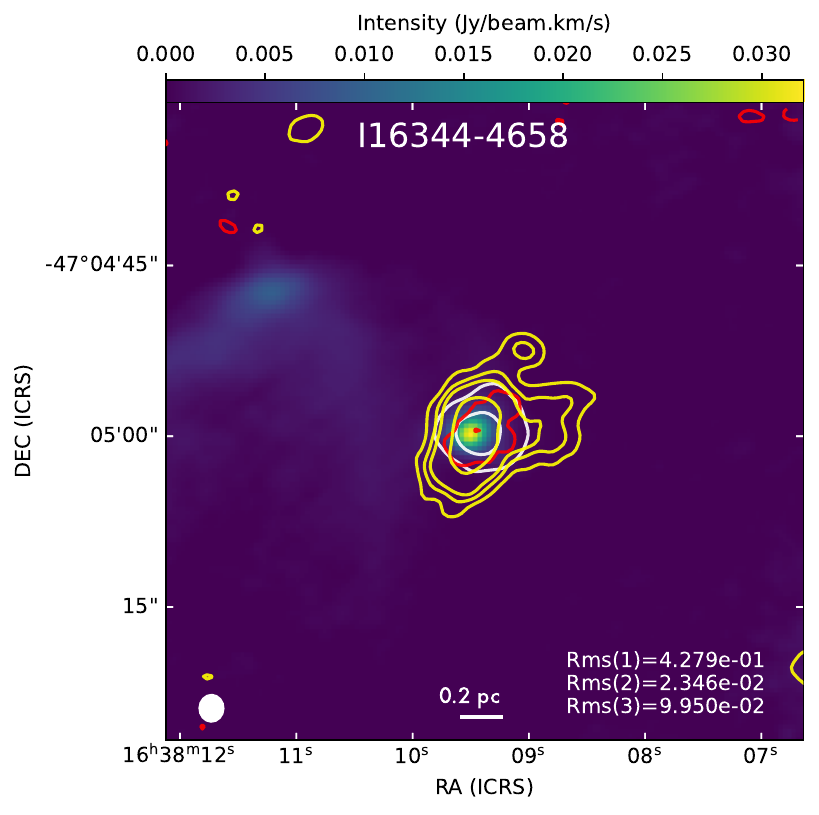}}
\quad
{\includegraphics[width=6.5cm,height=6.5cm]{H13CO_I16348-4654.pdf}}
\caption{Continued.}
\end{figure}

\clearpage
\setcounter{figure}{\value{figure}-1}
\begin{figure}
\centering
{\includegraphics[width=6.5cm,height=6.5cm]{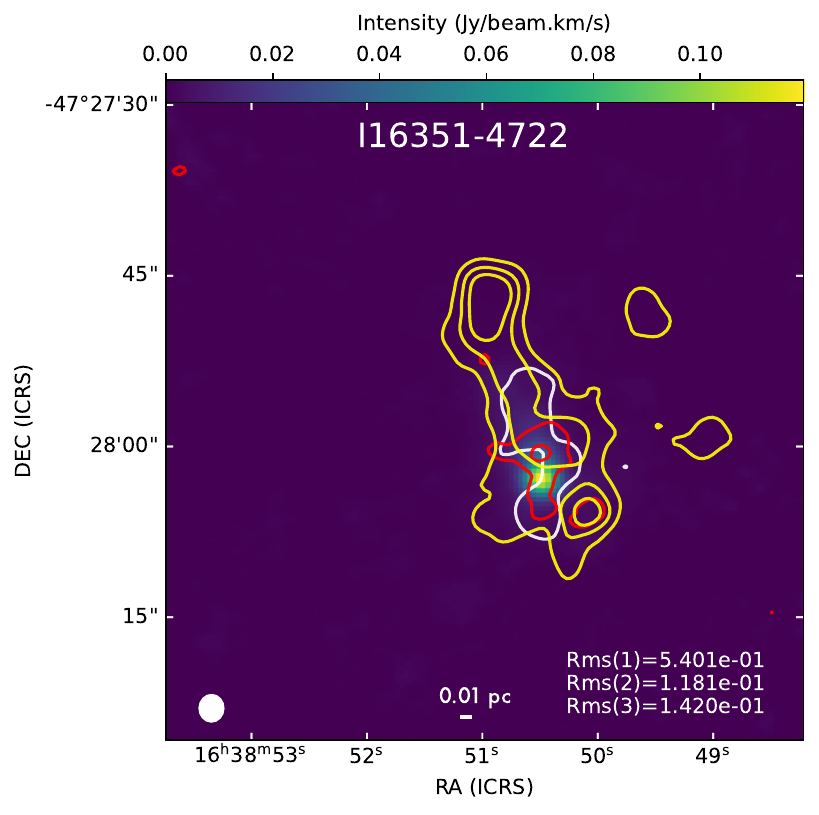}}
\quad
{\includegraphics[width=6.5cm,height=6.5cm]{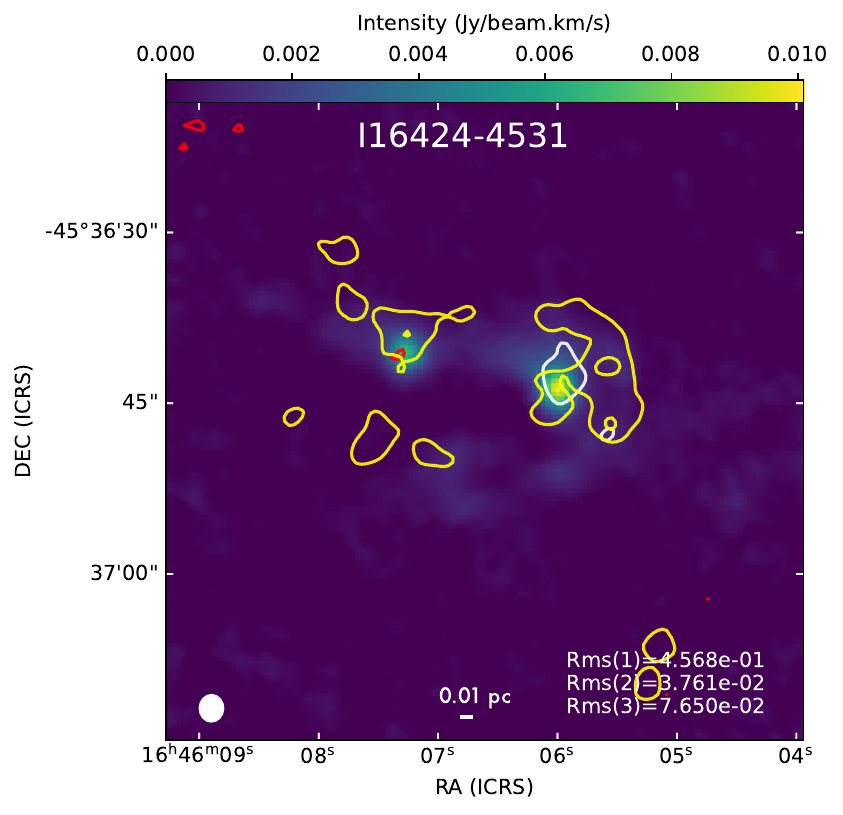}}
\quad
{\includegraphics[width=6.5cm,height=6.5cm]{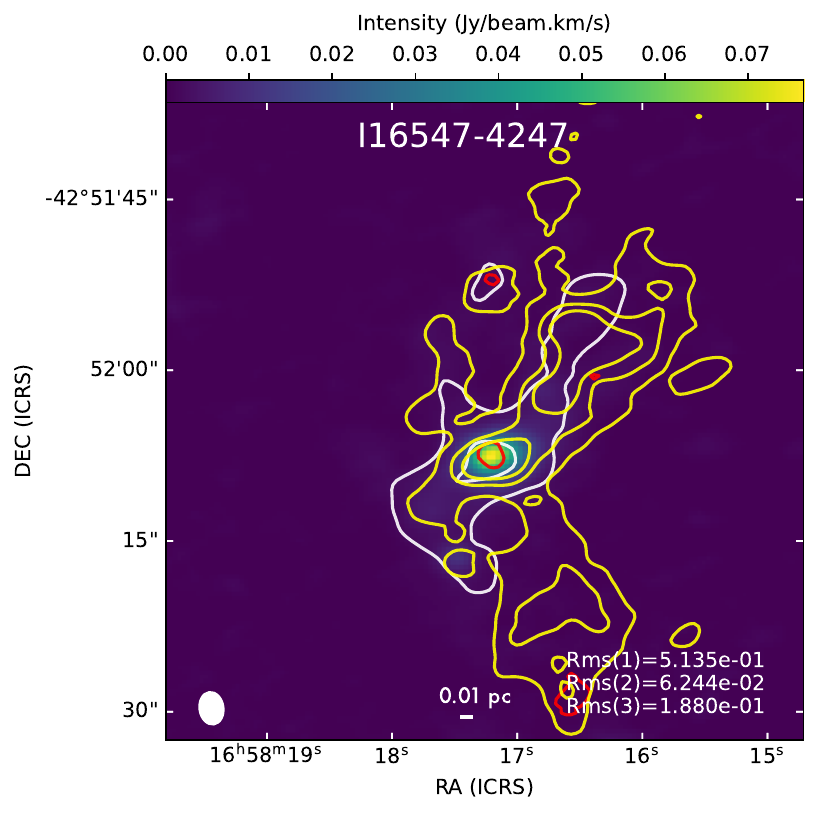}}
\quad
{\includegraphics[width=6.5cm,height=6.5cm]{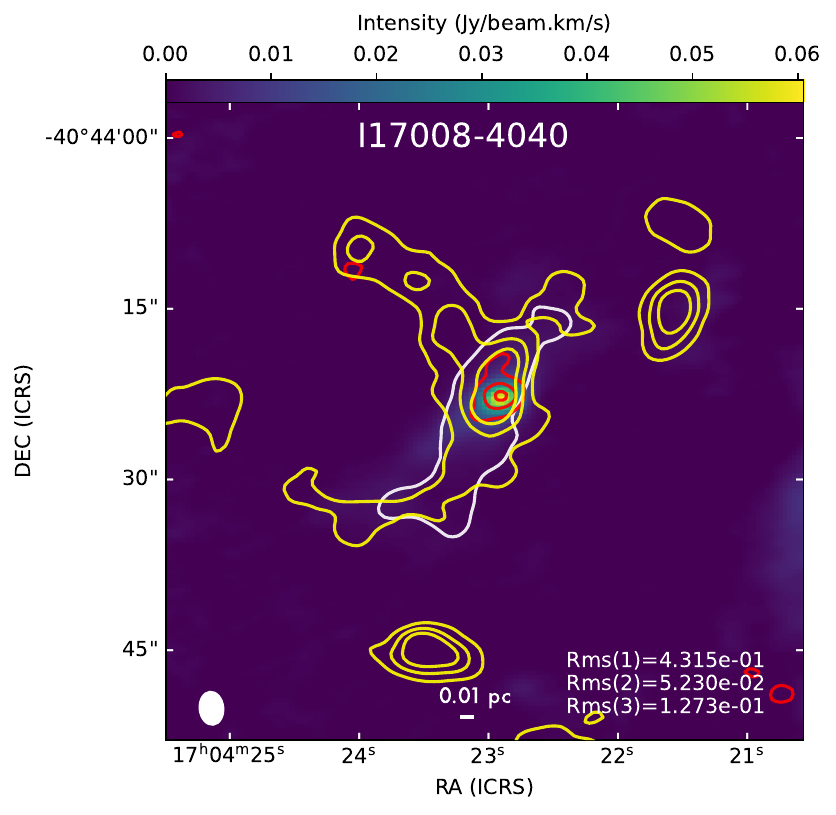}}
\quad
{\includegraphics[width=6.5cm,height=6.5cm]{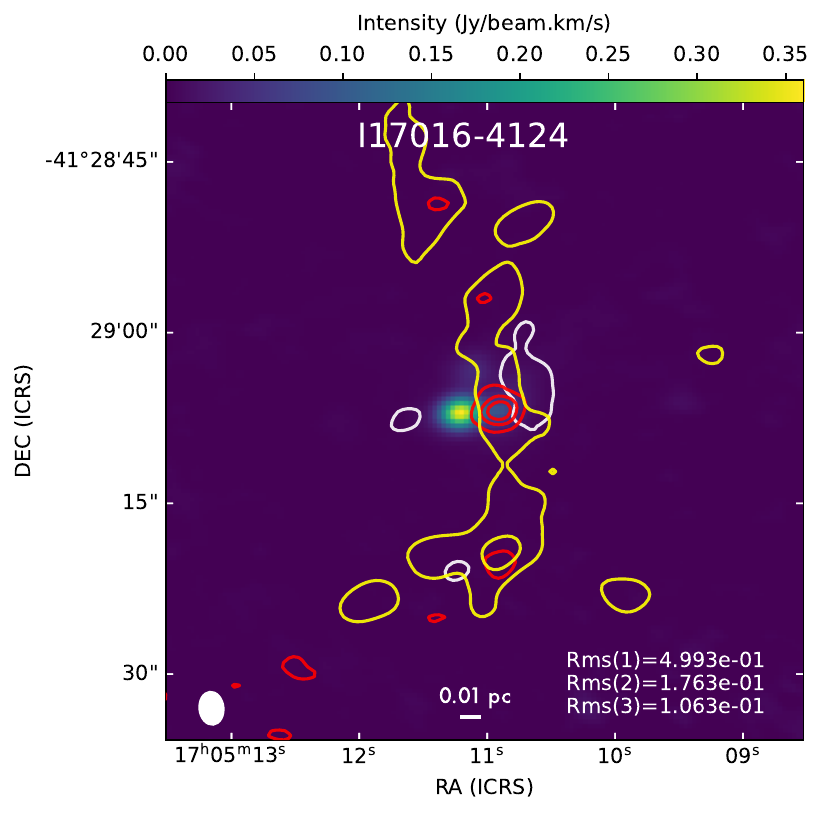}}
\quad
{\includegraphics[width=6.5cm,height=6.5cm]{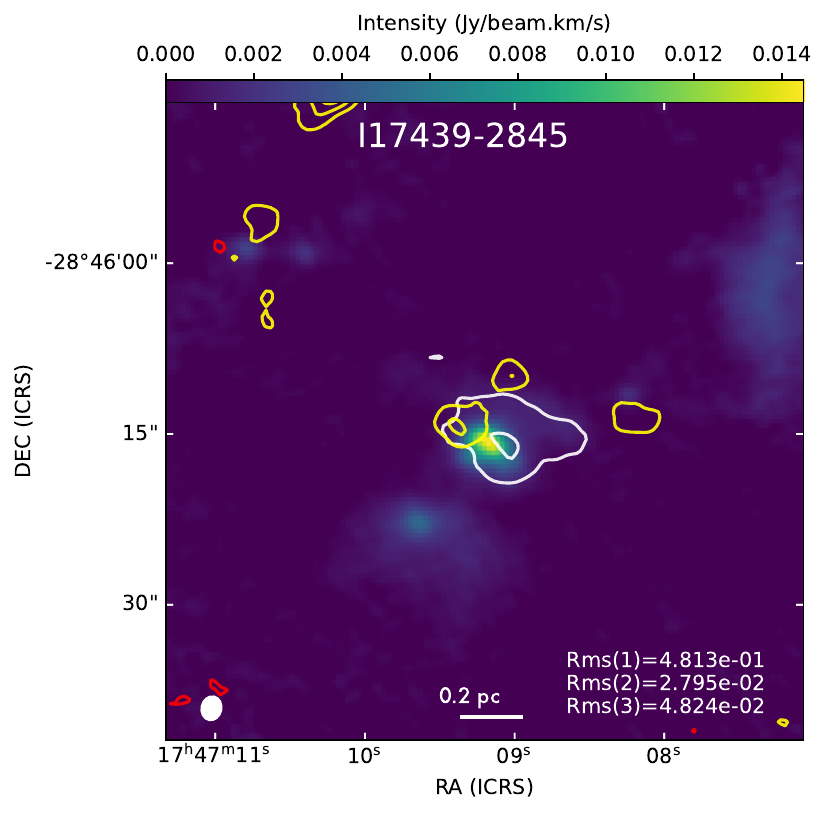}}
\caption{Continued.}
\end{figure}

\clearpage
\setcounter{figure}{\value{figure}-1}
\begin{figure}
\centering
{\includegraphics[width=6.5cm,height=6.5cm]{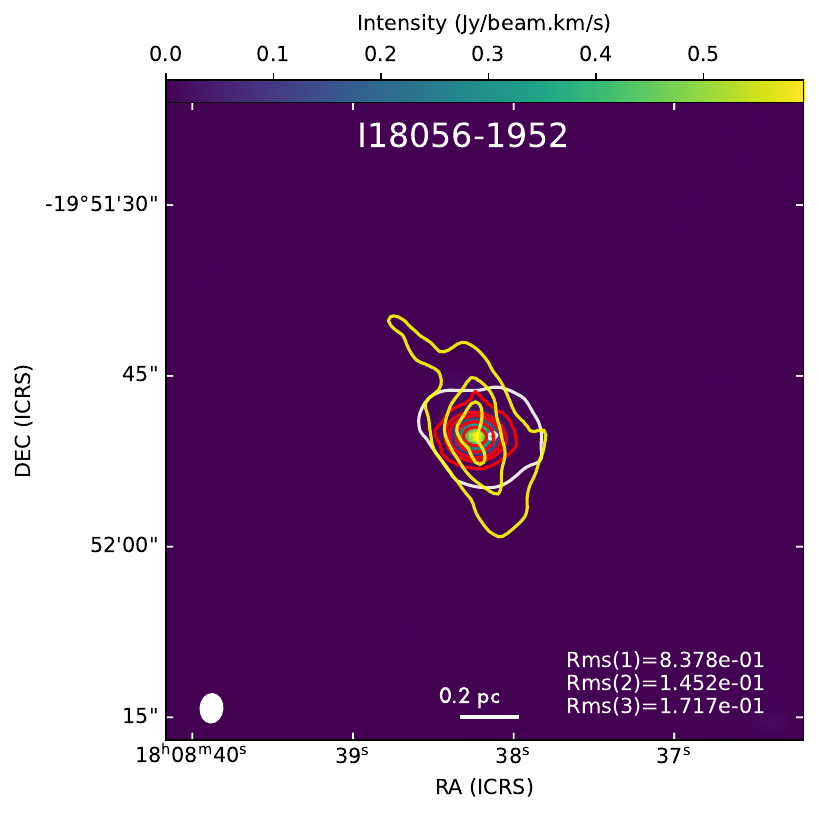}}
\quad
{\includegraphics[width=6.5cm,height=6.5cm]{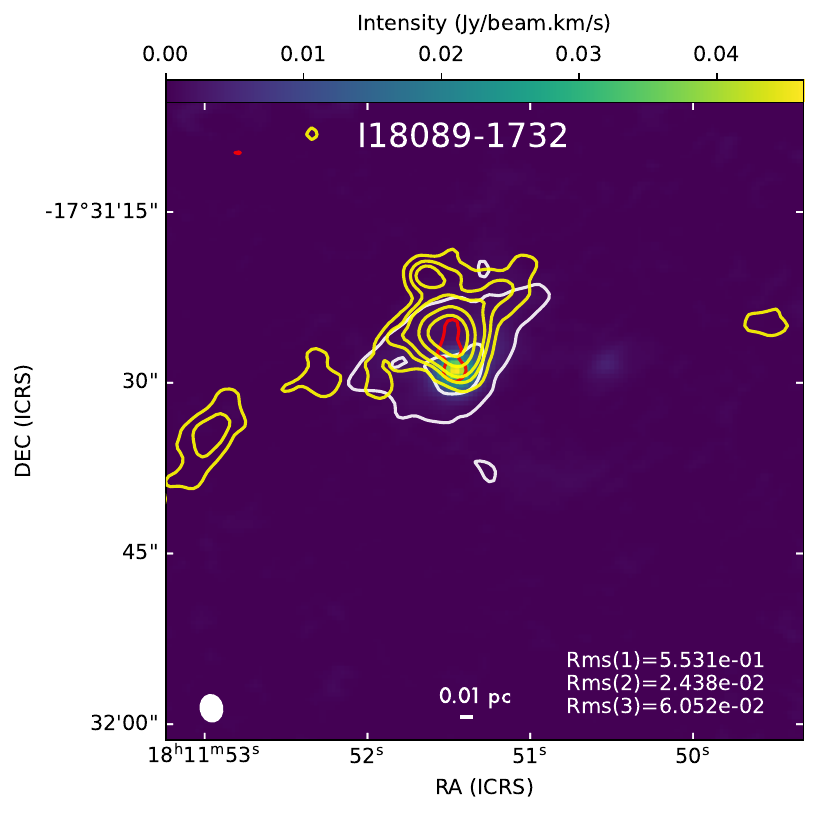}}
\quad
{\includegraphics[width=6.5cm,height=6.5cm]{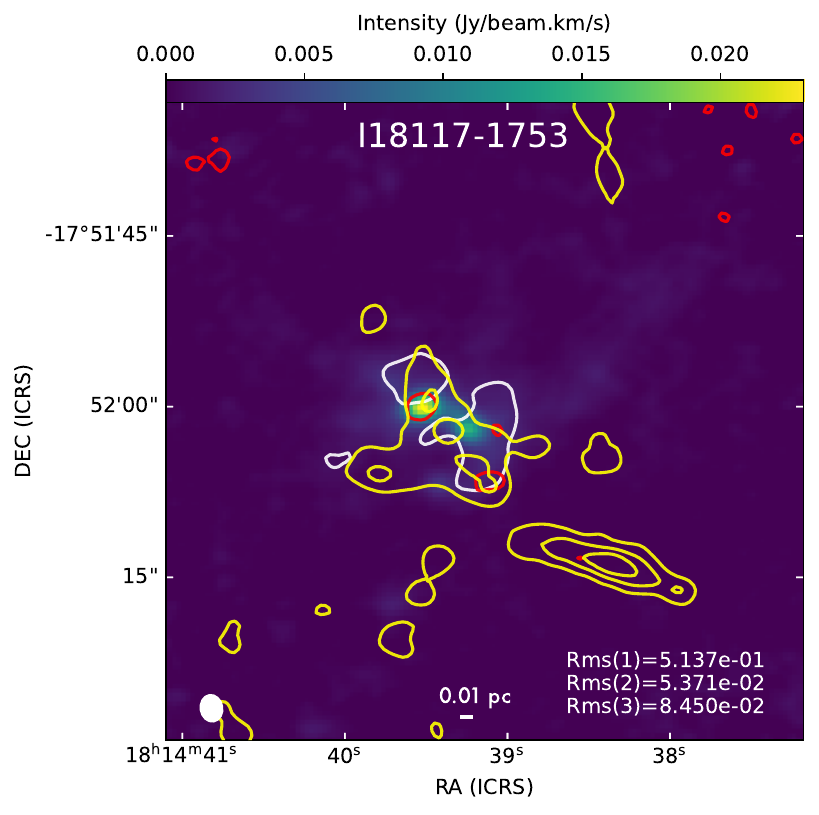}}
\quad
{\includegraphics[width=6.5cm,height=6.5cm]{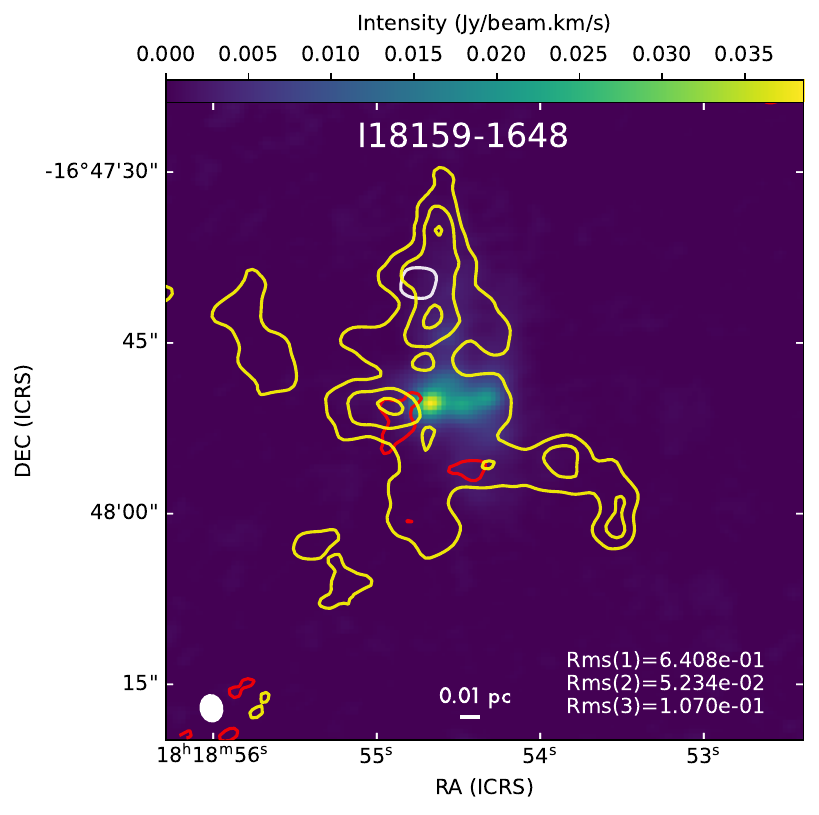}}
\quad
{\includegraphics[width=6.5cm,height=6.5cm]{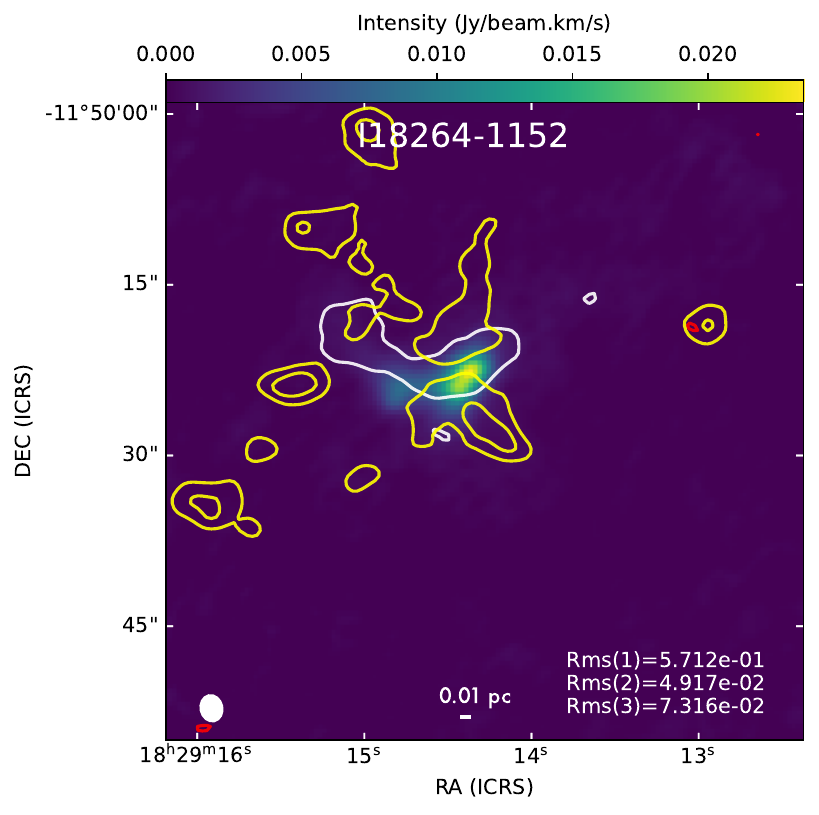}}
\quad
{\includegraphics[width=6.5cm,height=6.5cm]{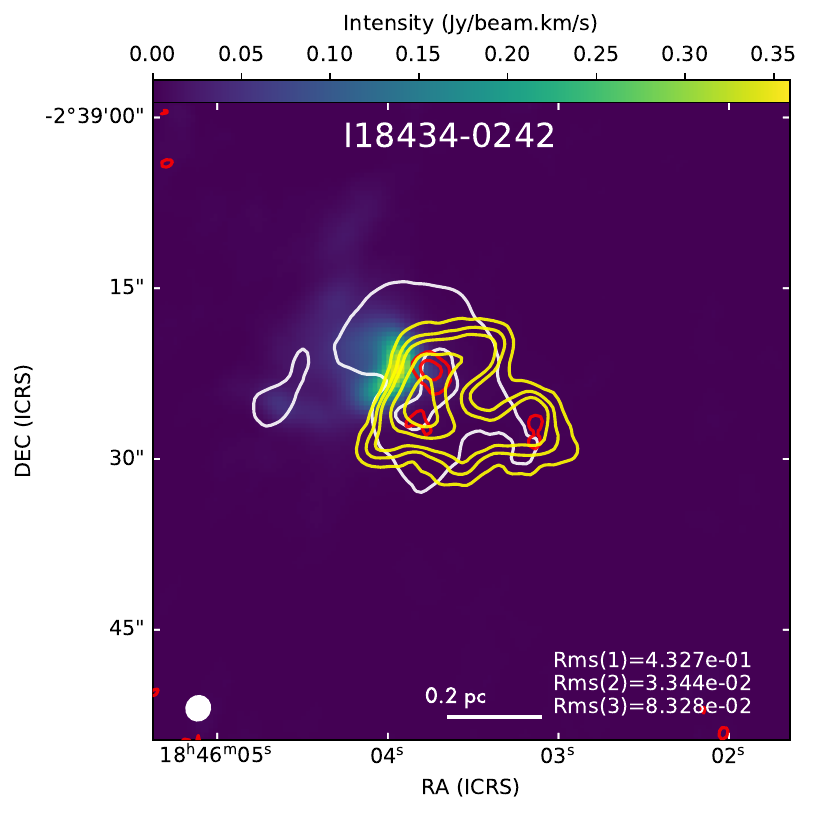}}
\caption{Continued.}
\end{figure}

\clearpage
\setcounter{figure}{\value{figure}-1}
\begin{figure}
  \centering 
{\includegraphics[width=6.5cm,height=6.5cm]{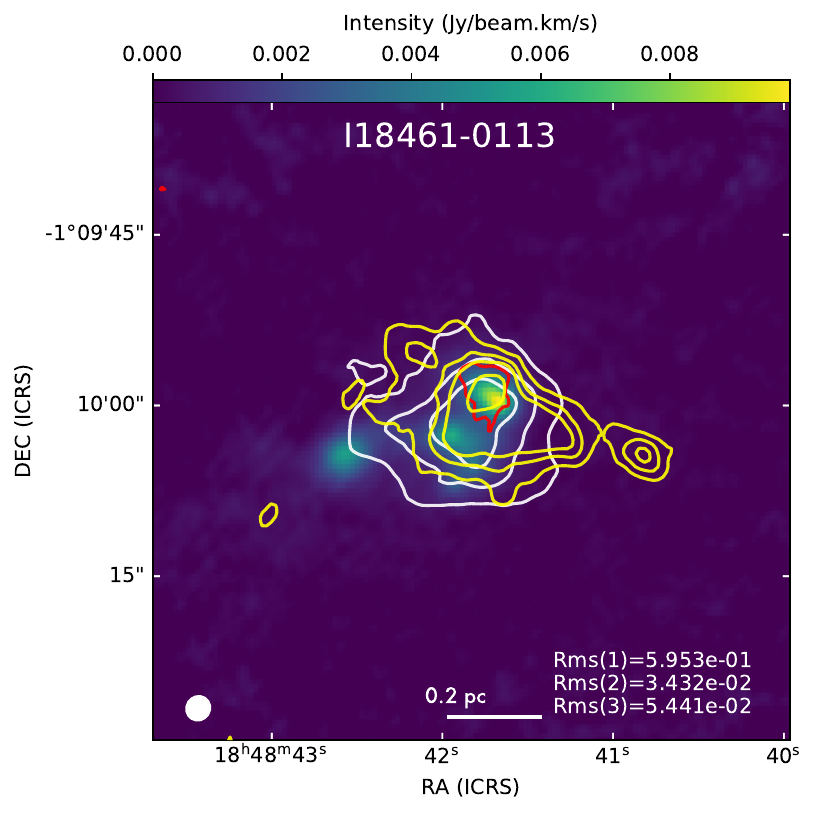}}
\quad
{\includegraphics[width=6.5cm,height=6.5cm]{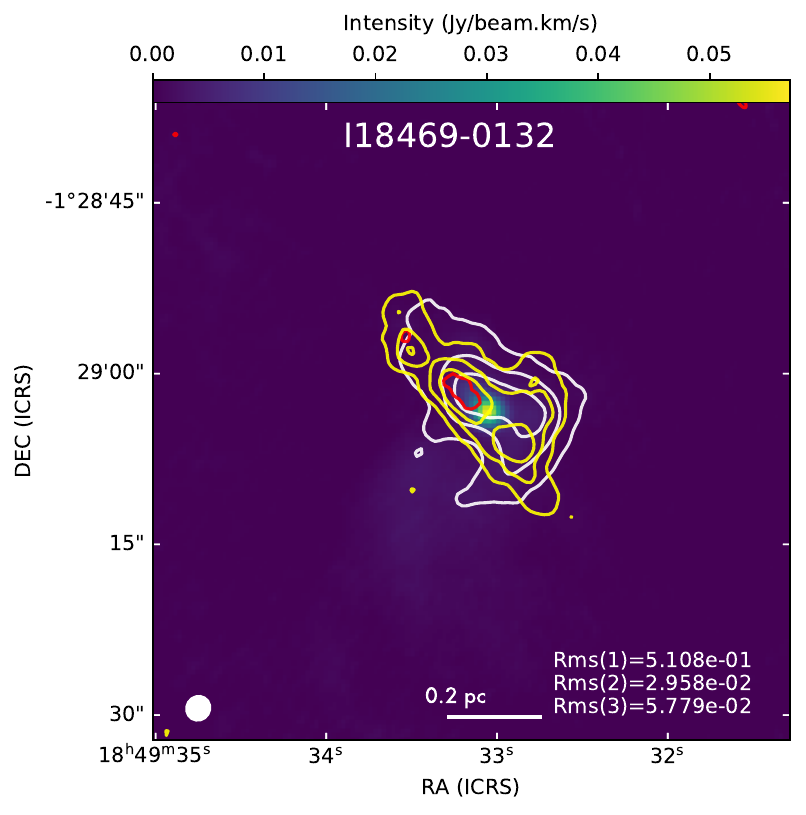}}
\quad
{\includegraphics[width=6.5cm,height=6.5cm]{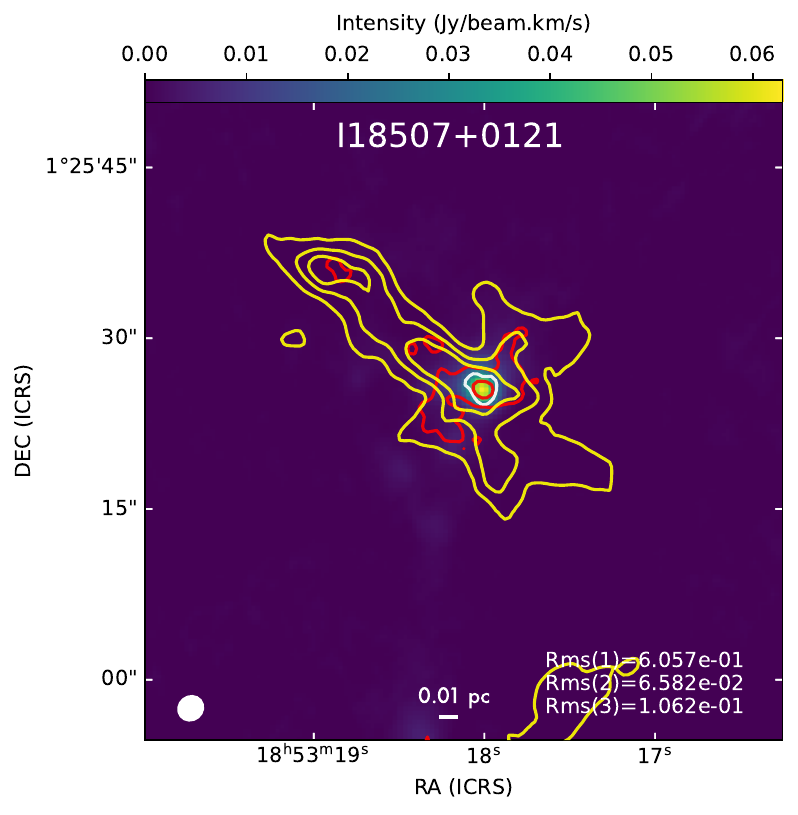}}
\quad
{\includegraphics[width=6.5cm,height=6.5cm]{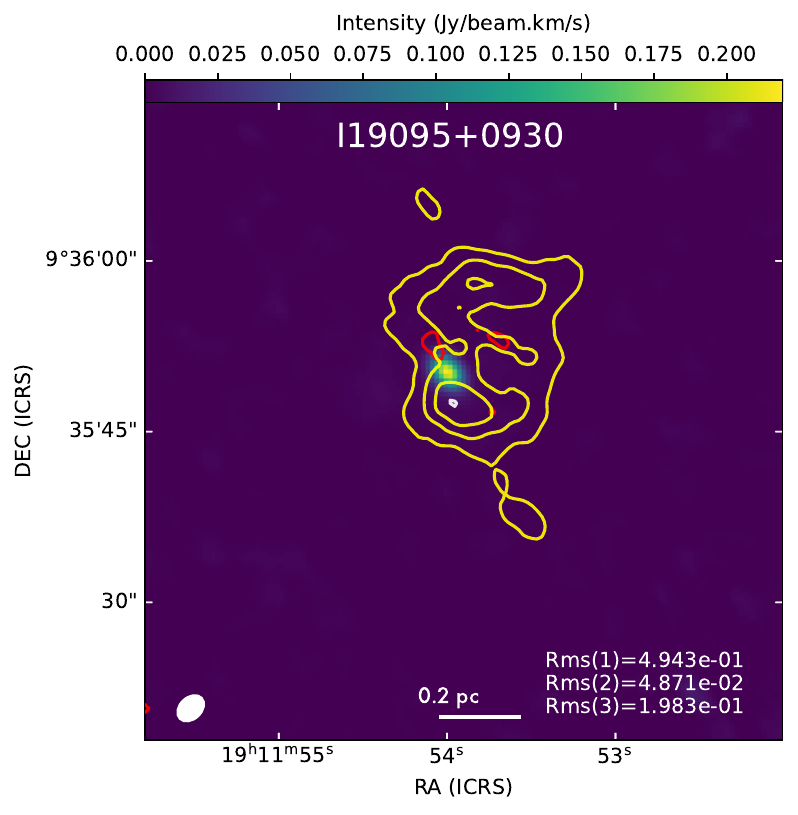}}
\quad
{\includegraphics[width=6.5cm,height=6.5cm]{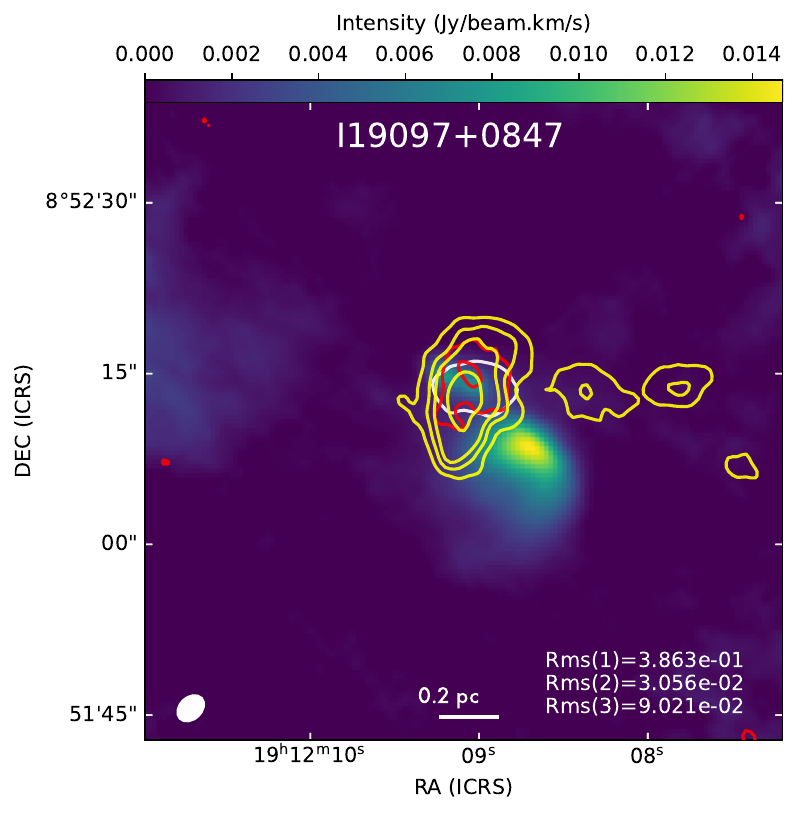}}
\caption{Continued}
\end{figure}

\section{Moment maps}  \label{app:C}

\begin{figure}
  \centering
{\includegraphics[height=0.1\textheight]{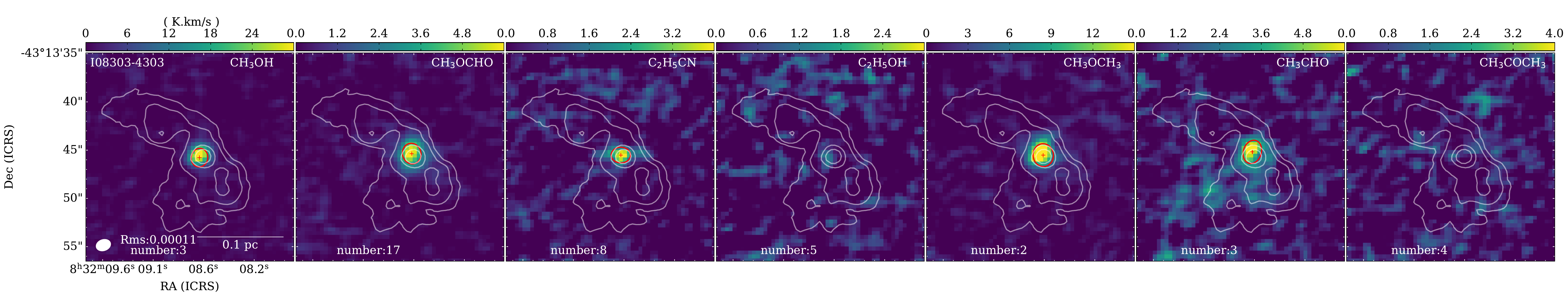}}
\quad 
{\includegraphics[height=0.1\textheight]{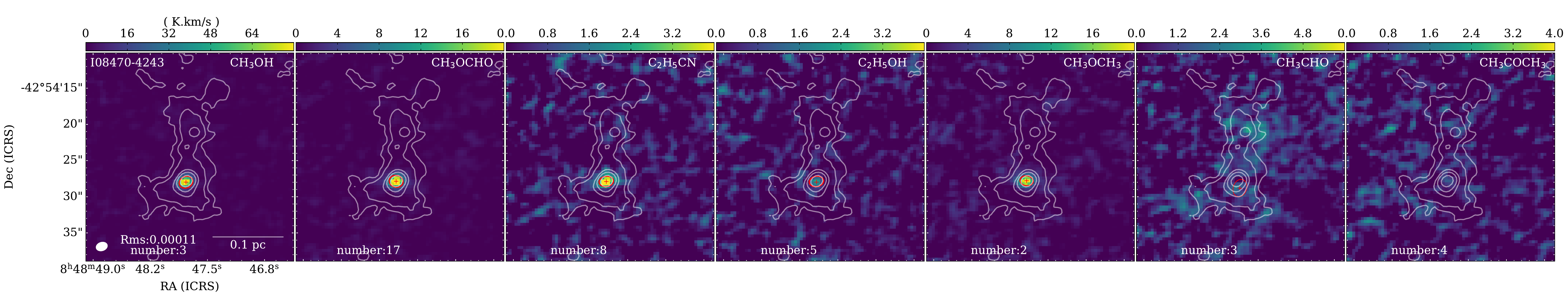}}
\quad 
{\includegraphics[height=0.1\textheight]{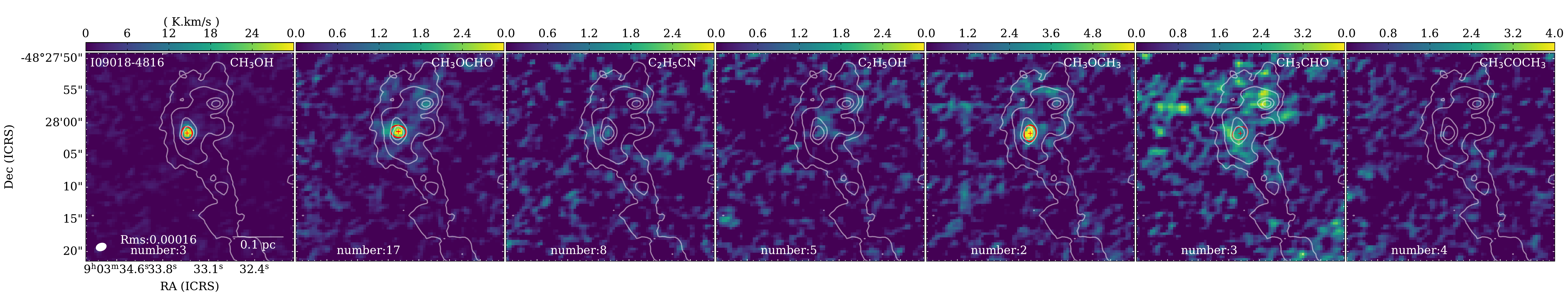}}
\quad 
{\includegraphics[height=0.1\textheight]{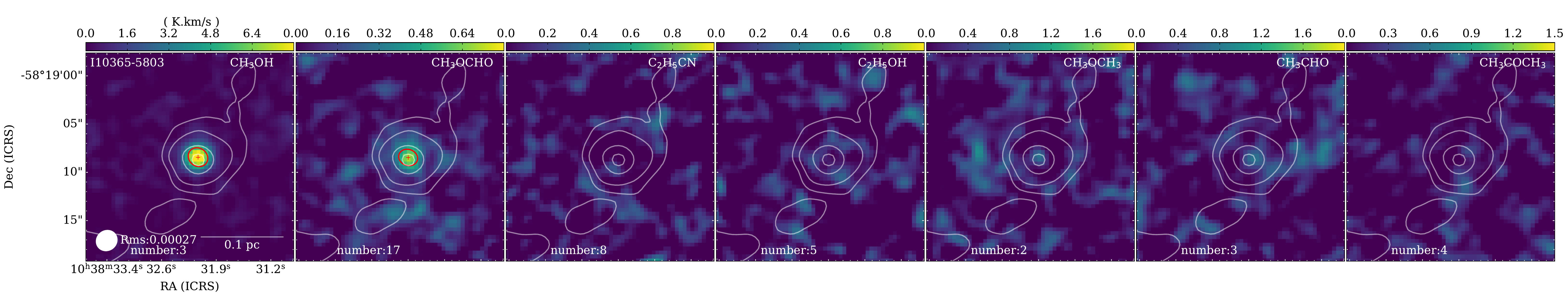}}
\quad
{\includegraphics[height=0.1\textheight]{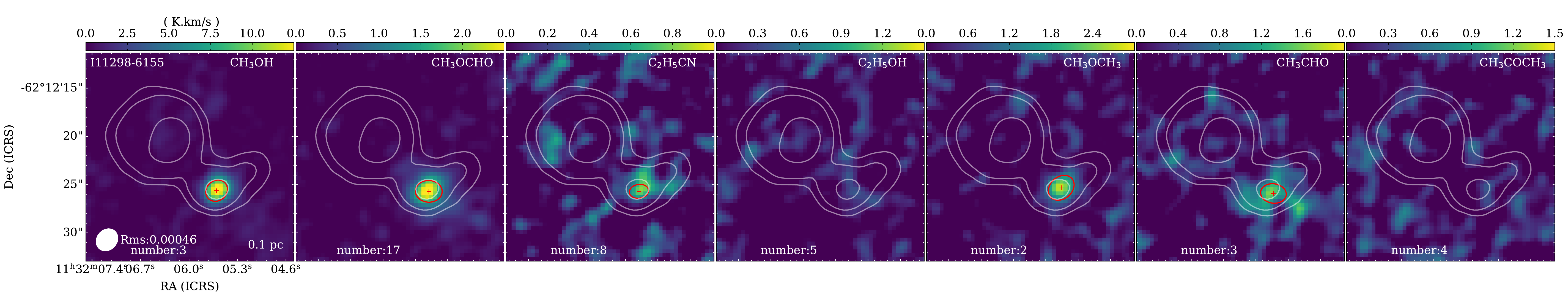}}
\quad 
{\includegraphics[height=0.1\textheight]{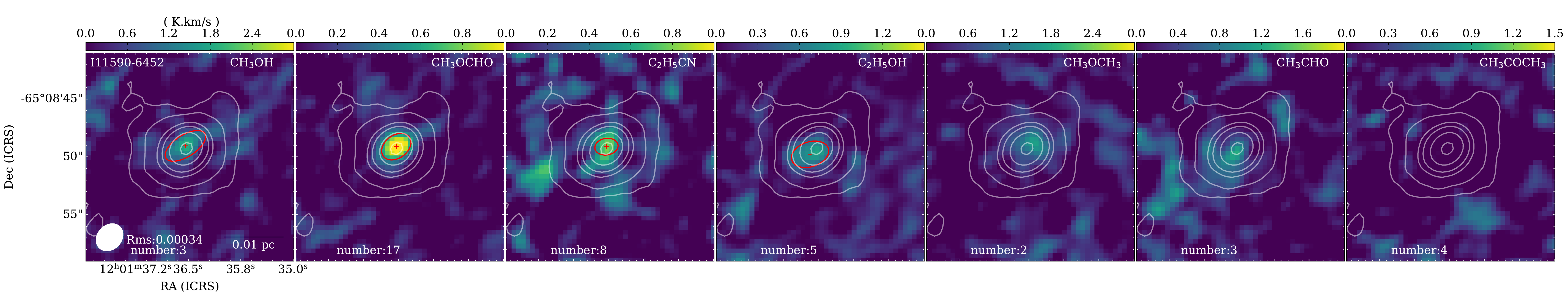}}
\quad
{\includegraphics[height=0.1\textheight]{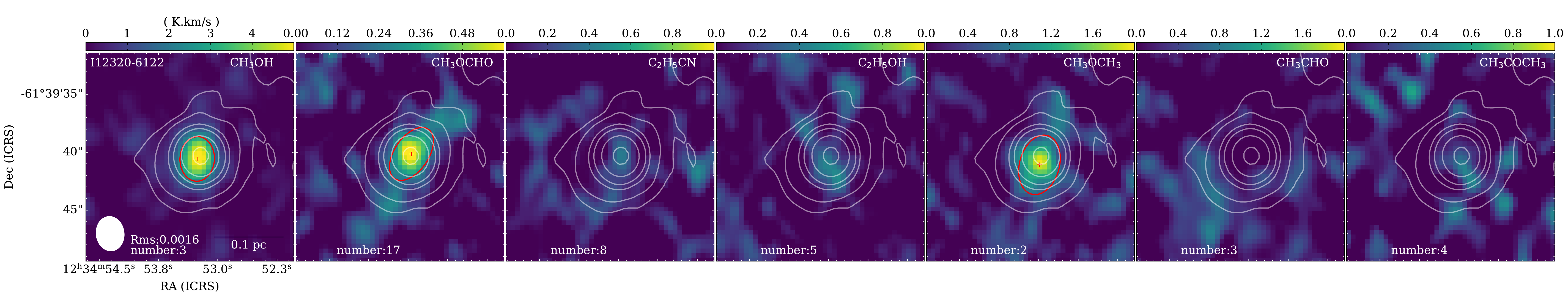}}
\quad
{\includegraphics[height=0.1\textheight]{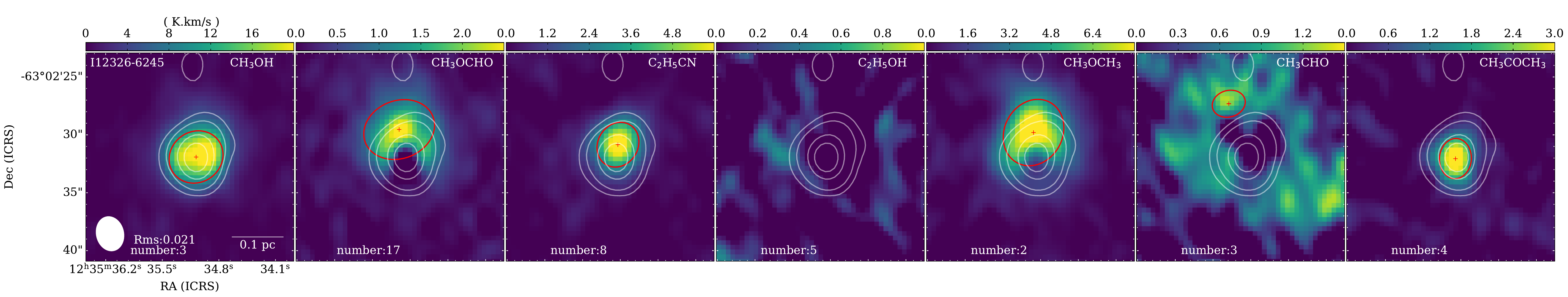}}
\quad 
{\includegraphics[height=0.1\textheight]{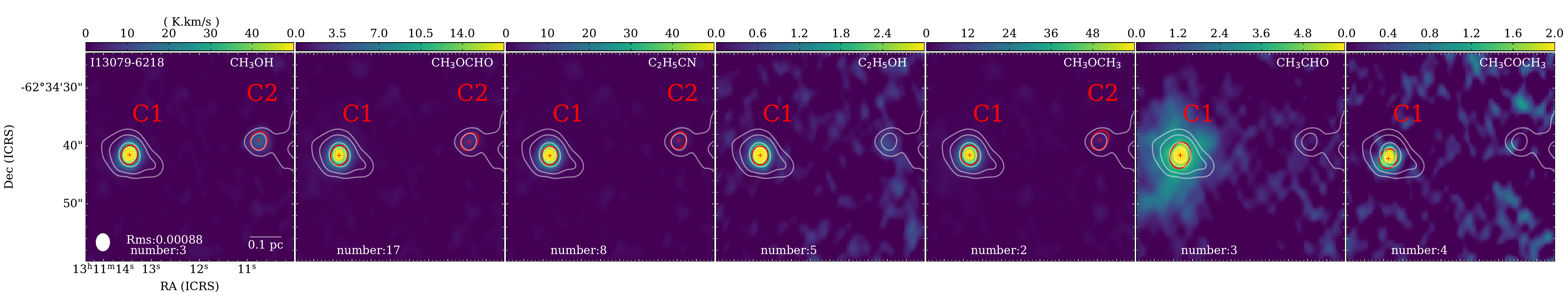}}
\caption{The figure displays the moment 0 map obtained by stacking seven molecules. The contour levels represent the continuum emission, with values of [5, 10, 30, 50, 100, 200] times the RMS. The red ellipses represent the deconvolved sizes from the two-dimensional Gaussian fits to the moment 0 maps, which were derived by integrating the molecular emission within a ±5 km/s velocity range using the CASA FITS functionality.}
\label{fig_allthecores}
\end{figure}

\clearpage
\setcounter{figure}{\value{figure}-1}
\begin{figure}
\centering 
{\includegraphics[height=0.1\textheight]{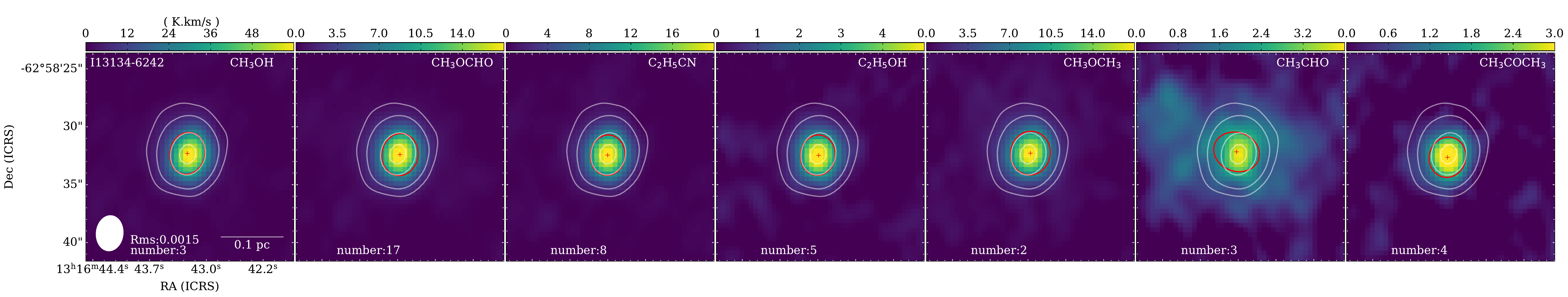}}
\quad 
{\includegraphics[height=0.1\textheight]{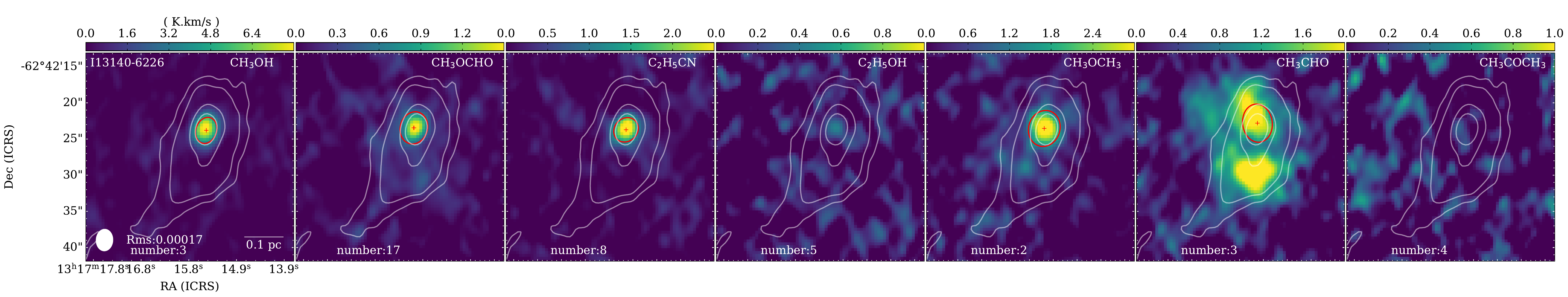}}
\quad 
{\includegraphics[height=0.1\textheight]{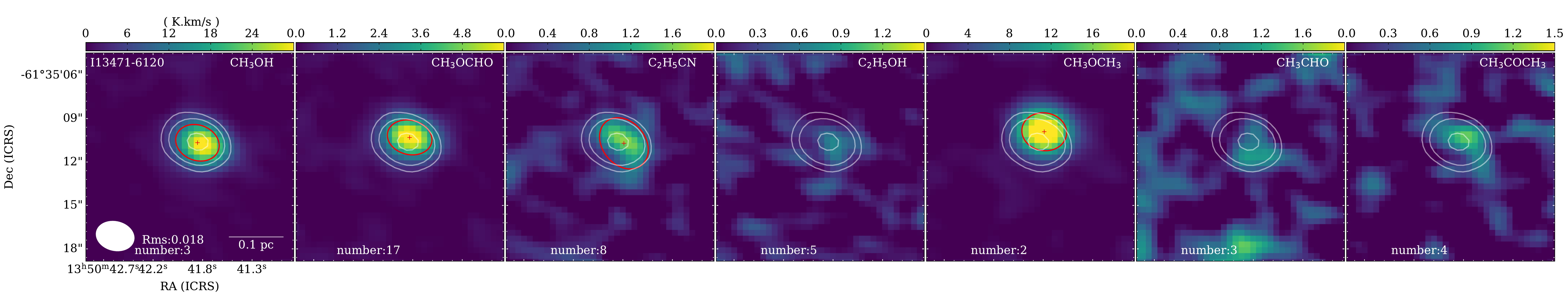}}
\quad
{\includegraphics[height=0.1\textheight]{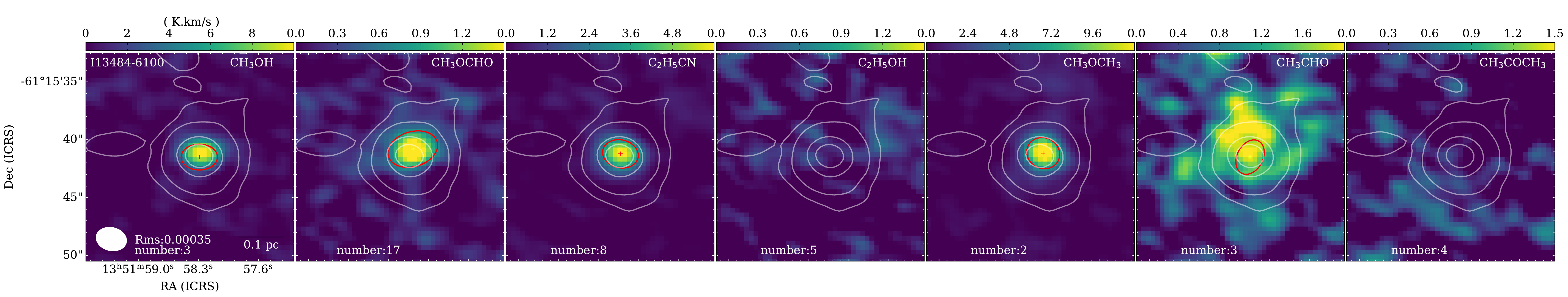}}
\quad 
{\includegraphics[height=0.1\textheight]{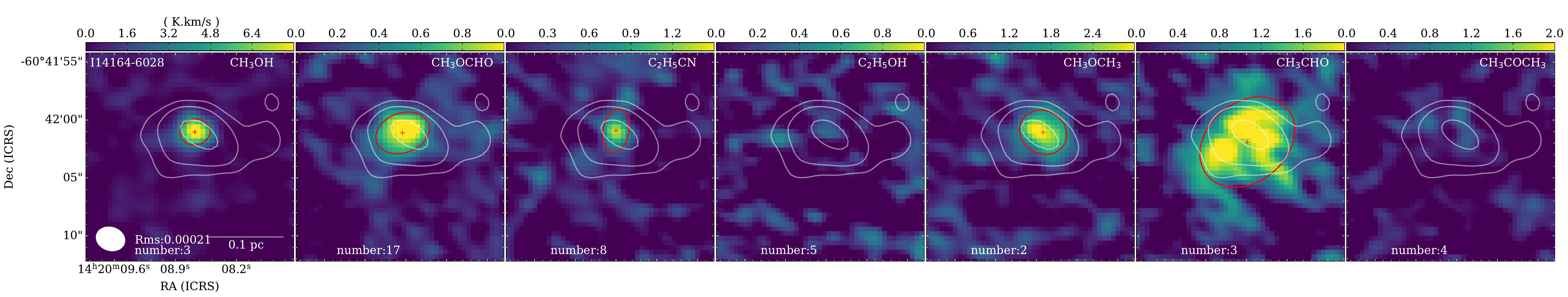}}
\quad
{\includegraphics[height=0.1\textheight]{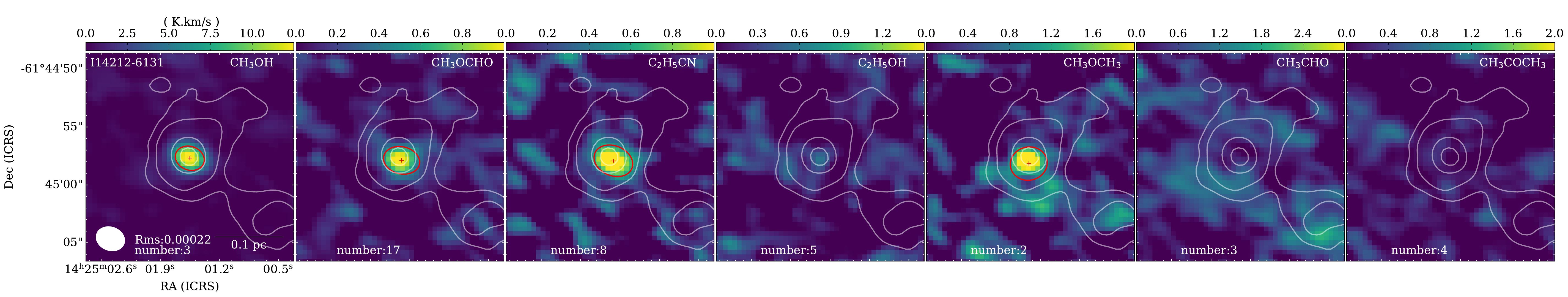}}
\quad
{\includegraphics[height=0.1\textheight]{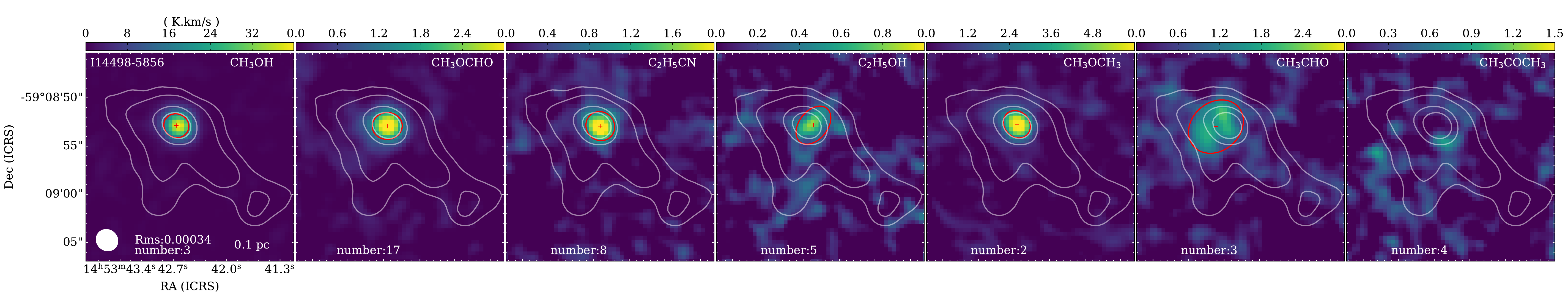}}
\quad 
{\includegraphics[height=0.1\textheight]{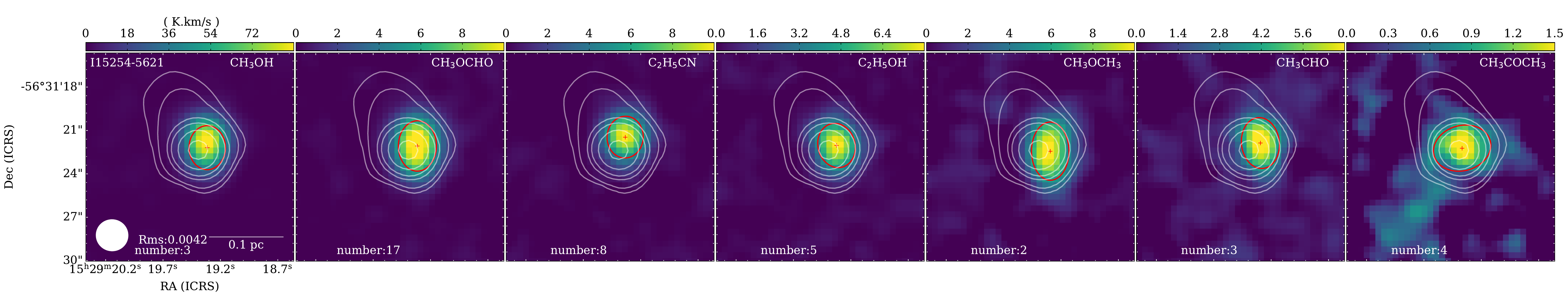}}
\quad
{\includegraphics[height=0.1\textheight]{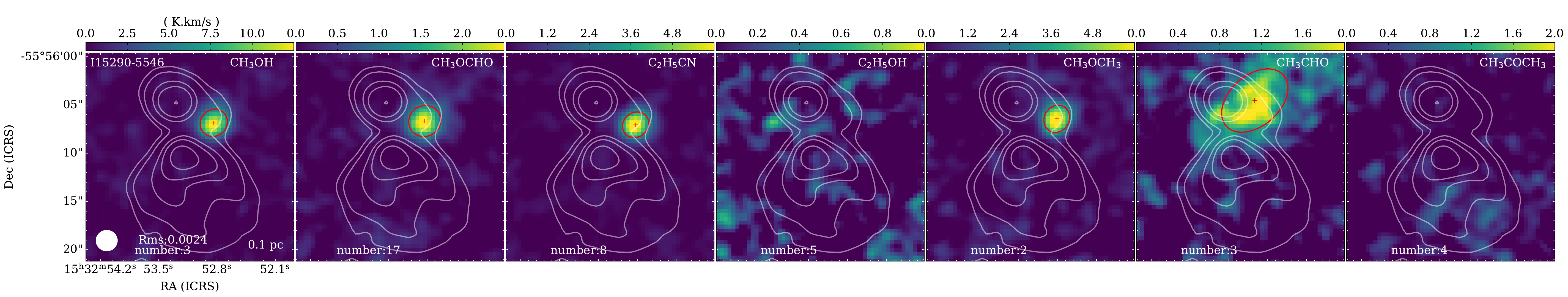}}
\caption{Continued.}
\end{figure}

\clearpage
\setcounter{figure}{\value{figure}-1}
\begin{figure}
\centering 
{\includegraphics[height=0.1\textheight]{stacked_mom0_I15394-5358.pdf}}
\quad
{\includegraphics[height=0.1\textheight]{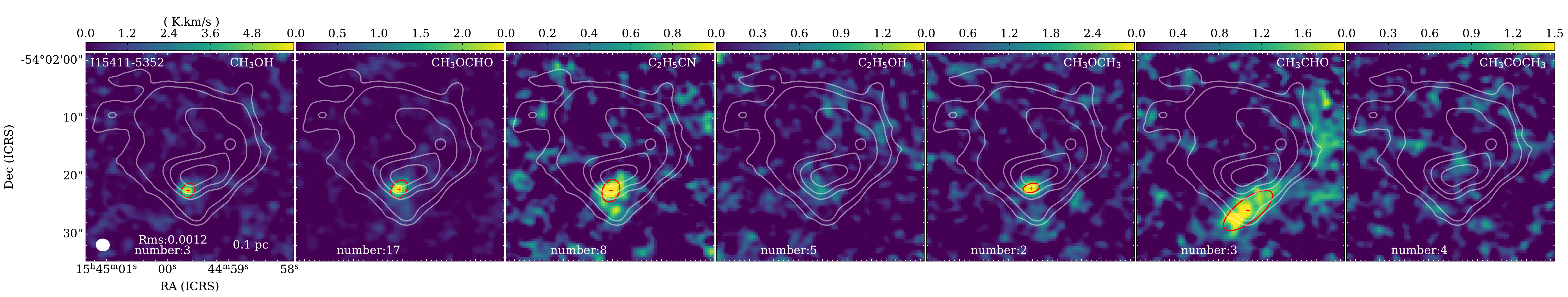}}
\quad
{\includegraphics[height=0.1\textheight]{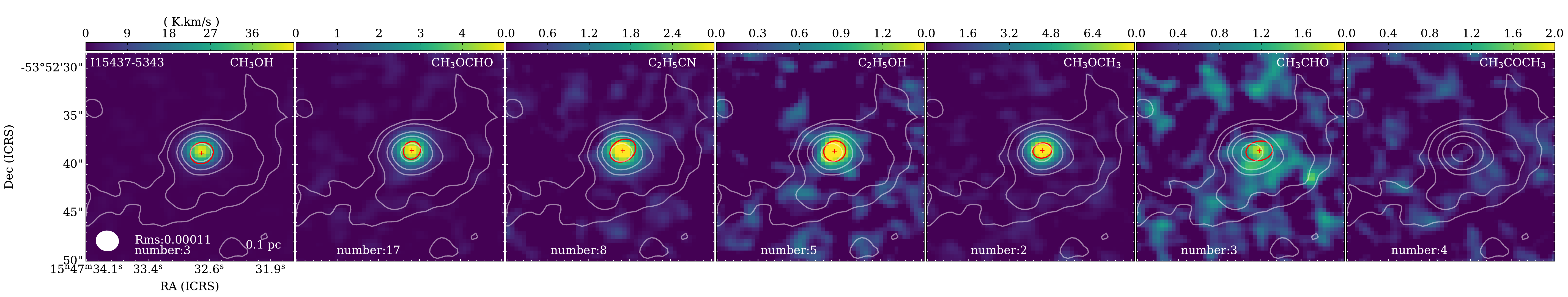}}
\quad 
{\includegraphics[height=0.1\textheight]{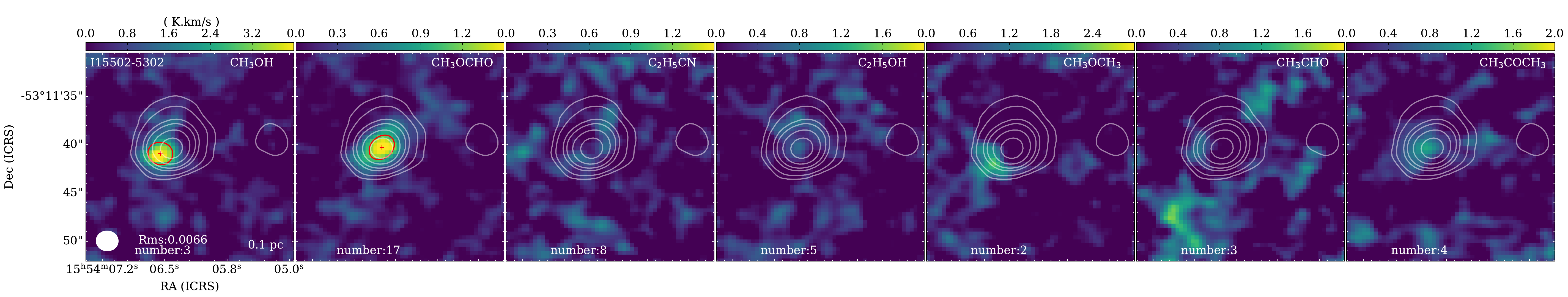}}
\quad
{\includegraphics[height=0.1\textheight]{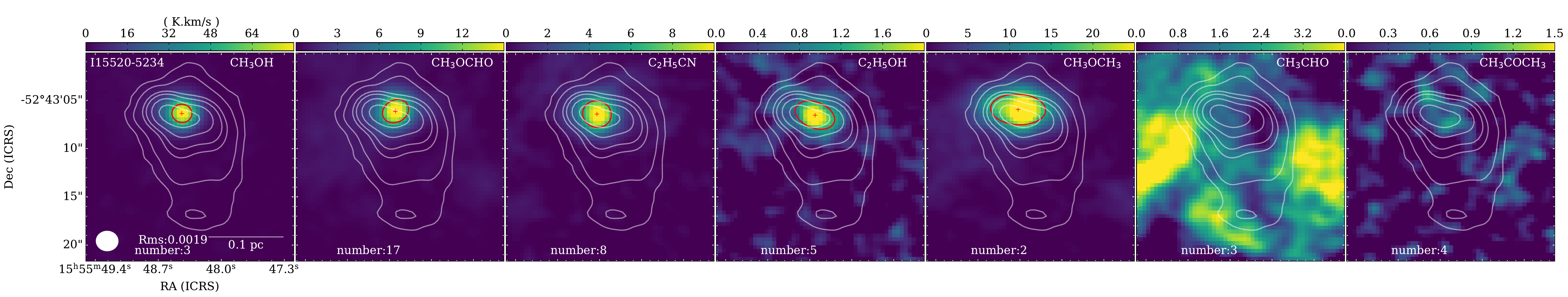}}
\quad
{\includegraphics[height=0.1\textheight]{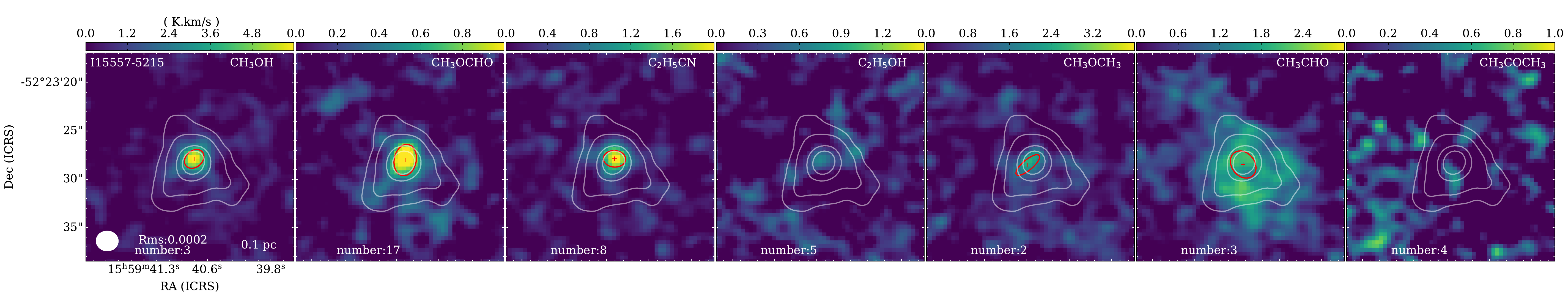}}
\quad
{\includegraphics[height=0.1\textheight]{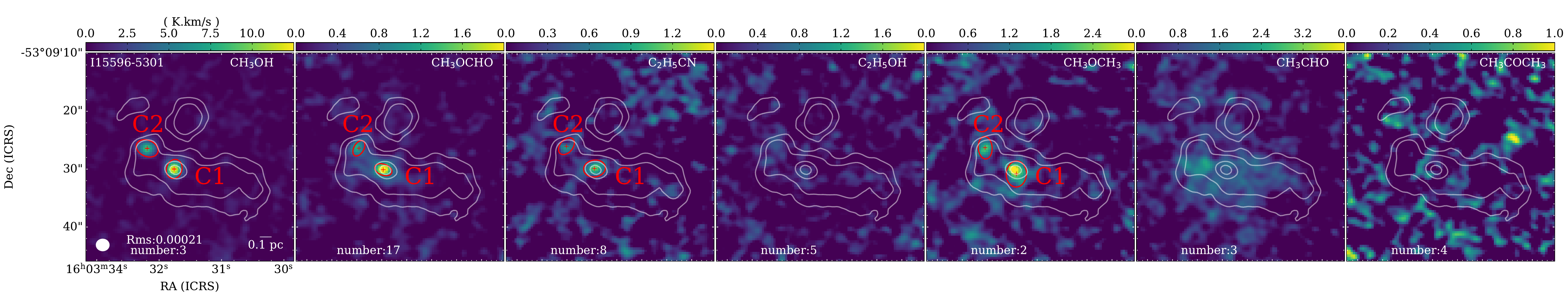}}
\quad
{\includegraphics[height=0.1\textheight]{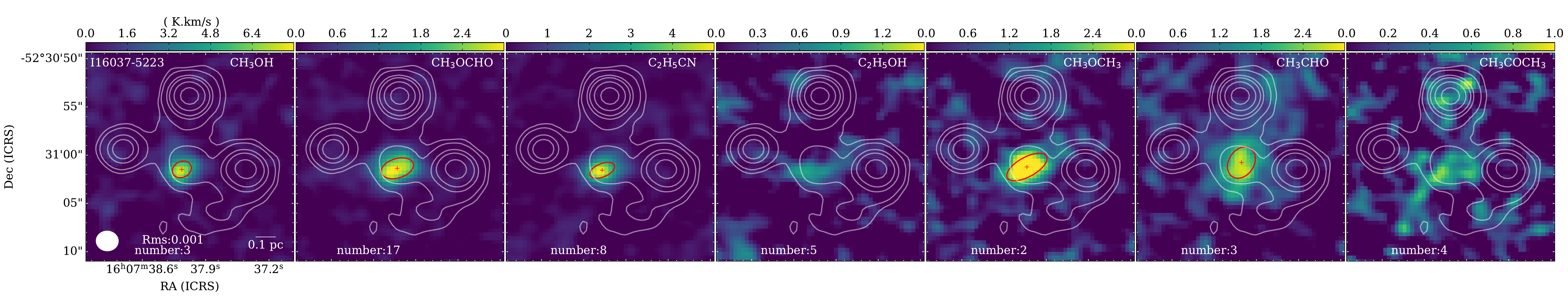}}
\quad 
{\includegraphics[height=0.1\textheight]{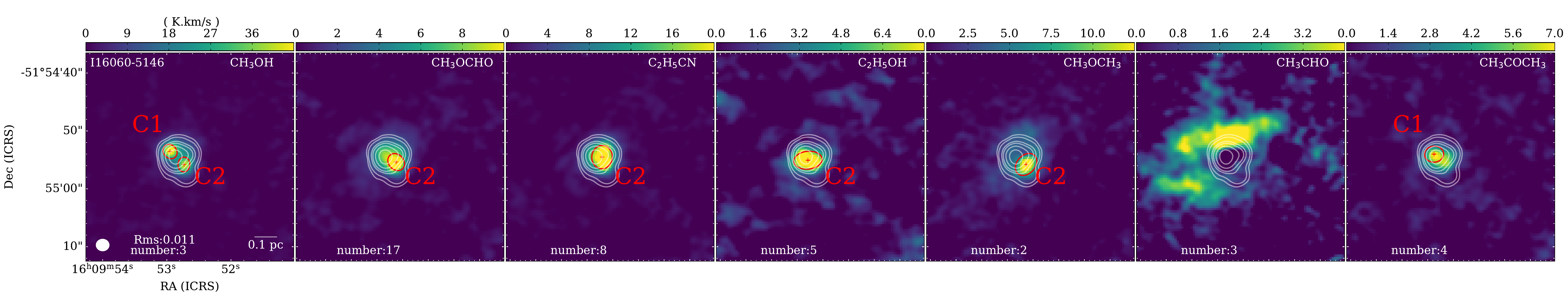}}
\caption{Continued.}
\end{figure}

\clearpage
\setcounter{figure}{\value{figure}-1}
\begin{figure}
\centering 
{\includegraphics[height=0.1\textheight]{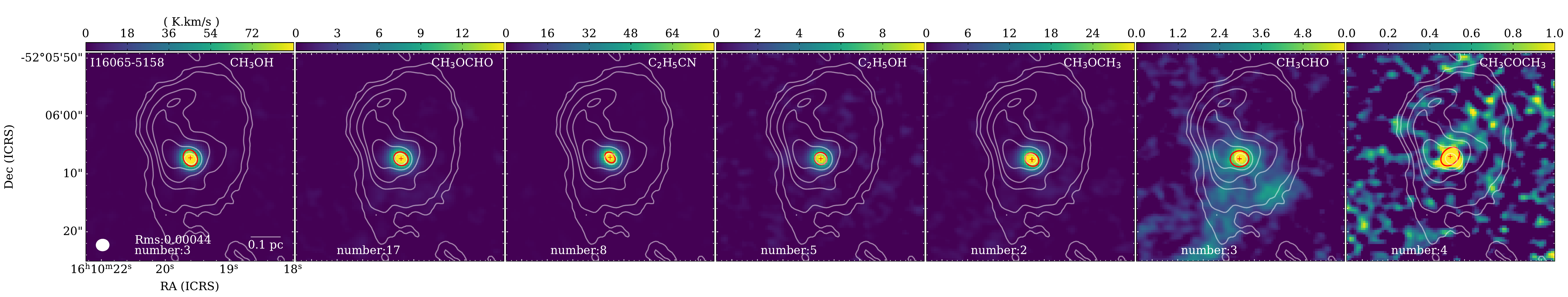}}
\quad 
{\includegraphics[height=0.1\textheight]{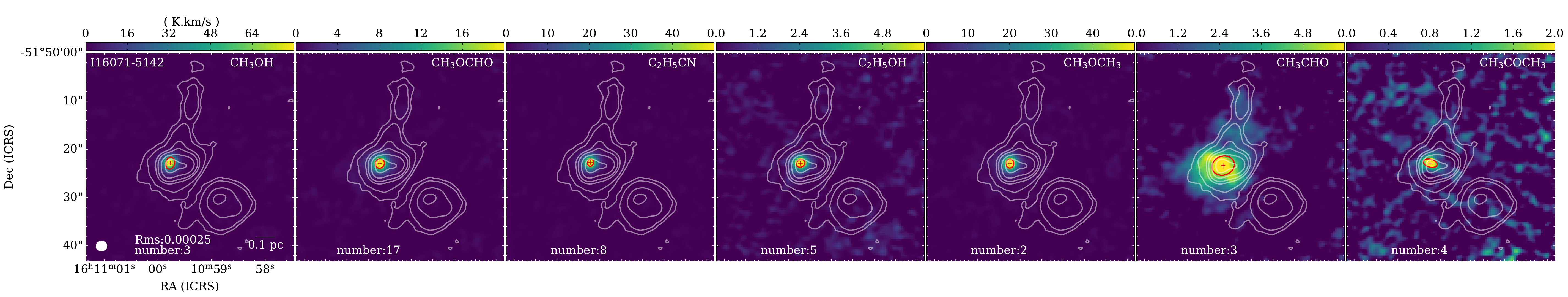}}
\quad 
{\includegraphics[height=0.1\textheight]{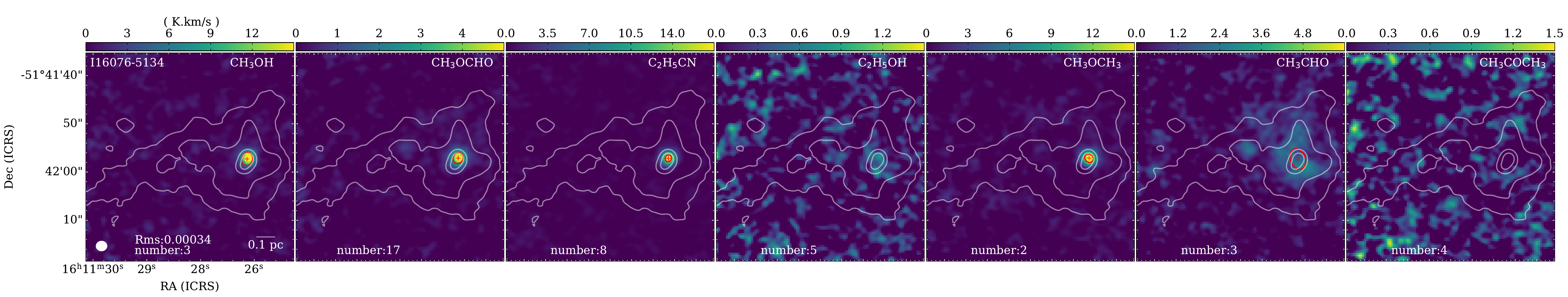}}
\quad
{\includegraphics[height=0.1\textheight]{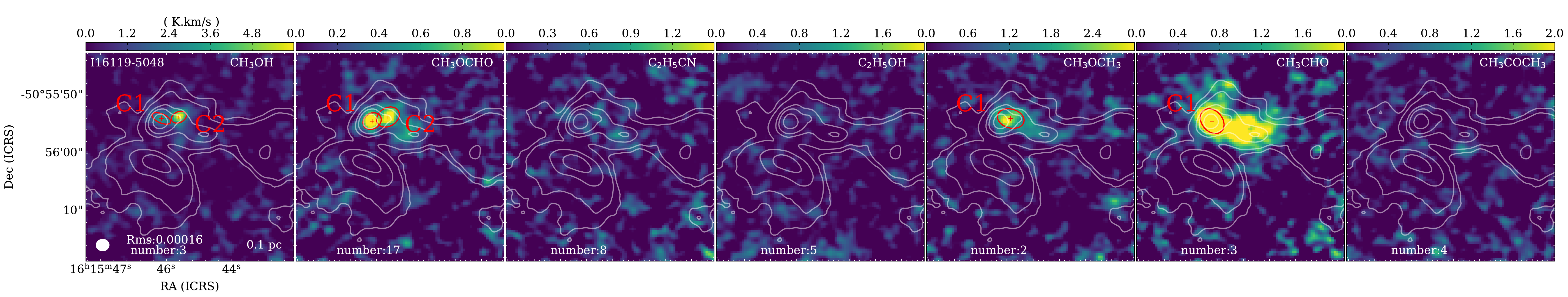}}
\quad 
{\includegraphics[height=0.1\textheight]{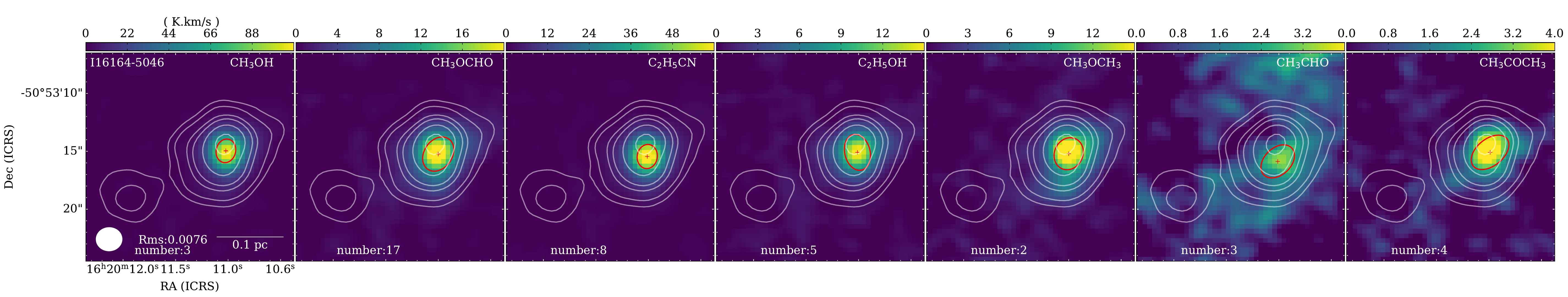}}
\quad 
{\includegraphics[height=0.1\textheight]{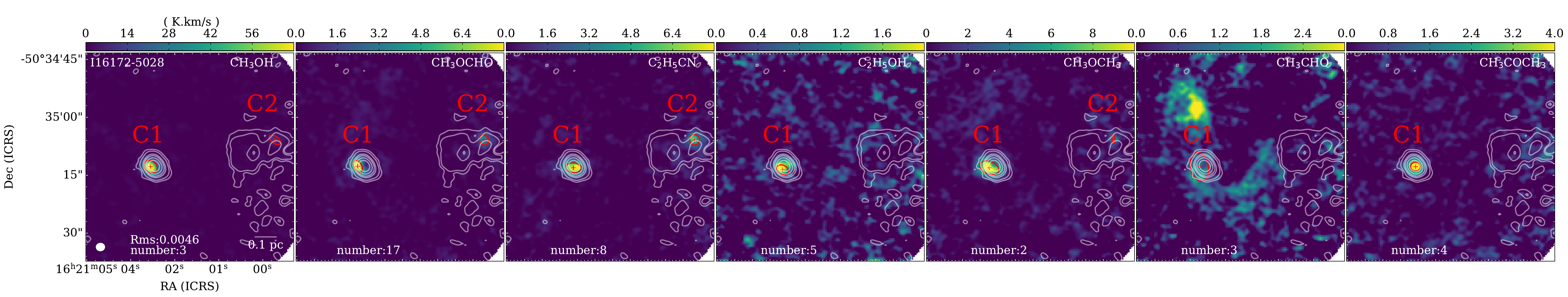}}
\quad 
{\includegraphics[height=0.1\textheight]{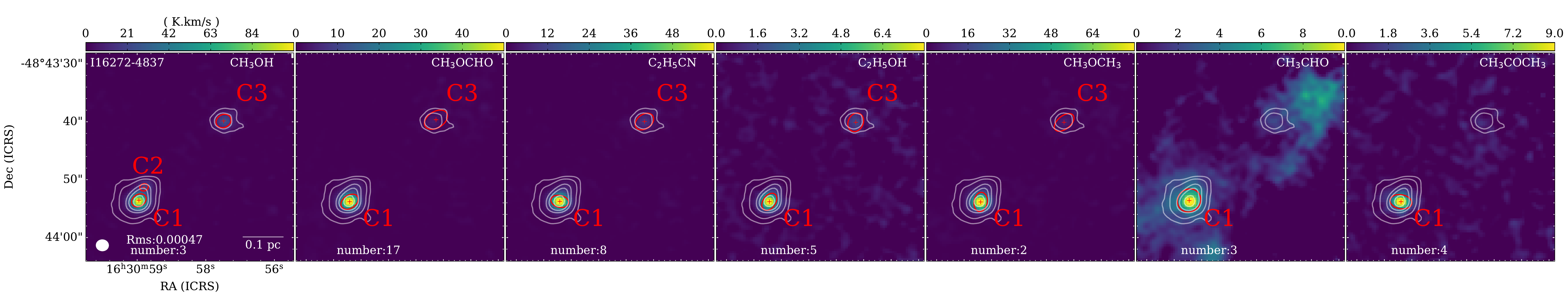}}
\quad 
{\includegraphics[height=0.1\textheight]{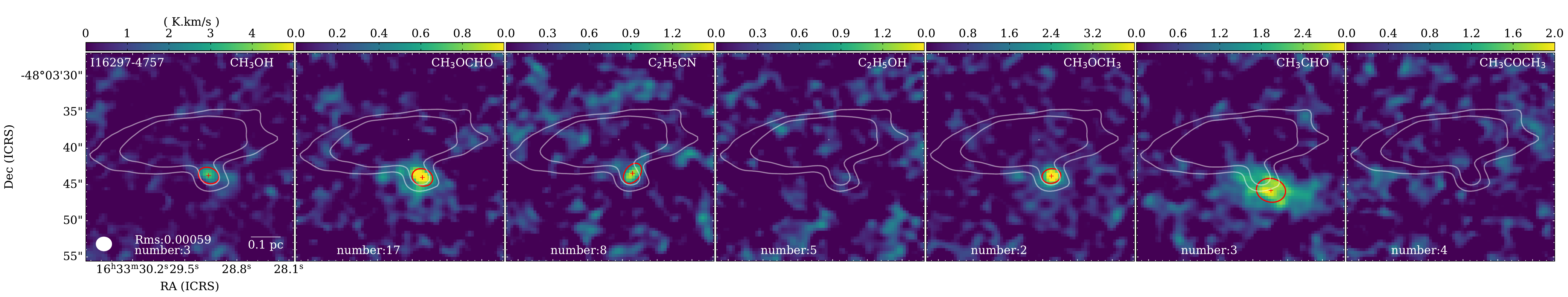}}
\quad 
{\includegraphics[height=0.1\textheight]{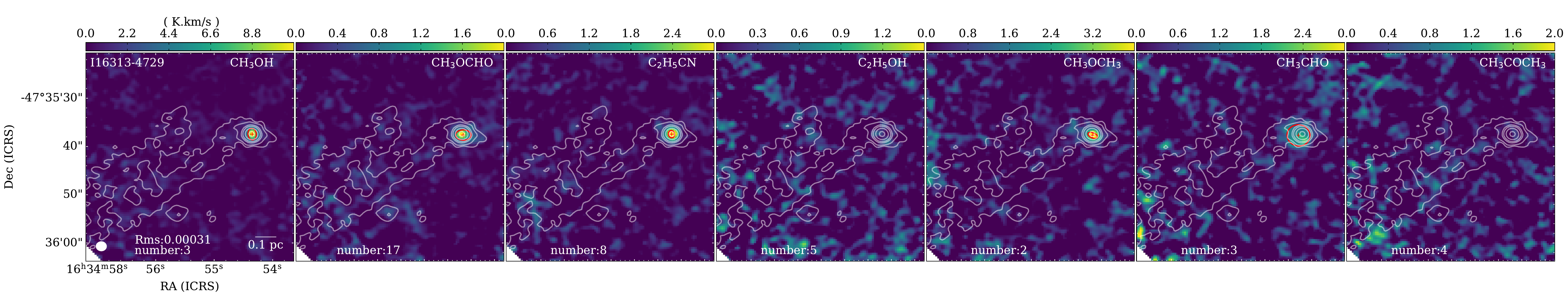}}
\caption{Continued.}
\end{figure}

\clearpage
\setcounter{figure}{\value{figure}-1}
\begin{figure}
\centering 
{\includegraphics[height=0.1\textheight]{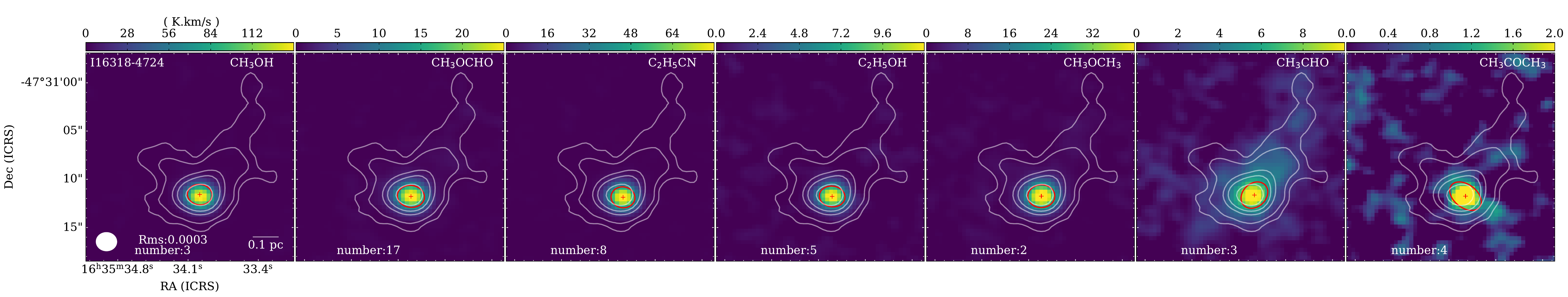}}
\quad 
{\includegraphics[height=0.1\textheight]{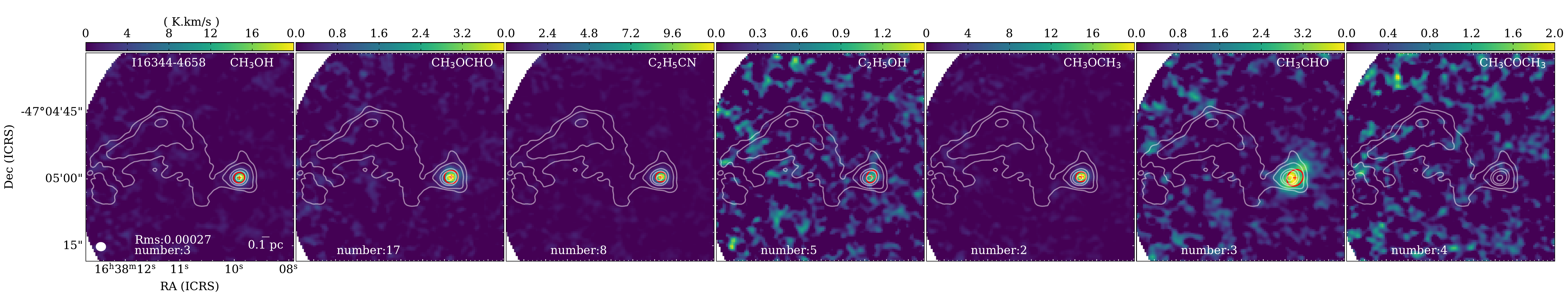}}
\quad 
{\includegraphics[height=0.1\textheight]{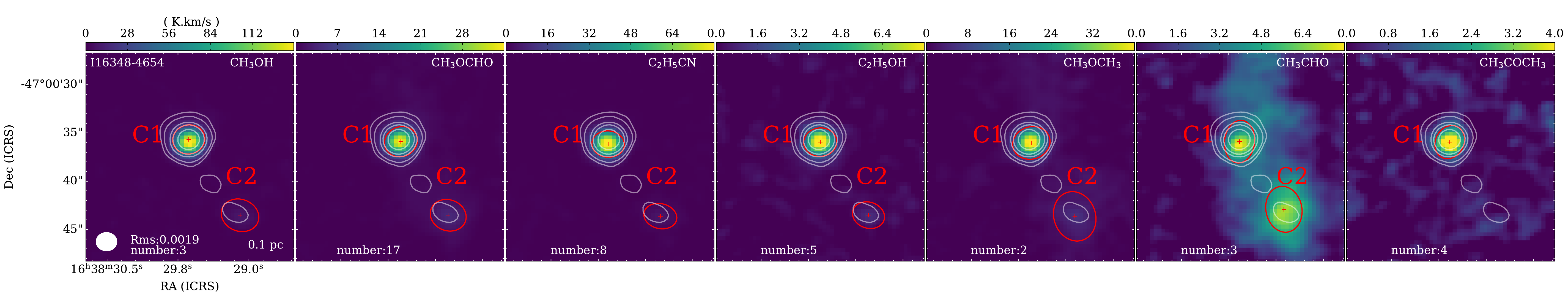}}
\quad 
{\includegraphics[height=0.1\textheight]{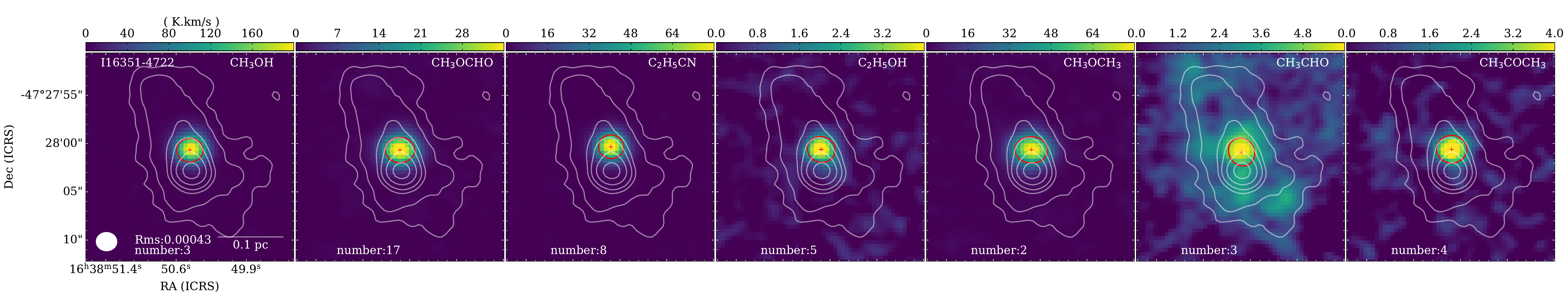}}
\quad
{\includegraphics[height=0.1\textheight]{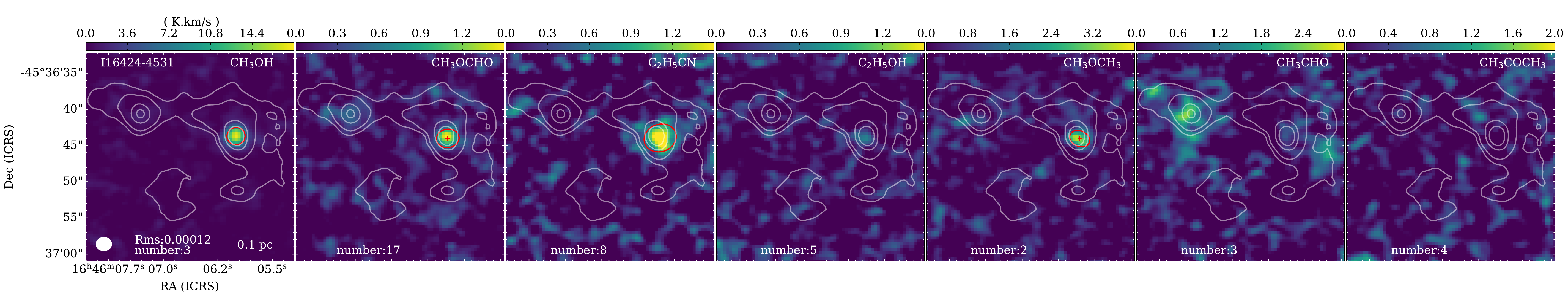}}
\quad
{\includegraphics[height=0.1\textheight]{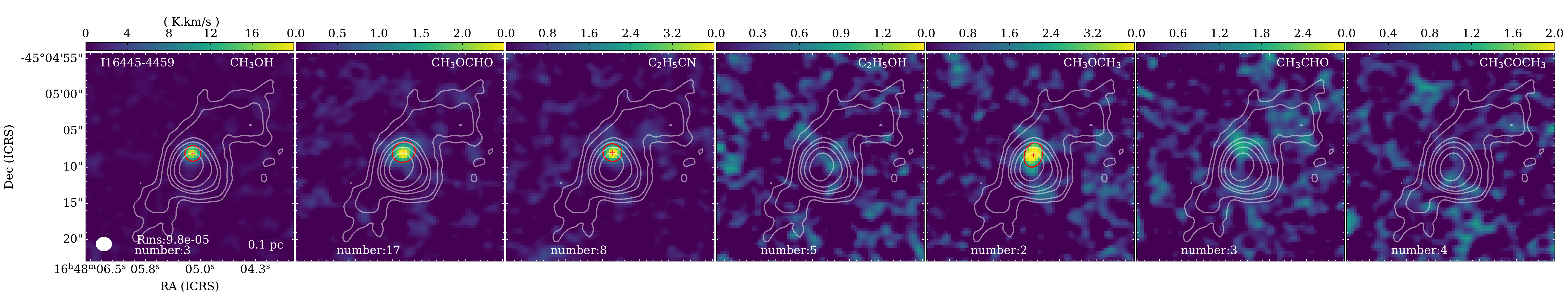}}
\quad
{\includegraphics[height=0.1\textheight]{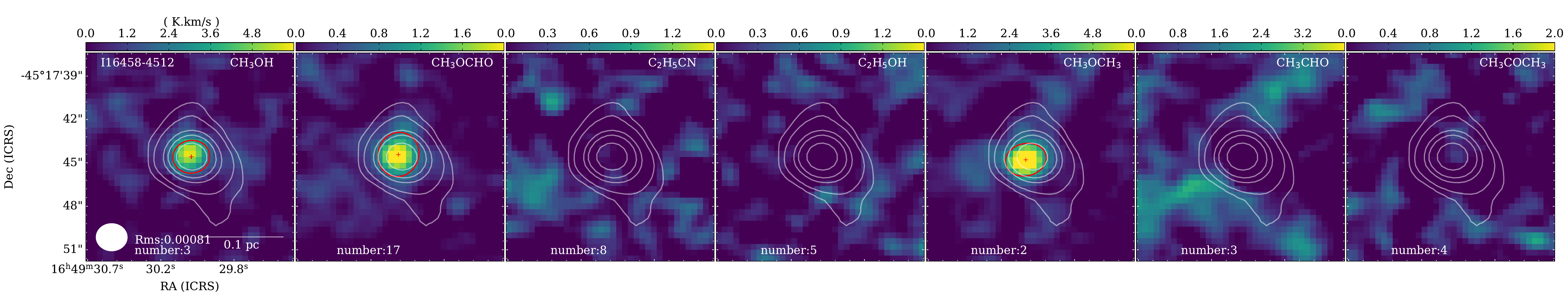}}
\quad 
{\includegraphics[height=0.1\textheight]{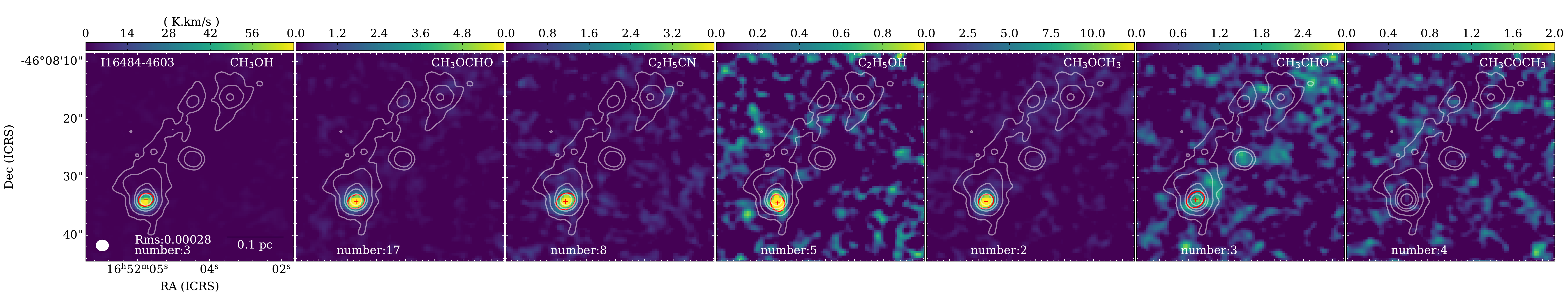}}
\quad 
{\includegraphics[height=0.1\textheight]{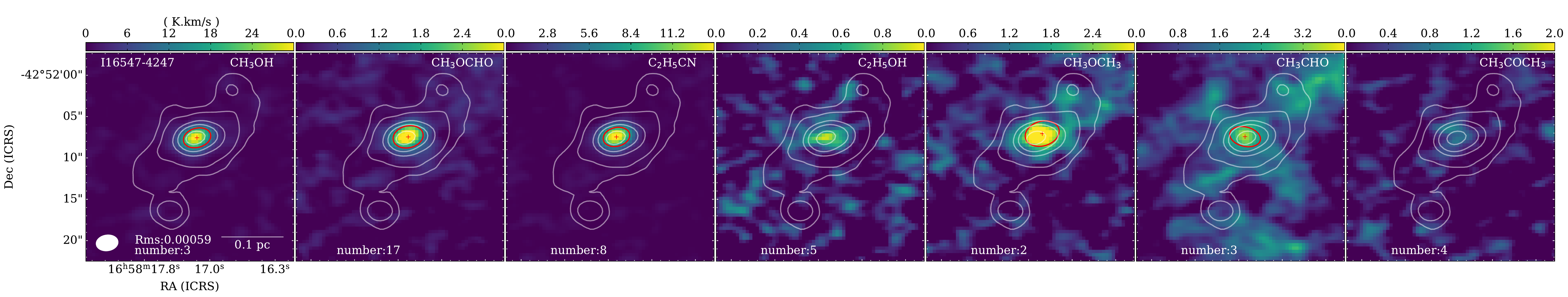}}
\caption{Continued.}
\end{figure}

\clearpage
\setcounter{figure}{\value{figure}-1}
\begin{figure}
\centering 
{\includegraphics[height=0.1\textheight]{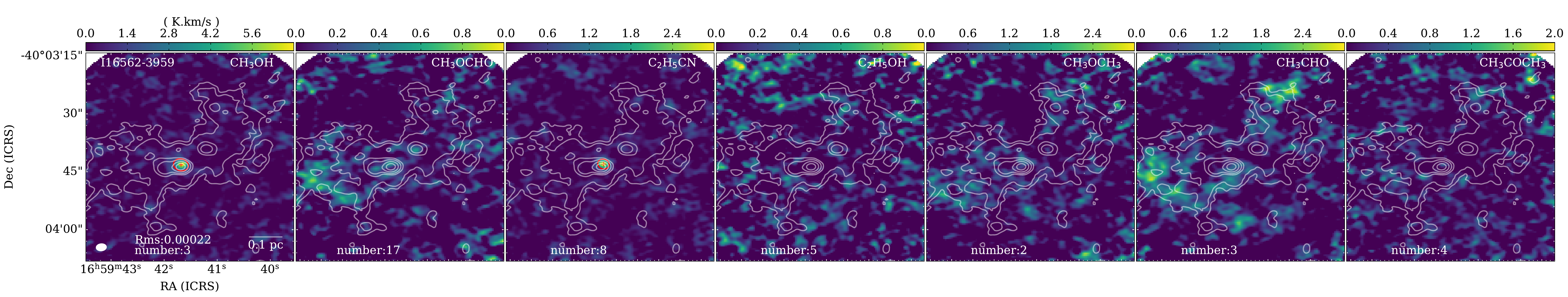}}
\quad
{\includegraphics[height=0.1\textheight]{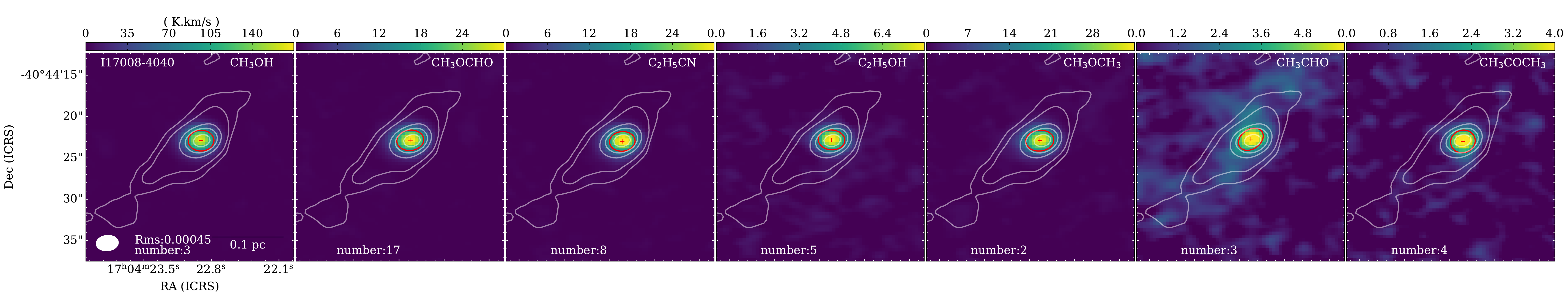}}
\quad 
{\includegraphics[height=0.1\textheight]{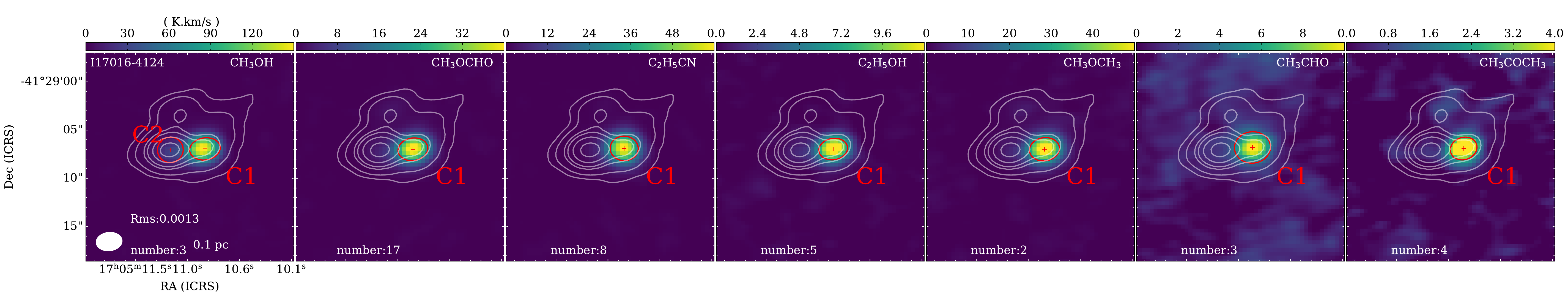}}
\quad 
{\includegraphics[height=0.1\textheight]{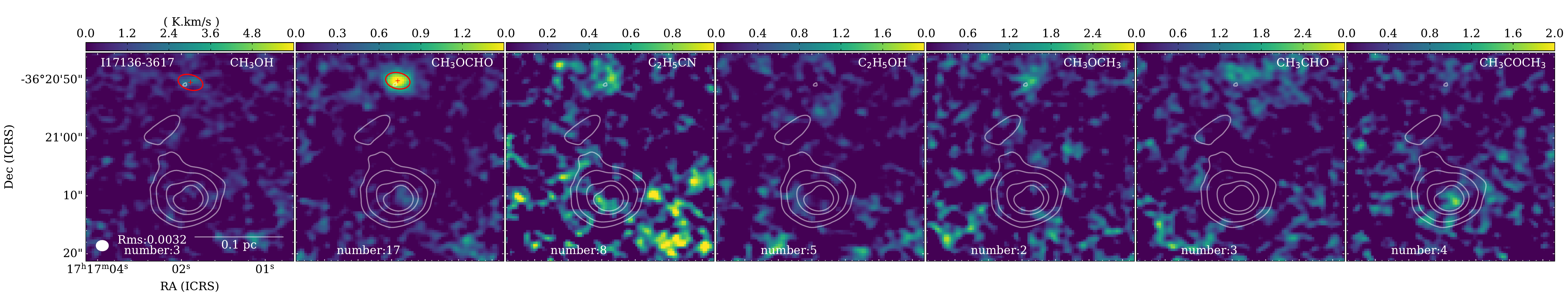}}
\quad
{\includegraphics[height=0.1\textheight]{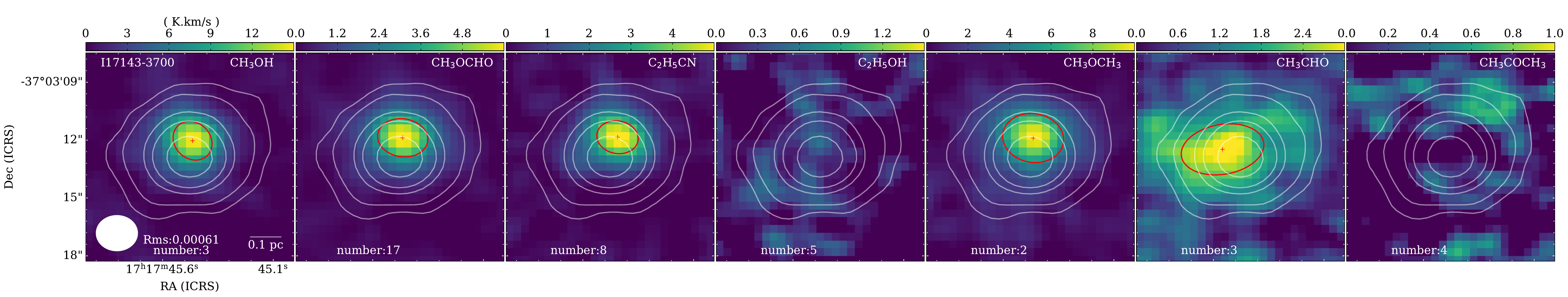}}
\quad
{\includegraphics[height=0.1\textheight]{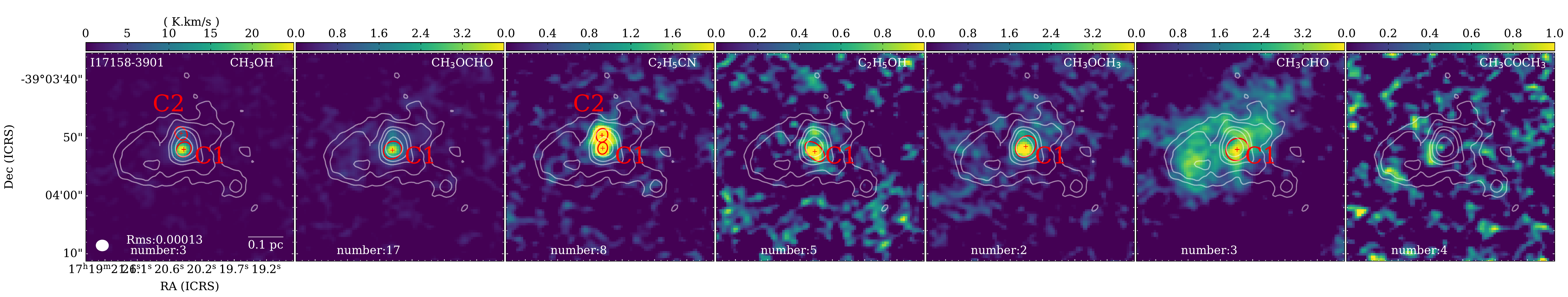}}
\quad  
{\includegraphics[height=0.1\textheight]{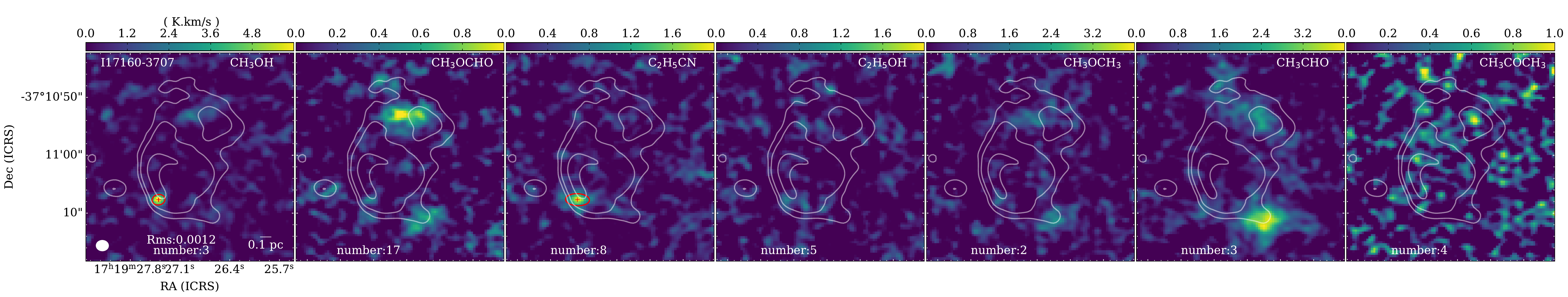}}
\quad 
{\includegraphics[height=0.1\textheight]{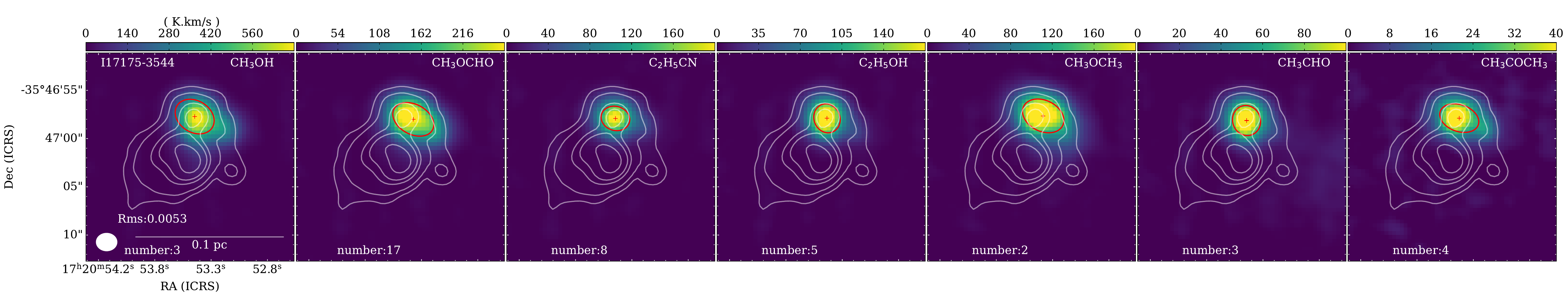}}
\quad 
{\includegraphics[height=0.1\textheight]{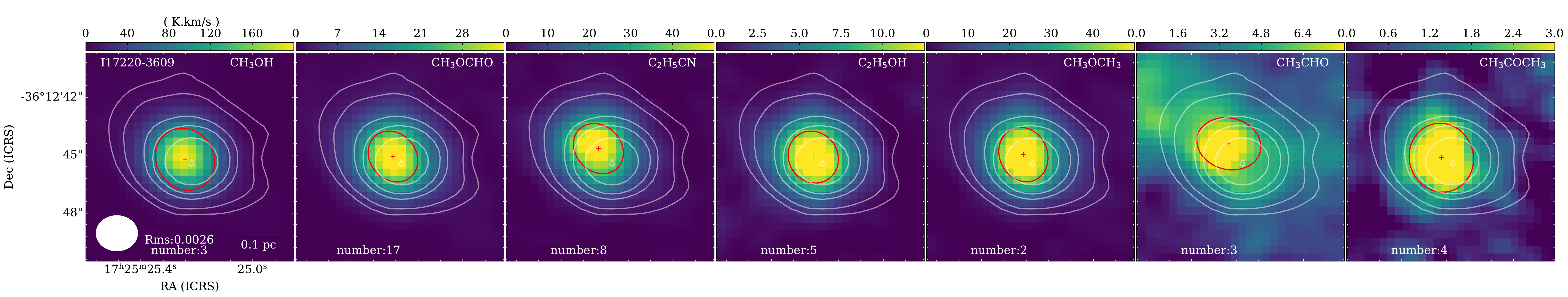}}
\caption{Continued.}
\end{figure}

\clearpage
\setcounter{figure}{\value{figure}-1}
\begin{figure}
\centering 
{\includegraphics[height=0.1\textheight]{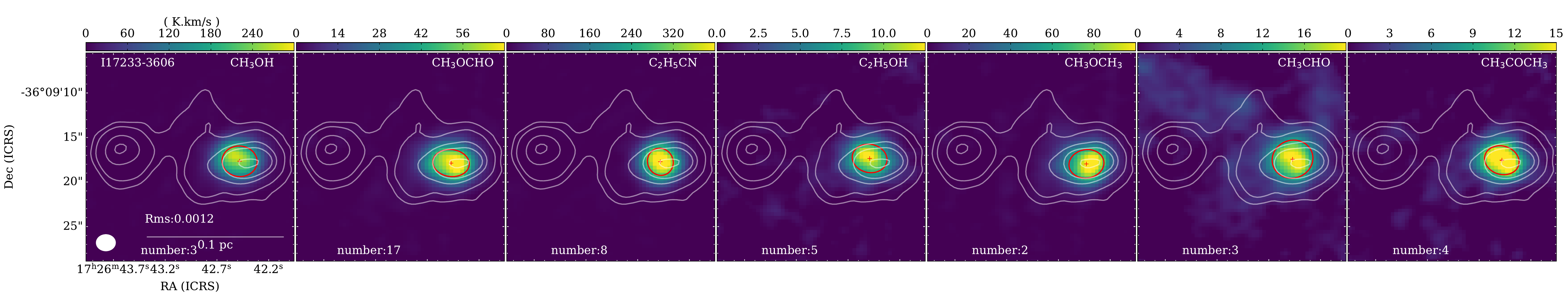}}
\quad 
{\includegraphics[height=0.1\textheight]{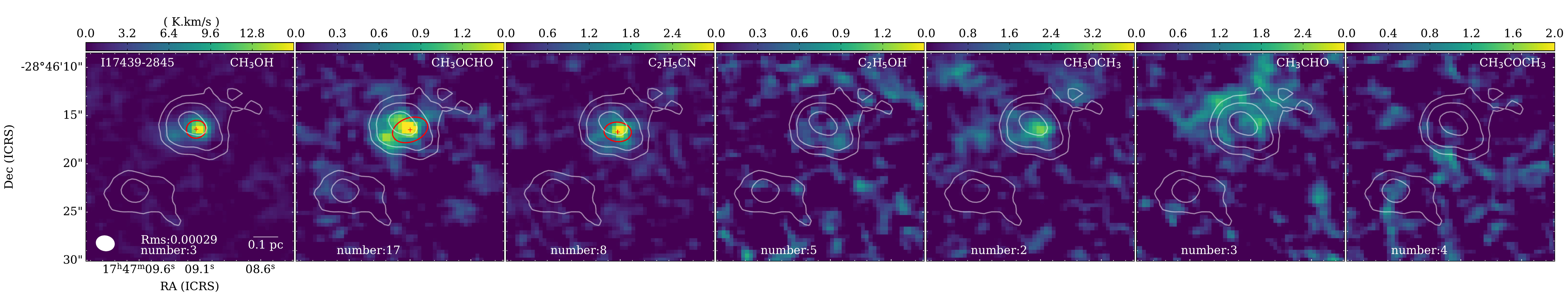}}
\quad 
{\includegraphics[height=0.1\textheight]{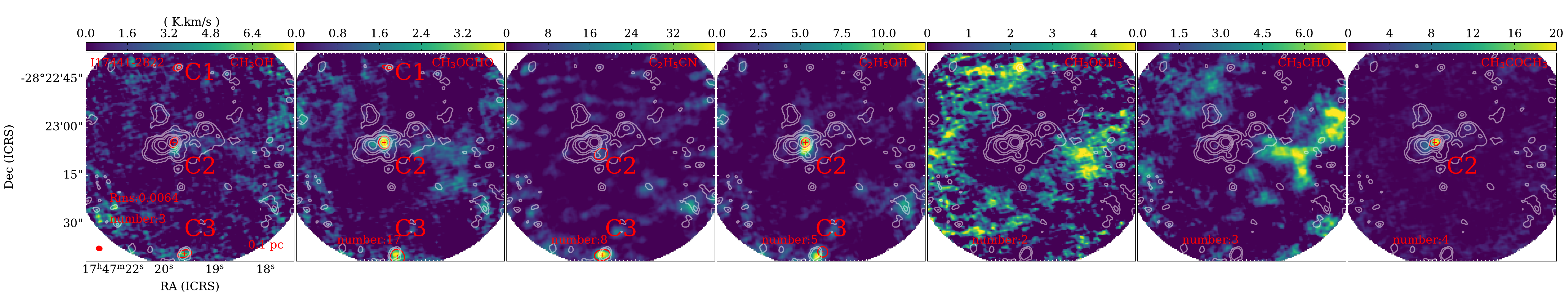}}
\quad 
{\includegraphics[height=0.1\textheight]{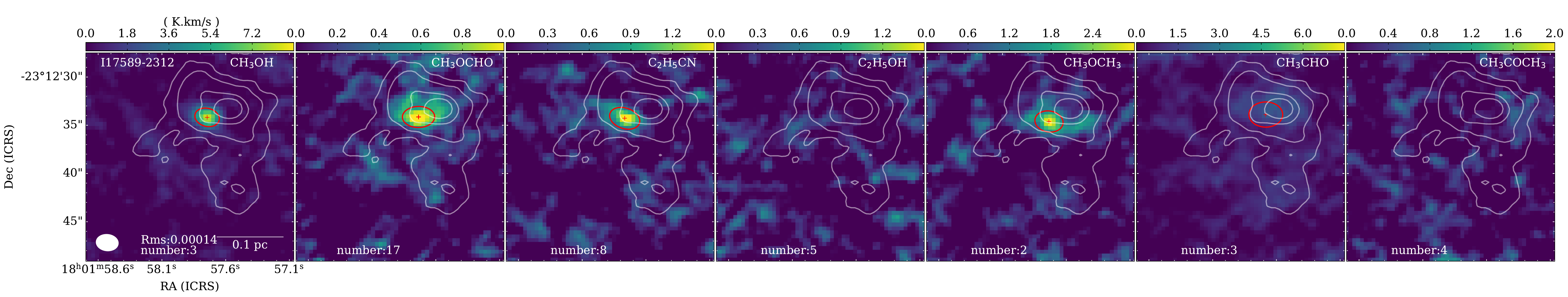}}
\quad 
{\includegraphics[height=0.1\textheight]{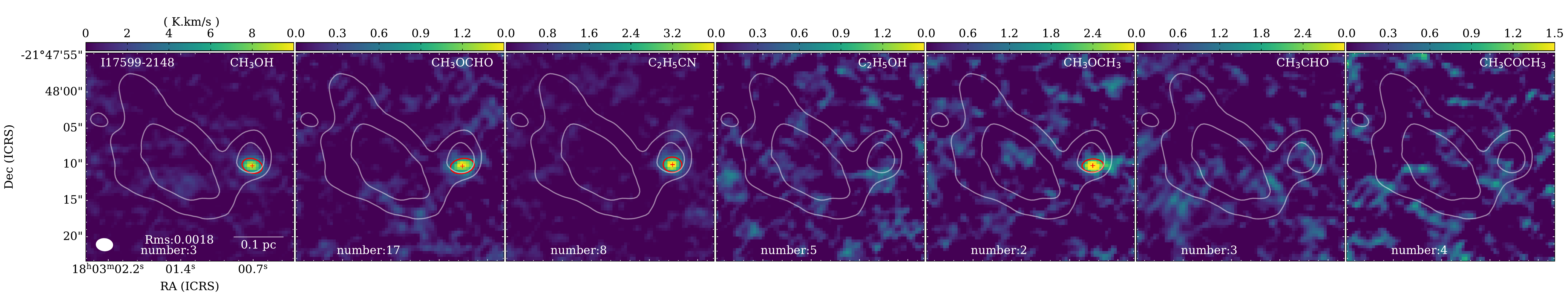}}
\quad
{\includegraphics[height=0.1\textheight]{stacked_mom0_I18032-2032.pdf}}
\quad 
{\includegraphics[height=0.1\textheight]{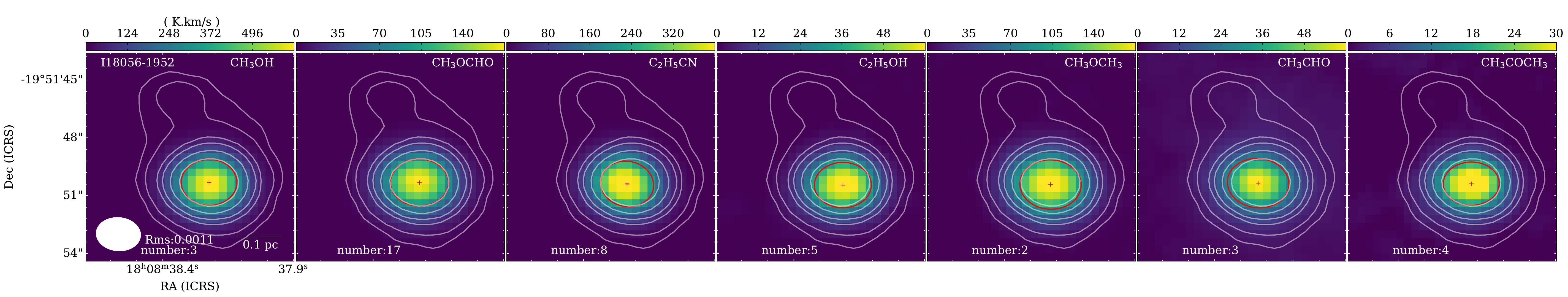}}
\quad
{\includegraphics[height=0.1\textheight]{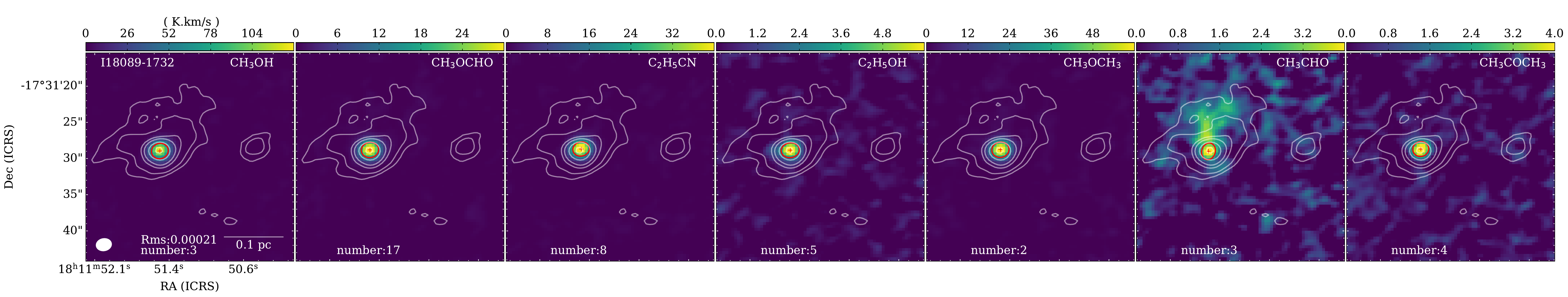}}
\quad
{\includegraphics[height=0.1\textheight]{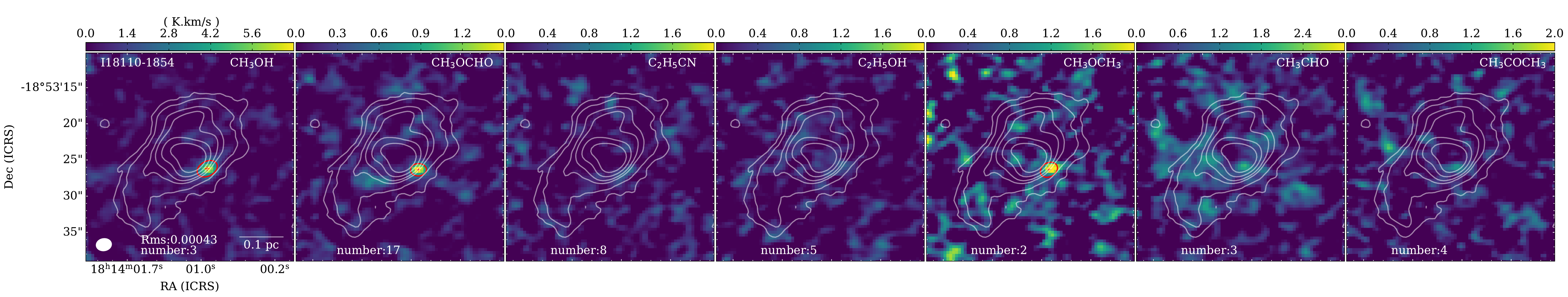}}
\caption{Continued.}
\end{figure}

\clearpage
\setcounter{figure}{\value{figure}-1}
\begin{figure}
\centering 
{\includegraphics[height=0.1\textheight]{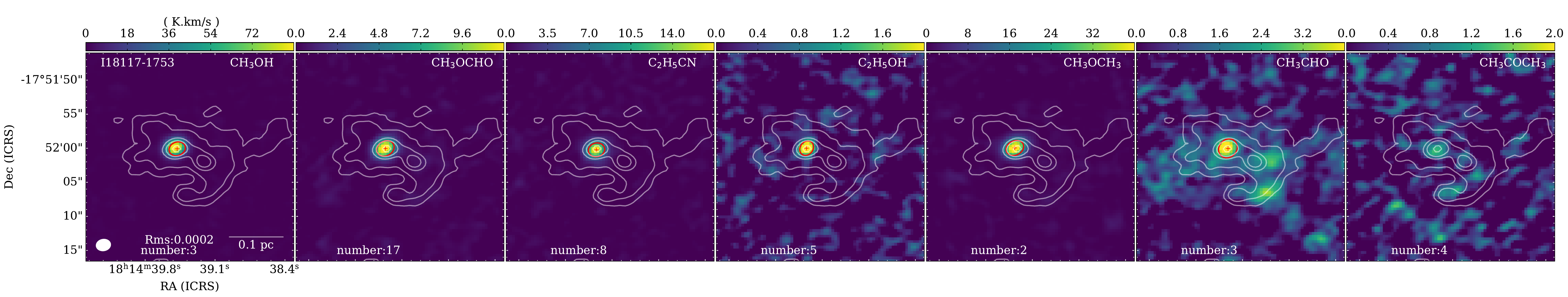}}
\quad 
{\includegraphics[height=0.1\textheight]{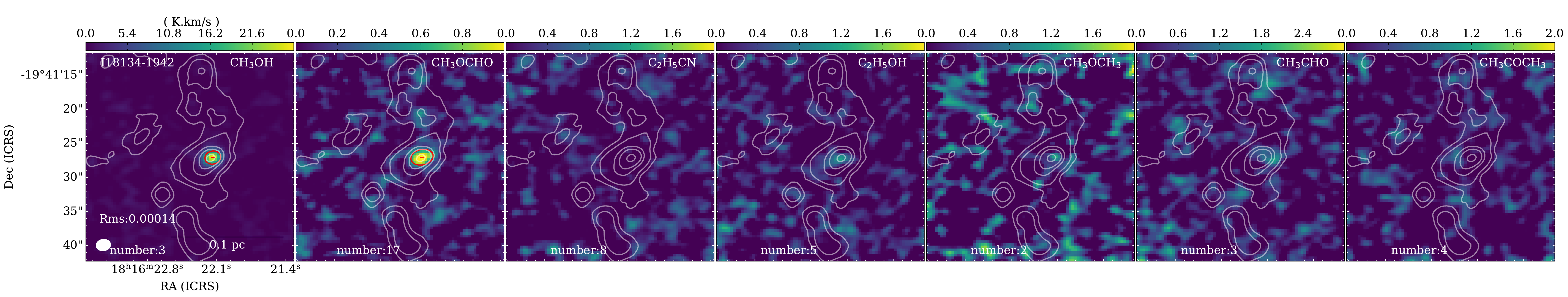}}
\quad 
{\includegraphics[height=0.1\textheight]{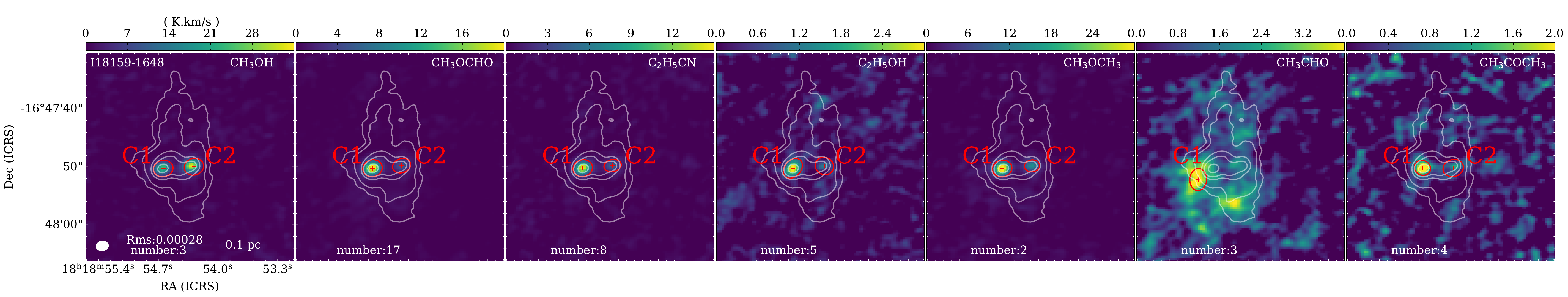}}
\quad 
{\includegraphics[height=0.1\textheight]{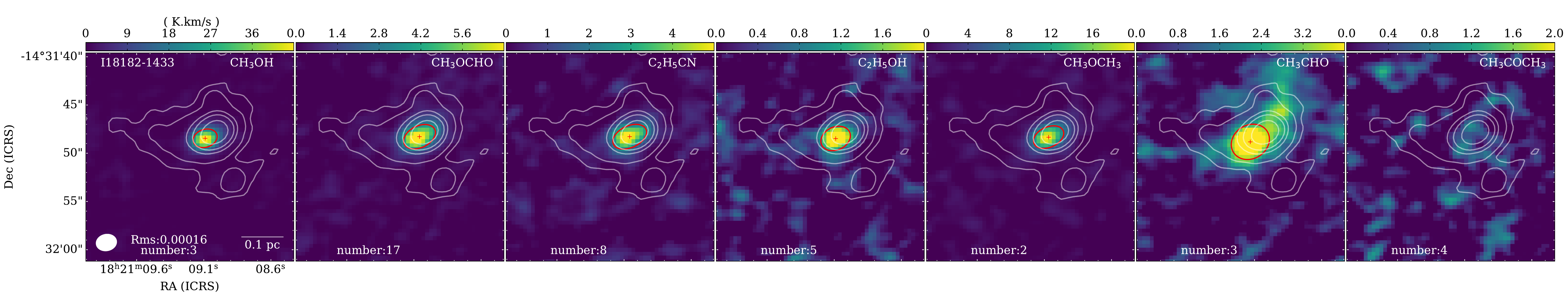}}
\quad 
{\includegraphics[height=0.1\textheight]{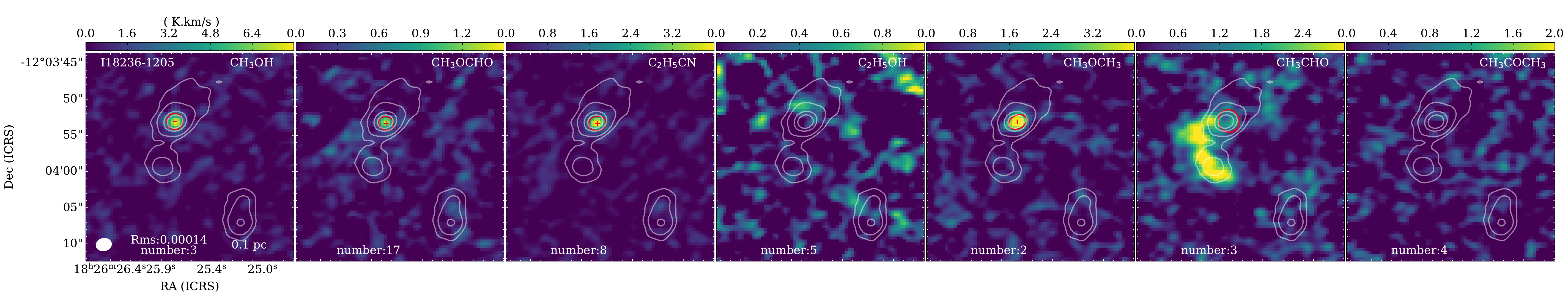}}
\quad 
{\includegraphics[height=0.1\textheight]{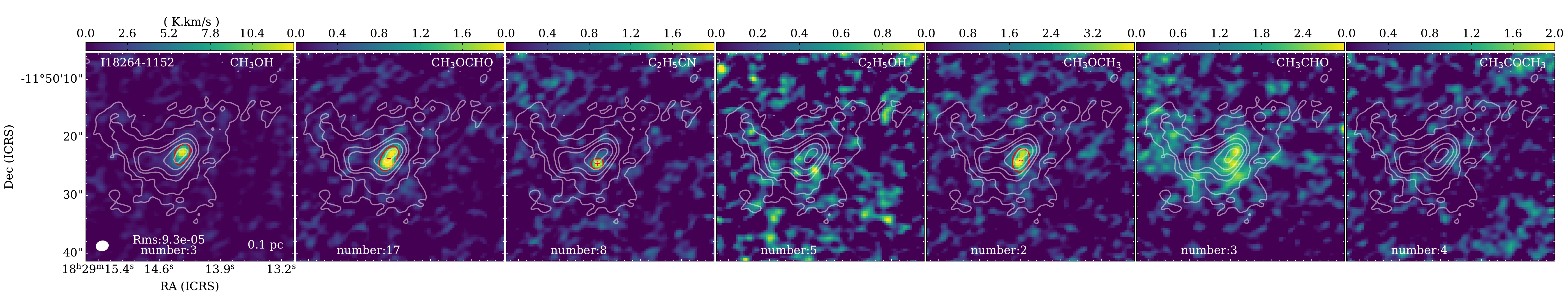}}
\quad
{\includegraphics[height=0.1\textheight]{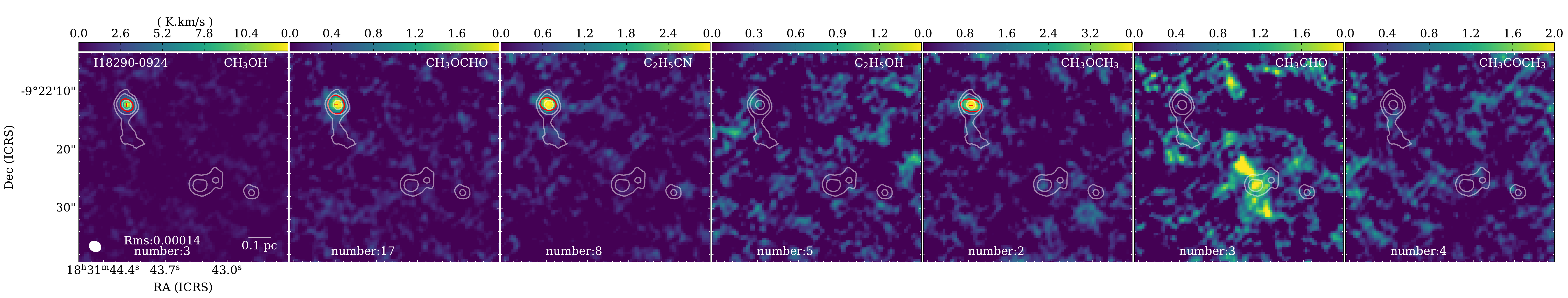}}
\quad
{\includegraphics[height=0.1\textheight]{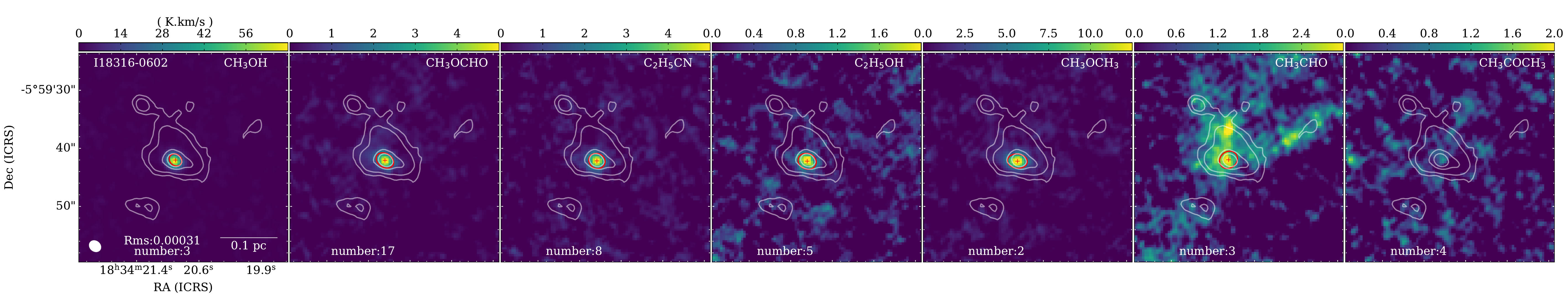}}
\quad 
{\includegraphics[height=0.1\textheight]{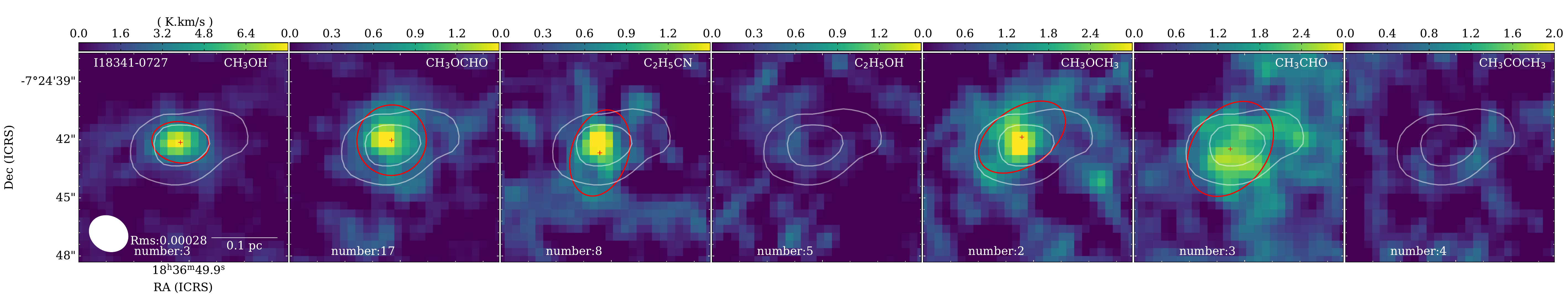}}

\caption{Continued.}
\end{figure}

\clearpage
\setcounter{figure}{\value{figure}-1}
\begin{figure}
\centering 
{\includegraphics[height=0.1\textheight]{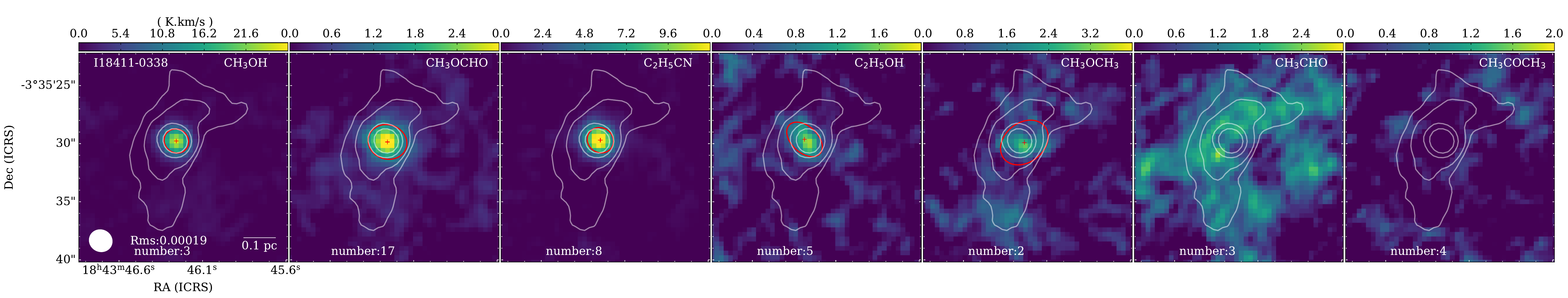}}
\quad 
{\includegraphics[height=0.1\textheight]{stacked_mom0_I18434-0242.pdf}}
\quad 
{\includegraphics[height=0.1\textheight]{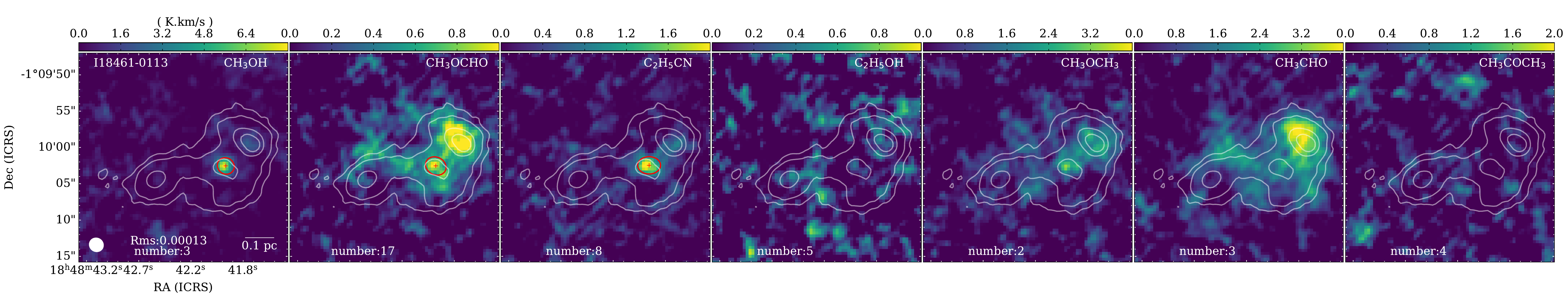}}
\quad 
{\includegraphics[height=0.1\textheight]{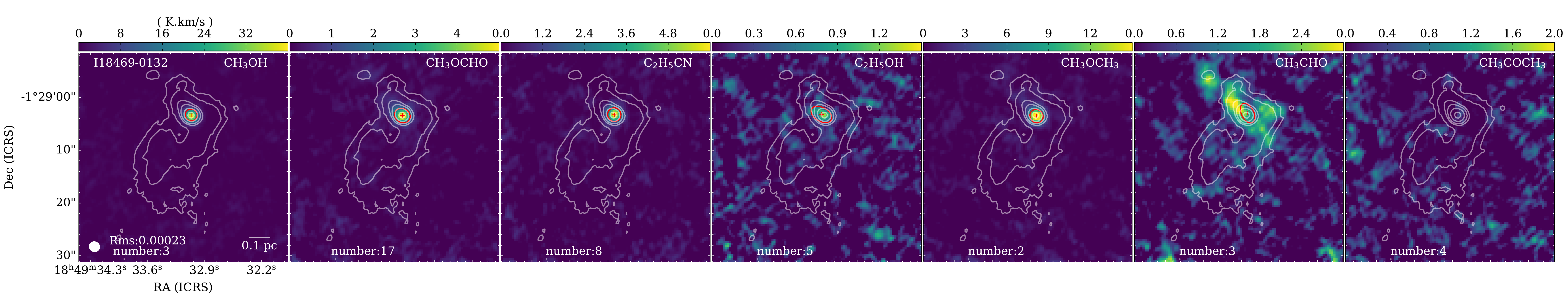}}
\quad 
{\includegraphics[height=0.1\textheight]{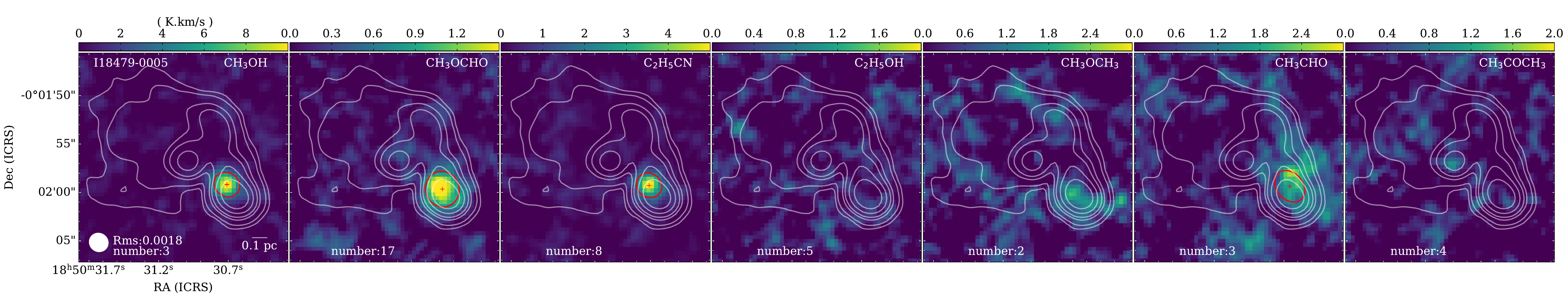}}
\quad 
{\includegraphics[height=0.1\textheight]{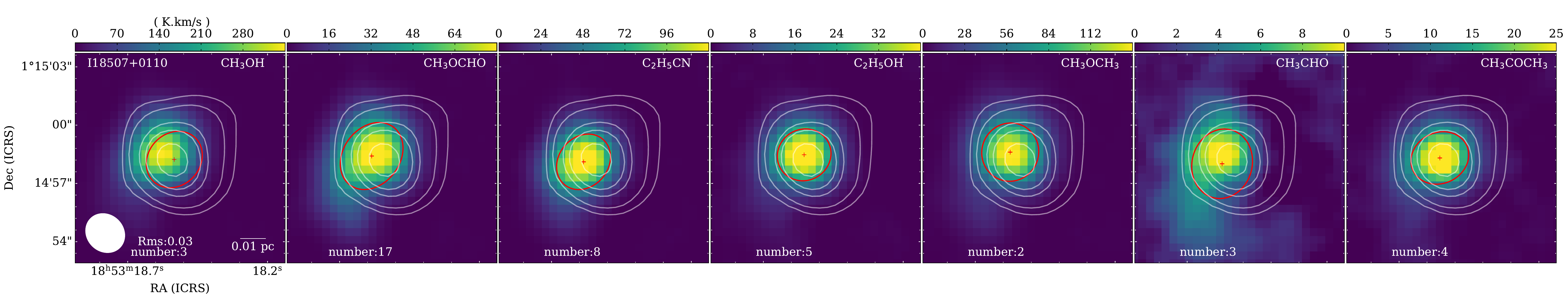}}
\quad 
{\includegraphics[height=0.1\textheight]{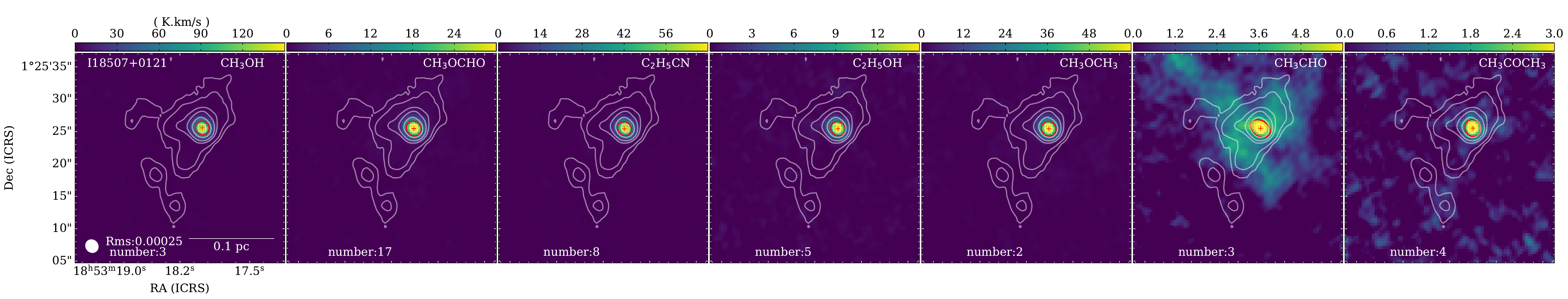}}
\quad 
{\includegraphics[height=0.1\textheight]{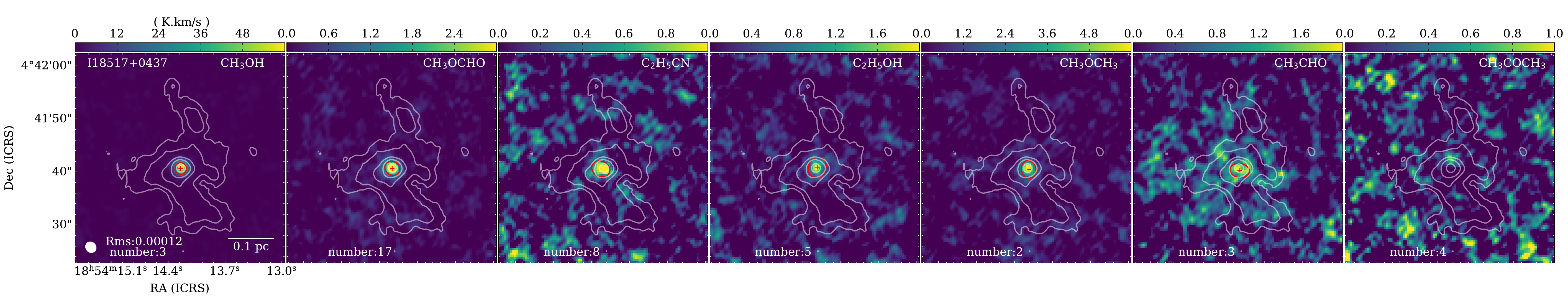}}
\quad 
{\includegraphics[height=0.1\textheight]{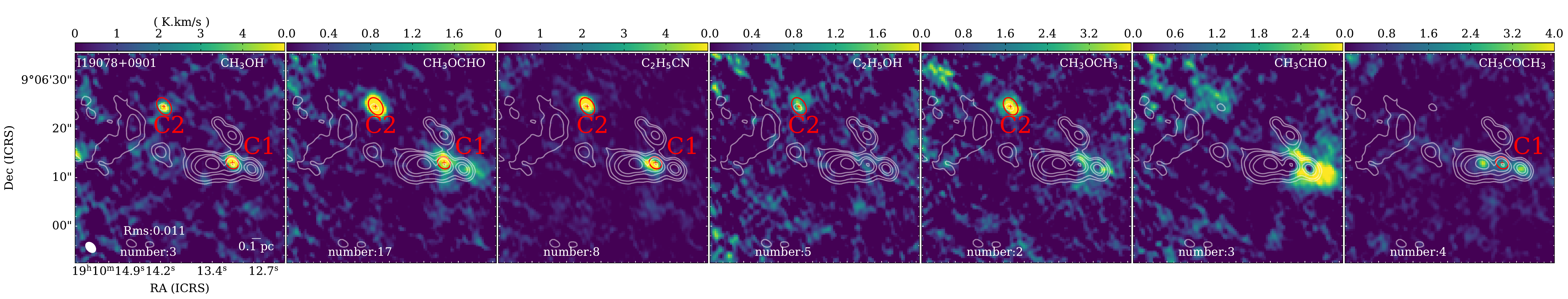}}
\caption{Continued.}
\end{figure}

\clearpage
\setcounter{figure}{\value{figure}-1}
\begin{figure}
\centering 
{\includegraphics[height=0.1\textheight]{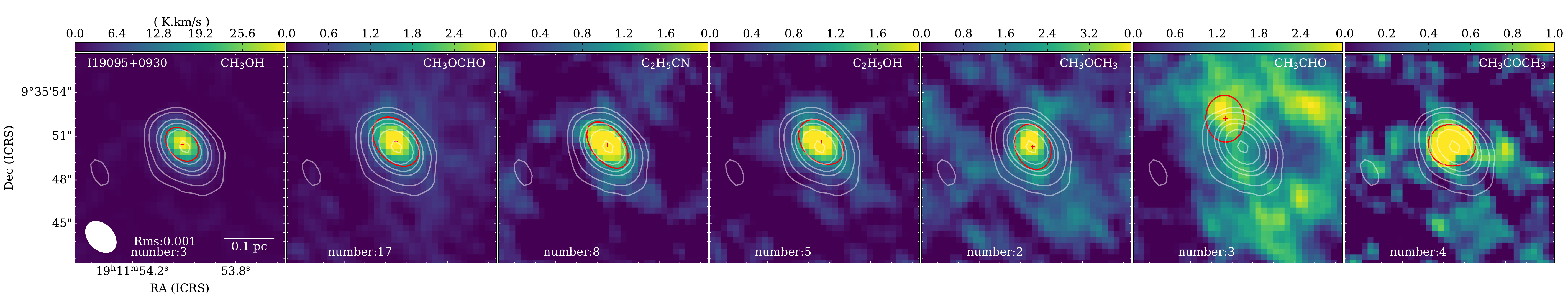}}
\quad 
{\includegraphics[height=0.1\textheight]{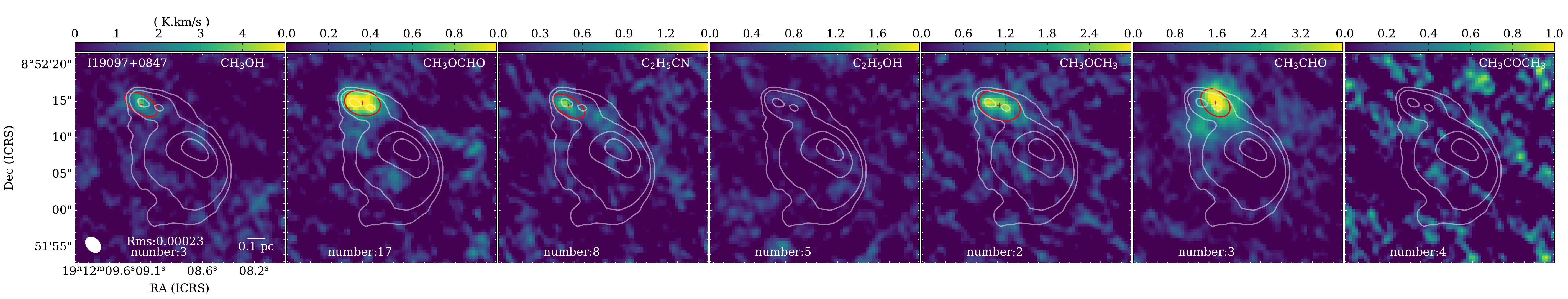}}
\caption{Continued}
\end{figure}

\end{document}